\documentclass[12pt,a4paper,leqno,openbib,oldfontcommands]{memoir}
\usepackage[T1]{fontenc}         
\usepackage[brazil,english,greek]{babel} 
\usepackage{indentfirst}         
\usepackage{amsmath,amssymb,bbold,mathrsfs} 
\usepackage{accents}             
\usepackage[or]{teubner}         
\usepackage{graphicx,color}      
\usepackage[all]{xy}             
\usepackage{fancybox}            
\usepackage{epipart}             
\usepackage{endnotes}            

\setpnumwidth{2.2em}
\setrmarg{3em}
\renewcommand{\bibname}{Bibliography}
\renewcommand{\theendnote}{\roman{endnote}}
\renewcommand{\makeenmark}{\textsuperscript{\theenmark}}
\renewcommand{\notesname}{Notes and translation of (some) epigraphs}
\renewcommand{\enoteheading}{\chapter*{\notesname}}

\newcommand{\lie}[1]{\pounds\!_{#1}}
\newcommand{\restr}[1]{\!\!\upharpoonright\!\!_{#1}}
\newcommand{\dlim}[1]{\lim_{\overrightarrow{\:#1\:}}}
\newcommand{\ilim}[1]{\lim_{\overleftarrow{\:#1\:}}}

\renewcommand{\beforepartskip}{\null\vspace*{\fill}}
\renewcommand{\afterpartskip}{\newpage}
\renewcommand{\partnamefont}{\normalfont\LARGE\scshape\raggedleft}
\renewcommand{\partnumfont}{\normalfont\LARGE\scshape\raggedleft}
\renewcommand{\parttitlefont}{\normalfont\LARGE\scshape\raggedleft}

\begin{document}

%

\newtheorem{theorem}{\mdseries\scshape Theorem}[chapter]
\newtheorem{lemma}[theorem]{\mdseries\scshape Lemma}
\newtheorem{proposition}[theorem]{\mdseries\scshape Proposition}
\newtheorem{corollary}[theorem]{\mdseries\scshape Corollary}
\newtheorem{scholium}[theorem]{\mdseries\scshape Scholium}
\newtheorem{definition}{\mdseries\scshape Definition}[chapter]
\newtheorem{remark}[definition]{\mdseries\scshape Remark}
\newtheorem{conjecture}{\mdseries\scshape Conjecture}[chapter]

%

\selectlanguage{english}
\pagenumbering{roman}
\pagestyle{empty}

\begin{center}
\sloppy
\begin{Large}
Universidade de S\~ao Paulo\\
Instituto de F\'{\i}sica\\
\end{Large}
\vspace*{1.6cm}
\begin{LARGE}
\textsc{Structural and Dynamical Aspects of the AdS/CFT Correspondence: a 
Rigorous Approach}\\
\end{LARGE}
\vspace*{1.6cm}
\begin{Large}
Pedro Lauridsen Ribeiro\\
\end{Large}
\vspace*{1.5cm}
\begin{flushright}
Advisor: Jo\~ao Carlos Alves Barata\\
\vspace{0.7cm}
\begin{minipage}[t]{8cm}
\hrulefill\vspace*{2mm}\\
Doctorate Thesis submitted to the Institute of Physics of the University 
of S\~ao Paulo for achievement of the title of Doctor in Sciences

\hrulefill\\
\end{minipage}
\end{flushright}
\vspace*{0.5cm}
\begin{flushleft}
\textbf{Examination Board:}\\
Prof. Dr. Jo\~ao Carlos Alves Barata (IF-USP)\\
Prof. Dr. \'Elcio Abdalla (IF-USP)\\
Prof. Dr. Frank Michael Forger (IME-USP)\\
Prof. Dr. Severino Toscano do R\^ego Melo (IME-USP)\\
Prof. Dr. Ricardo Schwarz Sch\"or (ICEx-UFMG)\\
\end{flushleft}
\vspace*{0.7cm}
\begin{large}
S\~ao Paulo\\
\vspace*{1.2ex}
2007
\end{large}
\end{center}

%

\newpage

\vspace*{\fill}
This is a (hopefully, apart from some corrections throughout the text) faithful English 
translation of my Ph.D. thesis, originally written in Portuguese, aiming at submission 
to arXiv's and LQP Crossroads' preprint servers for public use under these servers'
terms and conditions -- I intend to add some extra chapters in the near future with
material which has been excluded from the original Portuguese text due to the lack of 
spacetime.\\

I apologise the readers for not being able to provide the English translation of all 
epigraphs, more specifically regarding the texts of Guimar\~aes Rosa and Edith Pimentel 
Pinto -- not only because I couldn't find them (for the second author, it probably 
doesn't exist) but also because undertaking the translation task myself would go far 
beyond my literary capabilities (of course, if someone knows about an English translation, 
I'd be glad to be informed about it). Should you notice any typographic and/or language 
errors, I'll also be glad to know about them. Please contact me at
\texttt{pribeiro@fma.if.usp.br} and/or \texttt{pedro.lauridsen.ribeiro@desy.de}
regarding these matters.
\vspace*{0.5cm}
\begin{flushright}
The Author\\
Hamburg, December 2007
\end{flushright}
\vspace*{2cm}

\newpage

%

%

%
%

%

\chapterstyle{companion}

\vspace*{\fill}
\begin{flushright}
\emph{This thesis is dedicated to the memory of my grandmother, Maria de Lourdes 
Pimentel Pinto (${}^{\star\!}\mbox{1922}$ -- ${}^{\dagger\!}\mbox{1996}$) and my grand-aunt, 
Edith Pimentel Pinto (${}^{\star\!}\mbox{1921}$ -- ${}^{\dagger\!}\mbox{1992}$), for I
know that they watched over me (and still do!) all this time...}
\end{flushright}
\vspace*{2cm}

\cleardoublepage

\vspace*{\fill}
\epigraph{\emph{\selectlanguage{greek} \s{A}ut\`a t\`a pr\'agmata o\r{r}\~an,
diairo\~unta \s{e}ic \Ar{u}lhn, \s{a}\'ition, \'anafor\'an.\selectlanguage{english}}\endnotemark[1]}
{\textsc{Marcus Aurelius}\\ 
\emph{\selectlanguage{greek} T\`a \s{e}ic e\r{a}ut\'on \selectlanguage{english}}\endnotemark[2], 12.10}
\vspace*{0.5cm}
\epigraph{
\emph{Man h\"ort nur die Fragen, auf welche man imstande ist, eine Antwort zu finden.}\endnotemark[3]}
{\textsc{Friedrich Wilhelm Nietzsche}\\
\emph{Die fr\"ohliche Wissenschaft}\endnotemark[4], \S 196}

\cleardoublepage

\pagestyle{plain}

\begin{quotation}
\begin{center}
\textbf{Abstract.}
\end{center}
\vspace{2ex}
We elaborate a detailed study of certain aspects of (a version of) the \emph{AdS/CFT 
correspondence}, conjectured by Maldacena \cite{malda} and Witten \cite{witten1}, between 
quantum field theories in a gravitational background given by an asymptotically anti-de 
Sitter (AAdS) spacetime, and conformally covariant quantum field theories in the latter's 
conformal infinity (in the sense of Penrose), aspects such that: (a) are independent from 
(the pair of) specific models in Quantum Field Theory, and (b) susceptible to a recast in 
a mathematically rigorous mould. We adopt as a starting point the theorem demonstrated by 
Rehren \cite{rehren1} in the context of Local Quantum Physics (also known as Algebraic 
Quantum Field Theory) in anti-de Sitter (AdS) spacetimes, called \emph{algebraic holography} 
or \emph{Rehren duality}.\\

The main body of the present work consists in extending Rehren's result to a reasonably 
general class of $d$-dimensional AAdS spacetimes ($d>3$), scrutinizing how the properties 
of such an extension are weakened and/or modified as compared to AdS spacetime, and probing 
how non-trivial gravitational effects manifest themselves in the conformal infinity's 
quantum theory.\\

Among the obtained results, we quote: not only does the imposition of reasonably general 
conditions on bulk null geodesics (whose plausibility we justify 
through geometrical rigidity techniques) guarantee that our generalization is geometrically
consistent with causality, but it also allows a ``holographic'' reconstruction of the bulk
topology in the absence of horizons and singularities; the implementation of conformal 
symmetries in the boundary, which we explicitly associate to an intrinsically constructed
family of bulk asymptotic isometries, have a purely asymptotic character and is dynamically 
attained through a process of \emph{return to equilibrium}, given suitable boundary 
conditions at infinity; gravitational effects may cause obstructions to the reconstruction
of the bulk quantum theory, either by making the latter trivial in sufficiently small regions
or due to the existence of multiple inequivalent vacua, which on their turn lead to the 
existence of solitonic excitations localized around domain walls, similar to D-branes. The 
language employed for the quantum theories relevant for our generalization of Rehren 
duality follows the functorial formulation of Local Quantum Physics due to 
Brunetti, Fredenhagen and Verch \cite{bfv}, extended afterwards by Sommer \cite{sommer} 
in order to incorporate boundary conditions.
\end{quotation}

\cleardoublepage

\begin{quotation}

\selectlanguage{brazil}

\begin{center}
\textbf{Resumo.}
\end{center}
\vspace{2ex}
Elaboramos um estudo detalhado de alguns aspectos d(e uma vers\~ao d)a \emph{correspond\^encia 
AdS/CFT}, conjecturada por Maldacena \cite{malda} e Witten \cite{witten1}, entre teorias 
qu\^anticas de campo num fundo gravitacional dado por um espa\c{c}o-tempo assintoticamente anti-de 
Sitter (AAdS), e teorias qu\^anticas de campos conformalmente covariantes no infinito conforme 
(no sentido de Penrose) deste espa\c{c}o-tempo, aspectos estes: (a) independentes d(o par d)e 
modelos espec\'{\i}ficos em Teoria Qu\^antica de Campos, e (b) suscet\'{\i}veis a uma reformula\c{c}\~ao em 
moldes matematicamente rigorosos. Adotamos como ponto de partida o teorema demonstrado por 
Rehren \cite{rehren1} no contexto da F\'{\i}sica Qu\^antica Local (tamb\'em conhecida como Teoria 
Qu\^antica de Campos Alg\'ebrica) em espa\c{c}os-tempos anti-de Sitter (AdS), denominado 
\emph{holografia alg\'ebrica} ou \emph{dualidade de Rehren}. \\

O corpo do presente trabalho consiste em estender o resultado de Rehren para uma classe 
razoavelmente geral de espa\c{c}os-tempos AAdS $d$-di\-men\-sio\-nais ($d>3$), escrutinar como as 
propriedades desta extens\~ao s\~ao enfraquecidas e/ou modificadas em rela\c{c}\~ao ao espa\c{c}o-tempo 
AdS, e como efeitos gravitacionais n\~ao-triviais se manifestam na teoria qu\^antica no infinito 
conforme. \\

Dentre os resultados obtidos, citamos: condi\c{c}\~oes razoavelmente
gerais sobre geod\'esicas nulas no interior (cuja plausibilidade justificamos
por meio de resultados de rigidez geom\'etrica) n\~ao s\'o garantem que a nossa generaliza\c{c}\~ao
\'e geometricamente consistente com causalidade, como tamb\'em permite uma reconstr\c{c}\~ao
``hologr\'afica'' da topologia do interior na aus\^encia de horizontes e singularidades;
a implementa\c{c}\~ao das simetrias conformes na fronteira, que associamos explicitamente
a uma fam\'{\i}lia de isometrias assint\'oticas do interior constru\'{\i}da de maneira intr\'{\i}nseca, ocorre 
num car\'ater puramente assint\'otico e \'e atingida dinamicamente por um processo de \emph{retorno 
ao equil\'{\i}brio}, mediante condi\c{c}\~oes de contorno adequadas no infinito; efeitos gravitacionais
podem eventualmente causar obstru\c{c}\~oes \`a reconstru\c{c}\~ao da teoria qu\^antica no interior, ou por 
torn\'a-la trivial em regi\~oes suficientemente pequenas ou devido \`a exist\^encia de m\'ultiplos
v\'acuos inequivalentes, que por sua vez levam \`a exist\^encia de excita\c{c}\~oes solit\^onicas localizadas
ao redor de paredes de dom\'{\i}nio no interior, similares a D-branas. As demonstra\c{c}\~oes 
fazem uso extensivo de geometria Lorentziana global. A linguagem empregada para as teorias 
qu\^anticas relevantes para nossa generaliza\c{c}\~ao da dualidade de Rehren segue a formula\c{c}\~ao 
funtorial de Brunetti, Fredenhagen e Verch para a F\'{\i}sica Qu\^antica Local \cite{bfv}, estendida 
posteriormente por Sommer \cite{sommer} para incorporar condi\c{c}\~oes de contorno. 

\selectlanguage{english}

\end{quotation}

\cleardoublepage

\chapter*{Acknowledgements}

My sincere thanks to Prof. Jo\~ao Barata, for taking me in at the Department of
Mathematical Physics, for proposing the present theme of investigation and for
believing me even at the most difficult moments of the development of this project,
bearing thus a solid and essential support.\\

To my former, undergraduate research group, friends up to day I hold dearly:
Profs. Alinka L\'epine-Szily, Rubens Lichtenth\"aler Filho and Valdir Guimar\~aes,
for the support and enticement when I decided to migrate from experimental/computational
Nuclear Physics to Mathematical Physics, and above all for teaching me the first
steps of scientific research and a righteous attitude towards it.\\

To the (ex-)colleagues of the Department of Mathematical Physics (IFUSP) Alessandro Marques, 
Alex Dias, Am\'{\i}lcar Queiroz, Brenno Vallilo, Davi Giugno, Ivan Pontual, Karlucio 
Castello Branco, Leandro Bevil\'aqua, Marcelo Pires, ``dign\'{\i}ssimo'' Leonardo Sioufi and 
Patr\'{\i}cia Teles, for the priceless conversations (in both senses!) and for the company
along these more than four years.\\

To the masters and companions of the Institute of Mathematics and Statistics, for
being open to the enriching exchanges of ideas which contributed so much to my learning,
in particular the Operator Algebras group  (C\'{\i}ntia da Silva, 
David Dias, Marcela Merklen, Patr\'{\i}cia Hess and Profs. Cristina Cerri, Martha Salerno, 
Ricardo Bianconi and Severino Toscano), Prof. Michael Forger's group (Bruno Soares, 
Eder Annibale, Fernando Antoneli, Get\'ulio Bulh\~oes, Leandro Gomes, M\'ario Salles, Paola 
Gaviria, Sandra Yepes and Sandro Romero) and as well to Profs. Helena \'Avila, Paulo Cordaro 
to the colleagues Alexandre Kawano, Arthur Juli\~ao, Ant\^onio Ronaldo Garcia and Nestor Centuri\'on.\\

I'd also like to thank Profs. Bert Schroer, Christian J\"akel, Daniel Vanzella, Fernando 
Auil, Jens Mund and, once more, to Prof. Michael Forger for the patience and enthusiasm
during their participation at the series of informal seminars along the years 2004--2005
dedicated to the study of the reference \cite{bfv}, which constitutes a founding stone of
the conceptual framework underlying the present Thesis. I thank likewise these and Profs.
Aiyalam Balachandran, Detlev Buchholz, Henning Rehren, Klaus Fredenhagen, Kostas Skenderis, 
Sumati Surya e Walter Wreszinski, for the inspiring dialogs and/or email exchanges at 
crucial moments of the development of the present work, that, hence, constituted an 
invaluable aid with their suggestions and constructive criticism.\\

To the secretaries of the Department of Mathematical Physics: Am\'elia Genova, Elizabeth 
Vargas e Simone Shinomiya, for the solicitude and tireless sympathy.\\

To FAPESP, for the financial support (this project was graced with a Direct Doctorate
scholarship relative to the process 01/14360-1).\\

Well, it's a lot of people, and most certainly I forgot someone above... Thanks and 
my apologies to those!

And, lastly (and first), the more than special thanks:\\

To my parents, for this journey was also theirs, of us all!\\

To my brother Francisco, for the spiritual counterpoint.\\

To Clau, for keeping (enduring!) my mental (in)sanity, and for teaching me
the importance of the tings that must be said.\\

To God, Almighty Eternal Father, for showing me a path that fascinated me and
giving me strength to follow it. Let's hope it grants good fruits to His honour!

\cleardoublepage

\tableofcontents*

\cleardoublepage

\chapter*{\label{ch0} Preface}
\addcontentsline{toc}{chapter}{Preface}
\chaptermark{Preface}

\epigraph{\begin{center}\textsc{Sem\^antica}\end{center}
\emph{\begin{verse}[7cm] Eu sei l\'a o que querem dizer as palavras.\\[11pt]
Sou caminho \\ e se imanto as limalhas a meu jeito \\ o desenho
n\~ao \'e figurativo.\\[11pt]
E se empreendo, gaguejando,\\ a leitura do
meu campo de esfor\c{c}o \\ o feito \\ infante ainda \\ \'e mensagem
cifrada.\end{verse}}~\vspace*{-22pt}}{\textsc{Edith Pimentel Pinto} \\ 
``Normativa'' (\emph{Sinais e Conhecen\c{c}as}\endnotemark[5], 1986)}

Two of the greatest open problems in contemporary theoretical physics are the
lack of quantitative comprehension of the nonperturbative aspects of non Abelian 
gauge theories (quark confinement, etc.) and the quantization of the gravitational 
field, which makes itself phenomenologically necessary due to the limits of physical 
consistency of General Relativity, which describes classical gravity up to what we've
been able so far to test experimentally. One of the most tantalizing ideas in current
literature is that both problems are somehow related by a \emph{holographic} 
correspondence, in which gravity in a certain region of spacetime would admit a 
``dual'' description at the boundary of such a region by means of a gauge theory, in 
the same way that a bidimensional hologram reproduces a tridimensional image. Moreover, 
the energy regimes of the physical processes associated to both theories are of such 
nature that \emph{nonperturbative} aspects of the gauge theory would manifest themselves
at the \emph{(semi)classical} gravitational regime.\\

The history of how this idea arose goes back to the seventies. The four laws of black hole
dynamics, established by \textsc{Bardeen}, \textsc{Carter} and \textsc{Hawking} 
\cite{bardeench}, possess a strong analogy with the 4 laws of thermodynamics:

\begin{description}
\item[\mdseries\scshape Zeroth Law \upshape \cite{bardeench}:] The surface gravity $\kappa$ is 
constant throughout the event horizon of a stationary black hole \footnotemark[1] 
$\leftrightarrow$ a thermal bath in equilibrium possesses constant temperature $T$ throughout 
all its extension.
\item[\mdseries\scshape First Law \upshape \cite{bardeench}:] $\delta M=\Omega_H\delta J+\frac{1}{8\pi}
\kappa\delta A$, where $M$ is the mass of the black hole, $\Omega_H$ is the magnitude of the
axial component of the timelike \textsc{Killing} field tangent to the horizon, $J$ the angular
momentum of the black hole and $A$ the area of (a spatial section of) the event horizon. The variations
of the metric implicit in the formula are also (quasi)stationary \footnotemark[1] $\leftrightarrow$ 
$\delta Q=P\delta V+T\delta S$, where $Q$ is the amount of heat, $P$ the pressure, $V$ the volume 
and $S$ the entropy.
\item[\mdseries\scshape Second Law \upshape \cite{hawking1}:] \emph{Area Theorem} -- the area of 
(spatial sections of ) the event horizon cannot decrease throughout the \textsc{Cauchy} evolution
of a nonstationary black hole satisfying the null energy condition (NEC) \footnotemark[2] 
$\leftrightarrow$ the entropy of a closed system cannot decrease by means of any physical process.
\item[\mdseries\scshape Third Law \upshape \cite{israel}:] A nonstationary, sufficiently
regular black hole satisfying the weak energy condition cannot become \emph{extremal}, i.e., lose 
its trapped surfaces (i.e., acausal, codimension-two surfaces such that both of their normal null 
geodesic congruences possess negative expansion) external to its event horizon in finite
(retarded) time, which would imply $\kappa\rightarrow 0$ \footnotemark[3] $\leftrightarrow$ it's 
impossible to reach $T=0$ in finite time by means of any physical process.
\end{description}

\footnotetext[1]{\label{ch0fn1} Although the zeroth, first and second laws are stated
in \cite{bardeench}, only the zeroth and first laws are proved in this reference, 
for perturbations stationary black holes (the third law is only suggested in a rudimentary
form).}
\footnotetext[2]{\label{ch0fn2} The statement and proof of the second law were previously
elaborated by \textsc{Hawking} \cite{hawking1} in a nonstationary context.}
\footnotetext[3]{\label{ch0fn3} The third law was formulated in the present precise form 
and demonstrated by \textsc{Israel} \cite{israel}. One must recall that \textsc{Nernst}'s
formulation for the third law, i.e., $S=0$ at $T=0$, is neither universally valid (a typical example
is the ice, which possesses a residual entropy at $T=0$ due to the hydrogen bridges. Such a fact 
was established in an exact way in two dimensions by \textsc{Lieb} \cite{lieb1,lieb2}. Notice that 
the free \textsc{Bose} gas, which goes under \textsc{Bose-Einstein} condensation at $T=0$ and, hence, 
possesses degeneracy at the ground state, \emph{doesn't} constitute another (counter)example, for  
the asymptotic behaviour of such a degeneracy in the partition function as $T\searrow 0$ is just 
polynomial, and not exponential as in the case of ice, resulting in $S=0$ if we adopt 
\textsc{Boltzmann}'s definition for the entropy. I thank Prof. Walter Wreszinski for 
calling my attention to this point), nor is its ``natural'' analog for black holes \cite{wald4}.}

In particular, the \emph{area} of the event horizon of a black hole corresponds to the \emph{entropy}. 
\textsc{Bekenstein} \cite{bek} suggested that such an identification was more than just a mere analogy 
-- the entropy of an black hole in equilibrium would, indeed, be given by the formula \[ S=\frac{A_{hor}}{4},\]
and a generalization of the second law of thermodynamics, taking into account the matter entropy 
(outside the black hole) as well as the entropy of the black hole, should hold. This scenario, 
suggested by \textsc{Hawking}'s Area Theorem \cite{hawking1}, is not necessarily valid 
classically; however, the work of \textsc{Hawking} \cite{hawking2} showed \footnote{The 
geometrical optics approximation for the two point function near the horizon of a \textsc{Schwarzschild}, 
black hole, adopted by \textsc{Hawking} in \cite{hawking2}, is difficult to justify, for the 
behaviour of the metric in this region implies an index of refraction that varies very fast along the
relevant time coordinate. A conceptually precise calculation of \textsc{Hawking} radiation was obtained
by \textsc{Fredenhagen} and \textsc{Haag} \cite{frehaag2}.} that a \emph{quantum} field coupled to
the gravitational field of a black hole \emph{thermalizes}, after a sufficiently long time interval, 
\emph{precisely at temperature} $T=\frac{\kappa}{2\pi}$, \emph{suggested by the first law of black hole
dynamics!} This result is surprising, for the latter is a purely geometrical quantity, which indicates 
an \emph{universal} (i.e., independent of the quantum field model we couple to the classical
gravitational field) character for the laws of black hole thermodynamics.\footnote{From the viewpoint 
of the related \textsc{Unruh} \emph{effect}, this is not so surprising, if one has in mind 
the results of \textsc{Bisognano} and \textsc{Wichmann} in the context of \textsc{Wightman} quantum
fields \cite{bw1,bw2,sewell1}. However, such a comparison demands care, as the 
circumstances of the \textsc{Hawking} effect are different in several aspects \cite{wald3}.} Motivated 
by such a result, \textsc{'t Hooft} \cite{thooft1} proposed taking \textsc{Bekenstein}'s idea to 
the last consequences: \emph{the microscopic degrees of freedom of a quantum theory of gravity are
completely encoded in the boundary of the spacetime's volume where such theory is 
defined.} This principle, formalized later by \textsc{Susskind} \cite{suss1}, is named 
\emph{holographic principle}.\\

Within the integrated description of gauge theories and gravity proposed by string theory, 
two results suggest that this theory may provide a realization of the holographic principle: 
the microscopic calculation of \textsc{Bekenstein} entropy made by \textsc{Strominger} and
\textsc{Vafa} \cite{strovafa}, and, above all, the work of \textsc{Maldacena} \cite{malda}, which, 
by means of a(n effective limit of a) string model on the background spacetime $AdS_5\times S^5$, 
where the $AdS_5$ factor is the anti-\textsc{de Sitter} (AdS) spacetime in five dimensions, obtained
the field multiplet of a supersymmetric, conformally invariant gauge theory with structural group 
$SU(N)$ in the $N\rightarrow\infty$ limit, living at the conformal boundary of $AdS_5$ (the $S^5$ 
factor becomes a global group of internal (R-)symmetries $SU(4)\sim SO(6)$). \textsc{Maldacena} 
conjectured that this relation establishes a one-to-one correspondence between both theories, even 
outside the effective limit. The formulation of this correspondence in terms of $k$-point functions 
was obtained by \textsc{Gubser}, \textsc{Klebanov} and \textsc{Polyakov} \cite{gubskp} in a particular
case, and more in general by \textsc{Witten} \cite{witten1}. In these two works, it was shown that the 
(\textsc{Schwinger}) $k$-point functions of the gauge theory in the above limit could be obtained from
variations of the \emph{classical} (Euclidean) supergravity action around its value at the background
geometry under variation of the boundary conditions at conformal infinity. Moreover, \textsc{Witten}'s
prescription for the generating functional of the conformal field theory allowed one to rephrase 
\textsc{Maldacena}'s statement uniquely in terms of usual field theories, without reference to strings. 
Such a correspondence became known in literature as \emph{AdS/CFT correspondence}. By means of this 
formulation, the \textsc{Maldacena} conjecture was successfully tested in a myriad of particular 
cases.\footnote{Unfortunately, the phenomenal research activity aroused by the fundamental works 
about the AdS/CFT correspondence, cited above, made practically impossible to elaborate even a
pretensely representative bibliography of its different directions (the work \cite{malda} alone 
possesses more than 4.700 citations by 30/07/2007, according to the SPIRES-HEP online database). Thus, 
we shall content ourselves in citing only works of direct relevance to our development, as needed, and 
forward the interested reader SPIRES-HEP's homepage (\texttt{http://www.slac.stanford.edu/spires/}) 
for other directions.}\\

\textsc{Witten}'s formulation for the AdS/CFT correspondence opened way to the
following question: might it be possible to establish a correspondence between quantum
field theories (QFT's) in AdS and conformally covariant QFT's in AdS's conformal infinity,
\emph{uniquely from the fundamental principles of QFT?} The question indeed happens to have a
\emph{positive} answer: within a version of \textsc{Wightman}'s axiomatic framework \cite{pct} 
for quantum fields in AdS, a correspondence between $k$-point functions in the (latter's) bulk
conformally invariant $k$-point functions in the conformal boundary was rigorously established 
by \textsc{Bertola}, \textsc{Bros}, \textsc{Moschella} and \textsc{Schaeffer} \cite{bertobms}. 
Moreover, a correspondence between bulk and boundary algebras of local observables was demonstrated 
by \textsc{Rehren} \cite{rehren1}, within the algebraic formalism of \textsc{Haag} and \textsc{Kastler} 
\cite{hkast,haag} for QFT (\emph{Local Quantum Physics}) -- the latter correspondence is named 
\emph{algebraic holography} or \textsc{Rehren} \emph{duality}. Although the relation between 
these two formulations and \textsc{Witten}'s prescription has been investigated in several 
aspects \cite{dure1,dure2,rehren5}, a fundamental question remained unanswered: \emph{how 
gravitational effects are encoded in these formulations?}\\

As both versions refer themselves uniquely to (pure) AdS spacetimes (for they depend
critically on its causal structure), and not to more general geometries possessing 
the same conformal infinity of AdS (i.e., \emph{asymptotically AdS} (AAdS) spacetimes), 
even the formulation \emph{per se} of such rigorous counterparts in a more general 
geometrical context becomes elusive. There are, inclusively, evidences \cite{asmolin} 
that, for nontrivial AAdS geometries, \textsc{Rehren} duality becomes 
physically incompatible with known tests of the \textsc{Maldacena} conjecture, 
and, as such, needs to be reformulated. Which takes us to a second question, intimately 
linked to the former: \emph{How much from the underlying framework of} \textsc{Rehren} 
\emph{duality and its properties does survive in ``generic'' AAdS spacetimes?}\\

The body of the present Thesis proposes a natural generalization of \textsc{Rehren} 
duality to AAdS spacetimes and investigates how the properties of the original version
are modified by gravitational effects in the bulk, seeking to elucidate both questions 
elaborated above.\\

One of the fundamental lessons of Local Quantum Physics is that the physical information
of a quantum field theory is contained not in the individual fields / observables, but in 
the relative inclusions of local algebras. In particular, such information is highly sensitive 
to the causal structure of spacetime. Hence, it's to be expected that the distortion of the 
causal structure of the bulk of an AAdS spacetime due to gravitational effects be ``felt'' by
the ``dual'' theory at the boundary. We'll see, after a qualitative, global study of this
deformation within a reasonably general set of hypotheses about the bulk geometry, that 
this indeed occurs. We shall show effects of two kinds: 

\begin{enumerate}
\item The implementation of asymptotic isometries, far from being an immediate fact as it is for 
AdS, is dynamically attained as a process of \emph{return to equilibrium}, which has as a 
consequence the potential breaking of conformal symmetry. The states over the boundary quantum
theory obtained by this process possess a rather rich structure, of which we've only scratched
the surface, and whose thermal properties deviating from equilibrium encode nontrivial details
of the bulk geometry.
\item The structure of superselection sectors is radically modified by gravitational effects. 
Given the hypothesis that arbitrarily small bulk regions possess nontrivial observables, plus other
natural ones within the context of the algebraic theory of superselection sectors \cite{araki2,haag}, 
we'll show that the bulk theory acquires soliton excitations localized around domain walls, similar to
the D-branes that occur in the original formulation of the AdS/CFT correspondence, signaling spontaneous
breaking of internal symmetries. The nontrivial increase of the local algebras at the boundary needed 
to the incorporation of the intertwiners among different partial vacua makes these algebras to acquire 
\emph{nonlocal} (i.e., extended) elements implementing internal symmetries in the region where each
partial vacuum differs from the total vacuum. And, precisely because of these elements, it becomes 
fundamentally impossible, unlike in pure AdS, to completely reconstruct the bulk quantum theory from 
the boundary theory alone (Subsection \ref{ch4-sector-rec}, page \pageref{ch4-sector-rec}).
\end{enumerate}

We've put emphasis, in this work, in robust, mathematically rigorous results which are independent from 
a specific quantum field model, by virtue of the universal character of several aspects of the 
AdS/CFT correspondence. We shall not investigate in detail, though, the relation of the formalism proposed
here to \textsc{Witten}'s prescription -- such a task lies beyond the scope of the present work.

\section*{\label{ch0-org}On the structure of this Thesis}

This Thesis has three parts. In Part I, which comprises Chapters \ref{ch1} and \ref{ch2}, we 
develop the geometrical part of the work. In Part II, which comprises Chapters \ref{ch3} 
and \ref{ch4}, we present our proposal of generalization of \textsc{Rehren} duality within 
a general formalism for QFT in curved spacetime which incorporates covariance and locality in 
functorial manner, and study universal (i.e., independent of specific models) consequences
from the dynamical and structural viewpoints. Part III consists only of the Coda, which concludes
the work.\\

In Chapter \ref{ch1}, we'll start a systematic treatment of anti-\textsc{de Sitter} (AdS) 
and asymptotically anti-\textsc{de Sitter} (AAdS) spacetimes. We shall investigate, within 
a set of hypotheses about the global behaviour of null geodesics, how the causal structure
is globally modified by gravitational effects consistent with the boundary conditions at
infinity (Proposition\ref{ch1p1}, page \pageref{ch1p1}) -- the effect is mainly due to
gravitational time delay of null geodesics traversing the bulk (Theorem \ref{ch1t4},
page \pageref{ch1t4}). Some of these results are not new \cite{woolgar,gaowald}, 
but the technique of gravitational time delay is fundamental to most of the geometrical
developments to follow. We shall define a simple generalization of \emph{wedge} regions, 
employed by \textsc{Rehren} in \cite{rehren1}, to AAdS spacetimes (Definition \ref{ch1d3}, 
page \pageref{ch1d3}) -- such regions generalize, in a certain sense, the \textsc{Rindler} 
wedge $\{x^1>|x^0|\}$ in \textsc{Minkowski} spacetime, and constitute the prototype of an
event horizon of an \emph{asymptotically} stationary black hole in a background with negative
cosmological constant. We'll show how it's possible to reconstruct the bulk topology by means
of intersections of wedges enveloping sufficiently small diamonds (Subsubsection 
\ref{ch1-aads-causal-loc}, page \pageref{ch1-aads-causal-loc}). This result is original 
\cite{ribeiro2}, and the proof involves an interesting application of the method employed
by \textsc{Penrose} \cite{hawkellis,wald2} to prove the existence of singularities as 
result of gravitational collapse, together with a compactness argument, to obtain a geometric 
maximum principle that guarantees the precise envelope (Theorem \ref{ch1t9}, page \pageref{ch1t9}). 
The presentation of this Chapter has several improvements with respect to \cite{ribeiro2}.\\

In Chapter \ref{ch2}, we'll investigate the dynamics of the \textsc{Einstein} equations
in AAdS spacetimes (Section \ref{ch2-fefgra}, page \pageref{ch2-fefgra}). Such a discussion 
aims at estimating in a more quantitative way nontrivial gravitational effects in terms of 
geometrical quantities defined in the conformal boundary. We'll also construct in an intrinsic 
manner families of asymptotic isometries (boosts) naturally associated to each AAdS wedge
(Proposition \ref{ch2p3}, page \pageref{ch2p3}), and, more in general, families of asymptotically 
conformal diffeomorphisms associated to any relatively compact diamond in causally simple
spacetimes (formulae (\ref{ch2e42})--(\ref{ch2e44}), page \pageref{ch2e42}). To such families, 
it's possible to associate a \emph{surface gravity} in the horizon of each wedge or diamond 
(Definition \ref{ch2d1}, page \pageref{ch2d1}) and, hence, formulate an asymptotic analog of
the zeroth law of black hole (thermo)dynamics (Theorem \ref{ch2t2}, page \pageref{ch2t2}), and
corresponding results for the second law (contained in Proposition \ref{ch1p1}, page \pageref{ch1p1}, 
and properly reinterpreted in the context of this Chapter) and the characterization of reversible
processes (Theorem \ref{ch2t3}, page \pageref{ch2t3}, which follows directly from Proposition
\ref{ch1p1}, \emph{ibid.} and Remark \ref{ch1r3}, page \pageref{ch1r3}). The presented construction 
is sufficiently robust to be implemented in situations not necessarily linked to the geometrical 
context of the AdS/CFT correspondence.\\

In Chapter \ref{ch3}, we'll introduce a generalization of the formalism of \textsc{Haag} and 
\textsc{Kast\-ler} to curved spacetimes, proposed by \textsc{Brunetti}, \textsc{Fredenhagen} 
and \textsc{Verch} \cite{bfv} and formulated in terms of the language of \emph{categories} 
and \emph{functors}. \\

In Chapter \ref{ch4}, we'll present in a precise way our proposal of generalization for 
\textsc{Rehren} duality. For such, it's necessary to adapt the constructions of Chapter \ref{ch3} 
to our present geometrical context, which demands the imposition of \emph{boundary conditions} in
a covariant manner. Thus, we shall initiate our discussion with the algebraic treatment of this
issue proposed by \textsc{Sommer} \cite{sommer}, which  investigates how it's possible to extend a 
locally covariant quantum theory in the sense of \textsc{Brunetti et al.} to non globally hyperbolic 
spacetimes (Definition \ref{ch4d1}, page \pageref{ch4d1}). One follows the example of local quantum
theories in anti-\textsc{de Sitter} spacetimes -- here, we sought to insert the formulation proposed
by \textsc{Buchholz}, \textsc{Florig} and \textsc{Summers} \cite{bfs} within a locally covariant
context. The advantage of this formulation, besides the minimal number of premises, is the imposition
of boundary conditions at conformal infinity on the ``elementary'' states in the form of a condition of
thermodynamic stability (formulae (\ref{ch4e8}) and (\ref{ch4e9}), page \pageref{ch4e8}) -- which, 
in the case of AAdS spacetimes, is generalized in the form of a condition of ``return to equilibrium'' 
(condition (d), page \pageref{ch4e19}). Employing the techniques of \textsc{Borchers} and
\textsc{Yngvason} \cite{boryng} (stated in Theorems \ref{ch4t1}, \ref{ch4t2} and \ref{ch4t3}, page 
\pageref{ch4t1}ff.), we shall show that such a condition can be understood as a prescription for the
scaling behaviour of observables in a neighborhood of infinity (Proposition \ref{ch4p1}, page 
\pageref{ch4p1}), which can be generalized to asymptotically anti-\textsc{de Sitter} spacetimes. 
With this framework at hand, the generalization of \textsc{Rehren} duality (Definition \ref{ch4d2}, 
page \pageref{ch4d2}) we seek emerges naturally (Definition \ref{ch4d4}, page \pageref{ch4d4}). 
We'll investigate the structure of states over the obtained boundary theory by means of this 
generalization, and we'll end up with an outline of a correspondence between superselection sectors of the
bulk quantum theory and its boundary dual, linking sectors localized in diamonds (i.e., in the sense
of \textsc{Doplicher}, \textsc{Haag} and \textsc{Roberts} \cite{haag}) at the boundary and soliton 
sectors in the bulk, localized around codimension-two domain walls, quite similar to D-branes.\\

In the Coda, we'll present our conclusions and list possible future directions of investigation 
suggests by the present work.\\

A pedagogical obstacle to the conception of the present Thesis was the broad spectrum of employed
mathematical techniques -- global Lorentzian geometry, operator algebras, categories and functors.
Hence, we've collected most of the needed mathematical concepts in four Appendices, in order not to 
deviate the reader's attention from the central ideas of the present work and make the text 
mathematically self contained.\\

Appendix \ref{ap1} summarizes the necessary notions of Lorentzian geometry and causal structure. \\

Appendix \ref{ap2} presents basic concepts of Operator Algebras (C*-algebras, \textsc{von Neumann} 
algebras and \textsc{Borchers-Uhlmann} algebras), including a minimal introduction to 
\textsc{Tomita-Takesaki} theory, employed in Chapter \ref{ch4}. This Appendix begins with a detailed
treatment of *-algebras, seeking to unify the presentation of the results and concepts of operator 
algebras that depend only on the algebraic structure.\\

Appendix \ref{ap3} condensates the basic concepts of categories and functors employed in Chapters 
\ref{ch3} and \ref{ch4}.\\

Appendix \ref{ap4} draws quick strokes of the concepts of homotopy needed in Chapter \ref{ch1}.\\

I wish you all a nice reading!

\vspace*{\fill}

\begin{flushright}
The Author\\
São Paulo and Hamburg, resp. September and December 2007
\end{flushright}

\cleardoublepage

\listoffigures

\cleardoublepage

\chapter*{Frequently used notation and conventions}
\addcontentsline{toc}{chapter}{Frequently used notation and conventions}
\chaptermark{Frequently used notation and conventions}

\begin{itemize}
\item $\mathbb{Z},\,\mathbb{R},\,\mathbb{C}$ -- respectively, the ring 
of integers and the fields of real and complex numbers.
$\mathbb{Z}_{+/-},\,\mathbb{R}_{+/-}$ -- respectively, the positive ($>0$) / negative 
($<0$) integers and the positive / negative reals. $\bar{\mathbb{Z}}_{+/-},\,\bar{\mathbb{R}}_{+/-}$ -- 
respectively, the non negative ($\geq 0$) / non positive ($\leq 0$) integers and 
the non negative / non positive reals.
\item The real and imaginary parts of $z=x+iy\in\mathbb{C}$ are respectively
denoted by $\Re z\doteq x$ and $\Im z\doteq y$, and the complex conjugate of $z$,
by $\bar{z}\doteq x-iy$.
\item The power set of $X$ is denoted by $P(X)\doteq\{U\subset X\}$.
\item Given any sets $A,S$, $A^S\doteq\{a:S\rightarrow A\}$ denotes the 
\emph{Cartesian power} of $A$ to $S$. In particular, if $S=\{1,\ldots,n\}$, 
we write $A^S\doteq A^n=\{(a_1,\ldots,a_n):a_i\in A,\forall i=1,\ldots,n\}$.
\item Given an open set $\mathscr{O}$ in a topological space $X$, $K\Subset\mathscr{O}$
denotes that $K$ possesses compact closure $\bar{K}$ and $\bar{K}\subset\mathscr{O}$.
\item Given a symmetric non degenerate bilinear form $g$ over $\mathbb{R}^d$, the
\emph{signature} of $g$ is given by the pair of integers $(p,q)$ or by their difference $p-q$, 
where $p+q=d$ and $p$ is the number of \emph{negative} eigenvalues of $g$, named 
\emph{index} of $g$. We associate to $g$ the symmetric non degenerate bilinear form $g^{-1}
\doteq[g_{ij}]^{-1}$, of same index, over the dual of $\mathbb{R}^d$, and the volume $d$-form
given by $\sqrt{|g|}\doteq\sqrt{((-1)^p\det[g_{ij}])^{\frac{1}{2}}}$, where $[g_{ij}]$ 
is the matrix associated to $g$ in some orthonormal basis.
\item Manifolds and their subsets are denotes by capital calligraphic letters (e.g. 
$\mathscr{M}$, $\mathscr{O}$, $\mathscr{I}$, etc.). All manifolds employed in the text are 
$\mathscr{C}^\infty$, $\sigma$-compact (and, thus, paracompact), connected and orientable, 
unless otherwise indicated (regarding the notions of topology involved, see \cite{dugundji}).
\item Given a manifold $\mathscr{M}$, we denote the tangent and cotangent spaces at 
$p\in\mathscr{M}$ respectively by $T_p\mathscr{M}$ and $T^*_p\mathscr{M}$, and the tangent and 
cotangent bundles, by $T\!\mathscr{M}$ and $T^*\!\mathscr{M}$.
\item More generally, we write any vector bundle over a $d$-dimensional $\mathscr{M}$ as 
$\mathscr{E}\stackrel{p}{\longrightarrow}\mathscr{M}$, where $p$ denotes the projection (surjective 
submersion) map, $\mathscr{E}$ the total $(D+d)$-dimensional space and $\mathscr{M}$ the basis.
The typical fibre is usually denoted by $E_x\doteq p^{-1}(x)\cong\mathbb{R}^D,\,\forall 
x\in\mathscr{M}$.
\item Given a vector field $X$, we denote the \textsc{Lie} derivative of functions 
/ sections along $X$ by $\lie{X}$.
\item Tensor indices are given by small Greek letters $\mu$, $\nu$, etc. when written
in terms of local coordinates, and by small Latin letters $a$, $b$, etc. when they only
indicate the rank of the tensor, without reference to coordinates (\textsc{Penrose}--\textsc{Rindler}
abstract index notation). In both cases, the contraction of indices follows \textsc{Einstein}'s 
convention.
\item We denote by $\delta^a_b$ the \textsc{Kronecker} delta (= identity matrix as 
a linear operator in $T^{(*)}\!\mathscr{M}$).
\item Given a rank-$(r,s)$ tensor $T^{\mu_1\cdots\mu_r}_{\nu_1\ldots\nu_s}$,
we denote respectively the symmetric antisymmetric parts of $T$ w.r.t.
contravariant indices $\mu_j,\ldots,\mu_k$, $1\leq j<k\leq r$ 
by \[T^{\mu_1\cdots(\mu_j\cdots\mu_k)\cdots\mu_r}_{\nu_1\cdots\nu_s}\doteq
\frac{1}{(k-j+1)!}\sum_{\pi\in S_{k-j+1}}T^{\mu_1\cdots\mu_{\pi(j)}\cdots
\mu_{\pi(k)}\cdots\mu_r}_{\nu_1\cdots\nu_s}\] and \[T^{\mu_1\cdots[\mu_j
\cdots\mu_k]\cdots\mu_r}_{\nu_1\cdots\nu_s}\doteq\frac{1}{(k-j+1)!}
\sum_{\pi\in S_{k-j+1}}\mbox{sgn}(\pi)T^{\mu_1\cdots\mu_{\pi(j)}\cdots\mu_{\pi(k)}
\cdots\mu_r}_{\nu_1\cdots\nu_s},\] where $S_l$ is the group of permutations
of $l$ elements and $\mbox{sgn}(\pi)=1$ in case $\pi$ involves an even number
of 2-element switchings, and $-1$ if it involves an odd number of them.
Analogously, we denote respectively the symmetric and antisymmetric parts of $T$ 
w.r.t. covariant indices $\nu_m,\ldots,\nu_n$, $1\leq m<n\leq s$, by $T^{\mu_1\cdots
\mu_r}_{\nu_1\cdots(\nu_m\cdots\nu_n)\cdots\nu_s}$ e $T^{\mu_1\cdots\mu_r}_{\nu_1\cdots[\nu_m\cdots
\nu_n]\cdots\nu_s}$. If we want to exclude the interval $\mu_j,\ldots,\mu_k$
from the symmetrization or antisymmetrization of the indices $\mu_1,\ldots,\mu_l$,
we'll write respectively $(\mu_1\cdots|\mu_j\cdots\mu_k|\cdots\mu_l)$
or $[\mu_1\cdots|\mu_j\cdots\mu_k|\cdots\mu_l]$. The corresponding formulas
for (anti)symmetrization of a tensor shall have the combinatorial 
factor $\frac{1}{l!}$ substituted by $\frac{1}{(l-k+j-1)!}$. 
All these conventions hold without modification for abstract indices.
\item $\mathbb{R}^{p,q}$ denotes the semi-Riemannian manifold $(\mathbb{R}^{p+q},
\eta)$ (see Appendix \ref{ap1}), where the metric $\eta$ is given by the symmetric
non degenerate bilinear form of index $p$ \[\eta=\mbox{diag}(\stackrel{p}{-\cdots-}
\stackrel{q}{+\cdots+})\doteq\left[\begin{array}{ccc|ccc} 
-1 & & 0 & & & \\ & \ddots & & & 0 & \\ 0 & & -1 & & & \\
\cline{1-6} & & & +1 & & 0 \\ & 0 & & & \ddots & \\ & & & 0 & & +1
\end{array}\right].\] In particular, $d$-dimensional \textsc{Minkowski} spacetime
is given by $\mathbb{R}^{1,d-1}$ -- that is, we adopt the signature convention
$(-+\cdots+)$ (index 1) for Lorentzian metrics.
\item Given $x,y\in\mathbb{R}^d$, the Euclidean norm of $x$ is denoted by $|x|$,
and the corresponding scalar product, by $\langle x,y\rangle$.
The \textsc{Lebesgue} measure of a \textsc{Borel} subset $A$ of $\mathbb{R}^d$ is
denoted by $|A|$ (for the relevant notions of measure theory, see \cite{munroe,rudinrc}).
\item $S^d$ denotes the $d$-dimensional sphere $\{x\in\mathbb{R}^{d+1}:|x|=1\}$, 
whose induced Euclidean metric is denoted by $d\Omega^2_d=(d\theta^1)^2+\sin^2\theta^1
((d\theta^2)^2+\sin^2\theta^2(\cdots((d\theta^{d-1})^2+\sin^2\theta^{d-1}(d\theta^d)^2)\cdots))
=(d\theta^1)^2+\sin^2\theta^1d\Omega^2_{d-1}$, where $\theta^I=(\theta^1,\ldots,\theta^d)$ are 
the angular (spherical) coordinates of $S^d$, i.e., $0\leq\theta^1,\ldots,\theta^{d-1}\leq\pi$
and $0\leq\theta^d<2\pi$.
\item The norm of a normed vector space is denoted by $\Vert.\Vert$, with
additional indication when needed.
\item The scalar product of a pre-Hilbert space (real or complex)
is denoted by $\langle .,.\rangle$. We set the convention that, in the complex case, 
the sesquilinear form $\langle .,.\rangle$ is linear in the \emph{second} 
variable and anti-linear in the \emph{first}.
\item *-algebras, in particular C*- and \textsc{von Neumann},
are denoted by Gothic capital letters (e.g. $\mathfrak{F}$, 
$\mathfrak{A}$, $\mathfrak{R}$, etc.). Such algebras, if unital, have
their identity element denoted by $\mathbb{1}$ (see Appendix \ref{ap2}).
\item Functors are also denoted by Gothic capital letters.
\item Asymptotic behaviour -- we say that $f(x)=O(g(x))$ as
$x\rightarrow x_0$ ($x_0$ can be $\pm\infty$) if there exists a neighbourhood
$\mathscr{U}\ni x_0$ and $C>0$ such that $|f(x)|\leq C|g(x)|$, $\forall x\in
\mathscr{U}$ and $f(x)=o(g(x))$ as $x\rightarrow x_0$ if for all
$\epsilon>0$ there exists a neighbourhood $\mathscr{U}\ni x_0$ such that $|f(x)|<
\epsilon |g(x)|$, $\forall x\in\mathscr{U}$; $f$ and $g$ are said to be 
\emph{equivalent} (notation: $f(x)\sim g(x)$) as $x\rightarrow x_0$
if there exists a neighbourhood $\mathscr{U}\ni x_0$ and $C>0$ such that $C^{-1}g(x)\leq 
f(x)\leq Cg(x)$, $\forall x\in \mathscr{U}$. Given a sequence of functions $\{g_k
\}_{k\in\mathbb{Z}_+}$, we say that $f$ is an \emph{asymptotic sum} of $g_k,\,k
\in\mathbb{Z}_+$ as $x\rightarrow x_0$ (notation: $f\sim_{x\rightarrow 
x_0}\sum^\infty_{k=1}g_k$) if $|f-\sum^K_{k=1}g_k|=O(g_{K+1})$, for all $K$.
\item $\mathscr{D}(\mathscr{O})$ denotes the space of $\mathscr{C}^\infty$ functions
of compact support in $\mathscr{O}$ (\emph{test functions}), and 
$\mathscr{S}(\mathbb{R}^d)$, the space of $\mathscr{C}^\infty$ functions whose derivatives
of order $\geq 0$ go to 0 faster tan any polynomial in $x$ as $|x|\rightarrow\infty$ 
(\emph{tempered} or \textsc{Schwartz} test functions). We also write, following \textsc{L. Schwartz}, 
$\mathscr{E}(\mathscr{O})=\mathscr{C}^\infty(\mathscr{O})$. We shall employ these
different notations alternately along the text. Respectively, the 
(general) distributions, tempered e compactly supported distributions 
are given by the topological duals $\mathscr{D}'(\mathscr{O}),\,\mathscr{S}'
(\mathbb{R}^d),\,\mathscr{E}'(\mathscr{O})$.
\item When $\mathscr{O}\subset\mathscr{M}$, where $(\mathscr{M},g)$ is a $d$-dimensional, 
orientable semi-Riemannian manifold, we use invariant volume element 
$\sqrt{|g|}$ to identify test function spaces with the corresponding test densities (= $d$-forms,
since we've assumed $\mathscr{M}$ orientable). Hence, the topological duals are distributions
and not distributional densities (i.e., by the \emph{``smearing''} $u(f)$ of a distribution 
$u$ with a test function $f$, one understands the expression $u(f\sqrt{|g|})$, for the metric
will always be clear in context).
\item Given a vector bundle $\mathscr{E}\stackrel{p}{\longrightarrow}\mathscr{M}$,
the space of \emph{sections} $\mathscr{C}^\infty$ of $\mathscr{E}$ is the $\mathscr{C}^\infty
(\mathscr{M})$-module denoted by $\Gamma^\infty(\mathscr{M},\mathscr{E})\doteq\{\phi:
\mathscr{M}\stackrel{\mathscr{C}^\infty}{\longrightarrow}\mathscr{E}:p\circ\phi=
\mbox{id}_{\mathscr{M}}\}$. Denoting the \emph{zero section} of $\mathscr{E}$ (i.e., which 
takes the value of the origin at each fibre) simply by $0$, we define the \emph{support}
$\mbox{supp}\phi$ of a section $\phi$ as the complement of the largest open set in $\mathscr{M}$ 
where $\phi=0$. We denote, then, space of compactly supported $\mathscr{C}^\infty$ sections
of $\mathscr{E}$ by $\Gamma^\infty_c(\mathscr{M},\mathscr{E})$.
\end{itemize}

\cleardoublepage

\pagenumbering{arabic}
\pagestyle{Ruled}

\thisfancyput(0in,-4.8in){\includegraphics{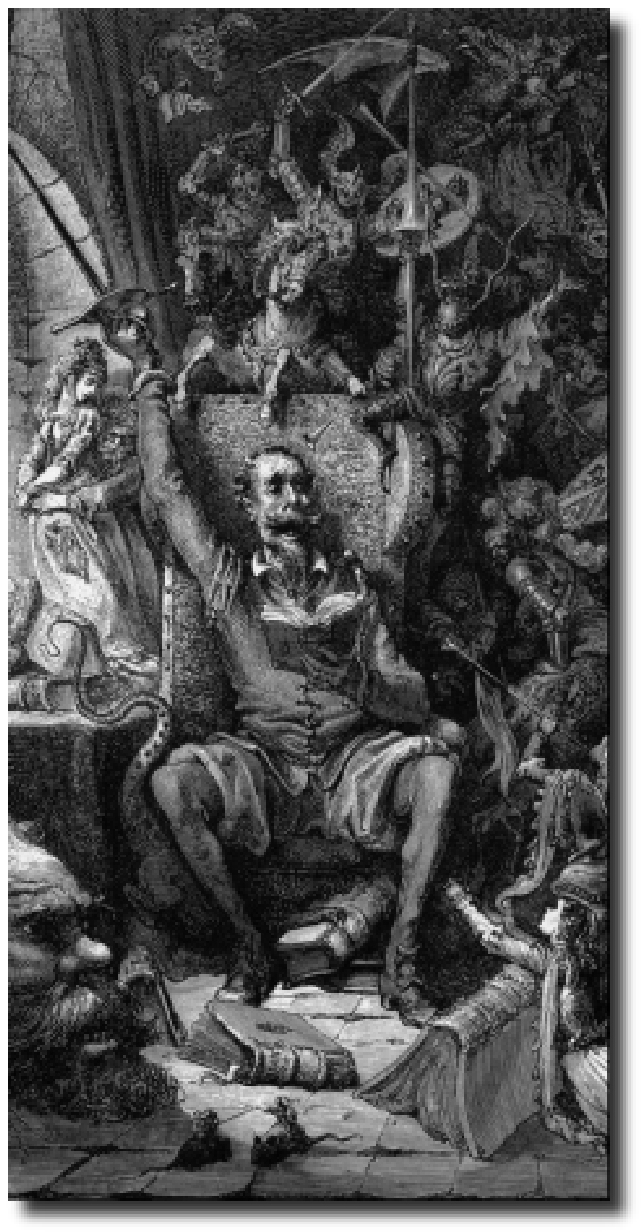}}

\epigraphhead[135]{\epigraph{\qquad En efeto, rematado ya su juicio, vino
a dar en el m\'as estra\~no pensamiento que jam\'as dio loco en el mundo, y fue
que le pareci\'o convenible y necesario, as\'{\i} para el aumento de su honra como
para el servicio de su rep\'ublica, hacerse caballero andante y irse por todo
el mundo con sus armas y caballo a buscar las aventuras y a ejercitarse en
todo aquello que \'el hab\'{\i}a le\'{\i}do que los caballeros andantes se ejercitaban,
deshaciendo todo g\'enero de agravio y poni\'endose en ocasiones y peligros donde,
acab\'andolos, cobrase eterno nombre y fama. Imagin\'abase el pobre ya coronado
por el valor de su brazo, por lo menos del imperio de Trapisonda; y as\'{\i}, con
estos tan agradables pensamientos, llevado del estra\~no gusto que en ellos 
sent\'{\i}a, se dio priesa a poner en efeto lo que deseaba. Y lo primero que hizo
fue limpiar unas armas que hab\'{\i}an sido de sus bisabuelos, que, tomadas de or\'{\i}n
y llenas de moho, luengos siglos hab\'{\i}a que estaban puestas y olvidadas en un
rinc\'on.\endnotemark[6]}
{\textsc{Miguel de Cervantes Saavedra}\\ \emph{El ingenioso hidalgo
don Quijote de la Mancha}\endnotemark[7], Primero Libro, Cap. I}}
\part{The AdS/CFT correspondence and its geometrical manifestations}

\chapter{\label{ch1}The geometry of asymptotically anti-\textsc{de Sitter} spacetimes}
\chaptermark{AAdS Geometry}

\epigraph{\emph{Devia ou n\~ao devia contar-lhe, por motivos de talvez.
Do que digo, descubro, deduzo. Ser\'a, se? Apalpo o evidente? Trebusco.
Ser\'a este nosso desengon\c{c}o e mundo o plano -- intersec\c{c}\~ao de planos --
onde se completam de fazer as almas?}}{\textsc{Jo\~ao Guimar\~aes Rosa}\\
``O espelho'' (\emph{Primeiras Est\'orias}\endnotemark[8])}

\section{\label{ch1-spacef}Generalities on space forms}

Among the $d$-dimensional Lorentzian metrics $g$ which solve the \textsc{Einstein} equations
without matter and with cosmological constant $\Lambda$
\begin{equation}\label{ch1e1}
\mbox{Ric}(g)-\frac{1}{2}R(g)g+\Lambda g=0,
\end{equation}
the simplest ones are those with constant sectional curvature.\footnote{For $d<4$, these are 
the \emph{only} solutions of (\ref{ch1e1})! See Appendix \ref{ap1}, formula (\ref{ap1e1}), 
page \pageref{ap1e1} discussion that follows.} More precisely, if $g$ is such that its 
sectional curvature
\begin{equation}\label{ch1e2}
K(g)(X,Y)\doteq\frac{g(\mbox{Riem}(g)(X,Y)X,Y)}{g(X,X)g(Y,Y)-g(X,Y)^2}=C
\end{equation}
for all pairs $X,Y$ of tangent vectors at a point $p$ generating a plane in which
$g$ is a nondegenerate bilinear form \footnote{The denominator in (\ref{ch1e2})
denotes the square of the Lorentzian area of the parallelogram with sides $X$ and $Y$.} 
($C$ may, em principle, depend on $p$), then $g$ satisfies (\ref{ch1e1}) if and only if 
$C=\frac{2}{(d-1)(d-2)}\Lambda$. In this case, we have:

\begin{equation}\label{ch1e3}
\mbox{Riem}(g)_{abcd}=\frac{2\Lambda}{(d-1)(d-2)}(g_{ac}g_{bd}-g_{ad}g_{bc})
\end{equation}
($g_{ac}g_{bd}-g_{ad}g_{bc}$ can be seen as the metric induced by $g$ in the tangent
space of 2-vectors; $\mbox{Riem}(g)_{abcd}$, on its turn, is a symmetric bilinear 
form in this space. The identity (\ref{ch1e3}) hence follows from (\ref{ch1e2}) by 
polarization). Conversely, any $d$-dimensional Lorentzian manifold with constant
sectional curvature $C$ satisfies (\ref{ch1e1}) with $\Lambda=\frac{(d-1)(d-2)}{2}C$. 
In these circumstances, we can invoke two results (see \cite{oneill} for the proof
of both):

\begin{theorem}[\cite{oneill}]\label{ch1t1} 
Let $(\mathscr{M},g)$ and $(\mathscr{M}',g')$ be two $d$-dimensional Lorentzian manifolds of
constant sectional curvature $K(g)=K(g')=C$. Then any points $p\in\mathscr{M}$, $p'\in\mathscr{M}'$ 
possess isometric neighbourhoods. If $\mathscr{M}'$ is geodesically complete, the isometry in 
question extends uniquely to a \emph{local isometry} $\phi:\mathscr{M}\rightarrow\mathscr{M}'$ 
(i.e., the tangent map $d\phi(p):T_p\mathscr{M}\rightarrow T_{p'}\mathscr{M}'$ is a isometry, 
for all $p\in\mathscr{M}$).\hfill~$\Box$
\end{theorem}

\begin{theorem}[\cite{oneill}]\label{ch1t2}
Let $(\mathscr{M},g)$ and $(\mathscr{M}',g')$ be two $d$-dimensional, geodesically complete 
and connected Lorentzian manifolds with constant sectional curvature $K(g)=K(g')=C$. 
Then, for any points $p\in\mathscr{M}$, $p'\in\mathscr{M}'$ and any isometry $L:T_p
\mathscr{M}\rightarrow T_{p'}\mathscr{M}'$, there exists a unique isometric covering map 
(see Definition \ref{ap4d3}, page \pageref{ap4d3}) $\phi:\mathscr{M}\rightarrow
\mathscr{M}'$ such that $d\phi(p)=L$.\hfill~$\Box$
\end{theorem}

\begin{definition}\label{ch1d1}
A geodesically complete Lorentzian manifold with constant sectional curvature is called 
a \emph{space form}.
\end{definition}

It follows from Theorem \ref{ch1t2} that any simply connected space form is uniquely 
determined by $\Lambda$. We list below such solutions.

\begin{description}
\item{$\Lambda>0$ --} \textsc{de Sitter} spacetimes (notation: $dS_d(\Lambda)$),
given by the spatially compact hyperboloid 
\begin{equation}\label{ch1e4}
dS_d(\Lambda)\doteq\{X\in\mathbb{R}^{1,d}:\eta(X,X)=R^2\},\,R=\sqrt{\frac{(d-1)
(d-2)}{2\Lambda}}.
\end{equation}
Its isometry group is the \textsc{de Sitter} \emph{group} $O(1,d)$, with component 
connected to identity \footnote{For orientable and time orientable Lorentzian manifolds, 
this is given by the \emph{proper} (preserving the orientation of the volume element, i.e., 
whose tangent map has determinant one) and \emph{orthochronous} (preserving time orientation)
isometries.} $SO_e(1,d)$.
\item{$\Lambda=0$ --} \textsc{Minkowski} spacetime $\mathbb{R}^{1,d-1}$. Its isometry group is
the \textsc{Poincar\'e} \emph{group} $O(1,d-1)\ltimes\mathbb{R}^d$, with component connected to 
identity $SO_e(1,d-1)\ltimes\mathbb{R}^d\doteq\mathscr{P}^\uparrow_{+,d}$.
\item{$\Lambda<0$ --} \emph{anti-}\textsc{de Sitter} spacetimes (notation: $AdS_d(\Lambda)$),
given by the universal covering (see Appendix \ref{ap4}, Definition \ref{ap4d3}, page 
\pageref{ap4d3}) of the timelike compact hyperboloid
\begin{eqnarray}\label{ch1e5}
 & AdS_d(\Lambda)\doteq\{X\in\mathbb{R}^{2,d-1}:(X^0)^2-(X^1)^2-\cdots-(X^{d-1})^2+(X^d)^2=R^2\},& \\
 & R=\sqrt{-\frac{(d-1)(d-2)}{2\Lambda}}. & \nonumber
\end{eqnarray}
Its isometry group is the \emph{anti-}\textsc{de Sitter} \emph{group}, given in the
fundamental domain (\ref{ch1e5}) by $O(2,d-1)$, with component connected to identity $SO_e(2,
d-1)$.
\end{description}

\begin{remark}
In a slight abuse of notation, we shall omit in general the cosmological constant when 
employing a notation above for the different simply connected space forms (i.e., we'll 
write $(A)dS_d$ instead of $(A)dS_d(\Lambda)$), whenever such a practice doesn't cause 
confusion. Unlike a recurrent practice in literature, we don't identify $AdS_d$ with the 
fundamental domain (\ref{ch1e5}), for the latter has a minimal role throughout this work, 
besides causing unnecessary technical complications.
\end{remark}

All solutions above are \emph{maximally symmetric:} the \textsc{Lie} algebra of
\textsc{Killing} fields of each space form has maximal dimension $\frac{d(d+1)}{2}$ 
(conversely, any connected Lorentzian manifold whose \textsc{Lie} algebra of \textsc{Killing} 
fields has maximal dimension possesses constant sectional curvature). By a standard result 
\cite{oneill}, every \textsc{Killing} field uniquely extends to a one-parameter group of 
isometries, whence it follows that the \textsc{Lie} group of isometries of a space form has 
also dimension $\frac{d(d+1)}{2}$. Moreover: let $p,q$ be two points of a space form $\mathscr{M}$. 
We can connect them by a finite sequence of geodesic segments $\gamma_i,i=1,\ldots,k$ with affine 
parameters $\lambda_i\in[0,1]$ -- in this case, $\gamma_i(1)=\gamma_{i+1}(0)$, $\gamma_1(0)=p$ and 
$\gamma_k(1)=q$ -- by covering any curve segment linking $p$ to $q$ with a finite number of normal
neighbourhoods. Consider the middle point $\bar{p}_i$ of $\gamma_i$, and the isometry $L_i:
T_{\bar{p}_i}\!\mathscr{M}\rightarrow T_{\bar{p}_i}\!\mathscr{M}$ given by $-\mbox{id}_{T_{\bar{p}_i}
\!\mathscr{M}}$. By Theorem \ref{ch1t1}, each $L_i$ determines an isometry $\phi_i:\mathscr{M}
\rightarrow\mathscr{M}$, in this case satisfying $\phi_i(\gamma_i(0))=\gamma_i(1)$. Finally, 
$\phi=\phi_k\circ\cdots\circ\phi_1$ is an isometry linking $p$ to $q$. That is, any two points 
$p,q$ are connected by an isometry, i.e., \emph{any space form is a homogeneous space} (i.e., 
the action of the isometry group is \emph{transitive}). In particular, the isometry group acts in
space forms \emph{without fixed points}, i.e., given any point $p$, there exists an isometry $\phi$ 
such that $\phi(p)\neq p$.\\

\section{\label{ch1-ads}Geometry of $AdS_d$: boundary conditions at infinity}

We can write a global system of spherically symmetric spatial coordinates for $AdS_d$. Consider 
$(\tau,r,\mathbf{e})\in\mathbb{R}\times\bar{\mathbb{R}}_+\times S^{d-2}$. Writing

\begin{equation}\label{ch1e6}
\left\{ \begin{array}{l@{\,=\,}l}
X^0 & \sqrt{R^2+r^2}\sin t \\
\mathbf{X} & r\mathbf{e} \\
X^d & \sqrt{R^2+r^2}\cos t
\end{array} \right.,\,R\mbox{ given by (\ref{ch1e5})},
\end{equation}
the $AdS_d$ metric becomes
\begin{equation}\label{ch1e7}
ds^2=-R^2\left(1+\frac{r^2}{R^2}\right)dt^2+\left(1+\frac{r^2}{R^2}\right)^{-1}dr^2
+r^2 d\Omega^2_{d-2}.
\end{equation}

The fundamental domain is obtained by the restriction $-\pi<t\leq\pi$. As the covering is, 
thus, obtained by ``unwrapping'' the time coordinate $\tau$, we shall denominate the global 
chart (\ref{ch1e6}--\ref{ch1e7}) the \emph{the covering chart}. A coordinate system
defined only in the domain $\{X:X^{d-1}+X^d>0\}$ (called \textsc{Poincar\'e} \emph{(fundamental) 
domain}), but far more convenient for the analysis of several geometrical aspects
of $AdS_d$, is given by $(x^\mu,z)\in\mathbb{R}^{1,d-2}\times\mathbb{R}_+$, if we write
\begin{equation}\label{ch1e8}
\left\{ \begin{array}{l@{\,=\,}l}
X^\mu & \frac{R}{z}x^\mu\;(\mu\,=\,0,\ldots,d-2) \\
X^{d-1} & R\left(\frac{1-z^2}{2z}+\frac{1}{2z}x_\mu x^\mu\right) \\
X^d & R\left(\frac{1+z^2}{2z}-\frac{1}{2z}x_\mu x^\mu\right)
\end{array} \right.
\end{equation}

In this case, (\ref{ch1e7}) acquires the following form:

\begin{equation}\label{ch1e9}
ds^2=\frac{R^2}{z^2}\left(\eta_{\mu\nu}dx^{\mu}dx^{\nu}+dz^2\right).
\end{equation}

This chart is denominated \emph{horocyclic} or a \textsc{Poincar\'e} \emph{chart}. One can see 
by formula (\ref{ch1e9}) that each timelike hypersurface in the \textsc{Poincar\'e} domain given
by $z=const.$ is conformal to $\mathbb{R}^{1,d-2}$ by a factor 

\begin{equation}\label{ch1e10}
(X^{d-1}+X^d)^2=\frac{R^2}{z^2}. 
\end{equation}

Hence, the \textsc{Poincar\'e} domain corresponds to the half $z>0$ of $\mathbb{R}^{1,d-1}\ni
(x^\mu,x^d-1\doteq z)$, up to the conformal factor (\ref{ch1e10}). We can obtain the action 
of several subgroups of $SO_e(2,d-1)$ on $AdS_d$ from the subgroup of (the component connected 
to identity of) the conformal group $SO_e(2,d)$ of $\mathbb{R}^{1,d-1}$ that individually 
preserves each half $\{(x^\mu,x^{d-1}):x^{d-1}\gtrless 0\}$ by using this parametrization, 
as follows:

\begin{itemize}
\item\textsc{Poincar\'e} \emph{subgroup}\emph{:}
\begin{equation}\label{ch1e11}(z,x^\mu)\,\mapsto\,(z,\Lambda^\mu_\nu x^\nu+a^\mu),\,\Lambda
\in SO_e(1,d-2),\,a\in\mathbb{R}_d.
\end{equation}
This subgroup preserves the $z=$constant hypersurfaces, and acts on each one as the
\textsc{Poincar\'e} group acts in $\mathbb{R}^{1,d-2}$.
\item\emph{Dilation subgroup:}
\begin{equation}\label{ch1e12}(z,x^\mu)\,\mapsto\,(\lambda z,\lambda x^\mu),\,\lambda\in
\mathbb{R}_+.
\end{equation} 
\end{itemize}

The remaining isometries, which do not preserve the fundamental \textsc{Poincar\'e} domain, will 
be presented in Subsection \ref{ch1-ads-infty}.

\subsection{\label{ch1-ads-infty}Structure of conformal infinity}

Let us consider the covering parametrization (\ref{ch1e6}). We'll show now that $AdS_d$ 
possesses a conformal infinity in the sense of Definition \ref{ap1d2}. Applying the change
of variables 
\begin{eqnarray}\label{ch1e13}
& \tau=t,\,u=2\left(\sqrt{1+\frac{r^2}{R^2}}\right)-\frac{r}{R},\,dr=-R\left(\frac{1}{u^2}+\frac{1}{4}
\right)du & \\ & r\in[0,+\infty)\,\leftrightarrow\,u\in (0,2], & \nonumber
\end{eqnarray} 
the metric (\ref{ch1e7}) becomes
\begin{equation}\label{ch1e14}
ds^2=\frac{R^2}{u^2}\left[-\left(1+\frac{u^2}{4}\right)^2dt^2+du^2+\left(1-\frac{u^2}{4}
\right)^2d\Omega^2_{d-2}\right].
\end{equation}

The spatial infinity of $AdS_d$ is obtained by taking $r\rightarrow\infty$, which corresponds to 
the limit $u\searrow 0$. The metric of the conformal completion $d\bar{s}^2=u^2ds^2$, in this 
limit, becomes 
\begin{equation}\label{ch1e15}
d\bar{s}^2=R^2(-dt^2+d\Omega^2_{d-2}+du^2)=R^2(du^2+ds^2_0),
\end{equation}
where $ds^2_0$ is the metric of \textsc{Einstein}\emph{'s static universe} in $(d-1)$ dimensions 
($ESU_{d-1}$), which is isometric to $\mathbb{R}\times S^{d-2}$. In order to see the conformal
boundary of the fundamental domain, let us return to coordinates $\tau$, $r$; one notices that 
the boundary is then conformal to $\mathbb{R}^{1,d-2}$ by a factor $(X^{d-1}+X^d)^{-2}$. 
Taking projective (\textsc{Dirac-Weyl}) coordinates, we have:

\begin{equation}\label{ch1e16}
x^\mu\,=\,\frac{X^\mu}{X^{d-1}+X^d}\;\stackrel{r\rightarrow\infty}{\longrightarrow}\;\left\{ 
\begin{array}{l@{\,=\,}l}x^0 & \frac{\sin\tau}{\cos\tau+\mathrm{e}^d} \\ \mathbf{x} & 
\frac{\mathbf{e}}{\cos\tau+\mathrm{e}^d} \end{array} \right..
\end{equation}

More precisely, the conformal boundary of the fundamental domain of $AdS_d$ corresponds to 
the conformal compactification of $\mathbb{R}^{1,d-2}$ (notation: $c\mathbb{R}^{1,d-2}$).

\begin{figure}[ht!]
\begin{center}
\input{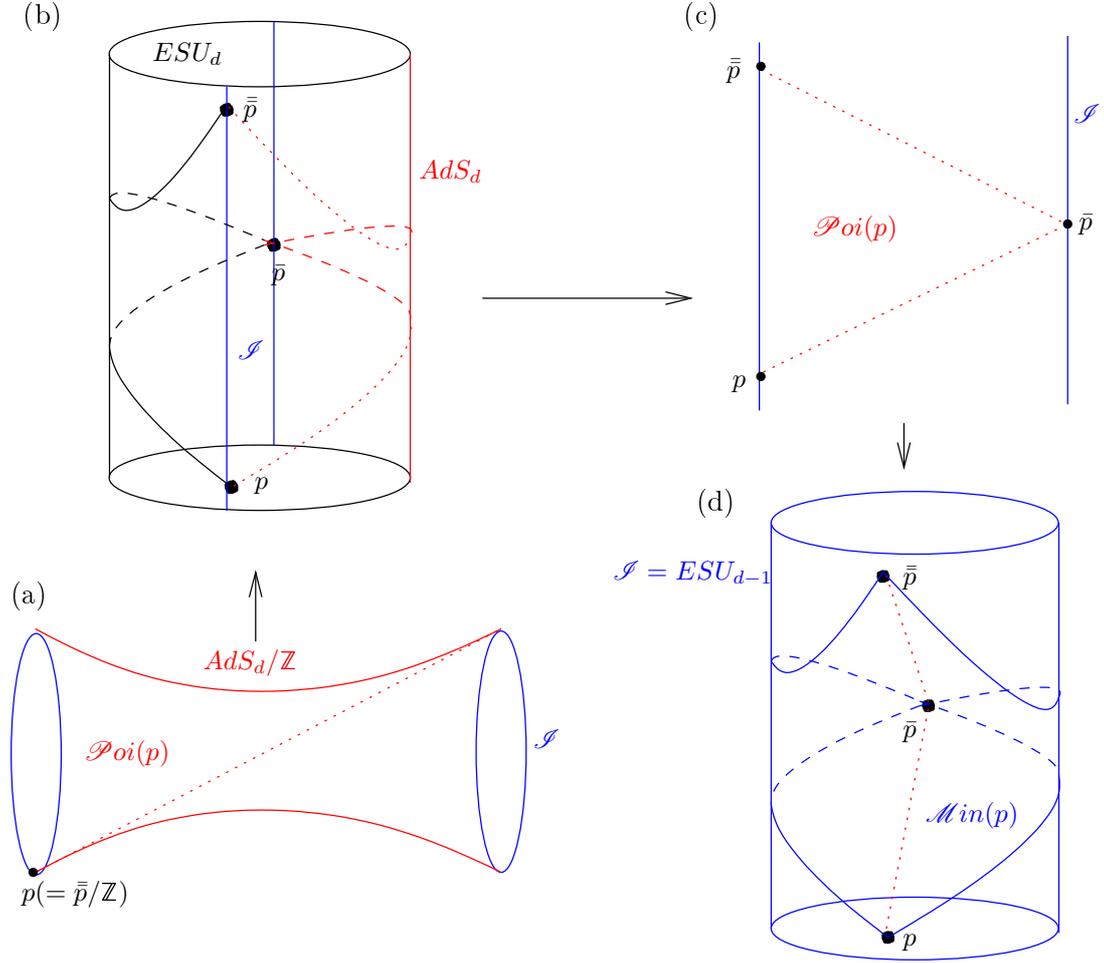} 
\end{center}
\caption{\label{ch1f1}\small \textbf{(a)} Fundamental domain of $AdS_d$ (red), given
by the quotient modulo $\mathbb{Z}$, where the spatial infinity is represented in blue and the 
(fundamental) \textsc{Poincar\'e} domain is the upper part of the dotted cut; 
\textbf{(b)} Image of the conformal embedding of $AdS_d$ into $ESU_d$ (red). The infinity 
$\mathscr{I}$ of $AdS_d$ is represented in blue and the \textsc{Poincar\'e} domain
$\mathscr{P}oi(p)$ is delimited by the dotted curves; \textbf{(c)} $\mathscr{P}oi(p)$ 
(side view) delimited by the dotted lines in red, with $\mathscr{I}$ 
represented in blue; \textbf{(d)} Visualization of $\mathscr{I}$ as $ESU_{d-1}$ 
(red) with the \textsc{Minkowski} domain $\mathscr{M}in(p)$, corresponding to the
image of the conformal embedding of $\mathbb{R}^{1,d-2}$ into $ESU_{d-1}$, in blue.}
\end{figure}

$ESU_{d-1}$, contrary to $c\mathbb{R}^{1,d-2}$, doesn't suffer from causality conflicts 
due to the action of the conformal group of $\mathbb{R}^{1,d-2}$, which extends to the
universal covering \footnote{\emph{Modulo} the group $\mathbb{Z}_2$ of reflections across 
the spatial infinity of $c\mathbb{R}^{1,d-2}$.} $\widetilde{SO}_e(2,d-1)$. Its global causal 
ordering is given by \cite{lm}
\begin{equation}\label{ch1e17}
(\tau,\mathrm{e})\ll_\mathscr{I}(\tau',\mathrm{e}')\;\mbox{if and only if}\;\tau-\tau'>2
\mbox{Arccos}(\mathrm{e}.\mathrm{e}'),
\end{equation} 
where Arccos$(a)$ corresponds to the principal branch of $\cos^{-1}(a)$ ($-\pi<a\leq\pi$). 
In terms of the \textsc{Poincar\'e} parametrization (\ref{ch1e8}), each $z=$const. hypersurface  
corresponds, in projective coordinates, exactly to $\mathbb{R}^{1,d-2}$, a result which extends
to $z=0$. Thus, conversely to (\ref{ch1e16}), we can obtain once more the global chart of 
$ESU_{d-1}$ by explicitly writing the conformal embedding of $\mathbb{R}^{1,d-2}$ into the latter. 
Namely, rewriting $\eta$ in ``spatially'' spherical coordinates
\begin{equation}
\eta=-(dx^0)^2+d\mathbf{x}.d\mathbf{x}=-(dx^0)^2+dr^2+r^2d\Omega^2_{d-3}\nonumber
\end{equation}
and moving to advanced ($v$) and retarded ($u$) radial coordinates in the light cone
\begin{equation}
\left.\begin{array}{c} v\doteq t+r \\ u\doteq t-r\end{array}\right\}\Rightarrow
\eta=-du\,dv+\frac{1}{4}(v-u)^2d\Omega^2_{d-3},\nonumber
\end{equation}
we see that, under the choice of conformal factor $\Omega^2=\frac{4}{(1+u^2)(1+v^2)}$
and the coordinate change $(u,v)\mapsto(U,V)$ (given by $U=2\mbox{arctan}u$ e
$V=2\mbox{arctan}v$) and $(U,V)\mapsto(T\doteq\frac{V+U}{2},R\doteq\frac{V-U}{2})$,
it follows that
\begin{equation}
ds_0^2=\Omega^2\eta=-dT^2+dR^2+\sin^2R d\Omega^2_{d-3}=-dT^2+d\Omega^2_{d-2},\nonumber
\end{equation}
according to (\ref{ch1e15}). That is, the coordinate change $(x^\mu)\mapsto(T,\theta^1
\doteq R,\theta^2,\ldots,\theta^{d-2})$ corresponds from the passive viewpoint to what, 
from the active viewpoint, is the conformal embedding $\mathbb{R}^{1,d-2}$ into $ESU_{d-1}$
($U$ and $V$ are the light-cone radial coordinates in $ESU_{d-1}$). One should also notice,
in the light of these considerations, that $AdS_d$, expressed in terms of the projective 
coordinates (\ref{ch1e16}) properly extended to the covering, corresponds to half of $ESU_d$. 
The \textsc{Poincar\'e} domain, on its turn, corresponds to the image of the conformal embedding 
of the half $x^{d-1}>0$ of $\mathbb{R}^{1,d-1}$ into $ESU_d$, and its infinity ($x^{d-1}=0$), 
to the image of the conformal embedding of $\mathbb{R}^{1,d-2}$ into $ESU_{d-1}$. The above 
analysis is pictorially synthesized in Figure \ref{ch1f1}.\\

Starting from what has been said above, it's immediate to identify the remaining subgroup of 
$\widetilde{SO}_e(2,d-1)$. The conjugation of the subgroup of translations along $x^\mu$ 
(see (\ref{ch1e11}), page \pageref{ch1e11}) by the \emph{relativistic ray inversion} map 
in $\mathbb{R}^{1,d-1}$ \[I:(x^\mu,x^{d-1})=(x^\mu,z)\mapsto-\frac{1}{z^2+x_\nu x^\nu}(x^\mu,z),
\,x_\nu x^\nu=\eta_{\rho\sigma}x^\rho x^\sigma\] results in the \emph{special conformal 
transformations} (also called \emph{conformal translations})
\begin{eqnarray}\label{ch1e18}(x^\mu,z)\,\mapsto\,I\circ(.+b^\mu)\circ I(x^\mu,z) =
\frac{1}{1-2b_\mu x^\mu+b^2(x^2+z^2)}.\\.(x^\mu-b^\mu(x^2+z^2),z),\,b\in\mathbb{R}^{1,
d-2}\ni a^\mu,\,a^2\,:=\,a_\mu a^\mu.\nonumber
\end{eqnarray} 

Such transformations, if conjugated by adequate elements of the subgroups 
(\ref{ch1e11}--\ref{ch1e12}) and extended to the parameter space of $\widetilde{SO}_e
(2,d-1)$, no longer preserve the \textsc{Poincar\'e} domain, but preserve the infinity
of $AdS_d$ as a whole. For $z=0$, (\ref{ch1e18}) reduces to the special conformal 
transformations in $\mathbb{R}^{1,d-2}$. The subgroups given by (\ref{ch1e11}) and (\ref{ch1e12}) 
in the projective coordinates $x^\mu$ correspond, respectively, to the actions of the 
\textsc{Poincar\'e} and dilation subgroups on $\mathbb{R}^{1,d-2}$. Summing up, the isometry
group of the fundamental domain of $AdS_d$ (resp. $AdS_d$) corresponds precisely to the 
conformal group of $(c)\mathbb{R}^{1,d-2}$ (resp. $ESU_{d-1}$).\\

\subsection{\label{ch1-ads-wedge}Wedges in AdS and diamonds in the boundary}

Let us consider, in $AdS_d$, the following causally complete region, called \emph{(standard)
wedge} (notation: $\mathscr{W}_0$):

\begin{eqnarray}\label{ch1e19}
\mathscr{W}_0 & = & \{X\in\mathbb{R}^{2,d-1}:\eta(X,X)=-R^2\mbox{ and }X^{d-1}>|X^0|\},
\end{eqnarray}
or, in \textsc{Poincar\'e} coordinates $(z,x^\mu)$:

\begin{equation}\label{ch1e20}
\mathscr{W}_0=\{(z,x^\mu):\sqrt{z^2+\mathbf{x}\cdot\mathbf{x}}<1-|x^0|,z<1\}.
\end{equation}

Taking the limit $z\rightarrow 0$, one notices that the intersection of $\mathscr{W}_0$ with 
the conformal boundary (notation: $\mathscr{K}_0$) is given by
\begin{equation}\label{ch1e21}
\mathscr{K}_0\,=\,\{x^\mu:|\mathbf{x}|<1-|x^0|\},
\end{equation}
that is, $\mathscr{K}_0$ corresponds to a \emph{diamond} (with spatial radius equal to 1) in 
$\mathbb{R}^{1,d-2}$.\\

A wedge in $AdS_d$ is the causal completion of the orbit of a uniformly accelerating observer. 
By formula \ref{ch1e19}, we see that such orbits are given by the restriction to the fundamental
domain of $AdS_d$ of the boosts in the $X^0-X^{d-1}$ plane:

\begin{eqnarray}
X^0 & \mapsto & X^0\cosh\lambda+X^{d-1}\sinh\lambda;\nonumber\\
\label{ch1e22} X^{d-1} & \mapsto & X^0\sinh\lambda+X^{d-1}\cosh\lambda;\\
\mathbf{X} & \mapsto & \mathbf{X};\;X^d\,\mapsto\,X^d,\,\lambda\in\mathbb{R}.\nonumber
\end{eqnarray}

In \textsc{Poincar\'e} coordinates, it's easier to visualize the transformation in the
conformal embedding of $AdS_d$ into $ESU_d$. $\mathscr{W}_0$ is the $x^{d-1}>0$ half of a
diamond contained in the conformal embedding of $\mathbb{R}^{1,d-1}$ into $ESU_d$, with 
vertices in $\mathscr{I}$. (\ref{ch1e22}) corresponds to the unique one-parameter subgroup of 
$\widetilde{SO}_e(2,d-1)$ which preserves this diamond. The former's action is symmetric under 
rotations around the axis determined by the vertices, and, restricted to the $x^0-z$ plane, becomes
\begin{equation}\label{ch1e23}
z_{\pm}\mapsto z_{\pm}(\lambda)\doteq\frac{(1+z_{\pm})-e^{-\lambda}(1-z_{\pm})}{(1+z_{\pm})
+e^{-\lambda}(1-z_{\pm})},\,\lambda\in\mathbb{R},
\end{equation}
where we've employed the light-cone coordinates $z_{\pm}\doteq x^0\pm z$. The corresponding isotropy 
subgroup for $\mathscr{D}_0$ is obtained, say, in the $x_0-x_1$ plane by substituting $x^1$ for $z$ and
the light-cone coordinates $x_\pm\doteq x^0\pm x^1$ for $z_\pm$ in (\ref{ch1e23}). The one-parameter 
subgroup of $SO_e(2,d-1)$ given by (\ref{ch1e23}) corresponds to the conjugation $(x,z)\mapsto (K^{-1}
\circ (e^\lambda.)\circ K)(x,z)$ of the dilation $(x,z)\mapsto (e^\lambda x,e^\lambda z)$ by the conformal
transformation $K$ in $\mathbb{R}^{1,d-1}$ given by the composition \[\mathscr{W}_0\ni(x,z)\doteq (x^0,
\mathbf{x},z)\mapsto K(x,z)\doteq I(x^0-1,\mathbf{x},z)-\left(\frac{1}{2},\mathbf{0},0\right),\] where 
$I$ is the relativistic ray inversion map. In this form, the rotational symmetry of the formula (\ref{ch1e23}) 
around the $x^0$ axis, invoked above, becomes evident.\\

Finally, as a prelude to Section \ref{ch1-aads}, we shall pass to proceed in a coordinate-free
manner. Denoting the vertices of $\mathscr{D}_0$ by $p_0=(-1,\mathbf{0})$ and $q_0=(1,\mathbf{0})$, 
we see that, if $\overline{AdS_d}$ is the conformal closure of $AdS_d$, then $(I^-(p_0,\overline{AdS_d})
\cap I^+(q_0,\overline{AdS_d}))\cap AdS_d=\mathscr{W}_0$ and $(I^-(p_0,\overline{AdS_d})\cap I^+(q_0,
\overline{AdS_d}))\cap\mathscr{I}=(I^-(p_0,\mathscr{I})\cap I^+(q_0,\mathscr{I}))=\mathscr{D}_0$. More
in general, let us define for $p,q\in\mathbb{R}^{1,d-2},p\ll_\mathscr{I} q$ the \emph{wedge}
\begin{equation}\label{ch1e24}
\mathscr{W}_{p,q}\doteq(I^-(p,\overline{AdS_d})\cap I^+(q,\overline{AdS_d}))\cap AdS_d
\end{equation}
and the \emph{diamond}
\begin{equation}\label{ch1e25}
\mathscr{D}_{p,q}\doteq(I^-(p,\overline{AdS_d})\cap I^+(q,\overline{AdS_d}))\cap{}\mathscr{I}=
I^-(p,\mathscr{I})\cap I^+(q,\mathscr{I})
\end{equation}
associated to $p,q$. In particular, $\mathscr{W}_0=\mathscr{W}_{p_0,q_0}$ and $\mathscr{D}_0=
\mathscr{D}_{p_0,q_0}$. Let us denote by $u^\lambda_{p,q}$ the one-parameter subgroup of isometries
of $AdS_d$ which preserve $\mathscr{W}_{p,q}$ and $\mathscr{D}_{p,q}$, given by (\ref{ch1e23}) 
in the case $p=p_0$, $q=q_0$. This subgroup is given by
\begin{eqnarray}\label{ch1e26}
u^\lambda_{p,q}(x,z) & = & K^{-1}_{p,q}(e^\lambda K_{p,q}(x,z)),\mbox{ where}\\
K_{p,q}(x,z)& \doteq & K\left(\Lambda_{p,q}\left(x-\frac{x(p)+x(q)}{2}\right),z\right),\nonumber
\end{eqnarray}
where $\Lambda_{p,q}$ is the \textsc{Lorentz} boost around the origin which makes the
direction $\overrightarrow{x(p)x(q)}$ parallel to the $x^0$ axis, and $K$ is the map used to
define the isotropy subgroup (\ref{ch1e23}) of $\mathscr{W}_0$ above. These one-parameter subgroups 
will play a key role in Chapters \ref{ch2} and \ref{ch4}.\\

Let us make now the description of the image of the conformal embedding of $\mathbb{R}^{1,d-2}$ 
into $ESU_{d-1}$ independent from the point chosen to represent the spatial infinity of the
former. Let $p\in\mathscr{M}in(r)$. All null geodesics emanating from $p$ will focus at a single
point of $\mathscr{I}$, which  constitutes the future endpoint of all achronal null generators of 
$\partial I^+(p,\mathscr{I})$. This point is denominated \emph{antipodal} of $p$, denoted by $\bar{p}$. 
The antipodal of $p$ has the following properties:

\begin{eqnarray}
\partial I^+(p,\mathscr{I})\smallsetminus\{p\} & = & \partial I^-(\bar{p},\mathscr{I})
\smallsetminus\{\bar{p}\};\label{ch1e27}\\
\partial I^+(p,\overline{AdS_d})\smallsetminus\{p\} & = & \partial I^-(\bar{p},\overline{AdS_d})
\smallsetminus\{\bar{p}\}.\label{ch1e28}
\end{eqnarray}

\begin{figure}[ht!]
\begin{center}
\input{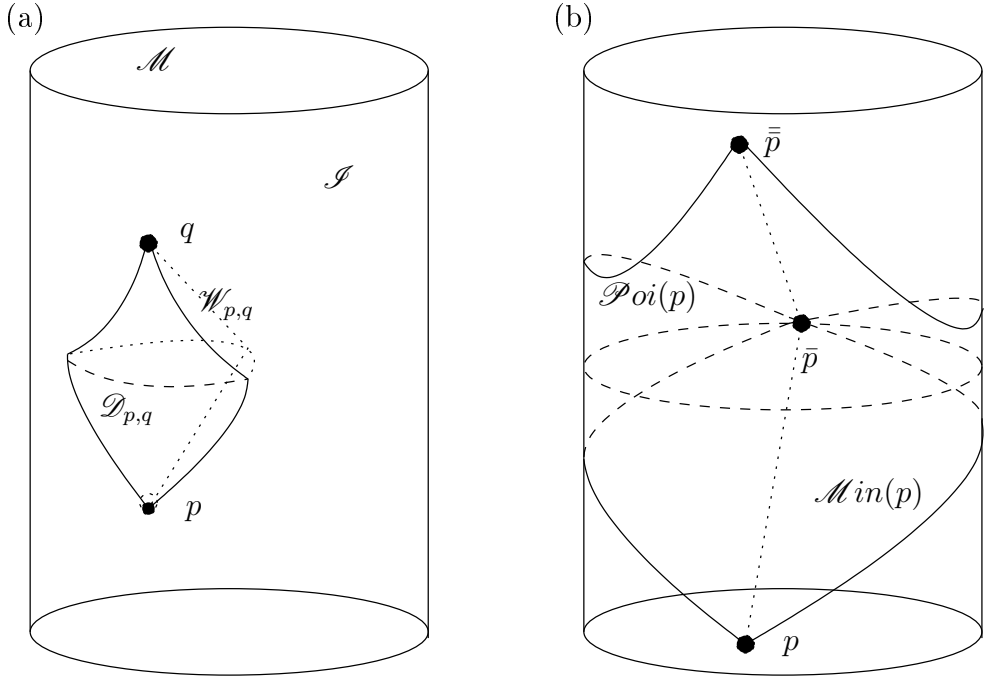} 
\end{center}
\caption{\label{ch1f2}\small \textbf{(a)} Correspondence between bulk wedges and boundary
diamonds. \textbf{(b)} A \textsc{Min\-kowski} domain and the corresponding \textsc{Poincar\'e} domain.}
\end{figure}

Let us define $\mathscr{M}in(p)\doteq\mathscr{D}_{p,\bar{\bar{p}}}$, the \textsc{Minkowski} 
\emph{domain} to the future of $p\in\mathscr{I}$. This region corresponds to the conformal
embedding of $\mathbb{R}^{1,d-2}$ into $\mathscr{I}$ such that $p$ is the future timelike
infinity of $\mathbb{R}^{1,d-2}$. $\mathscr{P}oi(p)\doteq\mathscr{W}_{p,\bar{\bar{p}}}$ corresponds 
to the domain of a \textsc{Poincar\'e} chart in $AdS_d$, hence called the \textsc{Poincar\'e} \emph{domain}
to the future of $p$. We shall extend the definitions (\ref{ch1e24}) of a wedge and (\ref{ch1e25}) of 
a diamond to all pairs of points $p\ll_\mathscr{I} q$ such that $p,q\in\overline{\mathscr{M}in(r)}$ for 
some $r\in\mathscr{I}$, with the same notation. 

\begin{remark}\label{ch1r1}
An equivalent way of characterizing wedges $\mathscr{W}_{p,q}$ and diamonds $\mathscr{D}_{p,q}$
such that $p,q\in\mathscr{M}in(r)$ for some $r\in\mathscr{I}$ is the following: as $ESU_{d-1}$ is 
globally hyperbolic, it follows that $\mathscr{D}_{p,q}$ is relatively compact and, thus,
globally hyperbolic as well. It's possible, then, to distinguish three possible situations:

\begin{enumerate}
\item There is no $r\in\mathscr{I}$ such that $p,q\in\overline{\mathscr{M}in(r)}$: in this case,
let us take $r=p$; it follows that $\bar{r}=\bar{p}\ll_{\mathscr{I}} q$ and, hence, $\mathscr{D}_{p,q}$
contains a \textsc{Cauchy} surface for $\mathscr{I}$. In particular, the \textsc{Cauchy} surfaces of
$\mathscr{D}_{p,q}$ are \emph{compact} and, more important, \emph{non contractible} (for the definition
of contractibility, see Section \ref{ap4-basics}, page \pageref{ap4-basics}). 
\item $p,q\in\mathscr{M}in(r)$ for some $r\in\mathscr{I}$: as here $\mathscr{D}_{p,q}$ can be understood
as a diamond in $\mathscr{M}in(r)$, the \textsc{Cauchy} surfaces of $\mathscr{D}_{p,q}$ are \emph{non 
compact} and \emph{contractible}.
\item $p,q\in\partial\mathscr{M}in(r)$ for some $r\in\mathscr{I}$: this borderline case
includes the \textsc{Minkowski} domains themselves, and, by definition, $\mathscr{D}_{p,q}$ can
then be obtained as the limit $\mathscr{D}_{p,q}=\lim_{n\rightarrow\infty}\mathscr{D}_{p,q_n}=
\bigcup^\infty_{n=1}\mathscr{D}_{p,q_n}$, where once more we choose $r=p$ and $\{q_n\}_{n\in\mathbb{Z}_+}
\subset\mathscr{M}in(r)$ is a sequence of points such that $q_n\stackrel{n\rightarrow
\infty}{\longrightarrow} q$ and $q_n\ll_{\mathscr{I}}q_{n+1}\ll_{\mathscr{I}}q$ for all $n$. 
In particular, $\mathscr{D}_{p,q_n}\in\mathscr{M}in(r_n)$, where $r_n\ll_{\mathscr{I}} r$ is 
sufficiently close to $r=p$. The \textsc{Cauchy} surfaces of $\mathscr{D}_{p,q}$ are also noncompact
and contractible.
\end{enumerate}

We emphasize that by no means can case 1 be incorporated by any limit of an increasing sequence 
of diamonds lying in case 2, as in case 3. Summing up, the collection of diamonds that interest us 
is the one composed of \emph{diamonds with contractible} \textsc{Cauchy} \emph{surface}. The 
diamonds within case 2 are, in a certain sense, ``dense'' in this collection -- given any 
$\mathscr{D}_{p,q}$ with contractible \textsc{Cauchy} surface and any $q'\ll_{\mathscr{I}}q$, 
no matter how close to $q$, we have that $\mathscr{D}_{p,q'}$ is a diamond within case 2.
\end{remark}
The structure of $ESU_{d-1}$ yet implies the following facts of great importance to us:

\begin{enumerate}
\item From (\ref{ch1e27}) and (\ref{ch1e28}), it follows that 
\begin{equation}\label{ch1e29}
\mathscr{W}_{q,\bar{p}}=(\mathscr{W}_{p,\bar{q}})'_{AdS_d}\mbox{ and }
\mathscr{D}_{q,\bar{p}}=(\mathscr{D}_{p,\bar{q}})'_\mathscr{I},\,\forall p,q\in
\mathscr{Min}(r)\mbox{ for some }r\mbox{ and }p\ll_\mathscr{I}\bar{q},
\end{equation} and, as a consequence,
\item\begin{equation}\label{ch1e30}
\mathscr{M}in(r)=(\{\bar{r}\})'_\mathscr{I}=(\{\bar{r}\})'_{\overline{AdS_d}}\cap{}
\mathscr{I},\,\mathscr{P}oi(r)=(\{\bar{r}\})'_{\overline{AdS_d}}\cap AdS_d.
\end{equation}
\item The bijection
\begin{equation}\label{ch1e31} \rho_{AdS_d}:\mathscr{W}_{p,q}\mapsto\rho_{AdS_d}
(\mathscr{W}_{p,q})\doteq\mathscr{D}_{p,q},\,p,q\in\mathscr{M}in(r)\mbox{ for some }r
\mbox{ and }p\ll_\mathscr{I}q,\end{equation} 
denominated \textsc{Rehren} \emph{bijection}, preserves causal complements.
\item The action of $\widetilde{SO}_e(2,d-1)$ on $\mathscr{I}$ is transitive and preserves the 
causal structure. Hence, it acts transitively on the set of diamonds. As 
$SO_e(2,d-1)$ is the conformal group of \textsc{Minkowski} spacetime, it follows that the 
former acts transitively on the collection of the diamonds contained in some \textsc{Minkowski}
domain. 
\item As $\widetilde{SO}_e(2,d-1)$ also preserves the causal structure of $AdS_d$, 
it also follows that this group acts transitively on the set of wedges.
\item Every element of $\widetilde{SO}_e(2,d-1)$ can be obtained from a finite sequence of
isometries of the form (\ref{ch1e26}) for a suitable family of diamonds / wedges within cases 
2 and 3 of Remark \ref{ch1r1}, page \pageref{ch1r1}. 
\end{enumerate}

The above construction, which makes use only of the causal relations in the conformal 
closure, constitutes the geometrical framework of \textsc{Rehren} duality \cite{rehren1}, and
was employed in this coordinate-free form by \textsc{Bousso} and \textsc{Randall} \cite{boura} 
for studying qualitative aspects of the AdS/CFT correspondence.

\section{\label{ch1-aads}AAdS spacetimes}

Having scrutinized with sufficient detail the structure of $AdS_d$, we shall now proceed with 
the general case of our interest. Following the nomenclature of Appendix \ref{ap1}, consider 
stably causal spacetimes $(\mathscr{M},g)$ endowed with conformal completion $(\overline{\mathscr{M}},
\bar{g})$, conformal infinity $(\mathscr{I},\bar{g}^{(0)})$ and conformal factor $z$ satisfying the 
conditions of Definition \ref{ap1d2} (page \pageref{ap1d2}).\\

\begin{definition}\label{ch1d2}
We say that $(\mathscr{M},g)$ is an \emph{AdS-type} spacetime id $(\mathscr{I},\bar{g}^{(0)})$ 
is a $(d-1)$-dimensional Lorentzian manifold (or, equivalently, the covector normal to $\mathscr{I}$, 
$dz\restr{z=0}$, is spacelike with respect to $\bar{g}$) and diffeomorphic to $\mathbb{R}\times S^{d-2}$. If, 
beyond that, $\bar{g}^{(0)}$ belongs to the conformal class of the metric of $ESU_{d-1}$, $(\mathscr{M},
g)$ is said to be \emph{asymptotically AdS} (AAdS).\\

Finally, a spacetime $(\mathscr{M},g)$ endowed with conformal infinity $(\mathscr{I},
\bar{g}^{(0)})$ is said to be \emph{locally AdS-type} (resp. \emph{locally AAdS}) if any
$p\in\mathscr{I}$ has an open neighbourhood $\overline{\mathscr{U}}$ in the conformal completion 
$(\overline{\mathscr{M}},\bar{g})$ such that $(\overline{\mathscr{U}},\bar{g}\restr{\overline{\mathscr{U}}})$ 
is isometric to a neighbourhood $\overline{\mathscr{V}}$ of $q\in\mathscr{J}$, where $\mathscr{J}$ is the 
conformal infinity of an AdS-type (resp. AAdS) spacetime.
\end{definition}

The definition of locally AdS-type ad locally AAdS spacetimes comprises the examples
of black holes negative cosmological constant (\textsc{Schwarzschild}-AdS, \textsc{Reissner-Nordstr\"om}-AdS, 
\textsc{Kerr-Newman}-AdS, etc.), not contemplated by the definition of AdS-type and AAdS spacetimes.\\

With Definition \ref{ch1d2} at hand, we define in AAdS spacetimes the antipodal $\bar{p}$ and the
\textsc{Minkowski} domain $\mathscr{M}in(p)$ to the future of $p\in\mathscr{I}$ exactly as in AdS, for
these definitions don't refer to the bulk geometry. More in general, we have the

\begin{definition}\label{ch1d3}
Let $(\mathscr{M},g)$ be a locally AdS-type spacetime with conformal infinity $(\mathscr{I},
\bar{g}^{(0)})$, and $p\ll_{\mathscr{I}}q\in\mathscr{I}$ such that $I^+(p,\overline{\mathscr{M}})\cap{}
I^-(q,\overline{\mathscr{M}})$ is relatively compact (hence, globally hyperbolic) and 
$I^+(p,\mathscr{I})\cap{}I^-(q,\mathscr{I})$ possesses contractible \textsc{Cauchy} surfaces. 
Let us denote the collection of pairs $(p,q)\in\mathscr{I}^2$ satisfying these conditions by $\mathscr{D}
(\mathscr{I})$.\\

The (bulk) \emph{wedge} associated to $p,q$ is the region \[\mathscr{W}_{p,q}=I^+(p,
\overline{\mathscr{M}})\cap{}I^-(q,\overline{\mathscr{M}})\cap\mathscr{M},\] and the 
(boundary) \emph{diamond} associated to $p,q$, the region \[\mathscr{D}_{p,q}=I^+(p,
\mathscr{I})\cap{}I^-(q,\mathscr{I}).\] The \emph{past horizon} $\partial_-\mathscr{W}_{p,q}$ 
(resp. \emph{future horizon} $\partial_+\mathscr{W}_{p,q}$) of $\mathscr{W}_{p,q}$ is given by 
\[\partial_-\mathscr{W}_{p,q}\doteq(\partial I^+(p,\overline{\mathscr{M}})\smallsetminus\{p\})\cap{}
I^-(q,\overline{\mathscr{M}})\cap{}\mathscr{M},\]\[\partial_+\mathscr{W}_{p,q}\doteq(\partial I^-(q,
\overline{\mathscr{M}})\smallsetminus\{q\})\cap{}I^+(p,\overline{\mathscr{M}})\cap{}\mathscr{M},\] and
the \emph{past horizon} $\partial_-\mathscr{D}_{p,q}$ (resp. \emph{future horizon} $\partial_+
\mathscr{D}_{p,q}$) of $\mathscr{D}_{p,q}$, by \[\partial_-\mathscr{D}_{p,q}\doteq(\partial I^+(p,
\mathscr{I})\smallsetminus\{p\})\cap{}I^-(q,\mathscr{I}),\,\partial_+\mathscr{D}_{p,q}\doteq(\partial 
I^-(q,\mathscr{I})\smallsetminus\{q\})\cap{}I^+(p,\mathscr{I}).\] We denote by $\mathscr{W}
(\mathscr{M},g) \doteq\{\mathscr{W}_{p,q}:(p,q)\in\mathscr{D}(\mathscr{I})\}$ the collection of wedges 
in $(\mathscr{M},g)$, and $\mathscr{D}(\mathscr{M},g)\doteq\{\mathscr{D}_{p,q}:(p,q)\in\mathscr{D}
(\mathscr{I})\}$, the collection of diamonds in $(\mathscr{I},\bar{g}^{(0)})$. The \textsc{Rehren} 
\emph{bijection} associated to $(\mathscr{M},g)$ is given by
 \begin{eqnarray}
\rho_{(\mathscr{M},g)}:\mathscr{W}(\mathscr{M},g) & \longrightarrow & \mathscr{D}(\mathscr{M},g)
\label{ch1e32}\\\mathscr{W}_{p,q} & \mapsto & \rho_{(\mathscr{M},g)}(\mathscr{W}_{p,q})\doteq
\mathscr{D}_{p,q}.\nonumber
\end{eqnarray}
\end{definition}

Notice that it's Remark \ref{ch1r1} which allowed us to naturally extend the definition of wedges
and diamonds to locally AdS-type spacetimes. If $(\mathscr{M},g)$ is an AAdS spacetime, the 
\textsc{Poincar\'e} domain to the future of $r\in\mathscr{I}$ coincides, as in the case 
of \textsc{Minkowski} domains, with the definition in the AdS case: $\mathscr{P}oi(r)\doteq
\mathscr{W}_{r,\bar{\bar{r}}}$.\\

The wedges in a locally AdS-type spacetime cover precisely the region \[I^-(\mathscr{I},
\overline{\mathscr{M}})\cap I^+(\mathscr{I},\overline{\mathscr{M}})\cap{}\mathscr{M}\doteq\mathscr{W}
(\mathscr{I}),\] called \emph{domain of outer communications} of $(\mathscr{M},g)$, which is 
identified in several examples with the exterior of AAdS black holes.\\

In Subsection \ref{ch1-aads-causal} below, we shall study the causal, topological and localization
properties of $\rho_{(\mathscr{M},g)}$, in terms of the null geodesics of AAdS spacetimes.

\subsection{\label{ch1-aads-causal}Global geometry and null geodesics}

\subsubsection{\label{ch1-aads-causal-delay}Causality (bulk to boundary)}

\textsc{Einstein}'s static universe $\mathscr{I}$ is globally hyperbolic, and the
orbits generated by the subgroup of $\widetilde{SO}_e(2,d-1)$ corresponding to time translations 
in $AdS_d$ (rotation of the points of the hyperquadric (\ref{ch1e5}), page \pageref{ch1e5}, 
in the $X^0-X^d$ plane) correspond to the lines parallel to the cylinder \footnote{The generator 
of this subgroup corresponds to $1/2(P^0+K^0)$ in a \textsc{Minkowski} domain, where $P^0$ is the 
generator of translations along the direction $x^0$, and $K^0$ the generator of special conformal 
transformations along the direction $x^0$.} $\mathbb{R}\times S^{d-2}$, constituting timelike 
geodesics which maximize the proper time between its points. Moreover, any point in $\mathscr{I}$ 
belong to a unique geodesic of this family, and we can choose a common parametrization to all these 
geodesics (denominated \emph{time generators} of $\mathscr{I}$) such that the union of the 
points corresponding to the same value of the affine parameter constitute a \textsc{Cauchy} surface. \\

With this, we can define the \emph{(gravitational) time delay} of a complete null geodesic
$\gamma$ in $\mathscr{M}$, with endpoints in $\mathscr{I}$: \emph{grosso modo}, 
it represents how much the future endpoint of $\gamma$ moves away from the antipodal of the past
endpoint of $\gamma$, which, in the AdS case, \emph{is} the future endpoint.\\

First, we'll prove that, given certain extra conditions, the gravitational time delay 
of inextendible null geodesics is always \emph{positive} (see Figure \ref{ch1f3} for an 
illustration of the result):

\begin{figure}[ht!]
\begin{center}
\input{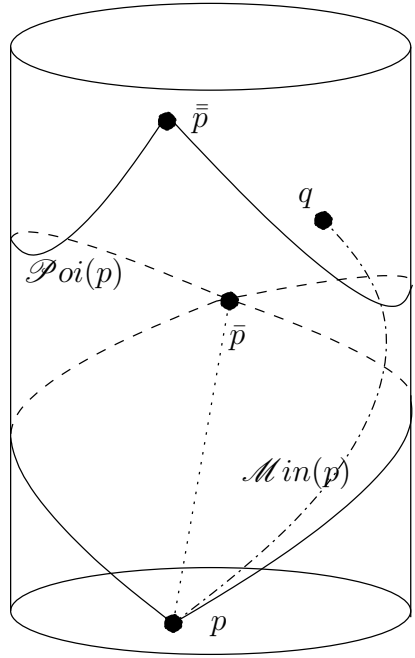} 
\end{center}
\caption{\label{ch1f3}\small Gravitational time delay. The null geodesics which emanate from 
$p\in\mathscr{I}$ and remain in $\mathscr{I}$ (full lines) focus at the antipodal $\bar{p}$.
Whereas in $AdS_d$ a null geodesic emanating from $p$ and traversing the bulk would also have 
$\bar{p}$ as its future endpoint (dotted line), in an AAdS spacetime satisfying the hypothesis
of global focusing of Theorem \ref{ch1t4} such a geodesic has future endpoint $q\in I^+(p,
\mathscr{I})$ (dashed/dotted line).}
\end{figure}

\begin{theorem}[positive gravitational time delay]\label{ch1t4}
Let $(\mathscr{M},g)$ be an AAdS \\ spacetime satisfying the following \emph{global focusing} 
hypothesis: any inextendible null geodesic possesses a pair of conjugate points (see in Appendix 
\ref{ap1} the discussion that follows equation (\ref{ap1e13}), page \pageref{ap1e13}). Then, given 
$p\in\mathscr{I}$, every inextendible null geodesic in $\mathscr{M}$ with past endpoint $p$ has, 
if any, future endpoint in $I^+(p,\mathscr{I})$. A similar result is valid if we exchange future
with past, and $I^+(p,\mathscr{I})$ with $I^-(p,\mathscr{I})$.
\begin{quotation}{\small\scshape Proof.\quad}
{\small\upshape Initially, we shall prove two Lemmata:

\begin{lemma}[Absence of causal shortcuts]\label{ch1l1}
Let $p,p'\in\mathscr{I}$. If $p\perp_\mathscr{I}p'$, then there cannot exist a causal curve 
in $(\overline{\mathscr{M}},\bar{g})$ linking $p$ to $p'$.
\begin{quote}{\small\scshape Proof.\quad}
{\small\upshape Suppose that $p'>_{\overline{\mathscr{M}}}p$ (the opposite case is treated in 
an analogous way). We'll prove that the gravitational time delay implied by the hypotheses 
of the Theorem contradicts the causal disjointness of $p$ and $p'$ with respect to $\mathscr{I}$. Denote 
by $T(p')$ the unique time generator of $(\mathscr{I},\bar{g}^{(0)})$ containing $p'$.

Notice that $\partial I^+(p,\mathscr{I})\doteq\Sigma$ is a closed, achronal hypersurface, 
which cuts $(\mathscr{I},\bar{g}^{(0)})$ in two disjoint open sets $I^+(p,\mathscr{I})
\doteq A$ and $\mathscr{I}\smallsetminus\overline{I^+(p,\mathscr{I})}$\\$\doteq B$, and intersects
each time generator of $\mathscr{I}$ precisely once, for any time generator has points in $I^+
(p,\mathscr{I})$ and $I^-(\bar{p},\mathscr{I})$. By hypothesis, $p'\in B$. Moreover, $T(p')$ must
cross $\Sigma$ at some instant. Hence, there exists a $p''\in T(p')$ such that $p''\gg_\mathscr{I} 
p'$ and $p''\in\Sigma$. Let $\eta$ be the null generator of $\Sigma$ which contains $p''$. As the 
segment of $\eta$ which links $p$ to $p''$ is null and achronal, $\eta$ is necessarily the fastest 
curve $(\mathscr{I},\bar{g}^{(0)})$ linking $p$ to $T(p')$.

Now, consider the achronal boundary $\partial I^+(p,\overline{\mathscr{M}})=\partial J^+(p,
\overline{\mathscr{M}})\doteq\overline{\Sigma}$. $\overline{\Sigma}\cap{}\mathscr{I}$ is 
closed, achronal and intersects each time generator of $\mathscr{I}$ at precisely one 
point, for every time generator possesses points in $I^+(p,\overline{\mathscr{M}})$ and $I^-
(p,\overline{\mathscr{M}})$, and $\overline{\Sigma}$ splits $\overline{\mathscr{M}}$ in two 
disjoint open sets (from the viewpoint of a manifold with boundary, of course) $I^+(p,\overline{
\mathscr{M}})\doteq\dot{A}$ and $\overline{\mathscr{M}}\smallsetminus\overline{I^+(p,
\overline{\mathscr{M}})}\doteq\dot{B}$. Thus, $T(p')$ needs to cross $\overline{\Sigma}$ 
at, say, $p'''$. Once that $p<p'$ from the viewpoint of $(\overline{\mathscr{M}},\bar{g})$, 
we must have $p'''\ll_\mathscr{I} p'$ or $p'''=p'$. In both cases, we have $p'''\ll_\mathscr{I}p''$, 
which implies that the null generator $\bar{\eta}$ of $\overline{\Sigma}$ containing $p'''$ is 
strictly faster than $\eta$. As $\eta$ was the fastest curve in $(\mathscr{I},\bar{g}^{(0)})$ 
linking $p$ to $T(p')$, $\bar{\eta}$ necessarily traverses $\mathscr{M}$, which contradicts the 
hypotheses of the Theorem, for $\bar{\eta}$ must be necessarily achronal and, hence, cannot have a 
pair of conjugate points \cite{beemee,hawkellis,wald2}.~\hfill~$\Box$}
\end{quote}
\end{lemma}
}\end{quotation}

\hfill\vspace*{-5.5ex}

\begin{quotation}{\small\upshape
\begin{lemma}\label{ch1l2}
Let $p,p'\in\mathscr{I}$. If $p'\in\partial I^+(p,\mathscr{I})$ and $p'\neq\bar{p}$, then there cannot
exist a null geodesic segment in $(\overline{\mathscr{M}},\bar{g})$ which doesn't belong to $\mathscr{I}$ 
linking $p$ to $p'$.
\begin{quote}{\small\scshape Proof.\quad}
{\small\upshape Let $\eta$ the (necessarily unique) null generator of $\partial I^+(p,
\mathscr{I})$ linking $p$ to $p'$. Suppose that there exists another null geodesic $\bar{\eta}$ 
traversing $mathscr{M}$ and liking $p$ to $p'$. As $(\mathscr{I},\bar{g}^{(0)})$ is totally 
geodesic by construction, it follows that $\bar{\eta}$ necessarily incides at $p'$ transversally 
to $\mathscr{I}$. Hence, if we choose any point $p''$ in $\eta$ after $p'$, then there exists a 
\emph{broken} future directed null geodesic segment linking $p$ to $p''$, which implies, on its turn, 
the existence of a timelike curve in $(\overline{\mathscr{M}},\bar{g})$ linking $p$ to $p''$
\cite{hawkellis}.

Let $T(p'')$ the time generator of $(\mathscr{I},\bar{g}^{(0)})$ containing $p''$. Now, 
consider $\overline{\Sigma}$ as in the previous Lemma. Once again, $T(p'')$ must cross
$\overline{\Sigma}$, say, at $p'''$. Thus, necessarily $p'''\ll_\mathscr{I} p''$. But 
this implies that, given that $(\overline{\mathscr{M}},\bar{g})$ is (strongly) causal, $p
\perp_\mathscr{I}p'$. This contradicts the result of Lemma \ref{ch2l1}.~\hfill~$\Box$}
\end{quote}
\end{lemma}

Going back to the proof of the Theorem, let $\gamma$ a null geodesic segment emanating from $p$, 
traversing $\mathscr{M}$ and with future endpoint $p'\in\mathscr{I}$. As we've assumed that 
$(\mathscr{M},g)$ is (strongly) causal, one sees that $p'\notin\overline{I^-(p,\overline{
\mathscr{M}})}$. The Lemmata \ref{ch1l1} and \ref{ch1l2} imply that, if $p'$ is not dragged 
into $I^+(p,\mathscr{I})$ by gravitational time delay, then $p'=\bar{p}$, exactly as in the
case of $AdS_d$. However, even in this case, the presence of a pair of conjugate points 
in $\gamma$ implies that there exists a timelike curve traversing the bulk and linking $p$ to 
$p'=\bar{p}$. Repeating the argument of Lemma \ref{ch1l2}, the result follows. The proof 
for the case with reverse time orientation is the same.
\hfill$\Box$}
\end{quotation}
\end{theorem}

\begin{remark}\label{ch1r2} 
The method used in the proof is similar to the one of the positive mass theorem for 
asymptotically flat spacetimes due to \textsc{Penrose, Sorkin} and \textsc{Woolgar} \cite{psorkinw}, 
and for AAdS spacetimes by \textsc{Woolgar} \cite{woolgar} and \textsc{Page, Surya} and \textsc{Woolgar} 
\cite{pagesw}. A proof of a result similar to Theorem \ref{ch1t4}, employing a somewhat different
strategy, was proposed by \textsc{Gao} and \textsc{Wald} \cite{gaowald}.
\end{remark}

The condition stated in Theorem \ref{ch1t4} about the global focusing of null geodesics exerts 
a crucial role in the latter's proof, for the omnipresence of pairs of conjugate points exclude 
presence of \emph{null lines}, i.e., complete \emph{achronal} null geodesics. This comes from the
fact that, given a pair of conjugate points $p<q$ in a null geodesic $\gamma$, any point $r>q$, 
resp. $r<p$ satisfies $p\ll r$, resp. $q\gg r$. More precisely, the proof of Theorem \ref{ch1t4} 
consists in showing that, in all possible instances of violation of the assertion, it's possible 
to construct a null line traversing the bulk. Such a situation sharply contrasts with the $AdS_d$
case, once that in the latter \emph{all} null geodesics emanating from $p$ and traversing the bulk 
meet at $\overline{p}$.\\

This focusing condition is valid if, for instance, the \textsc{Einstein} equations without matter 
(\ref{ch1e1}) are satisfied everywhere, for these imply the validity of the \emph{null energy
condition (NEC)} $\mbox{Ric}(g)_{ab}k^ak^b\geq 0$ for all lightlike $k^a$, \emph{and} if the 
so-called \emph{null generic condition} is satisfied by all inextendible null geodesics (see 
\cite{beemee} for a discussion on this latter condition). Roughly, this last condition 
is necessary so that the null expansion of $\partial I^+(p,\overline{\mathscr{M}})\cap\mathscr{M}$ 
eventually becomes negative at each one of its generators -- the \textsc{Raychaudhuri} equation 
(\ref{ap1e6}) and NEC, then, take care of showing that there exist two conjugate points in the
maximal extension of each generator (see in Appendix \ref{ap1} the discussion following the 
formulae (\ref{ap1e10}), page \pageref{ap1e10}, and (\ref{ap1e22}), page \pageref{ap1e22}).\\

The violation of the null generic condition suggests a certain geometrical rigidity for null
hypersurfaces having such a property. That this intuition is justified at global level, it can be
seen in the following results:

\begin{theorem}[Geometrical Maximum Theorem \upshape\cite{angalho,galloway1}]\label{ch1t5}
Let $\mathscr{S}_1$ \\ be a \emph{future} null hypersurface (i.e., whose (future) null generators 
are future inextendible) and $\mathscr{S}_2$ a \emph{past} null hypersurface (i.e., whose (past) 
null generators are past inextendible). Suppose that:
\begin{enumerate}
\item $\mathscr{S}_1$ and $\mathscr{S}_2$ possess a common point $p\in\mathscr{M}$ and $\mathscr{S}_2$ 
finds itself in the future side of $\mathscr{S}_1$ in a neighbourhood of $p$; 
\item $\mathscr{S}_1$ has null expansion $\theta_1>0$ (see in Appendix \ref{ap1} the discussion 
immediately preceding formula (\ref{ap1e8}), page \pageref{ap1e8}) in the sense of supports
\footnote{\upshape Given a future null $\mathscr{C}^0$ hypersurface $\mathscr{S}$ and $p\in\mathscr{S}$, 
a \emph{future (resp. past) support hypersurface} for $\mathscr{S}$ at $p$ is a future (resp. past)
null $\mathscr{C}^\infty$ hypersurface $\mathscr{S}_p$ possessing a null geodesic segment, containing 
$p$ in its interior and localized in the future (resp. past) side of $\mathscr{S}$ in a neighbourhood 
of $p$. We analogously define support hypersurfaces for past null $\mathscr{C}^0$ hypersurfaces. We
say that a future (resp. past) null $\mathscr{C}^0$ hypersurface $\mathscr{S}$ possesses \emph{expansion} 
$\theta\geq 0$ (resp. $\leq 0$) \emph{in the sense of supports} if, for any $p\in\mathscr{S}$, $\epsilon>0$, 
there exists a past (resp. future) support hypersurface $\mathscr{S}_{p,\epsilon}$ for $\mathscr{S}$ at 
$p$ such that the expansion $\theta_{p,\epsilon}$ of $\mathscr{S}_{p,\epsilon}$ satisfies $\theta_{p,
\epsilon}(p)\geq-\epsilon$ (resp. $\leq+\epsilon$). Support hypersurfaces constitute a way of studying 
the expansion of non differentiable null hypersurfaces due to the presence of caustics, as in for instance 
achronal boundaries, black hole horizons, etc..} such that the set of second null fundamental forms of the 
support hypersurfaces $\mathscr{S_{p,\epsilon}}$ is bounded below; 
\item $\mathscr{S}_2$ possesses expansion $\theta_2\leq 0$ in the sense of supports,
\end{enumerate}
then $\mathscr{S}_1$ and $\mathscr{S}_2$ coincide in a neighbourhood $\mathscr{O}$ of $p$. Moreover, 
$\mathscr{S}_1\cap{}\mathscr{O}=\mathscr{S}_2\cap{}\mathscr{O}$ is a null $\mathscr{C}^\infty$ hypersurface 
with expansion $\theta=0$.~\hfill~$\Box$
\end{theorem}

The proof of Theorem \ref{ch1t5} consists in denoting $\mathscr{S}_1$ and $\mathscr{S}_2$ locally 
by the graph of a \textsc{Lipschitz} function, and, in these coordinates, rewriting the null expansion as a
quasilinear elliptic operator of second order, similar to the mean curvature operator when the 
former is prescribed in a codimension-one Riemannian submanifold, and for which there are maximum 
principles (see, for instance, \cite{angalho,giltrud}). From Theorem \ref{ch1t5}, it follows the

\begin{theorem}[Null Splitting Theorem \upshape\cite{galloway1}]\label{ch1t6}
Let $(\mathscr{M},g)$ be a spacetime endowed with null geodesic completeness and whose \textsc{Ricci} 
tensor satisfies NEC everywhere. If $\mathscr{M}$ has a null line $\gamma$, then $\gamma$ belongs to 
an achronal, edgeless and \emph{totally geodesic} null hypersurface, i.e., any geodesic tangent to 
this hypersurface at any point belongs to the latter in a neighbourhood of this point.~\hfill~$\Box$
\end{theorem}

\begin{figure}[ht!]
\begin{center}
\input{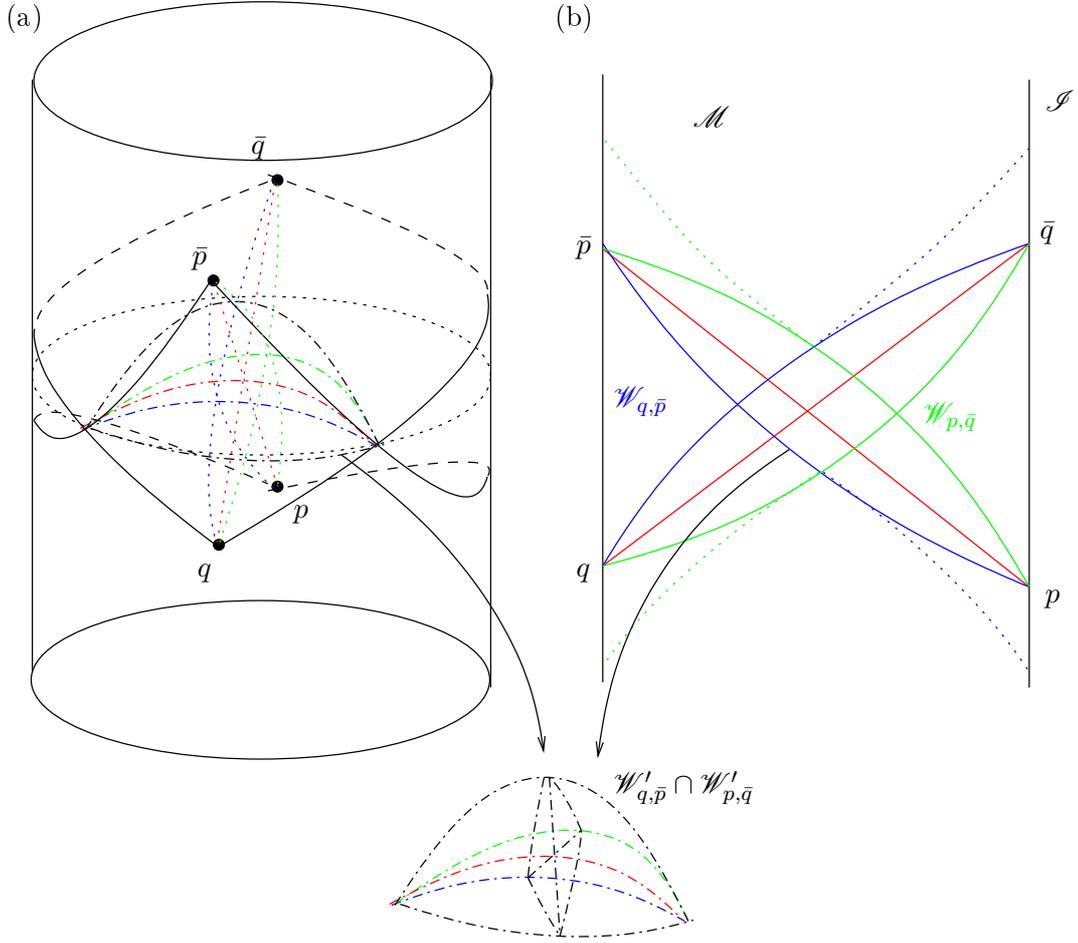} 
\end{center}
\caption{\label{ch1f4}\small Effect of gravitational time delay (whose effect on null geodesics is 
exemplified by the dotted lines in the longitudinal cut \textbf{(b)} of \textbf{(a)}) on wedges and 
their causal complements in AAdS spacetimes. Whereas in $AdS_d$ the causal complement of the wedge
$\mathscr{W}_{p,\bar{q}}$ is the wedge $\mathscr{W}_{q,\bar{p}}$ (red), the causal complements of both 
wedges (respectively green and blue) are no longer wedges in AAdS spacetimes satisfying the
conditions of Proposition \ref{ch1p1} and, hence, of Theorem \ref{ch1t4}, and in such a way that the open
region $\mathscr{W}_{p,\bar{q}}'\cap{}\mathscr{W}_{q,\bar{p}}'\neq\varnothing$ is \emph{nonvoid}, which
cannot happen in $AdS_d$.}
\end{figure}

Returning to our study, another fundamental difference between AdS spacetimes and ``generic'' AAdS 
spacetimes is expressed by the following Proposition, which in particular implies that the collection 
of wedges in an AAdS spacetimes may not be closed under causal complements, albeit the \textsc{Rehren} 
bijection still preserves causality in this context. More precisely, we have
$\mathscr{W}_{p,\bar{q}}'\cap{}\mathscr{W}_{q,\bar{p}}'\neq\varnothing$ (see Figure \ref{ch1f4}),
due to the

\begin{proposition}[Second Law of Dynamics of AAdS Wedges]\label{ch1p1} 
Let $(\mathscr{M},g)$ be an AAdS spacetime satisfying the NEC $\mbox{Ric}(g)
k^a k^b\geq 0$ for any lightlike vector $k^a$ and the conditions of Theorem \ref{ch1t4}. Define $\Xi^+_p
\doteq\partial I^+(p,\overline{\mathscr{M}})\smallsetminus\{p,\bar{p}\}$ and $\Xi^-_{\bar{p}}\doteq
\partial I^-(\bar{p},\overline{\mathscr{M}})\smallsetminus\{p,\bar{p}\}$. Then: 
\begin{enumerate}
\item[(i)] $\Xi^+_p\cap I^-(\bar{p},\overline{\mathscr{M}})=\Xi^-_{\bar{p}}\cap I^+(p,\overline{
\mathscr{M}})=\varnothing$.
\item[(ii)] $\Xi^+_p\cap{}\Xi^-_{\bar{p}}\cap{}\mathscr{M}=\varnothing$.
\end{enumerate}
\begin{quote}{\small\scshape Proof.\quad}
{\small\upshape (i) We know that $\Xi^+_p\cap{}\mathscr{I}=\Xi^-_{\bar{p}}\cap{}\mathscr{I}$, then 
let's concentrate only at the bulk. Namely, suppose that there exists $q\in\mathscr{M}$ such that 
$q\in\Xi^-_{\bar{p}}$ and $q\notin\Xi^+_p$. If $q\gg p$, this contradicts the fact that there is no
timelike curve linking $p$ to $\bar{p}$. Repeat the argument exchanging past with future, and 
the roles of $p$ and $\bar{p}$. 

(ii) Suppose that $\Xi^+_p$ and $\Xi^-_{\bar{p}}$ coincide at some point $q\in\mathscr{M}$. We know 
that (a) $\mathscr{S}_1\doteq\Xi^-_{\bar{p}}\cap{}\mathscr{M}$ \emph{is a future null} 
$\mathscr{C}^0$ \emph{hypersurface, and} $\mathscr{S}_2\doteq\Xi^+_{p}$ \emph{is a past null} 
$\mathscr{C}^0$ \emph{hypersurface}. Taking any $q_2\leq q\leq q_1$ with $q_i\in\mathscr{S}_i$, 
$i=1,2$, we see that $\partial I^-(q_1,\mathscr{M})$ contains a past support hypersurface $\mathscr{S}_-$
for $\mathscr{S}_1$ at $q$ and $\partial I^+(q_2,\mathscr{M})$, a future support hypersurface 
$\mathscr{S}_+$ for $\mathscr{S}_2$ at $q$. Using the hypothesis that $\mbox{Ric}(g)_{ab}k^a k^b\geq 0$ 
for any null vector $k^a$, we can invoke the \textsc{Raychaudhuri} equation (\ref{ap1e6}) (page 
\pageref{ap1e6}) and, recalling that the null geodesic congruences given by $\mathscr{S}_-$ and 
$\mathscr{S}_+$ have torsion $\omega_{+ab}=\omega_{-ab}=0$, show that $-\frac{1}{\theta^2_-}\frac{d
\theta_-}{d\lambda_-}\leq-\frac{1}{d-2}$ and $-\frac{1}{\theta^2_+}\frac{d\theta_+}{d\lambda_+}\geq+
\frac{1}{d-2}$, where $\lambda_\pm$ is an affine parametrization of the null geodesics $\gamma_\pm$ 
respectively in $\mathscr{S}_2$ and $\mathscr{S}_1$ such that $\gamma_\pm(0)=q$, $\gamma_-(\lambda)=q_2$ 
and $\gamma_+(\lambda)=q_1$, for some $\lambda>0$. Hence it follows that $\theta_-(p)\geq-
\frac{d-2}{\lambda}$ and $\theta_+(p)\leq\frac{d-2}{\lambda}$. Taking $q_2$ sufficiently close to $p$ 
and $q_1$ sufficiently close to $\bar{p}$,we can take $\lambda$ arbitrarily large. Thus, (b) $\mathscr{S}_1$ 
\emph{and} $\mathscr{S}_2$ \emph{satisfy} $\theta_2\leq 0\leq\theta_1$ \emph{in the sense of supports}. 
Hence, we conclude from (a), (i) and (b) that $\mathscr{S}_1$ and $\mathscr{S}_2$ satisfy the hypotheses 
of Theorem \ref{ch1t5} and, as a consequence, both coincide in a neighbourhood of $q$, where they form
a null $\mathscr{C}^\infty$ hypersurface with zero expansion. This implies that $\Xi^+_p$ and $\Xi^-_{\bar{p}}$ 
must coincide along a null geodesic containing $q$ and hence traversing the bulk. In this case, 
$\gamma$ must possess $p$ as past endpoint and $\bar{p}$ as future endpoint. But this geodesic 
belongs to an achronal boundary, and thus it must be achronal, contradicting the hypotheses of Theorem 
\ref{ch1t4}.~\hfill~$\Box$}
\end{quote}
\end{proposition}

\begin{remark}\label{ch1r3}
Notice that, before we invoke the hypothesis of global focusing of Theorem \ref{ch1t4} in the 
proof of (ii), we've seen that, once more, one produced a null line $\gamma$ contained in 
$\Xi^+_p$ and $\Xi^-_{\bar{p}}$. In the moment that we invoked Theorem \ref{ch1t5}, if $(\mathscr{M},g)$ 
were \emph{asymptotically simple}, we could have gone even further: by the Null Splitting Theorem 
\ref{ch1t6}, the null hypersurface which contains the null line in $\mathscr{M}$ linking $p$ to 
$\bar{p}$ has no boundary, which would be constituted by the endpoints of $\partial I^+(p,
\mathscr{M})$ and $\partial I^-(\bar{p},\mathscr{M})$ if it existed. Thus, it follows from 
null geodesic completeness of $(\mathscr{M},g)$ that $\partial I^+(p,\mathscr{M})=\partial 
I^-(\bar{p},\mathscr{M})$ is $\mathscr{C}^\infty$ and totally geodesic with respect to $g$.  
\end{remark}

An important consequence of Proposition \ref{ch1p1} is that the focusing condition causes a 
nontrivial ``shrinking'' of AAdS wedges towards conformal infinity, as we can see in Figure \ref{ch1f4} 
-- see the discussion at the end of Chapter \ref{ch2} for a dynamical interpretation of this phenomenon, 
which explains the why of our denominating this Proposition \emph{Second Law of Dynamics of AAdS Wedges}.

\subsubsection{\label{ch1-aads-causal-top}Simple connectedness (bulk \emph{versus} boundary)}

We shall cite now, for completeness, two results which show that every wedge $\mathscr{W}_{p,q}\in
\mathscr{W}(\mathscr{M},g)$ in a locally AdS-type spacetime $(\mathscr{M},g)$, satisfying 
certain additional conditions, is \emph{simply connected} if $\mathscr{D}_{p,q}$ is (for instance, 
wedges contained in a \textsc{Poincar\'e} domain of an AAdS spacetime). More precisely, any causal 
curve in the bulk with endpoints $p$ and $q$ is homotopic to a causal curve \emph{in} 
$\mathscr{I}$ linking $p$ to $q$. Summing up, the topology of $\mathscr{W}_{p,q}$ is determined by 
the topology of $\mathscr{D}_{p,q}$, showing that \emph{any wedge} $\mathscr{W}_{p,q}\in\mathscr{W}
(\mathscr{M},g)$ \emph{is contractible}.\\

This is a particular case of \emph{topological censorship} -- the topology of the region exterior to the
event horizon of a black hole is determined by the topology of $\mathscr{I}$ -- proven in the present
context by \textsc{Galloway, Schleich, Witt} and \textsc{Woolgar} \cite{galloway3}.

\begin{theorem}[Galloway, Schleich, Witt and Woolgar \cite{galloway3}]\label{ch1t7}
Let \\$(\mathscr{M},g)$ be a locally AdS-type spacetime such that $(\overline{\mathscr{M}},
\bar{g})$ is globally hyperbolic and $\mathscr{I}$ possesses a connected component $\mathscr{I}_0$ 
with a compact spacelike cut.\footnote{\upshape Given a (connected) manifold $\mathscr{M}$, 
a \emph{cut} is a hypersurface without boundary $\Sigma$ such that $\mathscr{M}\smallsetminus\Sigma$ 
has two connected components.} Suppose that any future complete null geodesic $\gamma$ in 
$(\overline{\mathscr{M}},\bar{g})$ emanating from some $p=\gamma(0)$ in a neighbourhood of 
$\mathscr{I}_0$ satisfies $\int^{+\infty}_0\mbox{\upshape Ric}(\bar{g}(\gamma(\lambda)))_{ab}
\dot{\gamma}^a(\lambda)\dot{\gamma}^b(\lambda)d\lambda\geq 0$. Then, there is no future causal 
communication of $\mathscr{I}_0$ through $\mathscr{M}$ with any other component of $\mathscr{I}$, i.e., 
$J^+(\mathscr{I}_0,\overline{\mathscr{M}})\cap(\mathscr{I}\smallsetminus\mathscr{I}_0)=\varnothing$.
~\hfill~$\Box$
\end{theorem}

Theorem \ref{ch1t7} splendidly complements Lemma \ref{ch1l1}, employed in the proof
of Theorem \ref{ch1t4}. Using Theorem \ref{ch1t7} and covering space arguments \cite{greenhar,oneill}, 
cone can finally prove the following \emph{topological censorship} theorem:

\begin{theorem}[Galloway, Schleich, Witt and Woolgar \cite{galloway3}]\label{ch1t8}
Let \\$(\mathscr{M},g)$ be a locally AdS-type spacetime such that $(\mathscr{W}(\mathscr{I})\cup
\mathscr{I},\bar{g}\restr{\mathscr{W}(\mathscr{I})\cup\mathscr{I}})$ is globally hyperbolic and 
$\mathscr{I}$ possesses a compact spacelike cut. Suppose that any future complete null geodesic 
$\gamma$ in $(\overline{\mathscr{M}},\bar{g})$ emanating from some $p=\gamma(0)$ in a 
neighbourhood of $\mathscr{I}$ satisfies $\int^{+\infty}_0\mbox{\upshape Ric}(\bar{g}(\gamma(\lambda
)))_{ab}\dot{\gamma}^a(\lambda)\dot{\gamma}^b(\lambda)d\lambda\geq 0$. Then any causal curve
$\gamma:(0,1)\rightarrow\mathscr{W}(\mathscr{I})$ with endpoints $\gamma(0),\gamma(1)\in\mathscr{I}$ 
is fixed endpoint homotopic to a curve $\gamma_0:[0,1]\rightarrow\mathscr{I}$ (see Appendix\ref{ap4}). 
~\hfill~$\Box$
\end{theorem}

It follows immediately from Theorem \ref{ch1t8} the contractivity of $\mathscr{W}_{p,q}$ for $p,q
\in\mathscr{D}(\mathscr{I})$. In particular, if $p,q\in\mathscr{D}(\mathscr{I})$ are the endpoints
of $\gamma$ in Theorem \ref{ch1t8}, we can choose $\gamma_0$ \emph{chronological}.

\subsubsection{\label{ch1-aads-causal-loc}Localization (boundary to bulk)}

Throughout this Subsubsection, it'll be assumed that $(\mathscr{M},g)$ is an asymptotically simple AAdS 
spacetime satisfying the hypotheses of Theorem \ref{ch1t4}. \\

Knowing the localization of the physical procedures in wedges may not be enough for the complete 
reconstruction of the bulk quantum theory in the context of Chapter \ref{ch4}, only by employing 
boundary information and the \textsc{Rehren} bijection $\rho_{(\mathscr{M},g)}$ introduced by Definition 
\ref{ch1d3}. We need to be able to specify the localization of physical procedures in arbitrarily 
small open sets, or, which amounts to the same thing, its localization with respect to a base for the bulk topology. 
This can be achieved, in principle, by taking intersections of wedges, but it's by no means clear
whether this results in a base or not. One thus needs to make this rough idea more precise.\\

It's known that, for strongly causal spacetimes, the topology generated by diamonds (\textsc{Alexandrov}
topology) coincides with the manifold topology \cite{beemee}. Hence, in AdS, the above question has a 
positive answer, for \emph{any diamond in $AdS_d$ can be enveloped by wedges:} given
\begin{equation}\label{ch1e33}
\mathscr{O}_{p,q}\doteq I^+(p,AdS_d)\cap I^-(q,AdS_d),\,p\ll_{AdS_d}q,
\end{equation}
we can write
\begin{equation}\label{ch1e34}
\mathscr{O}_{p,q}=\bigcap_{{r\in\partial I^-(p,\overline{AdS_d})\cap{}\mathscr{I},}\atop{s\in\partial I^+(q,
\overline{AdS_d}),\,r,s\in\mathscr{M}in(u)}}\mathscr{W}_{r,s}.
\end{equation}

We shall see soon that the achronality of inextendible null geodesics in $AdS_d$ is tantamount for 
a precise envelope, whereas in an asymptotically simple AAdS spacetime satisfying the hypotheses of 
Theorem \ref{ch1t4}, the question is more delicate, because of the following

\begin{proposition}\label{ch1p2} 
Let $p\in\mathscr{M}$. Then, $\partial I^+(p,\overline{\mathscr{M}})$ intersects each time generator
of $(\mathscr{I},\bar{g}^{(0)})$ \emph{precisely once}.
\begin{quote}{\small\scshape Proof.\quad}
{\small\upshape Due to the achronality of $\partial I^+(p,\overline{\mathscr{M}})$, the latter intersects 
each time generator of $(\mathscr{I},\bar{g}^{(0)})$ \emph{at most once}. Suppose that the assertion
is false. Then, given a time generator $T$, we have the following possibilities:
\begin{enumerate}
\item[(i)] $T\subset I^+(p,\overline{\mathscr{M}})$ -- Consider a complete null geodesic 
$\gamma$ crossing $p$, and let $q$ be the \emph{past} endpoint of $\gamma$. Then there exists a value
$t$ of the affine parameter of $T$ such that $T(t)\ll_\mathscr{I}q$. Hence, $T(t)\ll_{\overline{\mathscr{M}}}p$, 
which is absurd since $(\overline{\mathscr{M}},\bar{g})$ is chronological.
\item[(ii)] $T\cap J^+(p,\overline{\mathscr{M}})=\varnothing$ -- Consider a complete null geodesic 
$\gamma$ crossing $p$, and let $r$ the \emph{future} endpoint of $\gamma$. Then there exists a 
value $t$ of the affine parameter of $T$ such that $T(t)\gg_\mathscr{I}r$. Hence, $T(t)\gg_{\overline{
\mathscr{M}}}p$, contradicting the hypothesis.
\end{enumerate}
~\hfill~$\Box$}\end{quote}
\end{proposition}

\begin{proposition}\label{ch1p3} 
Let $q,r\in\mathscr{M}$ such that $r\in\partial I^-(q,\mathscr{M})$, and $\gamma$ a null generator 
of $\partial I^-(q,\mathscr{M})$ to which $r$ belongs. Let $s_1(r),s_2(r),s_3(r)\in\mathscr{I}$ 
defined as follows:
\begin{itemize}
\item $s_1(r)$ is the future endpoint of $\gamma$;
\item $s_2(r)$ is the point where $\partial I^+(q,\overline{\mathscr{M}})$ intersects the time 
generator of $T(s_1(r))$ to which $s_1(r)$ belongs;
\item $s_3(r)$ is the point where $\partial I^+(r,\overline{\mathscr{M}})$ intersects $T(s_1(r))$.
\end{itemize}
Then:
\begin{enumerate}
\item[(i)] $s_3(r)\leq_\mathscr{I}s_2(r)\leq_\mathscr{I}s_1(r)$.
\item[(ii)] $s_3(r)=s_2(r)=s_1(r)$ if and only if the segment of $\gamma$ linking $r$ to $s_1(r)$ 
is achronal.
\end{enumerate}
\begin{quote}{\small\scshape Proof.\quad}
{\small\upshape (i) Immediate, as (ii) $\Rightarrow$. It remains only to prove (ii) $\Leftarrow$. 
Namely, suppose that $s_3(r)$ coincides with $s_2(r)$. Then the null geodesic segment linking $q$ to 
$s_2(r)$ must belong to $\gamma$, for otherwise there would exist a broken null geodesic segment 
linking $r$ to $s_3(r)$, contradicting the definition of the latter (this, in particular, proves that 
$s_1(r)=s_3(r)$ even if we only assume $s_2(r)=s_3(r)$). If $\gamma$ is not achronal, once again 
we have a contradiction with the definition of $s_3(r)$.~\hfill~$\Box$}\end{quote}
\end{proposition}

\begin{remark}\label{ch1r4}
Results analogous to Propositions \ref{ch1p2} and \ref{ch1p3} are valid if we exchange future with
past.
\end{remark}

Now, let $\mathscr{O}_{p,q}\subset\mathscr{M}$ be onwards a \emph{relatively compact diamond possessing 
a contractible} \textsc{Cauchy} \emph{surface} -- any sufficiently small diamond satisfies both 
conditions. Let us then consider the region
\begin{equation}\label{ch1e35}
\mathscr{Q}_{p,q}=\bigcap_{{r\in\partial I^-(p,\overline{\mathscr{M}})\cap{}\mathscr{I},}\atop{s\in
\partial I^+(q,\overline{\mathscr{M}})\cap T(r)}}\mathscr{W}_{r,s}.
\end{equation}

IT follows naturally from this definition that $\mathscr{Q}_{p,q}\supset\mathscr{O}_{p,q}$, is 
\emph{causally complete}, as it's an intersection of causally complete regions, and
\begin{equation}\label{ch1e36}
\mathscr{Q}_{p,q}\cap J^+(q,\overline{\mathscr{M}})=\mathscr{Q}_{p,q}\cap J^-(p,\overline{
\mathscr{M}})=\varnothing.
\end{equation}

In $AdS_d$, $\mathscr{Q}_{p,q}=\mathscr{O}_{p,q}$. For asymptotically simple AAdS spacetimes 
satisfying the as hypotheses of Theorem \ref{ch1t4}, however, it may happen that 
$\mathscr{Q}_{p,q}\supsetneqq\mathscr{O}_{p,q}$. Analogously, defining $\mathscr{E}_{p,q}\doteq
\partial I^+(p,\mathscr{M})\cap{}\partial I^-(q,\mathscr{M})$, let's start from
\begin{equation}\label{ch1e37}
\widetilde{\mathscr{Q}}_{p,q}=\bigcap_{r\in\mathscr{E}_{p,q}}\mathscr{W}_{s'_3(r),s_3(r)},
\end{equation}
where $s'_3(r)$ corresponds to $s_3(r)$ if we exchange future with past in the statement of
Proposition \ref{ch1p3}. Here, $\widetilde{\mathscr{Q}}_{p,q}\subset\mathscr{Q}_{p,q}$ is once more 
causally complete, \emph{if nonvoid}. However, is the metric deep inside the bulk is sufficiently 
``distorted'', making a sufficient number of null generators of, say, $\partial I^-(q,\mathscr{M})$ 
to acquire pairs of conjugate points between $\mathscr{E}_{p,q}$ and $\mathscr{I}$, for all we know 
(Proposition \ref{ch1p3}, page \pageref{ch1p3}) $\widetilde{\mathscr{Q}}_{p,q}$ can very well be empty. 
This is suggested by the following remarks:

\begin{enumerate}
\item In a causally simple spacetime, any relatively compact diamond $\mathscr{O}_{p,q}$ is a
globally hyperbolic region, for which any \textsc{Cauchy} surface possess boundary $\mathscr{E}_{p,q}$;
\item Any causally complete region $\mathscr{U}$ possesses the following property: if 
$\mathscr{S}\subset\mathscr{U}$ closed, achronal set with respect to $\mathscr{U}$, then $D(\mathscr{S})\subset
\mathscr{U}$. 
\end{enumerate}

Both remarks together show that, if $r\in\mathscr{E}_{p,q}$ is such that a null generator of, 
say, $\partial I^-(q,\mathscr{M})$ crossing $r$ acquires a pair of conjugate points between $r$ 
and $s_1(r)$, then, by causal simplicity, there exists a neighbourhood of $q$ causally 
disjoint from $s_3(r)$, and, thus, $I^-(s_3(r),\overline{\mathscr{M}})$ cannot contain a 
\textsc{Cauchy} surface for $\mathscr{O}_{p,q}$. Since, on the other hand, this doesn't 
exclude the possibility that $\widetilde{\mathscr{Q}}_{p,q}$ may contain points outside 
$\mathscr{O}_{p,q}$ either, it's by no means clear whether the collections of $\mathscr{Q}_{p,q}$'s and 
$\widetilde{\mathscr{Q}}_{p,q}$'s produce bases for the topology of $\mathscr{M}$ or 
not. \\

A way of circumventing these problems could be to restrict our considerations to sufficiently 
small diamonds, such that no null generator of $\partial I^+(q,\overline{
\mathscr{M}})$ can pursue long enough a path beyond $q$ which allows it to acquire a 
pair of conjugate points. There is, however, a situation in which no matter how small the 
extension, it will always cease to be achronal: it's when $q$ \emph{itself} is conjugate to $s_1(r)$. 
In this limit case, $s_1(r)=s_2(r)$ but $s_2(r)\neq s_3(r)$.\\

We'll show now that the key for the end of these problems is trying to build a region similar to
$\widetilde{\mathscr{Q}}_{p,q}$, but employing, instead of the points $s_3(r)$, $s'_3(r)$ for 
$r\in\mathscr{E}_{p,q}$, the points for which the problem above, implied by Proposition 
\ref{ch1p3}, is, in a certain sense, ``minimized''. In order to undertake such a task and hence 
optimize our construction, we shall start from a different viewpoint, which will end up 
showing that the critical situation mentioned in the former paragraph is excluded by null geodesic
completeness.\\

First, notice that, by means of an argument analogous to the one employed in \cite{hawkellis} and 
\cite{wald2} to prove the existence of a topological (\textsc{Lipschitz}) manifold structure
for achronal boundaries, one can show that $\mathscr{E}_{p,q}$ is locally the graph of a 
locally \textsc{Lipschitz} function with values in $\mathbb{R}$ with $d-2$ arguments, and thus
an achronal, compact topological (\textsc{Lipschitz}) submanifold embedded into $\mathscr{M}$, 
with codimension two. Notice as well that one can parametrize in a $\mathscr{C}^\infty$ way the family 
of time generators of $(\mathscr{I},\bar{g}^{(0)})$ by means of a \textsc{Cauchy} surface
$\mathscr{S}$ of the latter, homeomorphic to $S^{d-2}$ and hence also compact. Let $t$ be the affine
parameter common to the time generators of $(\mathscr{I},b)$ mentioned above, and $\mathscr{F}_{p,q}$ 
some \textsc{Cauchy} surface for $\mathscr{O}_{p,q}$, whose boundary (edge) is of course $\mathscr{E}_{p,q}$. 
Define implicitly the functions
\begin{equation}\label{ch1e38}
\tau^\pm:\overline{\mathscr{F}_{p,q}}\times\mathscr{S}\ni(r,\theta)\mapsto\tau^\pm(r,\theta)\in\mathbb{R},
\end{equation}
where
\begin{equation}\label{ch1e39}
\partial I^\pm(R,\overline{\mathscr{M}})\cap T(\theta)=\{T(\theta)(\tau^\pm(r,\theta))\}.
\end{equation}

Proposition \ref{ch1p2} shows that the definition of $\tau^\pm$ is not empty. Moreover:

\begin{proposition}\label{ch1p4}
$\tau^+$ (resp. $\tau^-$) is upper (resp. lower) semicontinuous in $r$ at fixed $\theta$.
\begin{quote}{\small\scshape Proof.\quad}
{\small\upshape We'll just prove it for $\tau^+$ -- the result for $\tau^-$ follows analogously. Let 
$\epsilon>0$. $r$ belongs to the chronological past of the point $T(\theta)(\tau^+(r,\theta)+\epsilon)$, 
and thus there exists an open neighbourhood $U$ of $r$ in $\overline{\mathscr{F}_{p,q}}$ (with respect 
to the relative topology) which stays in the chronological past of $T(\theta)(\tau^+(r,\theta)+\epsilon)$. 
Hence, for all $r'\in U$, we must have $\tau^+(r',\theta)<\tau^+(r,\theta)+\epsilon$.~\hfill~$\Box$}
\end{quote}
\end{proposition}

One may yet prove that $\tau^\pm$ are locally \textsc{Lipschitz} in $\theta$ for fixed $r$, but this 
won't be used in what follows. The function $\tau^+(.,\theta)$ (resp. $\tau^-(.,\theta)$) will be called 
\emph{future} (resp. \emph{past}) \textsc{Fermat} \emph{potential} with respect to $\theta$.\footnote{Actually, 
$\tau^\pm$ also depend on the choice of foliation of $\mathscr{I}$ by \textsc{Cauchy} surfaces, but such a 
dependence is irrelevant for the use we'll make of $\tau^\pm$ and will be ignored.} The name is reminiscent 
from the \textsc{Huygens-Fermat} principle in geometrical optics (see, for instance, pages 249-250 of 
\cite{arnold} for a beautiful proof). Since $\mathscr{E}_{p,q}=\partial\mathscr{F}_{p,q}$ as well as 
$\overline{\mathscr{F}_{p,q}}$ are closed subsets of the compact set $\overline{\mathscr{O}_{p,q}}$, 
they are themselves compact. It then follows from a standard result in Analysis (see, for instance, 
pages 110-111 of \cite{fomin}) that $\tau^+(.,\theta)$ (resp. $\tau^-(.,\theta)$) possesses a maximum 
(resp. minimum) in $\overline{\mathscr{F}_{p,q}}$ as well as in $\mathscr{E}_{p,q}$. The next Theorem 
shows that $\pm\tau^\pm(.,\theta)$ has, indeed, a typical property of potentials:

\begin{theorem}[Maximum / minimum principle for Fermat potentials]\label{ch1t9} The 
maximum (resp. minimum) value of $\tau^+(.,\theta)$ (resp. $\tau^-(.,\theta)$) in 
$\overline{\mathscr{F}_{p,q}}$ is attained in $\mathscr{E}_{p,q}$.
\begin{quote}{\small\scshape Proof.\quad}
{\small\upshape As done in the proof of Proposition \ref{ch1p4}, we shall just prove the result for $\tau^+$. 
Let $r$ be a point of $\mathscr{E}_{p,q}$ where $\tau^+(.,\theta)$ attains its maximum in $\mathscr{E}_{p,q}$, 
and let $r'$ be a point of $\mathscr{F}_{p,q}$ such that $\tau^+(r',\theta)\geq\tau^+(r',\theta)$. In this case, 
it's obvious that $\mathscr{E}_{p,q}$ belongs to the causal past of $T(\theta)(\tau^+(r',\theta))$. Pick a 
curve segment in $\overline{\mathscr{F}_{p,q}}$ starting in $r'$, initially pointing outside $J^-(T(\theta)
(\tau^+(r',\theta)),\overline{\mathscr{M}})$ and terminating at some point of $\mathscr{E}_{p,q}$. Then any 
such curve segment must cross $\partial I^-(T(\theta)(\tau^+(r',\theta)),\overline{\mathscr{M}})$ at least 
once more after $r'$, and before (or when) hitting $\mathscr{E}_{p,q}$. This shows that $\partial I^-(T
(\theta)(\tau^+(r',\theta)),\overline{\mathscr{M}})\cap{}\overline{\mathscr{F}_{p,q}}$ contains an open 
subset $X$ of $\mathscr{F}_{p,q}$ outside the causal past of $T(\theta)(\tau^+(r',\theta))$.

The rest of the proof is analogous to the proof of \textsc{Penrose}'s singularity theorem
\cite{hawkellis,wald2}: namely, we'll show that the properties of $\partial X$ imply the 
existence of an incomplete null geodesic in $(\mathscr{M},g)$. First, we shall show that the 
closed, acausal set $\partial X=\partial I^-(T(\theta)(\tau^+(r',\theta)),\overline{
\mathscr{M}})\cap{}\overline{\mathscr{F}_{p,q}}$ is \emph{past trapped}, i.e., $\partial 
I^-(X,\mathscr{M})$ is \emph{compact}. The past null geodesics ``entering'' $\partial X$ 
constitute the past \textsc{Cauchy} horizon of $X$, which is hence contained in $\overline{
\mathscr{O}_{p,q}}$ and, thus, compact, for it's closed. The outgoing null geodesics are 
precisely the null generators of $\partial I^-(T(\theta)(\tau^+(r',\theta)),\overline{
\mathscr{M}})$ which cross $\partial X$. Let us adopt a common affine parametrization for the null
generators of $\partial I^-(T(\theta)(\tau^+(r',\theta)),\overline{\mathscr{M}})$ such that the zero  
of the affine parameter corresponds to $\partial X$. Then let $t_0$ the largest value of the affine 
parameter for which a past endpoint of $\partial I^-(T(\theta)(\tau^+(r',\theta)),\overline{\mathscr{M}})$  
is attained. This value must be finite, for any inextendible null geodesic must acquire a pair 
of conjugate point before hitting infinity, though the value of the affine parameter in a past 
endpoint of the null generator segment starting, say, in $r''\in\partial X$ can be zero if 
$r'$ happens to be a past endpoint himself. Anyway, the portion of $\partial I^-(T(\theta)(\tau^+(r',
\theta)),\overline{\mathscr{M}})$ in the causal past of $\partial X$, being closed, has a closed 
inverse image in the compact $[0,t_0]\times\partial X$ by the parametrization chosen for the null 
generators, and is thus compact. Hence, the set $\partial I^-(\partial X,\mathscr{M})=H^-(X)\cup{}
\partial X\cup(\partial I^-(T(\theta)(\tau^+(r',\theta)),\overline{\mathscr{M}})\cap J^-(\partial X,
\mathscr{M}))$ is a compact, achronal subset of $\mathscr{M}$, as stated. 

However, any causally simple spacetime is stably causal \cite{beemee}. That is, one can foliate
$\mathscr{M}$ in a $\mathscr{C}^\infty$ manner by ``constant-time'' spacelike, codimension-one
hypersurfaces. Due to the structure of conformal infinity, these surfaces cannot be compact. Moreover, 
each timelike \emph{orbit} of this foliation crosses an achronal set at most once. Following these 
orbits, one can continuously map $\partial I^-(\partial X,\mathscr{M})$ into a spacelike leaf. 
As the image of this map is compact, it must have a nonvoid boundary. It's, on the other hand, 
known that a set of the form $\partial I^-(Y,\mathscr{M}),Y\subset\mathscr{M}$ is a topological 
submanifold \emph{without boundary} of $\mathscr{M}$. This shows that some null generator of 
$\partial I^-(T(\theta)(\tau^+(r',\theta)),\overline{\mathscr{M}})$ must hit a singularity before
reaching its past endpoint. But this enters into conflict with the null geodesic completeness of 
$\widehat{\mathscr{M}}$, implied by asymptotic simplicity. Hence, no point of $\mathscr{F}_{p,q}$ 
can attain a maximum for $\tau^+(.,\theta)$ in $\overline{\mathscr{F}_{p,q}}$ -- the former is always
reached in $\mathscr{E}_{p,q}$.
\hfill$\Box$}\end{quote}
\end{theorem}

Proposition \ref{ch1p4} and Theorem \ref{ch1t9} together show that, for each $\theta$, there is always 
a $r\in\mathscr{E}_{p,q}$ such that, given \emph{any} \textsc{Cauchy} surface $\mathscr{F}_{p,q}$ 
for $\mathscr{O}_{p,q}$, the set $\overline{\mathscr{F}_{p,q}}$ always stays in the causal past of 
$T(\theta)(\tau^+(r,\theta))$. By Proposition \ref{ch1p3} and the above remarks, this can only happen 
is the achronal null geodesic segment $\gamma(r,\theta)$ linking $r$ to $T(\theta)(\tau^+(r,
\theta))$ crosses $q$. Hence, this maximum point is \emph{unique:} suppose otherwise. Then, there 
would exist another $r'\in\mathscr{E}_{p,q}$ such that there is an achronal null geodesic segment 
$\gamma(r',\theta)$ linking $r'$ to $T(\theta)(\tau^+(r',\theta))=T(\theta)(\tau^+(r,\theta))$ and crossing 
$q$. Now, consider the curve segment $\gamma'(r,\theta)$, which coincides with $\gamma(r,\theta)$ from 
$r$ to $q$, and coincides with $\gamma(r',\theta)$ from $q$ to $T(\theta)(\tau^+(r,\theta))$. This segment 
is necessarily broken, which conflicts with the achronality of $\gamma(r,\theta)$. Exchanging the 
roles of $r$ and $r'$, one can see that this argument also enters in conflict with the achronality of 
$\gamma(r',\theta)$.\\

Let us notice, however, that an arbitrary $r\in\mathscr{E}_{p,q}$ \emph{need not} maximize 
$\tau^+(.,\theta)$ for some $\theta$. The two situations where this indeed cannot occur are:

\begin{enumerate}
\item $r$ is conjugate to $q$ along a null generator of $\partial I^-(q,\mathscr{M})$ -- 
any future extension of this generator beyond $q$ won't be achronal;
\item $q$ is conjugate to $s_2(r)$ along a null generator of $\partial I^+(q,\overline{
\mathscr{M}})$, by the remarks made above.
\end{enumerate}

The second instance, on the other hand, is excluded by our line of reasoning, for it makes impossible, 
by Proposition \ref{ch1p3} and by Theorem \ref{ch1t9}, that $\tau^+(.,\theta)$ attains a maximum in 
$\mathscr{E}_{p,q}$. This cannot occur, since for each $\theta$ a maximum must exist by Proposition 
\ref{ch1p4}. The first instance can be circumvented if we pick $\mathscr{O}_{p,q}$ contained, say, in
a convex normal neighbourhood, which can always be done, as here $(\mathscr{M},g)$ is strongly causal. 
One can go beyond that and take $\mathscr{O}_{p,q}$ sufficiently small (albeit nonvoid) in a way that 
$r\in\mathscr{E}_{p,q}$ is a maximum of $\tau^+(.,\theta)$ for some $\theta$, for the only obstacle to
this would be the second instance above, excluded by the argument here presented. All these results 
possess a past counterpart, if we switch $q$ with $p$ and $\tau^+$ with $\tau^-$.\\

Summing up, we've shown that \emph{any sufficiently small $\mathscr{O}_{p,q}$ can always be 
precisely enveloped by wedges.} In this case, any point that doesn't belong to $\overline{
\mathscr{O}_{p,q}}$ belongs either to the chronological future of $\partial I^-(q,\mathscr{M})$ 
or to the chronological past of $\partial I^+(p,\mathscr{M})$, and, as such, will fail at belonging to 
any wedges enveloping $\mathscr{O}_{p,q}$. As the points in $\partial\mathscr{O}_{p,q}$ are
automatically excluded from the intersection by construction, one concludes that $\mathscr{O}_{p,q}=
\mathscr{Q}_{p,q}$ for $\mathscr{O}_{p,q}$ sufficiently small. Moreover, in this situation, 
each wedge in the definition (\ref{ch1e35}) of $\mathscr{Q}_{p,q}$ is guaranteed to be contained in
some \textsc{Poincar\'e} domain. Notice yet that the procedure we've proposed allows as well to envelope
the \emph{regular diamonds} (if sufficiently small) defined in \cite{guilrv}.

\chapter{\label{ch2}Dynamics and time evolution}

\epigraph{\begin{verse}[7cm] \emph{Because I know that time is always time \\
And place is always and only place \\ And what is actual is actual only 
for one time \\ And only for one place} \end{verse}~\vspace*{-13pt}}
{\textsc{T. S. Eliot} \\ ``I. Because I do not hope to turn again'' 
(\emph{Ash-Wednesday}, 1930)}

One has made extensive use of the global structure of AAdS spacetimes in Chapter \ref{ch1}, but 
the \textsc{Einstein} equations have exerted a minor role in these developments. However, 
it has become clear along the previous Chapter that nontrivial geometries in the bulk of an AAdS 
spacetime provoke a distortion of the causal structure in the large, which, thus, must be 
noticeable from the viewpoint of a theory of local observables at the boundary. Hence, it's highly
desirable to establish a more quantitative connection between these effects and the very 
gravitational dynamics, since \textsc{Maldacena}'s conjecture and many of its consequences involve
gravitational effects in an essential way.

\section[Gravitation]{\label{ch2-grav}Variational approach to classical gravity. Conceptual
foundations}

The \textsc{Einstein} equations (\ref{ch1e1}) are obtained in the Lagrangian approach by means
of the \textsc{Einstein-Hilbert} variational principle: for any compact set $K\Subset
\mathscr{M}$ with piecewise $\mathscr{C}^\infty$ boundary (or, more generally, we assume that the set 
of non differentiable points of $\partial K$ have zero $(d-1)$-dimensional \textbf{Lebesgue} measure), 
the action functional
\begin{equation}\label{ch2e1}
S_K[g]\doteq\frac{1}{16\pi G_d}\int_K \underbrace{(R(g)+2\Lambda)}_{\doteq\mathscr{L}_\Lambda(g);}
\sqrt{|g|}dx\,{}^{(}\mbox{\footnotemark}{}^{)}
\end{equation}
\footnotetext{$G_d$ is the $d$-dimensional \textsc{Newton} constant; henceforth, we choose 
units $G_d=(16\pi)^{-1}$, so as the numerical factor multiplying the right hand side of (\ref{ch2e1}) 
becomes equal to one.} is stationary for arbitrary variations of $g$ around a solution of the 
variational problem, supported in the interior of $K$. Namely, a finite variation of $g$ is obtained 
from a $\mathscr{C}^\infty$ curve of metrics $g_\lambda$, $\lambda\in(-1,1)$, which differ from $g$ 
in a compact set $K_1\subset int(K)$ and such that $g_0=g$. The corresponding infinitesimal variation 
(= tangent vector) around $g$ is given by $\delta g=\frac{d}{d\lambda}|_{\lambda=0}g_\lambda$. The 
corresponding variation in (\ref{ch2e1}) is given by 
\begin{equation}\label{ch2e2}
(\delta S_K)[g,\delta g]=\int_K \left[\nabla^a\underbrace{\left(\nabla^b\delta g_{ab}-g^{cd}
\nabla_a\delta g_{cd}\right)}_{\doteq\theta_a(g,\delta g );}+\left(\mbox{Ric}_{ab}-\frac{1}{2}R
g_{ab}+\Lambda g_{ab}\right)\delta g^{ab}\right]\sqrt{|g|}dx,
\end{equation}

where the first term of the right hand side is the divergence of the \textsc{Hodge}-dual 1-form
to the $(d-1)$-form $\theta^{a_1}\sqrt{|\det g|}_{[a_1\cdots a_d]}$, which vanishes in $\partial K$ for
the class of variations considered. The second term is the \textsc{Einstein} tensor $G(g)_{ab}
\doteq \mbox{Ric}(g)_{ab}-\frac{1}{2}R(g)g_{ab}$, plus the cosmological constant part $\Lambda$. 
Hence, we have the \textsc{Einstein} equations (\ref{ch1e1}). Let us notice that the use of a compact
domain of integration $K$ reflects the fact that the variational principle that leads to \ref{ch1e1}
is \emph{inherently local}, as any field theory.\\

We end this Section here emphasizing that, for any vector $T^a$, the expression $\left(\mbox{Ric}
(g)_{ab}-\frac{1}{2}R(g)g_{ab}+\Lambda g_{ab}\right)T^b$ doesn't have second order derivatives in 
direction of $T^a$. If $T^a$ is timelike, thus determining (locally) a time flow, we see that the 
components $\left(\mbox{Ric}(g)_{ab}-\frac{1}{2}R(g)g_{ab}+\Lambda g_{ab}\right)T^b=0$ of the
\textsc{Einstein} equations are actually \emph{constraints}, expressing the symmetry of the latter
under general coordinate transformations (\emph{passive} viewpoint) or, equivalently in mathematical 
terms, under local diffeomorphisms (\emph{active} viewpoint). The physical meaning of this fact, 
according to \textsc{Einstein}, is that gravity doesn't have an \emph{intrinsic} notion of (dynamical)
time evolution, for the very choice of time coordinate is a symmetry -- such a choice must, hence, be
made by means of a concrete physical procedure, in such a manner that the physical laws
which rule such procedures in our gravitational background are independent of this choice.
There are two ways to do this: one local -- by the specification of a nontrivial energy-momentum tensor
-- and the other global, if the metric becomes close to some fixed background geometry at sufficiently
long distances, for which such choice can be made naturally. The latter circumstance includes precisely 
the case of AAdS spacetimes. \\

In the next Section, we'll make use of this fact to explore in detail the geometry of (locally) AAdS 
spacetimes near conformal infinity.

\section{\label{ch2-fefgra}The \textsc{Fefferman-Graham} expansion}

Let us consider the case of AdS-type spacetimes $(\mathscr{M},g)$ as in Definition \ref{ch1d2}, 
whence we adopt the notation. Let's now scrutinize the form of the solutions of (\ref{ch1d1}) for 
such spacetimes in a collar neighbourhood $\overline{\mathscr{U}}_\epsilon\cong\mathscr{I}
\times[0,\epsilon)\ni(x,z)$, $\epsilon>0$, of $\mathscr{I}$ in $\overline{\mathscr{M}}$. This 
will be made by means of the asymptotic expansion obtained by \textsc{Fefferman} and \textsc{Graham} 
\cite{fefferman}, whose deduction we present below. In the subsequent argument, we closely follow 
\cite{him1} and, in a smaller measure, \cite{graham,grahamlee,dehsks}.\\

An equivalent way of writing (\ref{ch1e1}) is directly in terms of the \textsc{Ricci} tensor, 
resulting in
\begin{equation}\label{ch2e3}
\mbox{Ric}(g)_{ab}=\frac{2\Lambda}{d-2}g_{ab}.
\end{equation}

Invoking now the formula (\ref{ap1e24}) of the conformal transformation of $\mbox{Ric}(g)$ into 
$\mbox{Ric}(\bar{g})$, $\bar{g}=z^2 g$
\[\mbox{Ric}(g)_{ab}=\mbox{Ric}(\bar{g})_{ab}+\frac{d-2}{z}\bar{\nabla}_a\bar{\nabla}_bz+\bar{g}_{ab}
\bar{g}^{cd}\left(\frac{1}{z}\bar{\nabla}_c\bar{\nabla}_dz-\frac{d-1}{z^2}\bar{\nabla}_cz
\bar{\nabla}_dz\right),\]
we have the expression of (\ref{ch2e3}) in terms of $\bar{g}$
\begin{equation}\label{ch2e4}
\mbox{Ric}(\bar{g})_{ab}+\frac{d-2}{z}\bar{\nabla}_a\bar{\nabla}_bz+\bar{g}_{ab}\bar{g}^{cd}\left(
\frac{1}{z}\bar{\nabla}_c\bar{\nabla}_dz-\frac{d-1}{z^2}\bar{\nabla}_cz\bar{\nabla}_dz\right)=
\frac{2\Lambda}{(d-2)z^2}\bar{g}_{ab}.
\end{equation}

Notice, in particular, that if we multiply (\ref{ch2e4}) by $z^2$ and take $z=0$, it follows that 
$dz$ necessarily satisfies
\begin{equation}\label{ch2e5}
\bar{g}^{-1}(dz,dz)\restr{\mathscr{I}}=-\frac{2\Lambda}{(d-1)(d-2)}=\frac{1}{R^2},
\end{equation}
where $R$ is the AdS radius of $(\mathscr{M},g)$, defined in Chapter \ref{ch1}.\\

Let us now restrict (\ref{ch2e4}) to the timelike hypersurfaces $\mathscr{I}_\epsilon'\doteq 
z^{-1}(\epsilon')$, $0\leq\epsilon'<\epsilon$. First, we'll use the freedom we have in the
choice of $z$ to define a system of Gaussian normal coordinates in $\mathscr{U}_\epsilon$:

\begin{lemma}\label{ch2l1}
Given any AdS-type metric $g$ satisfying the \textsc{Einstein} equations (\ref{ch2e3}) in 
$\mathscr{U}_\epsilon$ for some $\epsilon>0$, we can choose $z$ and $\epsilon$ in a way that 
${}^{(*)}\,\bar{g}^{-1}(dz,dz)\doteq\zeta^2$ is constant in $\mathscr{U}_\epsilon$. In particular, 
$\zeta^2=R^{-2}$.
\begin{quote}{\small\scshape Proof.\quad}
{\small\upshape Let $z'=e^\sigma z$ and $\bar{g}'=e^{2\sigma}\bar{g}$; then, $dz'=e^\sigma(dz+zd
\sigma)$, and hence \[\bar{g}'^{-1}(dz',dz')=\bar{g}^{-1}(dz+zd\sigma,dz+zd\sigma)=\]\[=\bar{g}^{-1}
(dz,dz)+2z\bar{g}^{-1}(dz,d\sigma)+z^2\bar{g}^{-1}(d\sigma,d\sigma).\] Thus, the condition ${}^{(*)}$ 
for $dz'$ is equivalent to the first-order nonlinear partial differential equation \[2z\bar{g}^{-1}
(dz,d\sigma)+z^2\bar{g}^{-1}(d\sigma,d\sigma)=\frac{\zeta^2-\bar{g}^{-1}(dz,dz)}{z},\] where the
right hand side extends in a$\mathscr{C}^\infty$ way to $z=0$ due to (\ref{ch2e5}), which, on its
turn, fixes $\zeta^2=R^{-2}$. By the method of characteristics \cite{john} (see also Subsection 6.4 
of \cite{hormander3}), there exists $\epsilon>0$ and a unique $\mathscr{C}^\infty$ solution
of this equation in $\mathscr{U}_{2\epsilon}$ for arbitrary values of $\sigma\restr{\mathscr{I}}$. 
The result follows globally by employing a partition of unity subordinated to the covering 
$\{\mathscr{U}_{2\epsilon},\mathscr{M}\smallsetminus\overline{\mathscr{U}_\epsilon}\}$ of 
$\overline{\mathscr{M}}$.~\hfill~$\Box$}
\end{quote}
\end{lemma}

Applying the choice of $z$ dictated by Lemma \ref{ch2l1} to (\ref{ch2e4}), we see that the last term  
of the left hand side cancels the right hand side, resulting in
\begin{equation}\label{ch2e6}
\mbox{Ric}(\bar{g})_{ab}+\frac{d-2}{z}\bar{\nabla}_a\bar{\nabla}_bz+\bar{g}_{ab}\bar{g}^{cd}\frac{1}{z}
\bar{\nabla}_c\bar{\nabla}_dz=0.
\end{equation}

Notice yet that, by formulae (\ref{ap1e21}--\ref{ap1e23}) (pages \pageref{ap1e21}--\pageref{ap1e23}), 
the multiplication of $\bar{g}$ by a constant $K=\kappa^2>0$ doesn't modify neither the \textsc{Levi-Civita}
connection, nor the \textsc{Riemann} and \textsc{Ricci} tensors. Hence, we can make the substitution 
$g\mapsto-\frac{(d-1)(d-2)}{2\Lambda}g=R^2g$ and, simultaneously, take $\Lambda=-\frac{(d-1)(d-2)}{2}$, 
which results in $\bar{g}^{-1}(dz,dz)=1$ in $\mathscr{U}_\epsilon$. In this case, denoting by $Z_a\doteq
\bar{\nabla}_az$ the (unit) normal to the foliation of $\mathscr{U}_\epsilon$ induced by our choice of $z$, 
it follows that $K_{ab}\doteq-\bar{\nabla}_aZ_b$ is the extrinsic curvature of $\mathscr{I}_z$. We can, thus, 
write $\bar{g}$ as \[\bar{g}(x,z)_{ab}=\bar{g}^{(0)}(x,z)_{ab}+Z_aZ_b,\] where $\bar{g}^{(0)}(z,.)$ is the 
metric induced in each of these hypersurfaces ($\bar{g}^{(0)}(x,0)\doteq\bar{g}^{(0)}(x)$). Noticing that 
\[Z^aK_{ab}=-(\bar{\nabla}^az)\bar{\nabla}_a\bar{\nabla}_bz=-(\bar{\nabla}^az)\bar{\nabla}_b\bar{\nabla}_az
=-\frac{1}{2}\bar{\nabla}_b(\bar{\nabla}^az\bar{\nabla}_az)=0,\] we can project (\ref{ch2e6}) to its part 
tangent to the $\mathscr{I}_z$'s:

\begin{equation}\label{ch2e7}
\bar{g}^{(0)}(z)^c_a\bar{g}^{(0)}(z)^d_b\mbox{Ric}(\bar{g})_{cd}-\frac{d-2}{z}K_{ab}-\frac{1}{z}
\bar{g}^{(0)}(z)_{ab}\mbox{Tr}K=0.
\end{equation}
where $\mbox{Tr}K\doteq\bar{g}^{(0)}(z)^{cd}K_{cd}$ is the mean curvature. Finally, applying  
the \textsc{Gauss} equation (\ref{ap1e2}) to $\mbox{Ric}(\bar{g})$
\begin{equation}\label{ch2e8}
\bar{g}^{(0)}(z)^c_a\bar{g}^{(0)}(z)^d_b\mbox{Ric}(\bar{g})_{cd}=\mbox{Riem}(\bar{g})_{acbd}Z^cZ^d+
\mbox{Ric}(\bar{g}^{(0)})_{ab}+K^e_aK_{eb}-(\mbox{Tr}K)K_{ab},
\end{equation}
and making use of the identity (analogous to the \textsc{Riccati} equation (\ref{ap1e5}))
\begin{equation}\label{ch2e9}
\mbox{Riem}(\bar{g})_{acbd}Z^cZ^d=Z^e\bar{\nabla}_eK_{ab}-K^e_aK_{eb},
\end{equation}
we arrive, after multiplication of both sides by $z$, at 
\begin{equation}\label{ch2e10}
z\underbrace{Z^e\bar{\nabla}_eK_{ab}}_{=-\frac{1}{2}\partial^2_z\bar{g}^{(0)}(z)}+z\mbox{Ric}
(\bar{g}^{(0)})_{ab}-z(\mbox{Tr}K)K_{ab}-(d-2)K_{ab}-\bar{g}^{(0)}(z)_{ab}\mbox{Tr}K=0,
\end{equation}
which is the equation of evolution we sought for.\footnote{This equation is slightly different from 
the equation of evolution derived in \cite{fefferman} and studied in \cite{graham,grahamlee,dehsks}. 
In fact, \textsc{Fefferman} and \textsc{Graham} depart from considerations different from ours in
\cite{fefferman} to arrive at their analog of (\ref{ch2e10}).} Notice, however, that (\ref{ch2e10})
possesses constraints, for the contraction of the \textsc{Einstein} equations (\ref{ch2e3}) with any 
vector $X^a$ doesn't have second-order derivatives in the direction of $X^a$. The contraction of 
(\ref{ch2e6}) and (\ref{ch2e8}) with $\bar{g}^{(0)ab}$, plus the \textsc{Codazzi-Mainardi} equation
(\ref{ap1e3}) together with (\ref{ch2e6}), result respectively in the constraint equations 
(\ref{ch2e11}) and (\ref{ch2e12}) for the evolution of $(\bar{g}^{(0)},K)$ dictated by (\ref{ch2e10}):

\begin{eqnarray}
\label{ch2e11} R(\bar{g}^{(0)})+K_{ab}K^{ab}-(\mbox{Tr}K)^2+2\frac{d-2}{z}\mbox{Tr}K & = & 0,\\
\label{ch2e12} \bar{\nabla}^{(0)}_aK^a_b-\bar{\nabla}^{(0)}_b\mbox{Tr}K & = & 0.
\end{eqnarray}

The conservation of the constraints (\ref{ch2e11}--\ref{ch2e12}) along the evolution is guaranteed 
by the contracted \textsc{Bianchi} identities $\nabla^aG_\Lambda(g)_{ab}=0$, re-expressed in terms of 
$\bar{g}$.\\

(\ref{ch2e10}) constitutes a nonlinear, second-order Fuchsian system (see, for instance, \cite{kiche} 
for a definition of a Fuchsian system in the nonlinear case), which can be formally solved by 
\emph{initially} adopting the \textsc{Taylor}-series \emph{ansatz} 
\begin{equation}\label{ch2e13}
\bar{g}^{(0)}(z)=\bar{g}^{(0)}+\sum^\infty_{j=1}z^j\bar{g}^{(j)}.
\end{equation}

Notice that by taking $z=0$ in (\ref{ch2e10}), we have $K_{ab}\restr{\mathscr{I}}=0$, 
thus fixing our initial data $(g^{(0)},0)$. The coefficients $g^{(j)}$ are obtained 
by applying $\partial^{j-1}_z$ to both sides of (\ref{ch2e10}), resulting in 
\begin{eqnarray}\label{ch2e14}
\frac{d-1-j}{2}\partial^j_z\bar{g}^{(0)}(z)_{ab}+\frac{1}{2}\bar{g}^{(0)}(z)_{ab}\bar{g}^{(0)}
(z)^{cd}\partial^j_z\bar{g}^{(0)}(z)_{cd} & = & \\ =(\mbox{just terms with }\partial^k_z\bar{g}^{(0)}
(z),\,k<j). & & \nonumber
\end{eqnarray}

For future use, it's convenient to separate (\ref{ch2e12}--\ref{ch2e14}) into the trace part 
$\mbox{Tr}K$ and the trace-free part $P_{ab}\doteq K_{ab}-\frac{1}{d-1}\bar{g}^{(0)}(z)_{ab}\mbox{Tr}K$
\begin{eqnarray}
\label{ch2e15} (d-2-j)P^{(j)a}_b & = & \mbox{Ric}(\bar{g}^{(0)})^{(j-1)a}_b-\frac{1}{d-1}R
(\bar{g}^{(0)})^{(j-1)}\delta^a_b+ \\ & & -\sum^{j-1}_{m=0}(\mbox{Tr}K)^{(m)}P^{(j-1-m)a}_b,\nonumber\\
\label{ch2e16} (2d-3-j)(\mbox{Tr}K)^{(j)} & = & R(\bar{g}^{(0)})^{(j-1)}-\sum^{j-1}_{m=0}(\mbox{Tr}
K)^{(m)}(\mbox{Tr}K)^{(j-1-m)},
\end{eqnarray}
whence we can obtain by invoking the identity $\partial_z\bar{g}^{(0)}(z)_{ab}=-2\partial_z
\bar{g}^{(0)}(z)_{bc}K^c_a$ the explicit form of (\ref{ch2e13})
\begin{equation}\label{ch2e17}
j\bar{g}^{(j)}=-2\sum^{j-1}_{m=0}\left(\bar{g}^{(m)}_{bc}P^{(j-1-m)c}_a+\frac{1}{d-1}\bar{g}^{(m)}_{ab}
K^{(j-1-m)}\right).
\end{equation}

The ``Hamiltonian'' constraint (\ref{ch2e11}) guarantees the invariance of the solution of (\ref{ch2e10}), 
up to diffeomorphisms, under reparametrizations of the variable $z$, and doesn't exert any direct role 
in the determination of the coefficients of the expansion (\ref{ch2e13}).\\

Hence, we see that $\bar{g}^{(j)}$, $j<d-1$, is uniquely determined by $\bar{g}^{(0)}$ and its 
tangential derivatives \emph{at} $\mathscr{I}$ (in particular, if $\bar{g}^{(0)}$ is the 
ESU metric, these coefficients must match the ones of the pure AdS metric in a similar chart 
in a neighbourhood of $\mathscr{I}$). In particular, as (\ref{ch2e10}) is invariant under the 
parity transformation $z\mapsto -z$, it follows that $\bar{g}^{(j)}=0$ for $j$ odd, $j<d-1$. For 
$j=d-1$, the recursion relations given by (\ref{ch2e14}) are truncated, and, in this case, 
we have two possible scenarios:

\begin{itemize}
\item[(i) $d$ \textsc{even}: ] Once more, parity invariance demands $\mbox{Tr}\bar{g}^{(d-1)}
=0$, but the trace-free part of $\bar{g}^{(d-1)}$ is only subject to the conservation law 
$\bar{\nabla}^{(0)a}\bar{g}^{(d-1)}_{ab}=0$ coming from (\ref{ch2e12}).
\item[(ii) $d$ \textsc{odd}: ] (\ref{ch2e14}) nontrivially fixes the trace of $\bar{g}^{(d-1)}$, 
and (\ref{ch2e12}) results in $\bar{\nabla}^{(0)a}\bar{g}^{(d-1)}_{ab}=\bar{\nabla}^{(0)}_b\mbox{Tr}
\bar{g}^{(d-1)}$, thus prescribing the divergence of the trace-free part.
\end{itemize}

In the case of $d$ \emph{odd}, it's necessary to follow the usual strategy of solution of Fuchsian
systems, by modifying the initially adopted \emph{ansatz} so as to include terms in the form $z^k\log 
z,\,k\geq d-1$ (such terms don't exist for $d$ even). Such an analysis was originally sketched by 
\textsc{Fefferman} and \textsc{Graham} \cite{fefferman} for  $\bar{g}^{(0)}$ with arbitrary signature, 
having been analyzed in the case of Euclidean signature by several authors, among which we cite in 
particular the works of \textsc{Graham} and \textsc{Lee} \cite{grahamlee} and \textsc{Graham} 
\cite{graham}. Of great importance was the calculation of the coefficients $\bar{g}^{(j)}$, $j<d-1$, 
and the coefficient $\bar{h}^{(d-1)}$ of the term proportional to $z^{d-1}\log z$, quite involved and 
realized in detail for certain values of $d$ by \textsc{de Haro, Skenderis} and \textsc{Solodukhin} 
\cite{dehsks}. Thus proceeding, the expansion of $\bar{g}$ around $z=0$, suitably corrected by the 
above considerations, becomes
\begin{eqnarray}
\mbox{(i) $d$ \textsc{even:}}\quad \bar{g}(z,x)&\!\sim_{z\searrow 0}\!&dz^2+\bar{g}^{(0)}(x)+(z^2)
\bar{g}^{(2)}(x)+\cdots+(z^2)^{\left(\frac{d-2}{2}\right)}\bar{g}^{(d-2)}(x)+\nonumber\\ & & +z^{d-1}
\bar{g}^{(d-1)}+\cdots;\label{ch2e18}\\
\mbox{(ii) $d$ \textsc{odd:}}\quad \bar{g}(z,x)&\!\sim_{z\searrow 0}\!&dz^2+\bar{g}^{(0)}(x)+(z^2)
\bar{g}^{(2)}(x)+\cdots+(z^2)^{\left(\frac{d-3}{2}\right)}\bar{g}^{(d-3)}(x)+\nonumber\\ & & +(z^{d-1}
\log z)\bar{h}^{(d-1)}+z^{d-1}\bar{g}^{(d-1)}+\cdots.\label{ch2e19}
\end{eqnarray}
(we understand (\ref{ch2e18}) and (\ref{ch2e19}), for now, in a purely asymptotic way -- the question  
of convergence will be discussed later). The coefficient $\bar{h}^{(d-1)}$, denominated 
\textsc{Fefferman-Graham} tensor, possesses remarkable properties: as $\bar{g}^{(j)}$ with $j<d-1$, it 
only depends on $\bar{g}^{(0)}$ and its derivatives tangential to $\mathscr{I}$; it's conformally 
invariant; and vanishes if and only if the conformal class of $g^{(0)}$ has a representative which 
solves the \textsc{Einstein} equations without matter and zero cosmological constant (as, for instance, 
$ESU_{d-1}$). In the latter case, there are no coefficients in the form $z^j\log z$ in (\ref{ch2e19}), 
but this clearly ceases to be true for perturbations of $g^{(0)}$ with compact support and not coming
from local conformal rescaling, which affects the calculation of energy-momentum tensors for field
theories at the boundary obtained by the AdS/CFT correspondence, causing an anomaly in trace of this 
tensor to appear and, hence, breaking scale invariance. Such an anomaly, in this context, is called 
\emph{holographic} \textsc{Weyl} \emph{anomaly}.\cite{hesken1,hesken2}\\

Here, instead of detailing these calculations (for which we shall have no use in the remaining of the 
present work, and which can be found in the above cited references), we shall explore in more detail the
remaining components of $\mbox{Riem}(\bar{g})$, so as to make more precise the part of $\bar{g}^{(d-1)}$ 
not determined by the recursion relations (\ref{ch2e14}). First, we'll simplify (\ref{ch2e6}) 
in terms of $K_{ab}$. For such, we introduce \textsc{Schouten} \emph{tensor} associated to $\bar{g}$ 
\[S(\bar{g})_{ab}\doteq\frac{1}{d-2}\left(\mbox{Ric}(\bar{g})_{ab}-\frac{1}{2(d-1)}R(\bar{g})\bar{g}_{ab}
\right).\] This tensor has a simpler conformal transformation rule and a simpler relation involving 
$\mbox{Riem}(\bar{g})$ and $C(\bar{g})$ than the \textsc{Ricci} tensor; the latter becomes 
\begin{equation}\label{ch2e20}
\mbox{Riem}(\bar{g})_{acbd}=C(\bar{g})_{acbd}+2\bar{g}_{a[b}S(\bar{g})_{c]d}-2\bar{g}_{c[b}S(\bar{g})_{d]a},
\end{equation}
whereas (\ref{ch2e6}) becomes 
\begin{equation}\label{ch2e21}
S(\bar{g})_{ab}=\frac{1}{z}K_{ab}.
\end{equation}

Gathering (\ref{ch2e9}), (\ref{ch2e20}) and (\ref{ch2e21}) together, we have
\begin{equation}\label{ch2e22}
C(\bar{g})_{acbd}Z^cZ^d=\left(\partial_z-\frac{1}{z}\right)K_{ab}-K^c_aK_{cb}.
\end{equation}

Using th expansion of $\bar{g}^{(0)}(z)$ (\ref{ch2e13}) and denoting by $K^{(j)}_{ab}\doteq-\frac{j+1}{2}
\bar{g}^{(j+1)}$ the $j$-th coefficient of the corresponding expansion for $K_{ab}$, we obtain the 
coefficients of the expansion of $C(\bar{g})_{acbd}Z^cZ^d$ around $z=0$:

\begin{equation}\label{ch2e23}
(C(\bar{g})_{acbd}Z^cZ^d)^{(j)}=jK^{(j+1)}_{ab}+\sum^j_{m=0}K^{(j-m)c}_a K^{(m)}_{cb}.
\end{equation}

(\ref{ch2e23}) explicitly shows that the information lost by the truncation of the recursion relations 
(\ref{ch2e14}) at $j=d-1$, referring to the trace-free part of $\bar{g}^{(d-1)}_{ab}$, is contained in 
$(C(\bar{g})_{acbd}Z^cZ^d)^{(d-3)}$. Indeed, if we are capable of determining $\bar{g}^{(d-1)}_{ab}$ by
other means, we can proceed to higher orders and \emph{completely} determine the coefficients of the 
expansion (\ref{ch2e14}). Thus, we can say that the``correct'' initial data for (\ref{ch2e10}) are 
$(\bar{g}^{(0)},\bar{g}^{(d-1)})$.\\

We shall now restrict ourselves to the class of AAdS metrics, i.e., $\bar{g}^{(0)}$ is the metric of 
$ESU_{d-1}$. In this case, the coefficients $\bar{g}^{(j)}$, $j<d-1$, must match, after an adequate
choice of coordinates in $\mathscr{I}$, with those of the metric of the conformal completion of $AdS_d$ 
(\ref{ch1e15}). Hence, a general AAdS metric has in $\mathscr{U}_\epsilon$, after normalization of the 
cosmological constant, the form
\begin{equation}\label{ch2e24}
ds^2=\frac{1}{z^2}\left[-\left(1+\frac{z^2}{4}\right)^2dt^2+dz^2+\left(1-\frac{z^2}{4}\right)^2d
\Omega^2_{d-2}+O(z^{d-1})\right].
\end{equation}

Thus we have from (\ref{ch2e20}), (\ref{ch2e21}) and (\ref{ch2e24}) the following rough ``peeling'' 
property for the \textsc{Weyl} tensor: $C(\bar{g})^{(j)}_{abcd}=0$ for $j<d-3$, and, due to the 
\textsc{Codazzi-Mainardi} equation (\ref{ap1e3}) (page \pageref{ap1e3}), $(\bar{g}^{(0)}(.)^e_a\bar{g}^{(0)}
(.)^f_b\bar{g}^{(0)}(.)^h_c Z^dC(\bar{g})_{abcd})^{(j)}=0$ for $j\leq d-3$. In $AdS_d$, $d\geq 6$, we know 
that $\bar{g}^{(d-1)}_{ab}=0$ and, as $AdS_d$ is a space form, $C(\bar{g})_{abcd}=0$. Therefore, the sum in
the right hand side of (\ref{ch2e23}) also vanishes. As it's the same for all AAdS metrics and all $j\leq 
d-3$, it vanishes for all these cases. Hence, 
\begin{equation}\label{ch2e25}
K^{(d-2)}_{ab}=\frac{1}{d-3}(C(\bar{g})_{acbd}Z^cZ^d)^{(d-3)},\,\forall d\geq 6.
\end{equation}

In the case $d=4$, the sum in the right hand side of (\ref{ch2e23}) reduces to $2K^{(1)c}_a K^{(0)}_{cb}=0$,  
and for $d=5$, $K^{(1)c}_aK^{(1)}_{cb}=\frac{1}{8}\bar{g}^{(0)}_{ab}$, by virtue of (\ref{ch2e15}--\ref{ch2e17}). 
Thus, we conclude that
\begin{equation}\label{ch2e26}
\bar{g}^{(d-1)}=\left\{\begin{array}{lr} -\frac{2}{d-1}E_{ab} & (d=4\mbox{ or }d\geq 6)\\ -\frac{1}{2}
E_{ab}+\frac{1}{16}\bar{g}^{(0)}_{ab} & (d=5) \end{array}
\right.,
\end{equation}
where $E_{ab}=\frac{1}{d-3}\lim_{z\searrow 0}z^{3-d}C(\bar{g})_{acbd}Z^c Z^d$ is the rescaled electric part
of the \textsc{Weyl} tensor of $\bar{g}$ (the limit exists and is $\mathscr{C}^\infty$, for we've seen that 
the coefficients of order inferior to $d-3$ of the expansion of $C(\bar{g})_{acbd}Z^c Z^d$ vanish), which
leaves a general AAdS metric satisfying (\ref{ch2e3}) in the form
\begin{eqnarray}\label{ch2e27}
\bar{g}^{(0)}(z) & = & -\left(1+\frac{z^2}{4}\right)^2dt^2+dz^2+\left(1-\frac{z^2}{4}\right)^2d
\Omega^2_{d-2}+\\ & & +\left\{\begin{array}{lr}-\frac{2}{d-1}E_{ab}z^{d-1} & (d=4\mbox{ or }d\geq 6)\\
\left(-\frac{1}{2}E_{ab}+\frac{1}{16}\bar{g}^{(0)}_{ab}\right)z^4 & (d=5)\end{array}\right\}
+O(z^d).\nonumber
\end{eqnarray}

\begin{remark}\label{ch2r1}
The problem of convergence of the expansions (\ref{ch2e18}--\ref{ch2e19}) for $z<\epsilon$ sufficiently 
small was affirmatively solved, for $\bar{g}^{(0)}_{ab}$ and $E_{ab}$ \emph{real analytic}, 
by \textsc{Kichenassamy} \cite{kiche}. In the non analytic case, it's not possible, in the present state 
of the art, to proceed by means of analytic approximations, due to the lack of sufficiently strong
estimates of the solution in terms of these boundary data. This, at a first glance, was to be expected, 
for the \textsc{Cauchy} problem for hyperbolic equations is know to be ill posed for initial data at 
timelike hypersurfaces \cite{hadamard}. However, such an intuition is possibly only partially correct, 
for the ``true'' initial data for our problem (i.e., $\bar{g}^{(0)}_{ab}$ \emph{and} $E_{ab}$) make the latter 
considerably different from the usual \textsc{Cauchy} problem for second-order hyperbolic equations 
(remember that $\partial_z\bar{g}^{(0)}(z)\restr{z=0}=0$!) -- indeed, weighted estimates involving the 
\textsc{Weyl} tensor exert a fundamental role in the proof global nonlinear stability of \textsc{Minkowski} 
spacetime \cite{christoklai}. Partial results in this direction can be found in \cite{anderson3,anderson4,
anderson5,anderson6} and the references cited therein. Hence, in non analytic cases, we should understand 
the expansions (\ref{ch2e18}--\ref{ch2e19}) only as asymptotic expansions as $z\searrow 0$.
\end{remark}

Notice, finally, that, by virtue of the purely local character of the analysis made above, it remains 
valid without change for any \textsc{Poincar\'e} domain in an AAdS spacetime and, more in general, for any
locally AAdS spacetime (the whole discussion preceding formula (\ref{ch2e24}) is valid even for general 
locally AdS-type spacetimes). In the case of \textsc{Poincar\'e} domains, we must substitute the coordinates 
(\ref{ch1e8}) for the coordinates (\ref{ch1e6}), and the metric (\ref{ch1e9}) multiplied by $z^2$ for the 
metric (\ref{ch1e15}). An advantage of this chart in the present context is that $\bar{g}^{(j)}_{ab}=0$ for
$1\leq j\leq d-2$, whence it follows that $\bar{g}^{(d-1)}_{ab}=\frac{2}{d-1}E_{ab}$ for all $d\geq 4$, 
unlike (\ref{ch2e26}). The asymptotic form of an AAdS metric in the neighbourhood of a \textsc{Minkowski} 
domain then becomes
\begin{equation}\label{ch2e28}
\bar{g}^{(0)}(z)_{ab}=\left(\eta_{ab}-\frac{2}{d-1}z^{d-1}E_{ab}\right)+O(z^d).
\end{equation}

\section{\label{ch2-thermo}Gravitational ``thermodynamics'' in AAdS wedges}

We'll show how to associate to each wedge $\mathscr{W}_{p,q}$ of a locally AAdS spacetime 
$(\mathscr{M},g)$, $p,q\in\mathscr{D}(\mathscr{I})$, a flow of diffeomorphisms which extend 
to $\mathscr{D}_{p,q}$ in such a way that their action coincides in this region with the action
of the one-parameter subgroup (\ref{ch1e26}). In this case, it follows that these diffeomorphisms 
are \emph{asymptotic isometries}, i.e., if $T^a$ is the vector field tangent to the flow, then 
$\lie{T}g(x,z)=o(z)$ as $z\searrow 0$.\\

The construction we'll make is a particular case of a more general procedure, suitable to any
relatively compact diamond $\mathscr{O}_{p,q}=I^+(p)\cap{} I^-(q)$ in a causally simple spacetime 
$(\mathscr{M},g)$, closely related to the method of \textsc{Geroch} \cite{geroch1} for the construction 
of global time functions in globally hyperbolic spacetimes.

\subsection{\label{ch2-thermo-diam}Geometrical time evolution in diamonds}

We'll show initially how to associate to each diamond $\mathscr{D}_{p,q}$ at conformal infinity
of a locally AAdS spacetime $(\mathscr{M},g)$, $p,q\in\mathscr{D}(\mathscr{I})$ a global time 
function associated to the action of the one-parameter subgroup (\ref{ch1e26}). 

\begin{theorem}\label{ch2t1}
Consider a $d$-dimensional AAdS spacetime $(\mathscr{M},g)$, $r\in\mathscr{I}$ and $p_0,q_0\in
\mathscr{M}in(r)$ such that, if $\mathscr{M}in(r)\ni p\mapsto x^\mu(p)$ is the global Cartesian
chart of \textsc{Minkowski} spacetime, we have $x^\mu(p_0)=(-1,\mathbf{0})$ and $x^\mu(q_0)=(1,
\mathbf{0})$. Consider the one-parameter group $u^\lambda_{p_0,q_0}$ of conformal diffeomorphisms
(\ref{ch1e26}) of $(\mathscr{D}_{p_0,q_0},\eta\restr{\mathscr{D}_{p_0,q_0}})$, given for $p$ in the 
$x^0-x^1$ plane by \[x^{\pm}(u^\lambda_{p_0,q_0}(p))\doteq\frac{(1+x^{\pm})-e^{-\lambda}(1-x^{\pm})
}{(1+x^{\pm})+e^{-\lambda}(1-x^{\pm})},\,\lambda\in\mathbb{R}\] ($x^{\pm}=x^0\pm x^1$ are the light cone 
coordinates in the $x^0-x^1$ plane) and extended to $p\in\mathscr{D}_{p_0,q_0}$ by spatial rotations. 
Define the following foliation $\lambda\mapsto\Sigma^\lambda_{p_0,q_0}$ of $\mathscr{D}_{p_0,q_0}$ by
\textsc{Cauchy} surfaces:

\begin{itemize}
\item $\Sigma_{p_0,q_0}\doteq\Sigma^0_{p_0,q_0}=\mathscr{D}_{p_0,q_0}\cap{}\{x^0=0\}$;
\item $\Sigma^\lambda_{p_0,q_0}=u^\lambda_{p_0,q_0}(\Sigma^0_{p_0,q_0})$, i.e., the foliation is given 
by the diffeomorphism $F_{p_0,q_0}:\mathbb{R}\times\Sigma_{p_0,q_0}\ni(\lambda,p)\mapsto u^\lambda_{p_0,
q_0}(p)\in\mathscr{D}_{p_0,q_0}$.
\item $\lambda$ is, hence, the global time function associated to $F_{p_0,q_0}$.
\end{itemize}

Then, 
\begin{equation}\label{ch2e29}
\lambda=\frac{1}{d-1}\log\left[\frac{|\mathscr{D}_{p_0,u^\lambda_{p_0,q_0}(p)}|}{|
\mathscr{D}_{u^\lambda_{p_0,q_0}(p),q_0}|}\right]=\log\left[\frac{d_\eta(p_0,
u^\lambda_{p_0,q_0}(p))}{d_\eta(u^\lambda_{p_0,q_0}(p),q_0)}\right],\,\forall p\in
\Sigma_{p_0,q_0},
\end{equation} 
where $d_\eta$ is the Lorentzian distance associated to $\eta$ (see Appendix \ref{ap1}).
Recall that the \textsc{Lebesgue} measure is precisely the measure induced by the volume element
 $\sqrt{|\eta|}$ (we notice, for convenience, that $|\mathscr{D}_{p_0,q_0}|=2\int^1_0|B_r(0)|dr=2
\int^1_0\frac{r^{d-2}}{d-2}\mbox{\upshape Vol}S^{d-3}dr=\frac{2}{(d-1)(d-2)}\mbox{\upshape Vol}S^{d-3}$).
\begin{quote}{\small\scshape Proof.\quad}
{\small\upshape Due to the rotational symmetry of $u^\lambda_{p_0,q_0}$, we can restrict our 
considerations to the $x^0-x^1$ plane. In this case, let us write $x^\mu(p)=(0,x^1,0,\ldots,0)$, 
$x^1\in[-1,1]$, whence it follows that \[x^{\pm}(u^\lambda_{p_0,q_0}(p))=\frac{(1\pm x^1)-e^{-\lambda}
(1\mp x^1)}{(1\pm x^1)+e^{-\lambda}(1\mp x^1)},\] and, thus, 
\begin{eqnarray}
x^0(u^\lambda_{p_0,q_0}(p)) & = & \frac{1}{2}(x^+(u^\lambda_{p_0,q_0}(p))+x^-(u^\lambda_{p_0,q_0}
(p)))=\nonumber\\ & = & \frac{(1-(x^1)^2)(1-e^{-2\lambda})}{(1-(x^1)^2)(1+e^{-2\lambda})+2e^{-\lambda}
(1+(x^1)^2)}\nonumber
\end{eqnarray}
and 
\begin{eqnarray}
x^1(u^\lambda_{p_0,q_0}(p)) & = & \frac{1}{2}(x^+(u^\lambda_{p_0,q_0}(p))-x^-(u^\lambda_{p_0,q_0}
(p)))=\nonumber\\ & = & \frac{4x^1e^{-\lambda}}{(1-(x^1)^2)(1+e^{-2\lambda})+2e^{-\lambda}
(1+(x^1)^2)}.\nonumber
\end{eqnarray}

Let $q=(t,\mathbf{0})$, $t\in(-1,1)$. The diamond $\mathscr{D}_{p_0,q}$ is a translation
of $\frac{1+t}{2}\mathscr{D}_{p_0,q_0}$, and the diamond $\mathscr{D}_{q,q_0}$, a translation
of $\frac{1-t}{2}\mathscr{D}_{p_0,q_0}$ -- hence, we have $|\mathscr{D}_{p_0,q}|=\left(\frac{1+t}{2}
\right)^{d-1}|\mathscr{D}_{p_0,q_0}|$ and $|\mathscr{D}_{q,q_0}|=\left(\frac{1-t}{2}\right)^{d-1}
|\mathscr{D}_{p_0,q_0}|$. More in general, if $q=(t,r,0,$\\$\ldots,0)$, there exists a \textsc{Lorentz} 
boost in the $x^0-x^1$ plane around $p_0$ which takes $\mathscr{D}_{p_0,q}$ to $\mathscr{D}_{p_0,q^+}$, 
where $q^+=(((1+t)^2-r^2)^{\frac{1}{2}}-1,\mathbf{0})$, and a \textsc{Lorentz} boost in the $x^0-x^1$ 
plane around $q_0$ which takes $\mathscr{D}_{p_0,q}$ to $\mathscr{D}_{q^-,q_0}$, where $q^-=(1-((1-t)^2-
r^2)^{\frac{1}{2}},\mathbf{0})$. As \textsc{Lorentz} transformations preserve volume, we have 
\[|\mathscr{D}_{p_0,q}|=\left(\frac{((1+t)^2-r^2)^{\frac{1}{2}}}{2}\right)^{d-1}|\mathscr{D}_{p_0,q_0}|\] 
and \[|\mathscr{D}_{q,q_0}|=\left(\frac{((1-t)^2-r^2)^{\frac{1}{2}}}{2}\right)^{d-1}|\mathscr{D}_{p_0,q_0}|.\] 
Finally, taking $q=q(\lambda)=u^\lambda_{p_0,q_0}(p)$, it follows that \[(1+x^0(q(\lambda)))^2-x^1(q
(\lambda))^2=\frac{4(1-(x^1)^2)}{(1-(x^1)^2)(1+e^{-2\lambda})+2e^{-\lambda}(1+(x^1)^2)}\] and \[(1-x^0(q
(\lambda)))^2-x^1(q(\lambda))^2=\frac{4e^{-2\lambda}(1-(x^1)^2)}{(1-(x^1)^2)(1+e^{-2\lambda})+2e^{-\lambda}
(1+(x^1)^2)},\] and hence formula (\ref{ch2e29}). The second identity follows from the fact that 
$|\mathscr{D}_{p,q}|=\frac{1}{2^{d-2}(d-1)(d-2)}\mbox{\upshape Vol}S^{d-3}d_\eta(p,q)^{d-1}$.~\hfill~$\Box$}
\end{quote}

\end{theorem}

The importance of Theorem (\ref{ch2t1}) will soon become clear.\\

It's important to have in mind that, in $d$-dimensional, strongly causal spacetimes $(\mathscr{M},g)$, 
the Lorentzian distance $d_g$ coincides with the geodesic distance in convex normal neighbourhoods 
(see formula \ref{ch2e36}, page \pageref{ch2e36}, and the discussion that follows for more details). \\

Let us suppose now that $(\mathscr{M},g)$ is causally simple, so as any nonvoid, relatively compact 
diamond $\mathscr{O}_{p,q}$ is globally hyperbolic, whose \textsc{Cauchy} surfaces $\mathscr{S}_{p,q}$ are
acausal sets with edge $\partial I^+(p)\cap\partial I^-(q)$. Notice, however, that if 
$\overline{\mathscr{O}_{p,q}}$ is not contained in any geodesically convex neighbourhood, then 
$\mathscr{O}_{p,q}$ is \emph{not} equal to $D(\mathscr{S}_{p,q})$ \emph{in} $\mathscr{M}$, for in this 
case $\partial\mathscr{O}_{p,q}=(\partial I^+(p)\cap{}J^-(q))\cup(\partial I^-(q)\cap{}J^+(p))$ possesses 
geodesic segments with pairs of conjugate points, that is, there are points in $\partial\mathscr{O}_{p,q}$ 
which belong to null geodesics which cease to be achronal before reaching the edge of $\mathscr{S}_{p,q}$.
Hence, we have that $intD(\mathscr{S}_{p,q})\supsetneqq\mathscr{O}_{p,q}$ in this case (equality occurs 
only if the null geodesic segments which generate $\partial\mathscr{O}_{p,q}$ are achronal, as for 
example in the case that $\overline{\mathscr{O}_{p,q}}$ is contained in some geodesically convex
neighbourhood).\\

Another obstacle to showing that (\ref{ch2e29}) indeed generalises to a global time function for a 
relatively compact $\mathscr{O}_{p,q}$ whose level sets are \textsc{Cauchy} surfaces is that the \emph{second} 
identity in (\ref{ch2e29}) no longer holds if there is curvature present. The intuitive reason is 
that the ``packing'' number of small diamonds inside a larger one need not grow linearly with the 
volume of the latter (this argument can be made rigorous by employing semi-Riemannian volume 
comparison estimates \cite{es}). Hence, \textsc{Geroch}'s argument, which does yield a global
time function with \textsc{Cauchy} level sets, is not directly applicable. However, simple causality happens 
to be just strong enough to guarantee that the \emph{first} identity in (\ref{ch2e29}) does define 
a function with the desired properties.\\

The property we'll use is the reverse triangular inequality (see Appendix \ref{ap1}, page 
\pageref{ap1e22}) \[p\leq r\leq q\Rightarrow d_g(p,q)\geq d_g(p,r)+d_g(r,q).\] 

It follows from this inequality that, if $\gamma:[0,1]\rightarrow\mathscr{M}$ is an inextendible causal
curve in $\mathscr{O}_{p,q}$, then $\lambda\mapsto d_g(\gamma(\lambda),q)$ (resp. $\lambda\mapsto d_g(p,
\gamma(\lambda))$) is a function bounded by $d_g(p,q)$ ($<+\infty$ by virtue of the compactness of 
$\overline{\mathscr{O}_{p,q}}$ and the Definition \ref{ap1d1} of Lorentzian distance) and 
\emph{strictly} decreasing (resp. increasing) in $\lambda$ -- recall that any \emph{maximal} causal curve
(i.e., whose arc length between any two of its points is equal to the Lorentzian distance) is necessarily 
a geodesic, up to reparametrization. Thus, the level surfaces of $d_g(.,q)$ and $d_g(p,.)$ are \emph{acausal} 
for nonzero values. Moreover, thanks to causal simplicity, $d_g(p,.)$ (resp. $d_g(.,q)$) tends to zero 
along any past (resp. future) inextendible causal curve in $\mathscr{O}_{p,q}$. 
Employing the properties obtained above, we can repeat the proof of Proposition 3.1 of \cite{angalho2} 
within our context, arriving at the following

\begin{proposition}[\mdseries\upshape\cite{angalho2}]\label{ch2p1}
$d_g(p,.)$ and $d_g(.,q)$ are \emph{semiconvex}, i.e., for any $r\in\mathscr{O}_{p,q}$ there exists a 
neighbourhood  $\mathscr{U}\ni r$, a local chart $x:\mathscr{U}\rightarrow\mathbb{R}^d$ and $f\in
\mathscr{C}^\infty(\mathscr{U})$ such that $(d_g(p,.)\restr{\mathscr{U}}+f)\circ x^{-1}$ and $(d_g(.,q)
\restr{\mathscr{U}}+f)\circ x^{-1}$ are \emph{convex} in $x(\mathscr{U})$.~\hfill~$\Box$
\end{proposition}

Actually, the result we obtain is the following: there exist $\phi_{p,r},\phi_{r,q}\in\mathscr{C}^\infty(\mathscr{U}$
such that $d_g(p,r)=\phi_{p,r}(r)$ and $d_g(r,q)=\phi_{r,q}(r)$, $d_g(p,.)\geq\phi_{p,r}$ and $d_g(.,q)\geq\phi_{r,q}$
in $\mathscr{U}$ and the Hessians $D^2\phi_{p,r}$ and $D^2\phi_{r,q}$ are such that $D^2\phi_{p,r}(r)-c_{p,r}\mathbb{1}$
and $D^2\phi_{r,q}(r)-c_{r,q}\mathbb{1}$ are positive semidefinite matrices for $c_{p,r},c_{r,q}\in\mathbb{R}$,
which not only implies semiconvexity in the sense of Proposition \ref{ch2p1} \cite{angalho}, but also guarantees 
that the given definition is independent of coordinates. In these circumstances, we can invoke the classical
result of \textsc{Alexandrov}, which tells us that a convex function is not only locally \textsc{Lipschitz}, 
but is also twice differentiable almost everywhere with respect to \textsc{Lebesgue} measure (see \cite{evangar}
for a proof of this fact). Such a result obviously extends to semiconvex functions. As the restriction 
of $\mu_g$ to normal neighbourhoods is absolutely continuous with respect to \textsc{Lebesgue}
measure, we thus obtain the following

\begin{proposition}\label{ch2p2}
$d_g(p,.)$ and $d_g(.,q)$ are locally \textsc{Lipschitz} and twice differentiable almost everywhere.~\hfill~$\Box$
\end{proposition}

After this preparations, we can finally define our global time function in $\mathscr{O}_{p,q}$ 
associated to $d_g$:

\begin{equation}\label{ch2e35}
\mathscr{O}_{p,q}\ni r\mapsto\lambda^g_{p,q}(r)\doteq\log\left[\frac{d_g(p,r)}{d_g(r,q)}\right]\in\mathbb{R}.
\end{equation}

Our argument above shows that $\lambda^g_{p,q}$ is surjective and strictly increasing along any 
inextendible, future-directed causal curve $\gamma:(a,b)\rightarrow\mathscr{O}_{p,q}$ -- as a 
consequence, $\lambda^g_{p,q}(\gamma(t)):(a,b)\rightarrow\mathbb{R}$ is \emph{surjective} and
hence $(\lambda^g_{p,q})^{-1}(\lambda)$ is a \textsc{Cauchy} surface for all $\lambda$, as desired.\\

We'll show now that the asymptotic behaviour of $\lambda^g_{p,q}(r)$ in the case that $\mathscr{O}_{p,q}$ is
contained in a geodesically convex neighbourhood, as 
\begin{enumerate}
\item[(i)] $d_g(p,r)\rightarrow 0$ keeping $d_g(r,q)\neq 0$, and 
\item[(ii)] $d_g(r,q)\rightarrow 0$ keeping $d_g(p,r)\neq 0$. 
\end{enumerate}

The restriction on the size of $\mathscr{O}_{p,q}$ guarantees that $d_g(p,.)$, $d_g(.,q)$ and, hence, 
$\lambda^g_{p,q}$ are $\mathscr{C}^\infty$, but we emphasize that formulae that follow depend only 
on derivatives of order $\leq 2$ of $d_g(p,.)$ and $d_g(.,q)$. We shall later discuss other 
consequences of this hypothesis.\\

Let $\mathscr{U}$ be a geodesically convex neighbourhood. Let us then define the \textsc{Synge}
\emph{world function}

\begin{equation}\label{ch2e36}
\Gamma_g\mathscr{U}\times\mathscr{U}\ni(p,q)\mapsto\Gamma_g(p,q)\doteq-\frac{1}{2}\int^1_0 
g(\dot{\gamma}_{p,q}(s),\dot{\gamma}_{p,q}(s))ds,
\end{equation}
where $\gamma_{p,q}:[0,1]\rightarrow\mathscr{U}$ is the (unique) geodesic segment linking 
$p=\gamma_{p,q}(0)$ to $q=\gamma_{p,q}(1)$. $\Gamma_g$ has the following properties (see
\cite{friedlander} for the proofs):

\begin{itemize}
\item $\Gamma_g\in\mathscr{C}^\infty\left(\bigcup_{p\in\mathscr{M}}\{p\}\times\mathscr{U}_p\right)$, 
where $\mathscr{U}_p$ is an open, geodesically convex neighbourhood of $p$;
\item $\Gamma_g(p,q)=\Gamma_g(q,p)$;
\item $\nabla^a\Gamma_g(p,.)=-\dot{\gamma}^a_{p,.}(.)$ and $\nabla^a\Gamma_g(.,q)=-\dot{\gamma}^a_{.,q}
(.)$, where $.$ denotes the variable on which $\nabla$ acts. We immediately have the fundamental formula 
\begin{equation}\label{ch2e37}
g^{-1}(d_p\Gamma_g(p,q),d_p\Gamma_g(p,q))=g^{-1}(d_q\Gamma_g(p,q),d_q\Gamma_g(p,q))=-2\Gamma_g(p,q),
\end{equation}
where $d_p$ and $d_q$ denote respectively the differential with respect to the first and second variables.
\item $(\nabla_a\nabla_b\Gamma_g(p,.))(p)=-g_{ab}(p)$.
\end{itemize}

First, we must notice that $d_g$ and $2(\Gamma_g)^{\frac{1}{2}}$ coincide in $\mathscr{U}\times 
J^+(\mathscr{U},\mathscr{U})$, for $(\mathscr{M},g)$ is causally simple by hypothesis and, thus, 
strongly causal (see Theorem 4.27 in \cite{beemee}). $(\Gamma_g)^{\frac{1}{2}}$ is continuous but 
non differentiable in $d_g^{-1}(0)\cap{}(\mathscr{U}\times J^+(\mathscr{U},\mathscr{U})$, by virtue of 
the presence of the square root. Hence, we employ the equivalent formula
\begin{equation}\label{ch2e38}
\lambda^g_{p,q}(r)=\frac{1}{2}(\log(d_g(p,r)^2)-\log(d_g(r,q)^2))=\frac{1}{2}(\log(\Gamma_g(p,r))-
\log(\Gamma_g(r,q)))
\end{equation}
instead of (\ref{ch2e35}) in what follows. We assume, from now on, $p$ and $q$ \emph{fixed}
and $\Gamma_g(p,r)$ and $\Gamma_g(r,q)$ as functions of $r$ \emph{alone}, that is, differentials and 
covariant derivatives act \emph{only on} $r$. In this way, we can unclutter the notation quite a bit.\\

We'll first obtain the infinitesimal form of the diffeomorphism flow induced by the global time function
$\lambda^g_{p,q}$:

\begin{equation}\label{ch2e39}
\nabla_a\lambda^g_{p,q}(r)=\frac{1}{2}\left(\frac{\nabla_a\Gamma_g(p,r)}{\Gamma_g(p,r)}-\frac{\nabla_a
\Gamma_g(r,q)}{\Gamma_g(r,q)}\right).
\end{equation}

Let us now consider the vector field $T^a$ in $\mathscr{O}_{p,q}$ uniquely determined by the
following algebraic conditions:

\begin{itemize}
\item $T^a\nabla_a\lambda^g_{p,q}(r)=1$ for all $r\in\mathscr{O}_{p,q}$;
\item $g(T,X)=0$ for all $X^a$ tangent to $(\lambda^g_{p,q})^{-1}(t)$, $t\in\mathbb{R}$.
\end{itemize}

In other words, $T^a$ is the vector field which generates the flow of diffeomorphisms induced by
the foliation of $\mathscr{O}_{p,q}$ by the level hypersurfaces of $\lambda_{p,q}$. Hence, we have
\begin{equation}\label{ch2e40}
g(T,T)=\frac{1}{g^{-1}(d\lambda^g_{p,q},d\lambda^g_{p,q})},
\end{equation}
where
\begin{equation}\label{ch2e41}
g^{-1}(d\lambda^g_{p,q},d\lambda^g_{p,q})=-\frac{1}{2}\left(\frac{1}{\Gamma_g(p,r)}+\frac{1}{
\Gamma_g(r,q)}+\frac{g^{-1}(d\Gamma_g(p,r),d\Gamma_g(r,q))}{\Gamma_g(p,r)\Gamma_g(r,q)}\right).
\end{equation}

Thus,
\begin{eqnarray}\label{ch2e42}
T^a & = & \frac{1}{g^{-1}(d\lambda^g_{p,q},d\lambda^g_{p,q})}\nabla^a\lambda^g_{p,q}=\\
 & = & \frac{\Gamma_g(p,r)\nabla^a\Gamma_g(r,q)-\Gamma_g(r,q)\nabla^a\Gamma_g(p,r)}{\Gamma_g(p,r)+
\Gamma_g(r,q)+g^{-1}(d\Gamma_g(p,r),d\Gamma_g(r,q))}.\nonumber
\end{eqnarray}

It's clear from (\ref{ch2e42}) that, asymptotically as $r\rightarrow p$ and $r\rightarrow q$, that is, 
respectively (---) $d_g(p,r)\searrow 0$, $d_g(r,q)\nearrow d_g(p,q)$ and (+) $d_g(r,q)\searrow 0$, 
$d_g\nearrow d_g(p,q)$, $T^a$ approaches a conformal \textsc{Killing} field, i.e.,
\begin{eqnarray}\label{ch2e43}
\nabla_aT_b+\nabla_bT_a & = & f(.)g_{ab}+o(\min\{d_g(p,r),d_g(r,q)\})=\\ & = & \frac{2}{d}
(\nabla_cT^c)g_{ab}+o(\min\{d_g(p,r),d_g(r,q)\})\nonumber
\end{eqnarray}
as $r\rightarrow p$ or $r\rightarrow q$. More precisely, $T^a$ (resp. $-T^a$) approaches an 
infinitesimal \emph{dilation} around $p$ (resp. $q$). In geodesic normal coordinates $(x^\mu)$ 
around $p$ (resp. $q$), the components of $T^a$ become
\begin{equation}\label{ch2e44}
T^\mu=x^\mu-x^\mu(p)+O((x-x(p))^2)\mbox{ (resp. }x^\mu-x^\mu(q)+O((x-x(q))^2)).
\end{equation}

\subsection{\label{ch2-thermo-iso}Construction of asymptotic isometries}

Once more specializing ourselves to (locally) AdS-type spacetimes $(\mathscr{M},g)$, we see that, due to
the fact that $(\mathscr{I},\bar{g}^{(0)})$ is totally geodesic with respect to $(\overline{\mathscr{M}},\bar{g})$ 
(and, hence, $\frac{1}{2}d_{\bar{g}}^2\restr{\mathscr{I}^2\cap{}\mathscr{U}^2}=\Gamma_{\bar{g}}
\restr{\mathscr{I}^2\cap{}\mathscr{U}^2}=\Gamma_{\bar{g}^{(0)}}\restr{\mathscr{I}^2\cap{}\mathscr{U}^2}$, 
where $\mathscr{U}$ is a geodesically convex neighbourhood in $\overline{\mathscr{M}}$ -- see Proposition 
4.32 in \cite{beemee}), $\lambda^{\bar{g}}_{p,q}\restr{\mathscr{I}}\equiv\lambda^{\bar{g}^{(0)}}_{p,q}$ for
any pair $p\ll_{\mathscr{I}} q\in\mathscr{I}$ such that $\overline{\mathscr{W}_{p,q}}$ is contained in a 
geodesically convex neighbourhood in $(\overline{\mathscr{M}},\bar{g})$, for $\lambda^{\bar{g}}_{p,q}$ is 
constructed solely from $\Gamma_{\bar{g}}$. In particular, if $(\mathscr{M},g)$ is (locally) AAdS, and 
$x(p)=(-1,\mathbf{0})$ and $x(q)=(1,\mathbf{0})$ in a Cartesian chart $x:(\mathscr{M}in(r),\bar{g}^{(0)}
\restr{\mathscr{M}in(r)})\rightarrow(\mathbb{R}^{1,d-2},\eta)$ on a \textsc{Minkowski} domain $\mathscr{M}
in(r)$ form some $r\in\mathscr{I}$, it follows that $\lambda^{\bar{g}}_{p,q}\restr{\mathscr{I}}\equiv\lambda$, 
where $\lambda$ is given by formula (\ref{ch2e29}), which is the conclusion of Theorem \ref{ch2t1}. That
is, $T^a$ \emph{is not only tangent to} $\mathscr{I}$ \emph{as} $T^a\restr{\mathscr{I}}$ \emph{coincides 
with the vector field in} $\mathscr{D}_{p,q}$ \emph{generated by the one-parameter subgroup} $\lambda\mapsto 
u^\lambda_{p_0,q_0}$ \emph{of the conformal group of} $(\mathscr{I},\bar{g}^{(0)})$ \emph{given by 
(\ref{ch1e23})}. Extending the assertion of Theorem \ref{ch2t1} to $(p,q)\in\mathscr{D}(\mathscr{I})$ 
by the commutation relations of $SO_e(2,d-1)$ and recalling that (resp. the connected component to 
the identity of) the conformal group is precisely the set of diffeomorphisms which preserve the causal 
structure (resp., plus orientation and time orientation), we obtain:

\begin{proposition}\label{ch2p3}
Let $(\mathscr{M},g)$ be a locally AAdS spacetime, $(p,q)\in\mathscr{D}(\mathscr{I})$ such that 
$\overline{\mathscr{W}_{p,q}}$ is contained in a geodesically convex neighbourhood in $(\overline{\mathscr{M}},
\bar{g})$. Then $T^a$ is tangent to $\mathscr{I}$ and $\lambda^{\bar{g}}_{p,q}\restr{\mathscr{I}}=
\lambda^{\bar{g}^{(0)}}_{p,q}=\lambda$, where $u^\lambda_{p,q}$ is the isotropy subgroup (\ref{ch1e26}) of 
$\mathscr{D}_{p,q}$ of the conformal group of $(\mathscr{I},\bar{g}^{(0)})$.~\hfill~$\Box$
\end{proposition}

It follows from Proposition \ref{ch2p3} that $T^a$ generates a one-parameter group of \emph{asymptotic
isometries} of $(\mathscr{W}_{p,q},g\restr{\mathscr{W}_{p,q}})$, i.e., $\lie{T}g=o(z)$ as $z\searrow 0$.\\

It's important to notice that, \emph{in the case of} $AdS_d$, we see from the proof of Theorem \ref{ch2t1}  
and by employing \textsc{Poincar\'e} charts that $T^a$ is \emph{precisely} the \textsc{Killing} field
in $\mathscr{W}_{p,q}$ which generates the conformal isotropy subgroup of the latter, for any $(p,q)\in
\mathscr{D}(\mathscr{I})$. Hence, our construction naturally generalizes the \textsc{Killing} fields of
$AdS_d$.\\

Now what happens with ``large'' $\mathscr{W}_{p,q}$'s, i.e., such that $\overline{\mathscr{W}_{p,q}}$ 
is relatively compact in $\overline{\mathscr{M}}$, but \emph{not} contained in some geodesically convex
neighbourhood in $(\overline{\mathscr{M}},\bar{g})$? We have problems of two sorts, both stemming from 
the same source (namely, the critical points of the exponential map), but with different consequences,  
more or less severe for us:

\begin{enumerate}
\item $L_g$ ceases to be well defined, and $d_g$ ceases to be $\mathscr{C}^\infty$ in $d_g(p,.)^{-1}((0,
d_g(p,q)))$ and $d_g(.,q)^{-1}((0,d_g(p,q)))$. However, as we know (Proposition \ref{ch2p2}) that $d_g$ is 
twice differentiable $d_g(p,.)^{-1}((0,d_g(p,q)))$ and $d_g(.,q)^{-1}((0,d_g(p,q)))$ up to some set 
$\mathscr{U}$ satisfying $\mu_g(\mathscr{U})=\mu_{d_g}(\mathscr{U})=0$, all formulae we've obtained remain 
valid in the points where $d_g(p,.)^2$ and $d_g(.,q)^2$ are twice differentiable. We cannot, however, 
guarantee the validity of the limits of expressions involving derivatives of $d_g(.,q)^2$ as 
$r\rightarrow p$, or involving derivatives of $d_g(p,.)^2$ as $r\rightarrow q$, since we know nothing about 
the continuity of these outside geodesically convex neighbourhoods.
\item The second problem is specific of (locally) AAdS spacetimes. It may occur that (some) \emph{maximal} 
(with respect a $d_{\bar{g}}$) timelike geodesic linking $p$ to $q$ traverses $\mathscr{M}$ instead of
remaining at $\mathscr{I}$. This implies not only that $T^a$ is no longer tangent to $\mathscr{I}$, but also 
causes $T^a$ to acquire discontinuity points at $\mathscr{I}$. The hypothesis that such \emph{focusing} of
timelike geodesics doesn't occur is stronger and cannot be obtained as a consequence of all null geodesics 
in $\overline{\mathscr{M}}$ with endpoints in $\mathscr{I}$ and traversing $\mathscr{M}$ having a pair of 
conjugate points, employed for instance in Theorem \ref{ch1t4}. We can, of course, impose the hypothesis 
that maximal causal geodesics in $\overline{\mathscr{M}}$ connecting points $p$ and $q$ necessarily belong
to $\mathscr{I}$, along lines similar to what we've done for null geodesics along Chapter \ref{ch1}. However, 
this hypothesis has the annoying characteristic of depending globally on the conformal factor $z$, unlike
the weaker case of null geodesics, which depends only on the causal structure.
\end{enumerate}

A way of circumventing these problems is to \emph{regularize} $\lambda^g_{p,q}$ in a way that this 
regularization, here denoted $\bar{\lambda}^g_{p,q}$ for concreteness, satisfies $\bar{\lambda}^g_{p,q}
\restr{\mathscr{D}_{p,q}}=\lambda$, where $u^\lambda_{p,q}$ is given by (\ref{ch1e26}). Here the problem
is how to regularize a global time function defined in a spacetime \emph{with boundary} -- we shall solve
the problem in the following manner: let us define the \emph{double} ${}^2\overline{\mathscr{M}}$ of a 
locally AdS-type spacetime $\overline{\mathscr{M}}$ as the disjoint union of two copies of 
$\overline{\mathscr{M}}$ with the points in $\mathscr{I}$ identified. ${}^2\overline{\mathscr{M}}$ can
be endowed with a differentiable structure compatible with the one of each copy and such that the natural 
inclusion of each copy is a differentiable embedding \cite{hirsch}. The metric in ${}^2\overline{\mathscr{M}}$ 
induced by $\bar{g}$ (also denoted by $\bar{g}$, as there's no confusion) is $\mathscr{C}^\infty$ in the 
interior of each copy, and $\mathscr{C}^{d-1,1}$ (resp. $\mathscr{C}^\infty$) in $\mathscr{I}$ for $d$ odd 
(resp. $d$ even \footnote{The extra regularity owes itself to (i) the absence of log terms in the
\textsc{Fefferman-Graham} expansion and (ii) the parity symmetry in $z$ in formula (\ref{ch2e10}),
page \pageref{ch2e10}, which gives origin to the recursion relations.}), due to the \textsc{Fefferman-Graham} 
expansion for locally AdS-type spacetimes constructed in the previous Section -- the regularity 
is $\mathscr{C}^\infty$ for arbitrary $d$ if $(\mathscr{M},g)$ is locally AAdS or, more in general, 
if $\bar{g}^{(0)}$ possesses a representative $\bar{g}^{(0)'}$ of its conformal class which satisfies 
$\mbox{Ric}(\bar{g}^{(0)'})=0$. This construction is a generalization of the conformal identification 
of $AdS_d$ with half of $ESU_d$, seen in Chapter \ref{ch1}, page \pageref{ch1f1}. \\

We can extend $z$ a $z<0$ in ${}^2\overline{\mathscr{M}}$, thus defining a system of Gaussian normal 
coordinates around $\mathscr{I}$. The double $({}^2\overline{\mathscr{M}},\bar{g})$ has a group of \emph{global}
isometries, isomorphic to $\mathbb{Z}_2$, generated by the reflection $\zeta$ across $\mathscr{I}$ 
of a point in a copy of $\mathscr{M}$, having as image its counterpart in the other copy (locally close to 
$\mathscr{I}$, we have $z\circ\zeta=-z$). \\

We shall now define $\bar{\lambda}^{\bar{g}}_{p,q}$ \emph{in a causally simple AAdS spacetime} 
$(\mathscr{M},g)$ \emph{which satisfies the focusing hypothesis of Theorem \ref{ch1t4}, page 
\pageref{ch1t4}} (these requirements guarantee, as seen in Chapter \ref{ch1}, that $I^+(p,
\overline{\mathscr{M}})\cap{} I^-(q,\overline{\mathscr{M}})\cap\mathscr{I}=\mathscr{D}_{p,q}$), 
in the following steps:

\begin{itemize}
\item First, we define $\lambda^{\bar{g}}_{p,q}$ \emph{in} $({}^2\overline{\mathscr{M}},
\bar{g})$ (more precisely, in $\mathscr{O}_{p,q}=\mathscr{W}_{p,q}\cup\mathscr{D}_{p,q}\cup
\zeta(\mathscr{W}_{p,q})$), with the same formula.
\item Next, we regularize $\lambda^{\bar{g}}_{p,q}$ by employing the method of \textsc{Bernal}
and \textsc{Sánchez} \cite{bernsan1,bernsan2} (see \cite{sanchez} for an introductory sketch of
the method), obtaining a $\mathscr{C}^\infty$ global time function $\tilde{\lambda}^{\bar{g}}_{p,q}$
which foliates $\mathscr{O}_{p,q}$ by \textsc{Cauchy} surfaces. We assume yet that the obtained 
regularization satisfies $|\tilde{\lambda}^{\bar{g}}_{p,q}-\lambda^{\bar{g}}_{p,q}|<\delta$
for $\delta<1$ sufficiently small \cite{seifert}.
\item Let us define $\hat{\lambda}^{\bar{g}}_{p,q}=\frac{1}{2}(\tilde{\lambda}^{\bar{g}}_{p,q}+
\tilde{\lambda}^{\bar{g}}_{p,q}\circ\zeta)$. Since $\tilde{\lambda}^{\bar{g}}_{p,q}$ as well as
$\tilde{\lambda}^{\bar{g}}_{p,q}\circ\zeta$ are global time functions in $\mathscr{O}_{p,q}$
whose level hypersurfaces are \textsc{Cauchy} surfaces in $\mathscr{O}_{p,q}$, and the collection 
of such functions is closed under convex linear combinations \cite{seifert}, it follows that
$\hat{\lambda}^{\bar{g}}_{p,q}$ also satisfies these properties, plus the key additional property
(which soon will be crucial) $\partial_z\hat{\lambda}^{\bar{g}}_{p,q}\restr{\mathscr{I}}=0$ and the 
estimate in the previous item. Henceforth, the usefulness of the double is over -- we restrict 
$\hat{\lambda}^{\bar{g}}_{p,q}$ to $\overline{\mathscr{M}}$, and proceed only within this manifold.
\item We extend $\lambda^{\bar{g}^{(0)}}_{p,q}$ from $\mathscr{D}_{p,q}$ to a collar neighbourhood
$\mathscr{U}^{\delta'}_{p,q}\cong\mathscr{D}_{p,q}\times[0,0<\delta')\ni(x,w)$ of the latter in 
$\mathscr{W}_{p,q}$ by using the formula $\lambda^{\bar{g}^{(0)}}_{p,q}(x,w)=\lambda^{\bar{g}^{(0)}}_{p,q}(x)$, 
taking $\delta'$ small enough so that $\bar{\nabla}^a\lambda^{\bar{g}^{(0)}}$ remains timelike in 
$\mathscr{U}^{\delta'}_{p,q}$ and keeping the same notation for the extension (notice that such a 
restriction to $\delta'$ is independent of the restriction to $\delta$ in the first item).
\item We choose a partition of unity $\{\phi,1-\phi\}$ of $\mathscr{W}_{p,q}\cup\mathscr{D}_{p,q}$
subordinated to the covering $\{\mathscr{D}_{p,q}\times[0,\frac{3\delta'}{4}),\mathscr{W}_{p,q}\smallsetminus
(\mathscr{D}_{p,q}\times[0,\frac{\delta'}{2}])\}$, satisfying $|\partial_w\phi|\leq K(\delta')^{-1}$
\cite{whitney} and \emph{independent of} $x\in\mathscr{D}_{p,q}$ (as $\phi\equiv 0$ outside 
$\mathscr{U}^{\delta'}_{p,q}$, there are no problems with the second property). Let us define 
\begin{equation}\label{ch2e45}
\bar{\lambda}^{\bar{g}}_{p,q}\doteq\phi\lambda^{\bar{g}^{(0)}}_{p,q}+(1-\phi)\hat{\lambda}^{\bar{g}}_{p,q},
\end{equation}
which satisfies 
\begin{equation}\label{ch2e46}
d\bar{\lambda}^{\bar{g}}_{p,q}=(\lambda^{\bar{g}^{(0)}}_{p,q}-\hat{\lambda}^{\bar{g}}_{p,q})\partial_w\phi dw
+\phi d\lambda^{\bar{g}^{(0)}}_{p,q}+(1-\phi)d\hat{\lambda}^{\bar{g}}_{p,q}.
\end{equation}
\item We see immediately that we need to control the size of $|\lambda^{\bar{g}^{(0)}}_{p,q}-
\hat{\lambda}^{\bar{g}}_{p,q}|$ in $\mathscr{U}^{\delta'}_{p,q}$ so that $|\bar{\lambda}^{\bar{g}^{(0)}}_{p,q}-
\hat{\lambda}^{\bar{g}}_{p,q}|$ is as small as we want and, at the same time, $d\bar{\lambda}^{\bar{g}^{(0)}}_{p,q}$ 
is a future timelike covector field; the most severe obstacle to such requirements resides in the 
spacelike covector proportional to $\partial_w\phi$ in (\ref{ch2e46}), due to the \textsc{Whitney} 
estimate invoked in the former item, which is essentially optimal. We write, thus,
\begin{equation}\label{ch2e47}
|\lambda^{\bar{g}^{(0)}}_{p,q}(x,w)-\hat{\lambda}^{\bar{g}}_{p,q}(x,w)|\leq\underbrace{|\lambda^{\bar{g}^{(0)}
}_{p,q}(x)-\hat{\lambda}^{\bar{g}}_{p,q}(x,0)|}_{<\delta\,(a)}+\underbrace{|\hat{\lambda}^{\bar{g}}_{p,q}(x,w)
-\hat{\lambda}^{\bar{g}}_{p,q}(x,0)|}_{\leq K'\delta'^2\,(b)},
\end{equation}
where the estimate $(a)$ is valid by construction, and the estimate $(b)$ follows from the fact that 
$\hat{\lambda}^{\bar{g}}_{p,q}$ is $\mathscr{C}^\infty$ and $\partial_w\hat{\lambda}^{\bar{g}}_{p,q}
\restr{\mathscr{D}_{p,q}}\equiv 0$. 
\item Taking $\delta=K'(\delta')^2$ and $\delta'$ sufficiently small, it follows that 
$\bar{\lambda}^{\bar{g}}_{p,q}$ is a $\mathscr{C}^{\infty}$ global time function which: 
\begin{enumerate}
\item satisfies $|\bar{\lambda}^{\bar{g}}_{p,q}-\lambda^{\bar{g}}_{p,q}|<\delta''$, where $\delta''>0$ 
is as small as we like,
\item extends $\lambda^{\bar{g}^{(0)}}_{p,q}$ to $\mathscr{W}_{p,q}\cup\mathscr{D}_{p,q}$, and such that
\item $\bar{\nabla}^a\bar{\lambda}^{\bar{g}}_{p,q}$ is a future directed timelike vector field,
tangent to $\mathscr{I}$, and
\item $(\bar{\lambda}^{\bar{g}}_{p,q})^{-1}(t)$ is a \textsc{Cauchy} surface for $\mathscr{W}_{p,q}\cup
\mathscr{D}_{p,q}$ for all $t\in\mathbb{R}$ (it follows directly from the first property \cite{dieckmann} 
and the continuity of $\bar{\lambda}^{\bar{g}}_{p,q}$).
\end{enumerate}
\item By virtue of the results above, it follows that 
\begin{equation}\label{ch2e48}
\bar{T}^a\doteq\frac{1}{\bar{g}^{-1}(d\bar{\lambda}^{\bar{g}}_{p,q},d\bar{\lambda}^{\bar{g}}_{p,q})}
\bar{\nabla}^a\bar{\lambda}^{\bar{g}}_{p,q}
\end{equation}
is an asymptotic \textsc{Killing} field in $(\mathscr{W}_{p,q},g\restr{\mathscr{W}_{p,q}})$
satisfying $\bar{T}^a\restr{\mathscr{I}}=\frac{du^._{p,q}}{d\lambda}$, and, hence, the formulae
(\ref{ch2e43}) and (\ref{ch2e44}), as desired.
\end{itemize}

Although $\bar{\lambda}^{\bar{g}}_{p,q}$ all the properties we need, unfortunately it no longer 
possesses in $(\mathscr{M},g)$ the natural geometrical interpretation which $\lambda^{\bar{g}}_{p,q}$ 
does, though being as close to this as we desire. Nevertheless, we've shown that it's possible to 
explicitly build families of asymptotic isometries in (practically) any AAdS wedge. We shall make 
extensive use of this fact in Chapter \ref{ch4}.

\subsection{\label{ch1-thermo-eq}``Return to equilibrium'' in wedges}

In the work of \textsc{Martinetti} and \textsc{Rovelli} \cite{martrov}, it was proposed to associate
an ``\textsc{Unruh} effect'' to the group $u^\lambda_{p,q}$ associated to a diamond $\mathscr{D}_{p,q}$
in \textsc{Minkowski} spacetime, whose orbits were imagined as the worldlines of ``observers with 
finite lifetime''. For quantum field theories with conformally invariant vacuum, it's truly possible 
to prove such an association \cite{bglongo1}, but a proper discussion of this issue will have to
await Chapter \ref{ch4}. Nevertheless, formulae (\ref{ch2e43}) and (\ref{ch2e44}) make natural the 
following question: \emph{what if such a temperature were attained only in the limit} $\lambda
\rightarrow\pm\infty$ \emph{?} For quantum field theories which enjoy asymptotic freedom, scale 
invariance is indeed realized only in the short-distance limit, and the corrections to ``critical'' 
behaviour are given, for instance, by the operator product expansion (OPE). Thus, we can think that the
``thermal time hypothesis'' of \textsc{Martinetti} and \textsc{Rovelli}, cited above, is realized only 
in the weaker form of a \emph{return to equilibrium,} by means of an \emph{asymptotic} scale invariance
(we shall give a more precise sense to these considerations in Chapter \ref{ch4}).\\

We'll transplant this reasoning to AAdS wedges. From the viewpoint of the laws of black hole dynamics, 
if we imagine the boundary of a wedge (supposed geodesically convex, for simplicity) $\mathscr{W}_{p,q}$ 
as an ``asymptotically stationary'' \emph{event horizon}, and $\mathscr{W}_{p,q}$ as a ``domain of outer 
communications'', we can ask ourselves if there exists some asymptotic analog for these laws. It's now our 
objective to show that this indeed occurs, at least for the zeroth law.\\

First, we shall give the precise definition of \emph{surface gravity} within our context. Let $T^a$ be
as in (\ref{ch2e42}). Taking, for instance, $r\rightarrow r_-\in\partial_-\mathscr{W}_{p,q}$, we have 
$\Gamma_{\bar{g}}(p,r_-)=0$, $\Gamma_{\bar{g}}(r_-,q)\neq 0$ and $d\Gamma_{\bar{g}}(p,r_-)\neq 0$. Moreover, 
\begin{equation}\label{ch2e49}
\lim_{r\rightarrow r_-}\bar{g}(T,T)=-\lim_{r\rightarrow r_-}\frac{2\Gamma_{\bar{g}}(p,r)\Gamma_{\bar{g}}
(r,q)}{\Gamma_{\bar{g}}(p,r)+\Gamma_{\bar{g}}(r,q)+\bar{g}^{-1}(d\Gamma_{\bar{g}}(p,r),d\Gamma_{\bar{g}}
(r,q))}=0
\end{equation}
by construction. Equally, if $r\rightarrow r_+\in\partial_+\mathscr{W}_{p,q}$, we have $\Gamma_{\bar{g}}
(r_+,q)=0$, $\Gamma_{\bar{g}}(p,r_+)\neq 0$, $d\Gamma_{\bar{g}}(r_+,q)\neq 0$ and $\lim_{r\rightarrow r_+}
\bar{g}(T,T)=0$. In the case that $r\rightarrow r_0\in\partial I^+(p,\overline{\mathscr{M}})\cap{}\partial 
I^-(q,\overline{\mathscr{M}})$, we have $T^a\rightarrow 0$. Summing up, $T^a$ is, at the same time, 
\emph{tangent} and \emph{normal} to the past horizon $\partial_-\mathscr{W}_{p,q}$ and to the future 
horizon $\partial_+\mathscr{W}_{p,q}$ of $\mathscr{W}_{p,q}$ (notice that the definition of horizon we've 
given in page \pageref{ch1d3} excludes precisely the points of $\partial\mathscr{W}_{p,q}$ where 
$T^a\rightarrow 0$). Hence, we have
\begin{equation}\label{ch2e50}
\lim_{\rightarrow\partial_\pm\mathscr{W}_{p,q}}\bar{\nabla}_a\bar{g}(T,T)\doteq -2\bar{\kappa}_\pm
(\lim_{\rightarrow\partial_\pm\mathscr{W}_{p,q}}T_a),
\end{equation}
where $\bar{\kappa}_+$ (resp. $\bar{\kappa}_-$) is a $\mathscr{C}^\infty$ function on the submanifold 
$\partial_+\mathscr{W}_{p,q}$ (resp. $\partial_-\mathscr{W}_{p,q}$), given by
\begin{equation}\label{ch2e51}
\bar{\kappa}_\pm=-\lim_{\rightarrow\partial_\pm\mathscr{W}_{p,q}}\frac{T^a\bar{\nabla}_a\bar{g}(T,T)}{2
\bar{g}(T,T)}=\frac{g^{-1}(d\lambda^{\bar{g}}_{p,q},d\bar{g}^{-1}(d\lambda^{\bar{g}}_{p,q},
d\lambda^{\bar{g}}_{p,q}))}{2\bar{g}^{-1}(d\lambda^{\bar{g}}_{p,q},d\lambda^{\bar{g}}_{p,q})^2}.
\end{equation}

Let us suppose for the moment that $T^a$ is a conformal \textsc{Killing} field with respect to $\bar{g}$. In this 
case,
\begin{eqnarray}
\bar{\nabla}_a\bar{g}(T,T) & = & 2T^b\bar{\nabla}_a T_b=2T^b(\bar{\nabla}_a T_b+\bar{\nabla}_b T_a)-2T^b
\bar{\nabla}_b T_a=\nonumber\\ & = & \frac{4}{d}(\bar{\nabla}_b T^b)T_a-2T^b\bar{\nabla}_b T_a
\stackrel{\rightarrow\partial_\pm\mathscr{O}_{p,q}}{\longrightarrow}-2\bar{\kappa}_\pm T_a,\label{ch2e52}
\end{eqnarray}
whence it follows that 
\begin{equation}\label{ch2e53}
\kappa_\pm\doteq\lim_{\rightarrow\partial_\pm\mathscr{W}_{p,q}}\frac{T^aT^b\bar{\nabla}_bT_a}{\bar{g}(T,T)}
\doteq\lim_{\rightarrow\partial_\pm\mathscr{W}_{p,q}}\frac{T^aT^b\bar{\nabla}_aT_b}{\bar{g}(T,T)}+
\bar{\kappa}_\pm=\lim_{\rightarrow\partial_\pm\mathscr{W}_{p,q}}\left(\frac{2}{d}\bar{\nabla}_aT^a\right)+
\bar{\kappa}_\pm
\end{equation}
defines precisely the failure of the parametrization of the orbits of $T^a$ in $\partial_\pm\mathscr{D}_{p,q}$
to be an \emph{affine} parametrization, since $T^a$ is tangent to the null geodesics which generate
$\partial_+\mathscr{W}_{p,q}$ and $\partial_-\mathscr{W}_{p,q}$. More precisely, if $\lambda_-$
(resp. $\lambda_+$) is an affine parameter common to the future directed null geodesics which emanate from 
$p$ (resp. reach $q$) such that $\lambda_-=0$ at $p$ (resp. $\lambda_+=0$ at $q$), we have 
\begin{equation}\label{ch2e54}
\lambda_\pm=\mp e^{\kappa_\pm\lambda^g_{p,q}}.
\end{equation}

Another interpretation for $\kappa_\pm$, which follows from formulae (\ref{ch2e51})--(\ref{ch2e53}), is
the acceleration needed to maintain a test body at a fixed, but ``infinitesimally small'' distance, from 
the horizon, compensated by the redshift factor \emph{``redshift''} $(-\bar{g}(T,T))^{\frac{1}{2}}$ of the
orbit \cite{wald2,wald3}.

\begin{definition}\label{ch2d1}
$\kappa_-$ (resp. $\kappa_+$) is said to be the \emph{past} (resp. \emph{future}) m\emph{surface gravity}
of $\mathscr{O}_{p,q}$.
\end{definition}

For arbitrary $T^a$, we use the first (two) identity(ies) in (\ref{ch2e53}) to \emph{define} 
$\kappa_\pm$, but the last identity in formula (\ref{ch2e53}) is only \emph{asymptotically} valid, as 
$r_-\rightarrow p$, $r_+\rightarrow q$. With this caveat in mind, we can now formulate our ``zeroth law
of dynamics'':

\begin{theorem}[Zeroth Law of Dynamics of AAdS]\label{ch2t2}
Let $(\mathscr{M},g)$ be a (locally) \\AAdS spacetime, and $\mathscr{W}_{p,q}$ a wedge contained
in a geodesically convex neighbourhood in $(\overline{\mathscr{M}},g)$. Then $\lim_{\rightarrow p}
\kappa_-=-\lim_{\rightarrow q}\kappa_+=1$.
\begin{quotation}{\small\scshape Proof.\quad}
{\small\upshape A direct calculation employing the properties of $\Gamma_{\bar{g}}$ gives us
\[T^a\bar{\nabla}_a\bar{g}(T,T)=\frac{2\Gamma_{\bar{g}}(p,r)\Gamma_{\bar{g}}(r,q)}{(\Gamma_{\bar{g}}(p,r)+
\Gamma_{\bar{g}}(r,q)+\bar{g}^{-1}(d\Gamma_{\bar{g}}(p,r),d\Gamma_{\bar{g}}(r,q)))^2}.\]\[.\left[
(\Gamma_{\bar{g}}(r,q)-\Gamma_{\bar{g}}(p,r))\left(2+\frac{\bar{g}^{-1}(d\Gamma_{\bar{g}}(p,r),
d\Gamma_{\bar{g}}(r,q))}{\Gamma_{\bar{g}}(p,r)+\Gamma_{\bar{g}}(r,q)+\bar{g}^{-1}(d\Gamma_{\bar{g}}
(p,r),d\Gamma_{\bar{g}}(r,q))}\right)\right.+\]\[\left.-\frac{\Gamma_{\bar{g}}(r,q)\bar{\nabla}^a
\Gamma_{\bar{g}}(p,r)\bar{\nabla}^b\Gamma_{\bar{g}}(p,r)\bar{\nabla}_a\bar{\nabla}_b\Gamma_{\bar{g}}
(r,q)-\Gamma_{\bar{g}}(p,r)\bar{\nabla}^a\Gamma_{\bar{g}}(r,q)\bar{\nabla}^b\Gamma_{\bar{g}}(r,q)
\bar{\nabla}_a\bar{\nabla}_b\Gamma_{\bar{g}}(p,r)}{\Gamma_{\bar{g}}(p,r)+\Gamma_{\bar{g}}(r,q)+
\bar{g}^{-1}(d\Gamma_{\bar{g}}(p,r),d\Gamma_{\bar{g}}(r,q))}\right],\]
and, thus, 
\[-\frac{T^a\bar{\nabla}_a\bar{g}(T,T)}{2\bar{g}(T,T)}=-\frac{1}{2(\Gamma_{\bar{g}}(p,r)+\Gamma_{\bar{g}}
(r,q)+\bar{g}^{-1}(d\Gamma_{\bar{g}}(p,r),d\Gamma_{\bar{g}}(r,q)))}.\]\[.\left[(\Gamma_{\bar{g}}(r,q)
-\Gamma_{\bar{g}}(p,r))\left(2+\frac{\bar{g}^{-1}(d\Gamma_{\bar{g}}(p,r),d\Gamma_{\bar{g}}(r,q))}{\Gamma_{\bar{g}}
(p,r)+\Gamma_{\bar{g}}(r,q)+\bar{g}^{-1}(d\Gamma_{\bar{g}}(p,r),d\Gamma_{\bar{g}}(r,q))}\right)\right.+\]\[\left.
-\frac{\Gamma_{\bar{g}}(r,q)\bar{\nabla}^a\Gamma_{\bar{g}}(p,r)\bar{\nabla}^b\Gamma_{\bar{g}}(p,r)
\bar{\nabla}_a\bar{\nabla}_b\Gamma_{\bar{g}}(r,q)-\Gamma_{\bar{g}}(p,r)\bar{\nabla}^a\Gamma_{\bar{g}}
(r,q)\bar{\nabla}^b\Gamma_{\bar{g}}(r,q)\bar{\nabla}_a\bar{\nabla}_b\Gamma_{\bar{g}}(p,r)}{\Gamma_{\bar{g}}
(p,r)+\Gamma_{\bar{g}}(r,q)+\bar{g}^{-1}(d\Gamma_{\bar{g}}(p,r),d\Gamma_{\bar{g}}(r,q))}\right].\]
$\bar{\kappa}_+$ (resp. $\bar{\kappa}_-$) is obtained by taking the limit $\Gamma_{\bar{g}}(r,q)
\searrow 0$, $d\Gamma_{\bar{g}}(r,q)\neq 0$ (resp. $\Gamma_{\bar{g}}(p,r)\searrow 0$, 
$d\Gamma_{\bar{g}}(p,r)\neq 0$). Hence, we have \[\bar{\kappa}_+=\frac{1}{2(\Gamma_{\bar{g}}(p,r)
+\bar{g}^{-1}(d\Gamma_{\bar{g}}(p,r),d\Gamma_{\bar{g}}(r,q)))}.\]\[.\left[\Gamma_{\bar{g}}(p,r)\left(3-
\frac{\Gamma_{\bar{g}}(p,r)}{\Gamma_{\bar{g}}(p,r)+\bar{g}^{-1}(d\Gamma_{\bar{g}}(p,r),d\Gamma_{\bar{g}}(r,q))}
\right)+\right.\]\[\left.-\frac{\Gamma_{\bar{g}}(p,r)\bar{\nabla}^a\Gamma_{\bar{g}}(r,q)\bar{\nabla}^b
\Gamma_{\bar{g}}(r,q)\bar{\nabla}_a\bar{\nabla}_b\Gamma_{\bar{g}}(p,r)}{\Gamma_{\bar{g}}(p,r)+\bar{g}^{-1}
(d\Gamma_{\bar{g}}(p,r),d\Gamma_{\bar{g}}(r,q))}\right]\] and \[\bar{\kappa}_-=-\frac{1}{2(\Gamma_{\bar{g}}(r,q)
+\bar{g}^{-1}(d\Gamma_{\bar{g}}(p,r),d\Gamma_{\bar{g}}(r,q)))}.\]\[.\left[\Gamma_{\bar{g}}(r,q)\left(3-
\frac{\Gamma_{\bar{g}}(r,q)}{\Gamma_{\bar{g}}(r,q)+\bar{g}^{-1}(d\Gamma_{\bar{g}}(p,r),d\Gamma_{\bar{g}}(r,q))}
\right)+\right.\]\[\left.-\frac{\Gamma_{\bar{g}}(r,q)\bar{\nabla}^a\Gamma_{\bar{g}}(p,r)\bar{\nabla}^b
\Gamma_{\bar{g}}(p,r)\bar{\nabla}_a\bar{\nabla}_b\Gamma_{\bar{g}}(r,q)}{\Gamma_{\bar{g}}(r,q)+\bar{g}^{-1}
(d\Gamma_{\bar{g}}(p,r),d\Gamma_{\bar{g}}(r,q))}\right].\] }\end{quotation}

\hfill\vspace*{-4ex}

\begin{quotation}{\small\upshape The divergence of $T^a$, on its turn, leads to 
\[\bar{\nabla}_aT^a=\frac{1}{\Gamma_{\bar{g}}(p,r)+\Gamma_{\bar{g}}(r,q)+\bar{g}^{-1}(d\Gamma_{\bar{g}}(p,r),
d\Gamma_{\bar{g}}(r,q))}.\]\[\left[\Gamma_{\bar{g}}(p,r)\bar{\nabla}_a\bar{\nabla}^a\Gamma_{\bar{g}}(r,q)
-\Gamma_{\bar{g}}(r,q)\bar{\nabla}_a\bar{\nabla}^a\Gamma_{\bar{g}}(p,r)-\frac{1}{\Gamma_{\bar{g}}(p,r)+
\Gamma_{\bar{g}}(r,q)+\bar{g}^{-1}(d\Gamma_{\bar{g}}(p,r),d\Gamma_{\bar{g}}(r,q))}\right..\]\[.\left(
(\Gamma_{\bar{g}}(p,r)-\Gamma_{\bar{g}}(r,q))\bar{g}^{-1}(d\Gamma_{\bar{g}}(p,r),d\Gamma_{\bar{g}}(r,q))
+\Gamma_{\bar{g}}(r,q)\bar{\nabla}^a\Gamma_{\bar{g}}(p,r)\bar{\nabla}^b\Gamma_{\bar{g}}(p,r)\bar{\nabla}_a
\bar{\nabla}_b\Gamma_{\bar{g}}(r,q)+\right.\]\[\left.\left.-\Gamma_{\bar{g}}(p,r)\bar{\nabla}^a\Gamma_{\bar{g}}
(r,q)\bar{\nabla}^b\Gamma_{\bar{g}}(r,q)\bar{\nabla}_a\bar{\nabla}_b\Gamma_{\bar{g}}(p,r)\right)\right],\] 
which in the future horizon $\partial_+\mathscr{W}_{p,q}$ becomes \[\bar{\nabla}_aT^a\restr{\partial_+
\mathscr{W}_{p,q}}=\frac{1}{\Gamma_{\bar{g}}(p,r)+\bar{g}^{-1}(d\Gamma_{\bar{g}}(p,r),d\Gamma_{\bar{g}}(r,q))}.\]
\[\left[\Gamma_{\bar{g}}(p,r)\bar{\nabla}_a\bar{\nabla}^a\Gamma_{\bar{g}}(r,q)-\frac{1}{\Gamma_{\bar{g}}(p,r)
+\bar{g}^{-1}(d\Gamma_{\bar{g}}(p,r),d\Gamma_{\bar{g}}(r,q))}\right..\]\[.\left.\left(\Gamma_{\bar{g}}(p,r)
\bar{g}^{-1}(d\Gamma_{\bar{g}}(p,r),d\Gamma_{\bar{g}}(r,q))-\Gamma_{\bar{g}}(p,r)\bar{\nabla}^a\Gamma_{\bar{g}}
(r,q)\bar{\nabla}^b\Gamma_{\bar{g}}(r,q)\bar{\nabla}_a\bar{\nabla}_b\Gamma_{\bar{g}}(p,r)\right)\right],\] 
and in the past horizon $\partial_-\mathscr{W}_{p,q}$, \[\bar{\nabla}_aT^a\restr{\partial_-\mathscr{W}_{p,q}}
=-\frac{1}{\Gamma_{\bar{g}}(r,q)+\bar{g}^{-1}(d\Gamma_{\bar{g}}(p,r),d\Gamma_{\bar{g}}(r,q))}.\]\[\left[
\Gamma_{\bar{g}}(r,q)\bar{\nabla}_a\bar{\nabla}^a\Gamma_{\bar{g}}(p,r)-\frac{1}{\Gamma_{\bar{g}}(r,q)
+\bar{g}^{-1}(d\Gamma_{\bar{g}}(p,r),d\Gamma_{\bar{g}}(r,q))}\right..\]\[.\left.\left(\Gamma_{\bar{g}}(r,q)
\bar{g}^{-1}(d\Gamma_{\bar{g}}(p,r),d\Gamma_{\bar{g}}(r,q))-\Gamma_{\bar{g}}(r,q)\bar{\nabla}^a\Gamma_{\bar{g}}
(p,r)\bar{\nabla}^b\Gamma_{\bar{g}}(p,r)\bar{\nabla}_a\bar{\nabla}_b\Gamma_{\bar{g}}(r,q)\right)\right].\]
We obtain immediately from the formulae above that \[\lim_{\rightarrow q}\bar{\kappa}_+=1=-\lim_{\rightarrow p}
\bar{\kappa}_-,\,\lim_{\rightarrow q}\bar{\nabla}_aT^a=-d=-\lim_{\rightarrow p}\bar{\nabla}_aT^a.\]
The result hence follows from formula (\ref{ch2e53}) and the discussion after Definition \ref{ch2d1}.
~\hfill~$\Box$}
\end{quotation}
\end{theorem}

\begin{remark}
The result and the proof of Theorem \ref{ch2t2} hold for geodesically convex diamonds
$\mathscr{O}_{p,q}$ in arbitrary causally simple spacetimes with no change.
\end{remark}

There are two caveats about Theorem \ref{ch2t2} as far as the interpretation of the wedge 
$\mathscr{W}_{p,q}$ as the exterior region of an event horizon is concerned.

\begin{itemize}
\item All results of this Subsection involve the metric $\bar{g}$ of the conformal closure, and
not the physical metric $g$. We have, though, that
\begin{equation}\label{ch2e55}
\frac{T^a\bar{\nabla}_a\bar{g}(T,T)}{\bar{g}(T,T)}=\frac{T^a\nabla_a g(T,T)}{g(T,T)}+2z\langle
dz,T\rangle
\end{equation}
and
\begin{eqnarray}\label{ch2e56}
\bar{\nabla}_aT^a & = & \bar{g}^{ab}\bar{\nabla}_aT_b=\frac{1}{z^2}g^{ab}\bar{\nabla}_aT_b=\\
& = & \frac{1}{z^2}\nabla_aT^a-\frac{1}{z^2}g^{ab}C^c_{ab}T_c=\frac{1}{z^2}\nabla_aT^a+
\frac{d-2}{2z^3}\langle dz,T\rangle\,\Rightarrow\nonumber\\ & \Rightarrow & \nabla_a T^a=
z^2\bar{\nabla}_a T^a-\frac{d-2}{2z}\langle dz,T\rangle,\nonumber
\end{eqnarray}
where $C^c_{ab}$ is the tensor (\ref{ap1e26}) which relates the connections $\bar{\nabla}$ and $\nabla$ 
(see page \pageref{ap1e26}). As $T^a$ is tangent to $\mathscr{I}$ vanishes at $p$ and $q$ (see formula 
(\ref{ch2e44}), page \pageref{ch2e44}), it follows that $\langle dz,T\rangle=o(z)$ as we get close to $p$ 
or $q$. Taking into account that $T^a$ is an asymptotic \textsc{Killing} field, we obtain that $\kappa_\pm$ 
also defines the asymptotic past/future surface gravity with respect to $g$ as we get close to $p,q$. As the
asymptotic values of $\kappa_\pm$ are the same for all asymptotic \textsc{Killing} fields 
$\frac{du^\lambda_{p,q}}{d\lambda}$ in $\mathscr{W}_{p,q}$ as $z\searrow 0$, it follows that the conclusion 
of Theorem \ref{ch2t2} also holds for non geodesically convex wedges and the asymptotic \textsc{Killing} 
fields constructed in Subsection \ref{ch2-thermo-iso}.
\item The sign of $\kappa_+$ we've obtained is minus what it's expected from an event horizon, for the 
origin of our affine parameter for the (future oriented) generators of $\partial_+\mathscr{W}_{p,q}$ 
corresponds to $q$ and not to $\partial I^+(p,\overline{\mathscr{M}})\cap{}\partial I^-(q,\overline{\mathscr{M}})$, 
which would correspond to a ``bifurcation surface'' of the horizon (notice the change of sign in (\ref{ch2e54})).
A correction of this problem leads to the desired change of the sign of $\kappa_+$.
\end{itemize} 

Let us notice, finally, that there is a hypothesis about the normalization of $T^a$ implicit in the
definition of $\kappa_\pm$. More precisely, our definition is conditioned to the following fact: if 
$\Gamma_{\bar{g}^{(0)}}(p,r)=\Gamma_{\bar{g}^{(0)}}(r,q)$, i.e., $r$ is the middle point of the maximal
timelike geodesic in $\mathscr{M}in(r)\ni p,q$ linking $p$ to $q$, clearly implying that ${}^{(0)}
\bar{\nabla}^a\Gamma_{\bar{g}^{(0)}}(p,r)=-{}^{(0)}\bar{\nabla}^a\Gamma_{\bar{g}^{(0)}}(r,q)$, then $\bar{g}^{(0)}
(T,T)(r)=-\frac{d_{\bar{g}^{(0)}}(p,q)^2}{16}$. If we rescale $T^a$ by a factor $R\in\mathbb{R}$, $\kappa_\pm$ 
is rescaled by the same factor, by definition. This has, for instance, the consequence that, if we take 
$p_0=(-1,\mathbf{0})$, $q_0=(1,\mathbf{0})$ and rescale the $T^a$ associated to the diamond 
$\mathscr{D}_{Rp_0,Rq_0}$ by a factor $\frac{2}{R}$, so as to maintain $\bar{g}^{(0)}(T,T)(r)=\eta(T,T)(0)=-1$, 
it follows that $\kappa_\pm\stackrel{R\rightarrow+\infty}{\longrightarrow}0$, in a way consistent with the 
fact that $\frac{2}{R}T^a\stackrel{R\rightarrow+\infty}{\longrightarrow}(\partial_0)^a$ (physically, the 
``\textsc{Unruh} temperature'' associated to time translations in \textsc{Minkowski} spacetime is zero).\\

We can understand Proposition \ref{ch1p1} as a formulation of the second law of black hole dynamics 
for AAdS wedges, as the ``shrinking'' of AAdS wedges expressed there provokes a decrease of the area of
the transversal sections of the past and future horizons, the larger the deeper we penetrate the bulk, that
is, in the opposite orientation to the one of the time evolution (notice that, unlike \textsc{Hawking}'s
Area Theorem, we cannot express this result directly in terms of the physical spacetime $(\mathscr{M},g)$, 
for the transversal sections of horizons of AAdS wedges are noncompact and hence have always infinite area). 
The case in which there is no increase of ``entropy'', which characterizes \emph{reversible} processes, becomes:

\begin{theorem}[Characterization of reversible processes in AAdS wedges]\label{ch2t3}
Let $(\mathscr{M},g)$ be am asymptotically simple AAdS spacetime satisfying the hypotheses of
Proposition \ref{ch1p1}, and $(p,\bar{q})\in\mathscr{D}(\mathscr{I})$, $r\in\mathscr{D}_{p,\bar{q}}$. 
Suppose that, for $r$ sufficiently close to $q$, we have $\mathscr{W}_{r,\bar{q}}'\cap{}\mathscr{W}_{q,
\bar{r}}'=\varnothing$. Then $\partial_+\mathscr{W}_{s,\bar{q}}$ is a null $\mathscr{C}^\infty$ hypersurface
which is totally geodesic with respect to $g$ for all $s\in\mathscr{I}$ such that $(s,\bar{q})\in\mathscr{D}
(\mathscr{I})$. In particular, $\mathscr{W}_{s,\bar{q}}'\cap{}\mathscr{W}_{q,\bar{s}}'=\varnothing$ for all 
$s\in\mathscr{I}$ such that $(s,\bar{q})\in\mathscr{D}(\mathscr{I})$.
\begin{quote}{\small\scshape Proof.\quad}
{\small\upshape Follows immediately from Proposition \ref{ch1p1} and Remark \ref{ch1r3}.~\hfill~$\Box$}
\end{quote}
\end{theorem}

Notice that, by virtue of Theorem \ref{ch2t2}, it makes no sense to speak of a ``third law'' for the
dynamics of AAdS wedges and diamonds.

\cleardoublepage

\thisfancyput(0in,-1.8in){\includegraphics{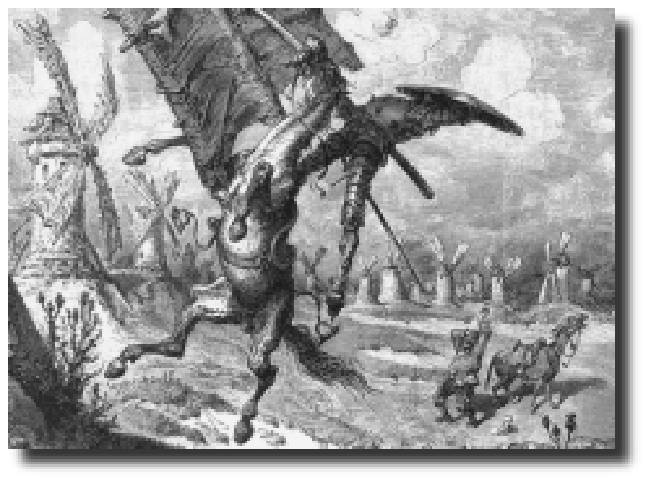}} 

\epigraphhead[30]{\epigraph{\qquad --- A lo que yo imagino --- dijo don
Quijote ---, no hay historia humana en el mundo que no tenga sus altibajos,
especialmente las que tratan de caballer\'{\i}as, las cuales nunca pueden estar
llenas de pr\'osperos sucesos.\endnotemark[9]}
{\textsc{Miguel de Cervantes Saavedra}\\ \emph{El ingenioso hidalgo 
don Quijote de la Mancha}, Segundo Libro, Cap. III}}
\part{Holographic avatars of Local Quantum Physics}

\chapter{\label{ch3}Locally Covariant Quantum Theories}

\epigraph{\textsc{(Hamlet.)}\quad --- \emph{O God! I could be bounded 
in a nutshell, and count myself a king of infinite space, were it 
not that I have bad dreams.}}{\textsc{William Shakespeare}\\
\emph{Hamlet, Prince of Denmark}, Act II, Scene II}

Having convinced ourselves about the need for rethinking the implementation of asymptotic isometries
of spacetime already in the context of classical gravity, we shall now introduce a proper 
formalism for the discussion of this question at the level of local quantum theories. Such a 
formalism was recently proposed by \textsc{Brunetti, Fredenhagen} and \textsc{Verch} \cite{bfv}, and 
one can base it on the observation, related to the remark at the beginning of Chapter \ref{ch2} (see
page \pageref{ch2e2}), that the specification of physical procedures and its relative ordering are
inherently \emph{local} -- these are insensitive to the structure of the Universe as a whole, into which
the localization region of these procedures is embedded. This, as we'll see, generalizes the notions
of isotony and covariance of physical procedures in a single blow. We can even raise this observation 
to the category of \emph{principle} --- say, of \emph{local covariance}.\\

On the other hand, we are immediately forced, in the light of these considerations, to consider 
\emph{all regions of all possible spacetimes on an equal footing.} More precisely, we must 
specify the possible physical procedures of a theory for each of these regions, in a coherent
manner. Mathematically, we must specify a \emph{functor} from the \emph{category} of ``admissible'' 
spacetime regions to the category of possible physical procedures (we advise the reader not familiar 
with the notions of category and functor to consult Appendix \ref{ap3}, page \pageref{ap3}). It's now 
our objective to make more precise the structure we've just sketched, and investigate some of its 
properties.

\section{\label{ch3-def}Basic definitions}

First, we shall consider a \emph{locally covariant quantum theory} on globally hyperbolic spacetimes, 
for this class is ``dynamically closed'' with respect to equations of motion which lead to causal 
propagation of \textsc{Cauchy} data. In this case, the causal commutation relations between observables 
can be uniquely determined (the necessary modifications of this concept in non globally hyperbolic 
cases will be exposed in Chapter \ref{ch4}, page \pageref{ch4}).

\begin{definition}\label{ch3d1} 
Let the categories
\begin{equation}\label{ch3e1}
\mathscr{G}lh_d=\left\{\begin{array}{rl} \mbox{objects:} & Obj\mathscr{G}lh_d=d\mbox{-dimensional,
globally hyperbolic}\\ & \mbox{spacetimes }(\mathscr{M},g);\\
\mbox{morphisms:} & \mbox{Hom}_{\mathscr{G}lh_d}((\mathscr{M},g),(\mathscr{M}',g'))=\{\psi:
(\mathscr{M},g)\rightarrow(\mathscr{M}',g')\\ & \mbox{isometric embeddings with open, 
causally}\\ & \mbox{convex image}\};
\end{array}\right.\nonumber
\end{equation}
and
\begin{equation}\label{ch3e2}
\mathscr{A}lg=\left\{\begin{array}{rl} \mbox{objects:} & Obj\mathscr{A}lg=\mbox{unital 
C*-algebras }\mathfrak{A};\\
\mbox{morphisms:} & \mbox{Hom}_{\mathscr{A}lg}(\mathfrak{A},\mathfrak{A}')=\{\alpha:\mathfrak{A}
\rightarrow\mathfrak{A}'\mbox{ unital *-monomorphisms}\}.
\end{array}\right.\nonumber
\end{equation}
A \emph{locally covariant quantum theory} is a covariant functor 
$\mathfrak{A}$ from $\mathscr{G}lh_d$ to $\mathscr{A}lg$.
\end{definition}

We say that $\mathfrak{A}$ is:

\begin{itemize}
\item \emph{locally causal} if, for any pair $(\mathscr{O}_i,g_i),(\mathscr{M},g)\in Obj
\mathscr{G}lh_d$, $\psi_i\in\mbox{Hom}_{\mathscr{G}lh_d}((\mathscr{O}_i,g_i),$\\$(\mathscr{M},g))$, 
$i=1,2$, such that $\psi_1(\mathscr{O}_1)$ and $\psi_2(\mathscr{O}_2)$ are causally disjoint in 
$(\mathscr{M},g)$, then \[[(\mathfrak{A}\psi_1)(\mathfrak{A}(\mathscr{O}_1,g_1)),(\mathfrak{A}
\psi_2)(\mathfrak{A}(\mathscr{O}_2,g_2))]=0\] in $\mathfrak{A}(\mathscr{M},g)$.
\item \emph{primitively causal} if, given $(\mathscr{O},g),(\mathscr{M},g)\in Obj
\mathscr{G}lh_d$, $\psi\in$$\mbox{Hom}_{\mathscr{G}lh_d}((\mathscr{O},g), (\mathscr{M},g))$, 
such that $\psi(\mathscr{O})$ contains a \textsc{Cauchy} surface for $(\mathscr{M},g)$, then 
$(\mathfrak{A}\psi)(\mathfrak{A}(\mathscr{O},g))=\mathfrak{A}(\mathscr{M},g)$, i.e., 
$\mathfrak{A}\psi$ is a \emph{C*-isomorphism}.
\item \emph{additive} if, given $(\mathscr{O}_i,g_i)\in Obj\mathscr{G}lh_d$, $\psi_i\in
\mbox{Hom}_{\mathscr{G}lh_d}((\mathscr{O}_i,g_i),(\mathscr{M},g))$, $i\in I$, such that the collection 
$\{\psi_i(\mathscr{O}_i)\}$ is an open covering of $(\mathscr{M},g)$, then 
\[\overline{\bigvee_{i\in I}(\mathfrak{A}\psi_i)(\mathfrak{A}(\mathscr{O}_i,g_i))}=\mathfrak{A}
(\mathscr{M},g)\], where $\vee_{i\in I}$ denotes the *-algebra generated by the *-algebras indexed 
by $i\in I$ and the above closure is in the C*-norm of $\mathfrak{A}(\mathscr{M},g)$.
\end{itemize}

Given $(\mathscr{M}_i,g_i)\in Obj\mathscr{G}lh$, $i=1,2$, such that $\mathscr{M}_1\subset
\mathscr{M}_2$ and $g_1=g_2\restr{\mathscr{M}_1}$, we denote, for future convenience, by 
$i_{\mathscr{M}_1,\mathscr{M}_2}$ the natural inclusion of $\mathscr{M}_1$ into $\mathscr{M}_2$.\\

The thoughtful reader will notice that we could generalize Definition \ref{ch3d1} so as to incorporate
any external classical bundle-valued field configuration over the objects of $\mathscr{G}lh$, and not 
only the metric. A particular case would be the dimensional parameters (masses, coupling constants, etc.) 
of the quantum field theory in question. This possibility will become more evident in Section \ref{ch3-field} 
(page \pageref{ch3-field}, where we define locally covariant quantum fields.

\section{\label{ch3-haag}Relation with the usual axioms}

We'll now show how one can recover the axiomatic scheme of \textsc{Haag} and \textsc{Kastler} 
\cite{hkast,haag} for theories of local observables on a \emph{fixed} spacetime starting from a 
locally covariant quantum theory $\mathfrak{A}$.\\

Consider a \emph{fixed} spacetime $(\mathscr{M},g)\in Obj\mathscr{G}lh_d$, with 
(possibly trivial) isometry group $G(\mathscr{M},g)$. Let 
\begin{equation}\label{ch3e3}
\mathscr{K}(\mathscr{M},g)\doteq\{\mathscr{O}\subset\mathscr{M}\mbox{ open, causally 
convex}\}.
\end{equation}

Notice that, by the definition (\ref{ch3e3}), $(\mathscr{O},g\restr{\mathscr{O}})$ is
automatically globally hyperbolic for all $\mathscr{O}\in\mathscr{K}(\mathscr{M},g)$, 
by virtue of the global hyperbolicity of $(\mathscr{M},g)$. Defining
\begin{equation}\label{ch3e4}
\mathfrak{A}_{\mathscr{M}}(\mathscr{O})\doteq\mathfrak{A}i_{\mathscr{O},\mathscr{M}}(\mathfrak{A}
(\mathscr{O},g\restr{\mathscr{O}})),
\end{equation}
it follows that
\begin{enumerate}
\item The correspondence $\mathscr{K}(\mathscr{M},g)\ni\mathscr{O}\mapsto\mathfrak{A}_{\mathscr{M}}
(\mathscr{O})$ is \emph{isotonous}, i.e., given $\mathscr{O}_1,\mathscr{O}_2\in\mathscr{K}
(\mathscr{M},g)$ such that $\mathscr{O}_1\subset\mathscr{O}_2$, we have $\mathfrak{A}_{\mathscr{M}}
(\mathscr{O}_1)\subset\mathfrak{A}_{\mathscr{M}}(\mathscr{O}_2)$.

To show this, notice that 
\begin{eqnarray}
\mathfrak{A}i_{\mathscr{O}_1,\mathscr{O}_2}(\mathfrak{A}(\mathscr{O}_1,g\restr{\mathscr{O}_1}))
\subset\mathfrak{A}(\mathscr{O}_2,g\upharpoonright\!_{\mathscr{O}_2}) & \Rightarrow & \nonumber\\ 
\Rightarrow\,\mathfrak{A}i_{\mathscr{O}_2,\mathscr{M}}(\mathfrak{A}i_{\mathscr{O}_1,\mathscr{O}_2}
(\mathfrak{A}(\mathscr{O}_1,g\restr{\mathscr{O}_1}))) & = & \mathfrak{A}i_{\mathscr{O}_1,\mathscr{M}}
(\mathfrak{A}(\mathscr{O}_1,g\restr{\mathscr{O}_1}))\subset\nonumber\\ & \subset & \mathfrak{A}
i_{\mathscr{O}_2,\mathscr{M}}(\mathfrak{A}(\mathscr{O}_2,g\restr{\mathscr{O}_2})),\nonumber 
\end{eqnarray}
where the equality follows from the covariance of $\mathfrak{A}$.

In other words, $\mathscr{K}(\mathscr{M},g)\ni\mathscr{O}\mapsto\mathfrak{A}_{\mathscr{M}}(\mathscr{O})$
is a \emph{precosheaf} of C*-algebras over the causally convex open sets of $\mathscr{M}$, ordered
by inclusion (see page \pageref{ap3d3} in Appendix \ref{ap3} for the categorical definition of the concept). 
If $\mathscr{K}(\mathscr{M},g)$ is \emph{directed} (i.e., given $\mathscr{O}_1,\mathscr{O}_2\in\mathscr{K}
(\mathscr{M},g)$, there exists $\mathscr{O}_3\in\mathscr{K}(\mathscr{M},g)$ such that $\mathscr{O}_1,
\mathscr{O}_2\subset\mathscr{O}_3$), then $\mathscr{K}(\mathscr{M},g)\ni\mathscr{O}\mapsto
\mathfrak{A}_{\mathscr{M}}(\mathscr{O})$ is a \emph{net}.
\item There exists a representation of $G(\mathscr{M},g)\ni u$ by *-automorphisms $\alpha_u\doteq
\mathfrak{A}u$ of $\mathfrak{A}(\mathscr{M},g)$ coherent with the precosheaf (net) structure 
above, that is, $\alpha_{u_1}\circ\alpha_{u_2}=\alpha_{u_1u_2}$ and $\alpha_u(\mathfrak{A}_{\mathscr{M}}
(\mathscr{O}))=\mathfrak{A}_{\mathscr{M}}(u(\mathscr{O}))$, $\forall u,u_1,u_2\in G(\mathscr{M},g)$, 
$\mathscr{O}\in\mathscr{K}(\mathscr{M},g)$.

The first property follows immediately from the covariance of $\mathfrak{A}$. To show the second 
property, notice that $u\restr{\mathscr{O}}$ is a isometric embedding of $(\mathscr{O},g\restr{\mathscr{O}})$ 
into $(u(\mathscr{O}),g\restr{u(\mathscr{O})})$, and that $i_{\mathscr{O},\mathscr{M}}=i_{u(\mathscr{O}),
\mathscr{M}}\circ u$, for all $u\in G(\mathscr{M},g)$, $\mathscr{O}\in\mathscr{K}(\mathscr{M},g)$. The
result then follows, once more, from the covariance of $\mathfrak{A}$, as in item 1.
\item If $\mathfrak{A}$ is \emph{locally causal}, it follows that the precosheaf (net) $\mathscr{O}
\mapsto\mathfrak{A}_{\mathscr{M}}(\mathscr{O})$ is locally causal in the usual sense of \textsc{Haag}
and \textsc{Kastler}, i.e., given $\mathscr{O}_1,\mathscr{O}_2\in\mathscr{K}(\mathscr{M},g)$ such that
$\mathscr{O}_1\perp_{\mathscr{M}}\mathscr{O}_2$, then $[\mathfrak{A}_{\mathscr{M}}(\mathscr{O}_1),
\mathfrak{A}_{\mathscr{M}}(\mathscr{O}_2)]=\{0\}$ (follows immediately from Definition \ref{ch3d1}).
\item Analogously to item 3, if $\mathfrak{A}$ is \emph{primitively causal} (resp. \emph{additive}), the 
precosheaf (net) $\mathscr{O}\mapsto\mathfrak{A}_{\mathscr{M}}(\mathscr{O})$ is primitively causal (resp. 
additive) in the usual sense of \textsc{Haag} and \textsc{Kastler} (up to weak closure of the local
algebras, which we take for granted -- see the end of this Chapter for more on this issue).
\end{enumerate}

\begin{definition}\label{ch3d2}
Let $\mathfrak{A}$ a locally covariant quantum theory, and $(\mathscr{M},g)\in Obj\mathscr{G}lh$.
The precosheaf $\mathscr{K}(\mathscr{M},g)\ni\mathscr{O}\mapsto\mathfrak{A}_{\mathscr{M}}(\mathscr{O})$ 
(net, if $\mathscr{K}(\mathscr{M},g)$ is directed) is said to be a \emph{realization of $\mathfrak{A}$ in
$(\mathscr{M},g)$}.
\end{definition}

There is a detail on the realization of $\mathfrak{A}$ in $(\mathscr{M},g)$ which requires a certain 
care with respect to the concept of a \textsc{Haag-Kastler} precosheaf (net) of local observables. It 
concerns the definition of \emph{quasilocal algebra:} the latter is given by the inductive limit
(see Appendix \ref{ap3}, formula \ref{ap3e1}, page \pageref{ap3e1} for the categorical definition
of the concept)
\begin{equation}\label{ch3e5}
\mathfrak{A}_{\mathscr{M}}\doteq\overline{\bigcup_{\mathscr{O}\in\mathscr{K}(\mathscr{M},g)}\mathfrak{A}
(\mathscr{O})^{\Vert.\Vert}},
\end{equation}
where $\Vert.\Vert$ is the norm of $\mathfrak{A}(\mathscr{M},g)$. Although clearly $\mathfrak{A}_{\mathscr{M}}
\subset\mathfrak{A}(\mathscr{M},g)$, \emph{it's not necessarily true that} $\mathfrak{A}_{\mathscr{M}}
=\mathfrak{A}(\mathscr{M},g)$! Nevertheless, is we ignore $\mathfrak{A}(\mathscr{M},g)$ and understand
$\mathfrak{A}(\mathscr{O})$ as local C*-subalgebras of $\mathfrak{A}_{\mathscr{M}}$, then we see that
the realization of $\mathfrak{A}$ in $(\mathscr{M},g)$ indeed defines a \textsc{Haag-Kastler} precosheaf.
A generalization of the concept of quasilocal algebra that allows one to define locally covariant
quantum theories in non globally hyperbolic spacetimes and incorporate the specification of boundary 
conditions will be presented in Chapter \ref{ch4}, page \pageref{ch4-bound}.\\

We'll now investigate some dynamical properties of a \emph{primitively causal} $\mathfrak{A}$. 
Consider a foliation of $(\mathscr{M},g)$ by \textsc{Cauchy} surfaces $\mathbb{R}\ni t\mapsto 
\Sigma_t=\tau^{-1}(t)$, given by the global time function $\tau$ -- more precisely, such foliation is given  
by the diffeomorphism $F^\tau:\mathbb{R}\times\Sigma\ni(t,x)\mapsto F^\tau(t,x)\in\Sigma_t\subset\mathscr{M}$, 
where we make the convention $\Sigma\doteq\Sigma_0$ (i.e., $F^\tau(t,\Sigma)=\Sigma_t$ and $F^\tau(\mathbb{R},x)$ 
is the orbit of $x\in\Sigma_0$ induced by the foliation). Attribute to each $\Sigma_t$ an open, causally convex
neigbourhood $\mathscr{N}_t\supset\Sigma_t$,\footnote{Notice that we haven't first defined $\mathscr{N}_{t_0}$ 
for a fixed $t_0$ and later on defined $\mathscr{N}_t$ as the image of $\mathscr{N}_{t_0}$ under the flux 
from $t_0$ to $t$ induced by the foliation. As this family of diffeomorphisms doesn't necessarily
preserve the causal structure of $\mathscr{N}_{t_0}$ -- for such, it's necessary and sufficient that the 
diffeomorphisms here are \emph{conformal transformations} \cite{beemee} --, the image of $\mathscr{N}_{t_0}$
may not be causally convex. Moreover, it's sometimes convenient to treat two \emph{disjoint} \textsc{Cauchy} 
surfaces without necessarily having the flux of a foliation connecting both;
a situation when this occurs will be illustrated later on.} and define the two-parameter family
of *-isomorphisms \cite{brunfe2,verch1}
\begin{equation}\label{ch3e6}
\mathbb{R}^2\ni(t_1,t_2)\mapsto\alpha^\tau_{t_1,t_2}\doteq(\mathfrak{A}i_{\mathscr{N}_{t_2},\mathscr{M}})^{-1}
\circ\mathfrak{A}i_{\mathscr{N}_{t_1},\mathscr{M}}:\mathfrak{A}(\mathscr{N}_{t_1},g\restr{\mathscr{N}_{t_1}})
\rightarrow\mathfrak{A}(\mathscr{N}_{t_2},g\restr{\mathscr{N}_{t_2}}),
\end{equation}
denominated \emph{propagator} of $\mathfrak{A}$ in $(\mathscr{M},g)$ associated to $\tau$. The propagator
$\alpha^\tau_{.,.}$ satisfies
\begin{equation}\label{ch3e7}
\alpha^\tau_{t_2,t_3}\circ\alpha^\tau_{t_1,t_2}=\alpha^\tau_{t_1,t_3},\,\alpha^\tau_{t,t}=
\mbox{id}_{\mathfrak{A}(\mathscr{M},g)},\,\forall t_1,t_2,t_3,t\in\mathbb{R},
\end{equation} 
i.e., $\alpha^\tau_{.,.}$ implements the \emph{pair groupoid} $\mathscr{P}\mathbb{R}$ associated to 
$\mathbb{R}$.\footnote{Given a set $S$, the \emph{pair groupoid} $\mathscr{P}S$ associated to $S$ 
is defined in the following way (for the definition of groupoid, see Appendix \ref{ap3}, page \pageref{epimono}): 
$Obj\mathscr{P}S\doteq S$, $Arr\mathscr{P}S=S^2\ni(s_1,s_2)$, where $D((s_1,s_2))=s_1$, $CD((s_1,s_2))=s_2$, 
$\mathbb{1}_s=(s,s)$ and the composition law is given by $(s_2,s_3)(s_1,s_2)=(s_1,s_3)$.} It's important to 
notice that, although the propagator preserves, by definition, the localization of the physical procedures 
throughout time (i.e., of the orbits of the foliation), it doesn't implement in a \emph{geometrical} way the flux 
of diffeomorphisms $F(.+t,.)\circ F^{-1}$, $t\in\mathbb{R}$: consider, for instance, an open set $\mathscr{S}$
of $\Sigma$, and define $\mathscr{S}_t\doteq F^\tau(t,\mathscr{S})$. It's \emph{false} that $\alpha^\tau_{t_1,t_2}
(\mathfrak{A}_{\mathscr{N}_{t_1}}(D(\mathscr{S}_{t_1})\cap{}\mathscr{N}_{t_1},g\restr{D(\mathscr{S}_{t_1})
\cap{}\mathscr{N}_{t_1}}))=\mathfrak{A}_{\mathscr{N}_{t_2}}(D(\mathscr{S}_{t_2})\cap{}\mathscr{N}_{t_2},
g\restr{D(\mathscr{S}_{t_2})\cap{}\mathscr{N}_{t_2}})$ ($D(.)$ denotes the \textsc{Cauchy} development; notice
that the intersection of two causally convex open sets is causally convex), showing that $\alpha^\tau_{.,.}$ 
is, indeed, a \emph{dynamical} and not a kinematic object.\\

We also emphasize that the size of $\mathscr{N}_t$ is essentially irrelevant. Indeed, we can take 
the set $\mathfrak{N}_t$ of all open, causally convex neighbourhoods of $\Sigma_t$ and define
the projective limits (see Appendix \ref{ap3}, formula \ref{ap3e1}, page \pageref{ap3e1} for the 
categorical definition of the concept)
\begin{eqnarray}\label{ch3e8}
\mathfrak{A}_{\Sigma_t} & \doteq & \bigcup_{\mathscr{N}_t\in\mathfrak{N}_t}\mathfrak{A}(\mathscr{N}_t,g
\restr{\mathscr{N}_t})/\sim,\\\mathfrak{A}_{\Sigma_t}(\mathscr{S}_t) & \doteq & \bigcup_{\mathscr{N}_t\in\mathfrak{N}_t}
\mathfrak{A}_{\mathscr{N}_t}(D(\mathscr{S}_t)\cap{}\mathscr{N}_t,g\restr{D(\mathscr{S}_t\cap{}\mathscr{N}_t)})/\sim',
\,\mathscr{S}\subset\Sigma\mbox{ open},\nonumber
\end{eqnarray}
where the equivalence relation $\sim$ identifies the elements of the orbits of the family of *-isomorphisms
$\mathfrak{A}i_{\mathscr{N}^1_t,\mathscr{N}^2_t}$ and the equivalence relation $\sim'$ identifies the elements 
of the orbits of the family of *-isomorphisms $\mathfrak{A}i_{D(\mathscr{S}_t)\cap{}\mathscr{N}^1_t,D(\mathscr{S}_t)
\cap{}\mathscr{N}^2_t}$, $\mathscr{N}^i_t\in\mathfrak{N}_t$, $i=1,2$, $\mathscr{N}^1_t\subset\mathscr{N}^2_t$. 
This procedure defines the quasilocal algebra and the local subalgebras of \emph{germs} of physical procedures 
at instant $t$ with respect to the foliation induced by $\tau$. We also denote by $\alpha^\tau_{t_1,t_2}$ the propagator 
linking $\mathfrak{A}_{\Sigma_{t_1}}$ to $\mathfrak{A}_{\Sigma_{t_2}}$ induced by (\ref{ch3e8}). Notice that the 
projective limit is \emph{not} the ``restriction of $\mathfrak{A}(\mathscr{M},g)$ to $\Sigma_t$'' (such an 
identification is possible only for free quantum field theories, for which the restriction to spacelike 
hypersurfaces is a mathematically well defined operation).

\section{\label{ch3-state}Quantum states in curved spacetimes}

Now, the notion of \emph{states} requires more care in this formalism, for, unlike physical procedures, 
states \emph{are inherently nonlocal} -- they are sensitive to the global geometry of spacetime and can 
as well respond in a nonlocal manner to local perturbations of the metric. 

\begin{definition}\label{ch3d3}
Let the category
\begin{equation}\label{ch3e9}
\mathscr{S}ts=\left\{\begin{array}{rl} \mbox{objects:} & Obj\mathscr{S}ts=\mbox{sets }
\mathfrak{S}\mbox{ of states over unital}\\ & \mbox{C*-algebras;}\\
\mbox{morphisms:} & \mbox{Hom}_{\mathscr{S}ts}(\mathfrak{S},\mathfrak{S}')=\{\sigma:\mathfrak{S}
\rightarrow\mathfrak{S}'\mbox{ linear positive maps;}\}.
\end{array}\right.\nonumber
\end{equation}
Given a locally covariant quantum theory $\mathfrak{A}$, a \emph{(locally contravariant) state space} 
is a contravariant functor $\mathfrak{S}$ from $\mathscr{G}lh_d$ to $\mathscr{S}ts$, such that $\mathfrak{S}
\psi=(\mathfrak{A}\psi)^*$ is the \emph{dual (pullback)} of $\mathfrak{A}\psi$ for all $\psi\in 
Arr\mathscr{G}lh$, i.e., given $\omega\in\mathfrak{S}(\mathscr{M}',g')$ and $\psi\in\mbox{Hom}_{\mathscr{G}lh}
((\mathscr{M},g),$\\$(\mathscr{M}',g'))$, we have $\mathfrak{S}\psi\in\mbox{Hom}_{\mathscr{S}ts}(\mathfrak{S}
(\mathscr{M}',g'),\mathfrak{S}(\mathscr{M},g))$ is defined by $\mathfrak{S}\psi(\omega)=(\mathfrak{A}\psi)^*
(\omega)$\\$\doteq\omega\circ\mathfrak{A}\psi$.
\end{definition}

\section{\label{ch3-field}Locally covariant quantum fields}

In the developments presented so far, the topological structure of C*-algebras attributed to the objects 
of $\mathscr{A}lg$ have had only a secondary role (definition of quasilocal algebras, etc.). Indeed, 
we could have defined the objects of $\mathscr{A}lg$ as unital *-algebras, and attributed a topological
structure suited to convenience. Before proceeding further, let us define the category
\begin{equation}\label{ch3e10}
\mathscr{T}alg=\left\{\begin{array}{rl} \mbox{objects:} & Obj\mathscr{T}alg=\mbox{unital 
topological *-algebras }\mathfrak{F};\\
\mbox{morphisms:} & \mbox{Hom}_{\mathscr{T}alg}(\mathfrak{F},\mathfrak{F}')=\{\alpha:\mathfrak{F}
\rightarrow\mathfrak{F}'\mbox{ unital *-monomorphisms}\}.
\end{array}\right.\nonumber
\end{equation}
and the pushforward functor of test functions $\mathfrak{D}:\mathscr{G}lh_d\rightarrow
\mathscr{T}vs$ given by
\begin{eqnarray}\label{ch3e11}
\mathfrak{D}(\mathscr{M},g) & = & \mathscr{C}^\infty_c(\mathscr{M}),\\
\mathfrak{D}\psi(f)(p) & = & \psi_*(f)(p)\doteq\left\{\begin{array}{lr} f(q) & (p=\psi(q))\\
0 & \mbox{otherwise}\end{array}\right.,\nonumber
\end{eqnarray}
where $\mathscr{T}vs$ is the category of topological vector spaces (see Appendix \ref{ap3}).
We can also, sometimes, understand $\mathfrak{D}$ as a functor from $\mathscr{G}lh$ to the
category $\mathscr{T}op$ of topological spaces.

\begin{definition}\label{ch3d4}
A \emph{locally covariant quantum field theory} is a covariant functor $\mathfrak{F}:
\mathscr{G}lh\rightarrow\mathscr{T}alg$. A \emph{locally covariant scalar quantum field}
is a natural transformation $\Phi:\mathfrak{D}\rightarrow\mathfrak{F}$, where $\mathfrak{D}$ and
$\mathfrak{F}$ are here understood as functors from $\mathscr{G}lh$ to $\mathscr{T}op$. We say that
$\Phi$ is \emph{linear} if $\Phi_{(\mathscr{M},g)}\in Arr\mathscr{T}vs$ for all $(\mathscr{M},g)
\in Obj\mathscr{G}lh$.
\end{definition}

\begin{remark}\label{ch3r1}
Notice that the hypothesis of $\Phi_{(\mathscr{M},g)}$ being linear in the sense of Definition \ref{ch3d4}
has \emph{nothing} to do with the nature of its self interaction!  
\end{remark}

The notions of \emph{local causality, primitive causality, additivity}, \emph{realization} of 
$\mathfrak{F}$ in $(\mathscr{M},g)\in Obj\mathscr{G}lh$ and of \emph{state space} are 
naturally obtained as extensions of the corresponding concepts for  locally covariant quantum
theories (the only needed modification is the substitution of ``*-algebra'' for ``C*-algebra'' in 
formula (\ref{ch3e9}) of Definition \ref{ch3d3}, page \pageref{ch3e9}).\\

A typical example (albeit apparently trivial) of a locally covariant quantum field theory is the 
attribution, to each $(\mathscr{M},g)\in Obj\mathscr{G}lh$, of the \textsc{Borchers-Uhlmann} 
\emph{algebra} $\mathscr{F}(\mathscr{M})$ (see Appendix \ref{ap2-borchers}, Definition \ref{ap2d10}, 
page \pageref{ap2d10}) -- it's immediate to see that $\mathfrak{F}$ satisfies Definition \ref{ch3d4}. 
This locally covariant quantum field theory serves uniquely the purpose of defining the kinematics 
of an Hermitian scalar field, and it's onwards called the \textsc{Borchers-Uhlmann} \emph{functor}. 
The realization of this $\mathfrak{F}$ in $(\mathscr{M},g)\in Obj\mathscr{G}lh$ has as its quasilocal 
algebra $\mathfrak{F}_{\mathscr{M}}=\mathfrak{F}(\mathscr{M})$ (it suffices to apply partitions of unity), 
and possible examples of locally covariant scalar quantum fields associated to $\mathfrak{F}$ are:

\begin{itemize}
\item $\Phi_{(\mathscr{M},g)}=\mbox{id}_{\mathscr{C}^\infty_c(\mathscr{M})}$, i.e, $\Phi$ identifies
$f\in\mathscr{C}^\infty_c(\mathscr{M})$ with the corresponding element of $\mathfrak{F}(\mathscr{M})$;
\item $\Phi_{(\mathscr{M},g)}(f)=[f]$, where $[f]$ is the equivalence class of $f$ under the quotient
of $\mathfrak{F}(\mathscr{M})$ modulo a *-ideal $\mathfrak{I}(\mathscr{M},g)$. In order to $(\mathscr{M},g)
\mapsto\mathfrak{F}(\mathscr{M})/\mathfrak{I}(\mathscr{M},g)$ to define a locally covariant quantum
field theory, it's necessary that the correspondence $(\mathscr{M},g)\mapsto\mathfrak{I}
(\mathscr{M},g)$ defines a covariant functor from $\mathscr{G}lh$ to $\mathscr{T}alg$. In this case,
the quotient map defines a natural transformation, by virtue of the its universal property
\cite{jacobson} (see also Appendix \ref{ap3}, page \pageref{ap2d10}). The specification of *-ideals 
as $\mathfrak{I}(\mathscr{M},g)$ constitutes a way to impose a dynamics to the \textsc{Borchers-Uhlmann}
algebra. This example, as well as the previous one, result in linear fields.
\item Here we elaborate the previous example, basing on the considerations at the end of Section 
\ref{ap2-borchers}. Consider a state space $\mathfrak{S}$, where $\mathfrak{S}(\mathscr{M},g)$ consists 
of a set of hierarchies $\omega$ of $k$-point distributions $\omega_k\in\mathscr{D}'(\mathscr{M}^k)$ 
and $\mathfrak{S}\psi$ is the pullback $\psi^*$ of distributions under $\psi\in Arr\mathscr{G}lh$, 
such that \[\mathfrak{I}(\mathscr{M},g)\subset\bigcap_{\omega\in\mathfrak{S}(\mathscr{M},g)}\mbox{Ann}\omega.\] 
In this case, the locally covariant quantum field $\Phi$ of the former example naturally induces a set 
of states $\Phi^*_{(\mathscr{M},g)}\mathfrak{S}(\mathscr{M},g)$ over each $\widetilde{\mathfrak{F}}
(\mathscr{M},g)\doteq\mathfrak{F}(\mathscr{M})/\mathfrak{I}(\mathscr{M},g)$ \[(\Phi^*_{(\mathscr{M},g)}
\omega_k)([f_1],\ldots,[f_k])=\omega_k(f_1,\ldots,f_k),\] in a way that $\Phi^*_.\mathfrak{S}(.)$ is a
locally contravariant state space.
\end{itemize}

The possibility of proceeding as in the formalism of \textsc{Wightman} and defining quantum fields
starting from states by means of the \textsc{Wightman}-GNS representation can, of course, can, of course,
be conveyed for the realization of the \textsc{Borchers-Uhlmann} functor in some $(\mathscr{M},g)\in
\mathscr{G}lh$, although it should be considered with care as far as the Principle of Local Covariance
is concerned, for \emph{individual} states are not locally contravariant. We may, nevertheless, 
consider the following scenario -- by the way, already assessed by \textsc{Haag} and \textsc{Kastler} 
in \cite{hkast}, and which served them as a motivation for the introduction of the algebraic approach
to QFT -- under the light of the examples addressed above.\\

Let $\mathfrak{S}$ be a state space associated to $\mathfrak{F}$, such that
\begin{equation}\label{ch3e12}
\mbox{Ann}\omega_{(\mathscr{M},g)}=\mbox{Ann}\omega'_{(\mathscr{M},g)}\doteq\mathfrak{I}
(\mathscr{M},g),\,\forall\omega_{(\mathscr{M},g)},\omega'_{(\mathscr{M},g)}\in\mathfrak{S}
(\mathscr{M},g),\,(\mathscr{M},g)\in Obj\mathscr{G}lh. 
\end{equation}

Employing the contravariance of $\mathfrak{S}$, we can show that, in this case, the 
correspondence $(\mathscr{M},g)\mapsto\mathfrak{I}(\mathscr{M},g)$ defines a locally covariant quantum
field theory. Hence, as we've seen, $\Phi_{(\mathscr{M},g)}(f)=[f]=f/\mathfrak{I}(\mathscr{M},g)$ defines 
then a locally covariant scalar quantum field, thus realizing our \emph{desideratum} about incorporating 
the dynamical information of a locally covariant quantum field theory into the specification of a state 
space, showing at the same time that the quantum field obtained is independent of a choice of a 
representative from the set of states, by the First *-Isomorphism Theorem.\\

From the viewpoint of a locally covariant quantum field theory $\mathfrak{A}$, (\ref{ch3e12}) 
has an even deeper interpretation: the GNS representations of $\mathfrak{A}(\mathscr{M},g)$
associated to each $\omega_{(\mathscr{M},g)}\in\mathfrak{S}(\mathscr{M},g)$ are not only 
\emph{quasiequivalent} (i.e., the sets of normal states over $\pi_{\omega_{(\mathscr{M},g)}}
(\mathfrak{A}(\mathscr{M},g))''$ and $\pi_{\omega'_{(\mathscr{M},g)}}(\mathfrak{A}(\mathscr{M},g))''$
coincide, or, equivalently, these two \textsc{von Neumann} algebras are  *-isomorphic --
see Theorem 2.4.26 of \cite{bratteli1}), but also, as a consequence, the GNS representations
of the quasilocal algebra $\mathfrak{A}_{\mathscr{M}}$ of the realization of $\mathfrak{A}$ in any
$(\mathscr{M},g)$ are \emph{locally quasiequivalent}, that is, the sets of normal states over 
$\pi_{\omega_{(\mathscr{M},g)}}(\mathfrak{A}(\mathscr{O}))''$ and $\pi_{\omega'_{(\mathscr{M},g)}}(\mathfrak{A}
(\mathscr{O}))''$ coincide. Hence, local quasiequivalence becomes a highly desirable requirement for
physically relevant state spaces and essentially allows us to pass from local C*-algebras to local 
\textsc{von Neumann} algebras in the same manner as in the \textsc{Haag-Kastler} formalism. An even 
more detailed discussion about the role of local quasiequivalence in the context of the Principle of 
Local Covariance can be found in \cite{bfv}.

\chapter{\label{ch4}Implementing the Holographic Principle in AAdS spacetimes}
\chaptermark{Holographic Principle in AAdS}

\epigraph{\emph{\quad In due modi si raggiunge Despina: per nave o per cammello. La 
citt\`a si presenta differente a chi viene da terra e a chi dal mare.\\
\quad (...)\\
\quad Ogni citt\`a riceve la sua forma dal deserto a cui si oppone; e cos\`{\i} il cammelliere 
e il marinaio vedono Despina, citt\`a di confine tra due deserti.}\endnotemark[10]}
{\textsc{Italo Calvino}\\ ``Le citt\`a e il desiderio 3'' (\emph{Le citt\`a 
invisibili})\endnotemark[11]}

\section{\label{ch4-bound}Local covariance with boundary conditions}

In Chapter \ref{ch3}, we've built a general framework for the description of locally 
covariant quantum theories in globally hyperbolic spacetimes. Within such a class, commutation 
relations among local procedures (given by the elements of the local algebras) can be established 
without ambiguities, for globally hyperbolic spacetimes constitute ``closed'' systems with respect to 
\emph{causal} dynamical laws, i.e., which propagate local physical effects with speed inferior 
or equal of that of the light. However, we know that we can always cover \emph{any} spacetime with 
open, globally hyperbolic and causally convex neighbourhoods -- in the case of strongly causal 
spacetimes, such neighbourhoods even constitute a basis for the spacetime topology. And not even
for this should we expect that, in these ``open'' systems, we may assume local causality \emph{a 
priori}. \\

In the case of the realization of $\mathfrak{A}$ in a globally hyperbolic spacetime, 
local causality is coherent with the inclusions of the algebras, for by means of an
adequate choice of the family of globally hyperbolic neighbourhoods, it's always possible 
to find, given any two causally disjoint such neighbourhoods (one of them, eventually, being
sufficiently small), a third one which contains both.\footnote{See \cite{guilrv}. This fact if of
the utmost importance in the algebraic theory of superselection sectors.} It's precisely the violation 
of this property that causes difficulties in the case of non globally hyperbolic spacetimes. \\

To pinpoint such difficulties proper in the extension of the Principle of Local Covariance to
spacetimes such as (A)AdS, we shall now exhibit a more general way to construct the realization of 
a locally covariant quantum theory $\mathfrak{A}$, valid even for non globally hyperbolic spacetimes
$(\mathscr{M},g)$. This path was proposed by \textsc{Sommer} \cite{sommer} and makes more transparent 
the necessity and the role of boundary conditions on the imposition of local and primitive causality 
-- in order to make clearer the context of the latter property, we'll restrict ourselves to stably
causal spacetimes, for these, as seen in Appendix \ref{ap1}, can be foliated by ``constant time'' 
hypersurfaces, for which it's meaningful to formulate a (mixed boundary-)initial-value problem.\\

Let, then, $(\mathscr{M},g)$ be a stably causal spacetime, and $\mathscr{K}(\mathscr{M},g)$
the set of globally hyperbolic, causally convex open regions $(\mathscr{O},g\restr{\mathscr{O}})$, 
$\mathscr{O}\subset\mathscr{M}$ (we sometimes omit $g\restr{\mathscr{O}}$ whenever such procedure 
doesn't cause confusion). Let us consider initially, given a locally covariant quantum theory 
$\mathfrak{A}$,
\begin{equation}\label{ch4e1}
\mathfrak{A}_{\mathscr{M}}=\bigvee\{(A,\mathscr{O}):A\in\mathfrak{A}(\mathscr{O},g\restr{\mathscr{O}}),
\,\mathscr{O}\in\mathscr{K}(\mathscr{M},g)\},
\end{equation}
the \emph{free} *-algebra generated by the elements indicated above. Let now the *-ideal
$\mathfrak{I}_{\mathscr{M}}\doteq\mathfrak{A}_{\mathscr{M}}\mathfrak{E}_{\mathscr{M}}\mathfrak{A}_{\mathscr{M}}$, 
where $\mathfrak{E}_{\mathscr{M}}$ is the *-algebra generated by the relations
\begin{eqnarray}\label{ch4e2}
& (A,\mathscr{O})+(B,\mathscr{O})-(A+B,\mathscr{O}), & \\
& (A,\mathscr{O})(B,\mathscr{O})-(AB,\mathscr{O}), & \nonumber\\
& \lambda(A,\mathscr{O})-(\lambda A,\mathscr{O}),\,\forall\lambda\in\mathbb{C}\mbox{ and } & \nonumber\\
& (A,\mathscr{O}_1)-(\mathfrak{A}i_{\mathscr{O}_1,\mathscr{O}_2}(A),\mathscr{O}_2),\,
\forall\mathscr{O},\,\mathscr{O}_1,\,\mathscr{O}_2\in\mathscr{K}(\mathscr{M},g),\,\mathscr{O}_1
\subset\mathscr{O}_2. & \nonumber
\end{eqnarray}

We know that each element $(A,\mathscr{O})$ possesses the (C*-)norm $\Vert (A,\mathscr{O})\Vert=\Vert A\Vert$,
consistent with the relations (\ref{ch4e2}) (recall that $\mathfrak{A}i_{\mathscr{O}_1,\mathscr{O}_2}$
is a \emph{*-monomorphism} for $\mathscr{O}_1,\mathscr{O}_2\in\mathscr{K}(\mathscr{M},g)$, and, hence,
$\Vert(\mathfrak{A}i_{\mathscr{O}_1,\mathscr{O}_2}(A),\mathscr{O}_2)\Vert=\Vert(A,\mathscr{O}_1)\Vert
=\Vert A\Vert$!). Applying now the universal construction of \textsc{Blackadar} \cite{blackadar1,
blackadar2},\footnote{I thank Prof. Severino Toscano do R\^ego Melo for bringing the reference 
\cite{blackadar1} to my attention.} to the generators in (\ref{ch4e1}) and to the relations (\ref{ch4e2}), 
we define the collection $\Pi(\mathfrak{A}_{\mathscr{M}},\mathfrak{I}_{\mathscr{M}})$ of \emph{admissible 
*-representations} of the pair $(\mathfrak{A}_{\mathscr{M}},\mathfrak{I}_{\mathscr{M}})$, consisting in the 
*-representations $\pi$ of $\mathfrak{A}_{\mathscr{M}}$ such that $\pi(\mathfrak{I}_{\mathscr{M}})=\{0\}$.  
It thus follows that the \emph{universal} C*-seminorm $\Vert A\Vert_u=\sup_{\pi\in\Pi(\mathfrak{A}_{\mathscr{M}},
\mathfrak{I}_{\mathscr{M}})}\Vert\pi(A)\Vert$ is \emph{finite} for all $A$. Denoting the *-ideal 
$\bar{\mathfrak{I}}_{\mathscr{M}}\doteq\{A:\Vert A\Vert_u=0\}$, then we obtain the \emph{quasilocal} C*-algebra
$\mathfrak{A}(\mathscr{M},g)$ as the universal C*-algebra given by the generators $\mathfrak{A}_{\mathscr{M}}$ 
and relations $\mathfrak{I}_{\mathscr{M}}$:\footnote{Notice that we \emph{haven't} taken the quotient modulo
the *-ideal $\mathfrak{I}_{\mathscr{M}}$. Were there a way to guarantee the existence of an \emph{admissible} 
*-representation $\pi$ such that $\mbox{Ker}\pi=\mathfrak{I}_{\mathscr{M}}$, we could equivalently take such
a quotient. However, this is in general impossible, for $\mathfrak{I}_{\mathscr{M}}$ is not necessarily closed
in the C*-norm defined for the elements of $\mathfrak{A}_{\mathscr{M}}$.}
\begin{equation}\label{ch4e3}
\mathfrak{A}(\mathscr{M},g)\doteq\overline{(\mathfrak{A}_{\mathscr{M}}
/\bar{\mathfrak{I}}_{\mathscr{M}})^{\Vert.\Vert_u}}.
\end{equation}

Let's now see how the *-morphisms induced by isometric embeddings $\psi:(\mathscr{M}_1,g_1)
\rightarrow(\mathscr{M}_2,g_2)$ with causally convex image behave. Obviously, $\psi(\mathscr{K}(\mathscr{M}_1,
g_1))\subset\mathscr{K}(\mathscr{M}_2,g_2)$, and $\psi\restr{\mathscr{O}}$, $\mathscr{O}\in\mathscr{K}
(\mathscr{M}_1,g_1)$, defines and isometric embedding of $\mathscr{O}$ into $\psi(\mathscr{O})$, naturally
defining a *-monomorphism from $\mathfrak{A}_{\mathscr{M}_1}$ to $\mathfrak{A}_{\mathscr{M}_2}$. 
Let us define the following *-morphism:

\begin{eqnarray}\label{ch4e4}
\mathfrak{A}\psi:\mathfrak{A}(\mathscr{M}_1,g_1) & \rightarrow & \mathfrak{A}(\mathscr{M}_2,
g_2) \\ A+\bar{\mathfrak{I}}_{\mathscr{M}_1} & \mapsto & \mathfrak{A}\psi(A)+\bar{\mathfrak{I}}_{\mathscr{M}_2}.
\nonumber
\end{eqnarray}

Notice that, due to the considerations of the previous paragraph, if $[A]_{\mathscr{M}_i}$ is the equivalence 
class of $A\in\mathfrak{A}_{\mathscr{M}_i}$ (under the quotient) modulo $\bar{\mathfrak{I}}_{\mathscr{M}_i}$ 
(i.e., $[A]_{\mathscr{M}_i}=[A+J]_{\mathscr{M}_i}$ if $J\in\bar{\mathfrak{I}}_{\mathscr{M}_i}$), $i=1,2$, it 
follows that $\mathfrak{A}\psi([A+J]_{\mathscr{M}_1})=[\mathfrak{A}\psi\restr{\mathscr{O}}(A)]_{\mathscr{M}_2}$, 
for all $A\in\mathfrak{A}(\mathscr{O},g\restr{\mathscr{O}})$, showing that the definition (\ref{ch4e4}) isn't
empty and indeed defines a nontrivial *-morphism. $\mathfrak{A}\psi$ is unital, but \emph{not necessarily 
injective!} Hence, we must weaken the properties of the morphisms of the category of (C)*-algebras so as to 
$\mathfrak{A}$ to define a covariant functor from the (full super)category of the stably causal spacetimes 
to the latter -- the covariance of $\mathfrak{A}$ is guaranteed by the First *-Isomorphism Theorem (Theorem 
\ref{ap2t2}, page \pageref{ap2t2}). We'll now exhibit a situation which illustrates with perfection how this 
fact complicates the imposition of local causality. \\

Consider $\mathscr{O}_1,\,\mathscr{O}_2\in\mathscr{K}(\mathscr{M}_1,g_1)$ causally disjoint, such that there
is \emph{no} $\mathscr{O}_3\in\mathscr{K}(\mathscr{M}_1,$\\$g_1)$ such that $\mathscr{O}_1,\,\mathscr{O}_2\subset
\mathscr{O}_3$, but there exists an $\mathscr{O}_4\in\mathscr{K}(\mathscr{M}_2,g_2)$ such that $\psi
(\mathscr{O}_1),\,\psi(\mathscr{O}_2)$\\$\subset\mathscr{O}_4$. Let $\mathfrak{A}$ be locally causal (i.e., 
with respect to globally hyperbolic regions). Then we have \[[(\mathfrak{A}(i_{\psi(\mathscr{O}_1),\mathscr{O}_4}
\circ\psi)(A),\mathscr{O}_4),(\mathfrak{A}(i_{\psi(\mathscr{O}_2),\mathscr{O}_4}\circ\psi)(B),\mathscr{O}_4)]=0.\] 
However, there is nothing under our hypotheses preventing \[[(A,\mathscr{O}_1),(B,\mathscr{O}_2)]=C\neq 0,\] 
unless $A=0$ or $B=0$. Hence, for $A,B\neq 0$, it follows that $C\notin\mathfrak{I}_{\mathscr{M}_1}$ and 
$[C]_{\mathscr{M}_1}\in\mbox{Ker}\mathfrak{A}\psi$, potentially spoiling the injectivity of $\mathfrak{A}\psi$ 
(it may occur, though, that $C\in\bar{\mathfrak{I}}_{\mathscr{M}_1}$, but there is no way of guaranteeing this 
\emph{a priori}). Nevertheless, the *-morphisms $\mathfrak{A}(\psi\circ i_{\mathscr{O},\mathscr{M}_1})$ are always 
injective, for all $\mathscr{O}\in\mathscr{K}(\mathscr{M}_1,g_1)$.\\

The construction above allows one to write the following

\begin{definition}\label{ch4d1}
Let the categories
\begin{equation}\label{ch4e5}
\mathscr{S}tc_d=\left\{\begin{array}{rl} \mbox{objects:} & Obj\mathscr{G}lh_d=d\mbox{-dimensional, 
stably causal}\\ & \mbox{spacetimes }(\mathscr{M},g);\\
\mbox{morphisms:} & \mbox{Hom}_{\mathscr{S}tc_d}((\mathscr{M},g),(\mathscr{M}',g'))=\{\psi:
(\mathscr{M},g)\rightarrow(\mathscr{M}',g')\\ & \mbox{isometric embeddings with open, causally 
convex}\\ & \mbox{image}\};
\end{array}\right.\nonumber
\end{equation}
($\mathscr{G}lh$ is, hence, a full subcategory of $\mathscr{S}tc$) and
\begin{equation}\label{ch4e6}
\mathscr{A}lgb=\left\{\begin{array}{rl} \mbox{objects:} & Obj\mathscr{A}lg=\mbox{unital 
C*-algebras }\mathfrak{A};\\
\mbox{morphisms:} & \mbox{Hom}_{\mathscr{A}lg}(\mathfrak{A},\mathfrak{A}')=\{\alpha:\mathfrak{A}
\rightarrow\mathfrak{A}'\mbox{ unital *-morphisms}\\ & \mbox{(not necessarily injective)}\}.
\end{array}\right.\nonumber
\end{equation}
An \emph{extended locally covariant quantum theory} is a covariant functor $\mathfrak{A}$ from 
$\mathscr{S}tc_d$ to $\mathscr{A}lgb$, such that $\mathfrak{A}\restr{\mathscr{G}lh}$ is a locally covariant 
quantum theory in the sense of Definition \ref{ch3d1}, page \pageref{ch3d1}. Given $(\mathscr{M},g)\in Obj
\mathscr{S}tc_d$, the \emph{realization} of $\mathfrak{A}$ in $(\mathscr{M},g)$ is given by the precosheaf 
$\mathscr{K}(\mathscr{M},g)\ni\mathscr{O}\mapsto\mathfrak{A}(\mathscr{O})\doteq\overline{(\{(A,\mathscr{O}):
A\in\mathfrak{A}(\mathscr{O},g\restr{\mathscr{O}})\}/\bar{\mathfrak{I}}_{\mathscr{M}})^{\Vert.\Vert}}$, 
following \textsc{Blackadar}'s construction given above. We say yet that $\mathfrak{A}$ is \emph{regular} 
is each local algebra $\mathfrak{A}(\mathscr{M},g)$ coincides with the quasilocal algebra of the realization 
of$\mathfrak{A}$ in $(\mathscr{M},g)$, given by (\ref{ch4e3}).\footnotemark
\end{definition}
\footnotetext{Here, we indulge in a slight abuse of notation, for we'll only deal with the regular case
in what follows.}

It's immediate to see that the realization of $\mathfrak{A}$ in $(\mathscr{M},g)\in Obj\mathscr{G}lh_d$
in the sense of Definition \ref{ch4d1} coincides with Definition \ref{ch3d2} up to an *-isomorphism. The
loss of injectivity of the *-morphisms induced by $\mathfrak{A}$ translates itself to the freedom of choice
of boundary conditions. Before making this idea more precise, we'll show what becomes of the remaining 
properties of $\mathfrak{A}$. 

\begin{itemize}
\item If $(\mathscr{M},g)$ possesses a nontrivial group of isometries $G\ni\psi$, it becomes clear 
that $\psi$ induces a bijection of $\mathscr{K}(\mathscr{M},g)\ni\mathscr{O}$ onto itself. Thus, we see
that the *-morphism $\mathfrak{A}\psi$ is explicitly invertible, with inverse given by $\mathfrak{A}
(\psi^{-1})$, for all $\psi\in G$. Thus, the Principle of Local Covariance allows the implementation
of the action of isometry groups even in the non globally hyperbolic case.
\item \emph{Local causality} holds in the following sense: if the restriction of $\mathfrak{A}$ to 
the category of globally hyperbolic spacetimes is locally causal, and given, as in the example above,
$\mathscr{O}_1,\,\mathscr{O}_2\in\mathscr{K}(\mathscr{M}_1,g_1)$ causally disjoint with respect to $\mathscr{O}_3
\in\mathscr{K}(\mathscr{M}_1,g_1)$ (i.e., $\mathscr{O}_3\supset\mathscr{O}_1,\mathscr{O}_2$ and 
$\mathscr{O}_1\perp_{\mathscr{O}_3}\mathscr{O}_2$), it follows by local covariance that \[[\mathfrak{A}i_{
\mathscr{O}_1,\mathscr{M}}(A_1),\mathfrak{A}i_{\mathscr{O}_2,\mathscr{M}}(A_2)]=0,\] for all $A_i\in\mathfrak{A}
(\mathscr{O}_i,g\restr{\mathscr{O}_i})$, $i=1,2$.
\item \emph{Additivity} remains valid with no changes, as long as the covering consists in globally 
hyperbolic regions.
\item \emph{Primitive causality} has no immediate extension to non globally hyperbolic spacetimes. In the 
case of \emph{stably causal} spacetimes, a tentative definition could be the following: let $\tau$ be a 
global time function in $(\mathscr{M},g)$, and $\mathscr{N}\subset\mathscr{M}$ an open, causally convex
neighbourhood of, say, $\Sigma=\tau^{-1}(0)$. Then we say that $\mathfrak{A}$ is \emph{weakly primitively 
causal} if it's primitively causal with respect to globally hyperbolic regions and $\mathfrak{A}i_{\mathscr{N}\!,
\mathscr{M}}$ is a \emph{*-epimorphism}. Such a definition is convenient, for the ambiguity in the choice
of boundary conditions becomes expressed in a purely algebraic manner, as follows: consider any *-ideal 
$\mathfrak{J}$ of $\mathfrak{A}(\mathscr{N},g\restr{\mathscr{N}})$ containing $\mbox{Ker}\mathfrak{A}\psi$. 
If $\mathfrak{A}$ is weakly primitively causal, then the map $A+\mathfrak{J}\mapsto\mathfrak{A}\psi(A)+
\mathfrak{A}i_{\mathscr{N}\!,\mathscr{M}}(\mathfrak{J})$ is a *-isomorphism from $\mathfrak{A}(\mathscr{N},g
\restr{\mathscr{N}})/\mathfrak{J}$ onto $\mathfrak{A}(\mathscr{M},g)/\mathfrak{A}i_{\mathscr{N}\!,\mathscr{M}}
(\mathfrak{J})$, by the First *-Isomorphism Theorem. 
\end{itemize}

The specification of the ideal $\mathfrak{J}$ in the context of \emph{weak primitive causality} corresponds 
precisely to the choice of boundary conditions. That a ``minimal'' choice is always possible, is guaranteed 
by the definition of weak primitive causality (it suffices to take $\mathfrak{J}=\mbox{Ker}\mathfrak{A}
i_{\mathscr{N}\!,\mathscr{M}}$). A ``maximal'' choice would be the case in which $\mathfrak{J}$ is a 
\emph{maximal} ideal of $\mathfrak{A}(\mathscr{N},g\restr{\mathscr{N}})$ (the existence of maximal *-ideals
follows from the \textsc{Zorn} lemma; see Appendix \ref{ap3} for its statement in a categorical context) -- 
then the quotient *-algebra $\mathfrak{A}(\mathscr{N},g\restr{\mathscr{N}})/\mathfrak{J}$ is \emph{simple}, 
i.e., it has no nontrivial closed *-ideals, and any *-epimorphism with domain in $\mathfrak{A}(\mathscr{N},
g\restr{\mathscr{N}})/\mathfrak{J}$ is automatically a *-isomorphism. \\

It's important to choose \emph{boundary *-ideals} $\mathfrak{J}$ which are preserved by $\{\mathfrak{A}\psi:
\psi\in G(\mathscr{M},g)\}$, for only then is $G(\mathscr{M},g)$ implemented in $\mathfrak{A}(\mathscr{M},g)
/\mathfrak{A}i_{\mathscr{N}\!,\mathscr{M}}(\mathfrak{J})$ by a group *-automorphisms. Local covariance is 
preserved by the quotient, for $\mathfrak{A}i_{\mathscr{O},\mathscr{N}}(\mathfrak{A}(\mathscr{O},g
\restr{\mathscr{O}}))/(\mathfrak{J}\cap{}\mathfrak{A}i_{\mathscr{O},\mathscr{N}}(\mathfrak{A}(\mathscr{O},g
\restr{\mathscr{O}})))\cong(\mathfrak{A}i_{\mathscr{O},\mathscr{N}}(\mathfrak{A}(\mathscr{O},g
\restr{\mathscr{O}}))+\mathfrak{J})/\mathfrak{J}$, by the Second *-Isomorphism Theorem (Theorem \ref{ap2t3}, 
page \pageref{ap2t3}). \\

\begin{remark}\label{ch4r1}
As the constructions above make reference uniquely to *-algebraic properties, they naturally extend to 
\textsc{Borchers-Uhlmann} algebras and, as a consequence, to locally covariant quantum fields. Indeed, in
the case of free fields, the algebraic imposition of boundary conditions supersedes the corresponding 
functional analytic concept in the one-particle \textsc{Hilbert} space (i.e., self-adjoint extensions of 
the Hamiltonian generator of unitary time evolution) -- see \cite{sommer} for the particular case of the 
\textsc{Weyl} algebra of the \textsc{Klein-Gordon} field in $\{x\in\mathbb{R}^{1,d-1}:x^{d-1}>0\}$ with 
\textsc{Dirichlet} or \textsc{Neumann} boundary conditions at $x^{d-1}=0$. Both *-ideals in this example 
are maximal.
\end{remark}

\section{\label{ch4-ads}Locally covariant quantum theories in AdS}

We shall now consider the realization of a locally covariant quantum theory $\mathfrak{A}$ in $AdS_d$
in the spirit of Section \ref{ch4-bound}, following the work of \textsc{Buchholz, Florig} and \textsc{Summers} 
\cite{bfs}. (The connected component to identity of) the group of isometries of $AdS_d$ induces, as seen
in Section \ref{ch4-bound}, a group of *-automorphisms $\{\alpha_g:g\in SO_e(2,d-1)\}$ in $\mathfrak{A}(AdS_d)$ 
which preserves the local structure of $\mathfrak{A}$. \\

Of particular interest are the (quasi)local algebras associated to wedges $\mathscr{W}_{p,q}$, $p\ll 
q\in\mathscr{M}in(r)$, $r\in\mathscr{I}$. As seen in Subsection \ref{ch1-ads-wedge}, each $\mathscr{W}_{p,q}$ 
possesses a one-parameter subgroup of isometries $\{u^\lambda_{p,q},\lambda\in\mathbb{R}\}$ of $AdS_d$ which 
preserves the former, given by formula (\ref{ch1e26}). We emphasize here a property of $u^\lambda_{p,q}$ obtained
in the more general context of Chapter \ref{ch2} (see page \pageref{ch2e45}) and which will be of great 
importance in the present Chapter: the asymptotic behaviour of $g^\lambda_{p,q}$ as $\lambda\rightarrow\pm\infty$. 
It follows immediately from formulae (\ref{ch1e26}) (page \pageref{ch1e26}), (\ref{ch2e45}) (page \pageref{ch2e45})
and the adjoint action of $SO_e(2,d-1)$ that, in \textsc{Poincar\'e} coordinates,
\begin{equation}\label{ch4e7}
u^\lambda_{p,q}(x,z)\sim\left\{\begin{array}{lr}(e^{-\lambda}(x-x(q)),e^{-\lambda}z) & (\lambda\rightarrow
+\infty) \\ (e^\lambda(x-x(p)),e^\lambda z) & (\lambda\rightarrow-\infty) \end{array}\right.,\forall (x,z)
\in\mathscr{W}_{p,q},
\end{equation}
i.e., $u^\lambda_{p,q}$ acts asymptotically as scaling transformations around $p$ (resp. $q$) as $\lambda
\rightarrow-\infty$ (resp. $\lambda\rightarrow+\infty$). Gone through these geometrical preliminaries,
we'll specify what we assume from our physical models:

\begin{enumerate}
\item The model is given by the realization of a regular, extended locally covariant quantum theory
$\mathfrak{A}$ in the collection $\mathscr{W}(AdS_d)$ of the wedges of $AdS_d$, which, on its turn, provide 
by local covariance a realization of $\mathfrak{A}$ in $AdS_d$, according to the construction of Section 
\ref{ch4-bound}.
\item We suppose that the group $\mathfrak{A}u^\lambda_{p,q}$ of *-automorphisms of $\mathfrak{A}
(\mathscr{W}_{p,q})$ is strongly continuous for all $(p,q)\in\mathscr{D}(\mathscr{I})$.
\item $\mathfrak{A}$ is locally causal with respect to globally hyperbolic regions.\footnote{This condition isn't 
assumed in \cite{bfs}, which employs a minimal adaptation of the formalism of \textsc{Haag} and 
\textsc{Kastler} to AdS spacetimes.}
\end{enumerate}

We opt for realizing $\mathfrak{A}$ not only in $AdS_d$, but also in each wedge individually, for this 
allows us not only to introduce physically reasonable boundary conditions but also to extend a good 
deal of our following considerations to locally AAdS spacetimes. The condition on the ``elementary'' 
states will be discussed in Subsection \ref{ch4-ads-dyn} below. \\

We conclude this Subsection by outlining, for completeness, the original form of \textsc{Rehren} duality:
\cite{rehren1}. 

\begin{definition}\label{ch4d2}
The \textsc{Rehren}\emph{-dual} quantum theory to the realization of $\mathfrak{A}$ in $AdS_d$ is the
\emph{conformally covariant} precosheaf $\mathscr{D}_{p,q}\mapsto\mathfrak{A}(\mathscr{D}_{p,q})
\doteq\mathfrak{A}(\rho_{AdS_d}^{-1}(\mathscr{D}_{p,q}))=\mathfrak{A}(\mathscr{W}_{p,q})$ of 
C*-algebras, indexed by the collection of diamonds in $\mathscr{I}$. Here, $\rho_{AdS_d}$ is the
\textsc{Rehren} bijection (\ref{ch1e32}) (page \pageref{ch1e32}).
\end{definition}

\subsection{\label{ch4-ads-dyn}Boundary values as a dynamical condition}

It's natural to understand each $\mathfrak{A}(\mathscr{W}_{p,q})$ as a C*-dynamical system, whose 
group of *-automorphisms is given by $\mathfrak{A}u^\lambda_{p,q}$. The ``elementary'' states
$\omega$ in $\mathfrak{A}(AdS_d)$ should represent a situation of thermodynamic equilibrium,
since the dynamics is *-automorphic -- physically, the quantum dynamical system $(\mathfrak{A}
(\mathscr{W}_{p,q}),\mathfrak{A}u^._{p,q})$ is \emph{closed}. 

We demand that, for any uniformly accelerated observer associated to $\mathfrak{A}u^\lambda_{p,q}$, 
a state $\omega$ over $\mathfrak{A}(\mathscr{W}_{p,q})$ be \emph{passive}, i.e.,
\begin{equation}\label{ch4e8}
-i\omega(U^*\delta_{p,q} U)\geq 0
\end{equation}
for all $U\in\mathfrak{U}_1(\mathfrak{A}(\mathscr{W}_{p,q}))\cap D(\delta_{p,q})$, where $\delta_{p,q}$ 
is the derivation which generates $\mathfrak{A}u^\lambda_{p,q}$, with domain $D(\delta_{p,q})\subset
\mathfrak{A}(\mathscr{W}_{p,q})$, and $\mathfrak{U}_1(\mathfrak{A}(\mathscr{W}_{p,q}))$ is the connected 
component to identity of the group of unitary elements of $\mathfrak{A}(\mathscr{W}_{p,q})$. The requirement 
of passivity means that it's not possible to extract energy from $\omega$ by means of a cyclic process -- 
by this, one understands time-dependent perturbations internal to the system which are shut down outside a
finite time interval. Equivalently, $\omega$ satisfies the second law of thermodynamics in the formulation 
of \textsc{Kelvin} (in particular, $\omega$ is \emph{invariant} under $\mathfrak{A}u^\lambda_{p,q}$, 
guaranteeing the unitary implementability of $\mathfrak{A}u^\lambda_{p,q}$ in the GNS \textsc{Hilbert}
space $\mathscr{H}_\omega$), and the first member of (\ref{ch4e8}) corresponds to the \emph{(relative) entropy}
of the state $\omega (U^*.U)$ with respect to the state $\omega$ (see Appendix \ref{ap2}, page \pageref{ap2t10}). 
We also require that $\omega$ satisfies \emph{weak ergodicity (in mean):}
\begin{equation}\label{ch4e9}
\lim_{T\rightarrow+\infty}\frac{1}{T}\int^{T}_0 \omega(A\mathfrak{A}u^\lambda_{p,q}(B))d\lambda=\omega(A)
\omega(B),\,\forall A,B,
\end{equation}
which means that $\omega$ is an \emph{extremal} invariant state, that is, any order parameter (i.e., an 
element of the commutant of the GNS representation of $\mathfrak{A}$ associated to $\omega$ which is invariant 
under the action of $\mathfrak{A}u^\lambda_{p,q}$) is a $c$-number. In this case, it follows that $\omega$ is a 
KMS state at some inverse temperature $\beta\in[0,+\infty]$, i.e., for all $f\in\mathscr{S}(\mathbb{R})$ 
with $\hat{f}\in\mathscr{C}^\infty_c(\mathbb{R})$, we have
\begin{equation}\label{ch4e10}
\int^{+\infty}_{-\infty}\omega(A\mathfrak{A}u^\lambda_{p,q}(B))f(\lambda)d\lambda=\int^{+\infty}_{-\infty}
\omega(\mathfrak{A}u^\lambda_{p,q}(B)A)f(\lambda+i\beta)d\lambda,\,\forall A,B.
\end{equation}

The case $\beta=+\infty$, corresponding to the case where $\omega$ is a ground state, is excluded
by the following argument: considering, for instance, the wedge $\mathscr{W}=\{X:\eta(X,X)=-1,X^1>|X^0|,
X^d>0\}$ in the fundamental domain (\ref{ch1e5}) of $AdS_d$ (page \pageref{ch1e5}), we have that the generator 
of the isotropy subgroup of $\mathscr{W}$ is the generator $M_{01}$ of $SO_e(2,d-1)$. As $e^{-i\pi M_{12}}M_{01}
e^{-i\pi M_{12}}=-M_{01}$, it follows from the invariance of the spectrum under unitary transformations 
\cite{reedsimon1} that the Hamiltonian in the GNS \textsc{Hilbert} space $\mathscr{H}_\omega$ implementing 
$M_{01}$ cannot be a positive operator -- the same holds, thus, for \emph{all} wedges. \textsc{Buchholz, 
Florig} and \textsc{Summers} \cite{bfs} have shown that all $\omega$ satisfying (\ref{ch4e8})--(\ref{ch4e9}) presents
the \textsc{Unruh} \emph{effect}, i.e., they are KMS states at $\beta=\frac{1}{kT}=\frac{2\pi}{\kappa}=2\pi$, 
where $\kappa=1$ is the surface gravity at $\partial_\pm\mathscr{W}_{p,q}$ (Theorem \ref{ch2t2}, page 
\pageref{ch2t2}). Moreover, we obtain that the realization of $\mathfrak{A}$ in $AdS_d$ is $\omega$\emph{-weakly 
locally causal} with respect to $\mathscr{W}(AdS_d)$: given $A\in\mathfrak{A}(\mathscr{W}_{p,q})$ and $B\in
\mathfrak{A}(\mathscr{W}_{p',q'})$, $\mathscr{W}_{p',q'}\subset\mathscr{W}'_{p,q}$, we have $\omega(AB)=
\omega(BA)$, even if we don't demand local causality from $\mathfrak{A}$ in any way.\\

We'll now establish a sense in which conditions (\ref{ch4e8})--(\ref{ch4e9}) correspond to a choice of
boundary conditions at conformal infinity. If a state $\omega$ over $\mathfrak{A}(\mathscr{W}_{p,q})$ is
passive (\ref{ch4e8}) and weakly ergodic in mean (\ref{ch4e9}) with respect to the group of *-automorphisms 
$\mathfrak{A}u^\lambda_{p,q}$, then the cyclic vector $\Omega$ in the GNS \textsc{Hilbert} space $\mathscr{H}_\omega$ 
associated to $\omega$ is separating with respect to $\pi_\omega(\mathfrak{A}(\mathscr{W}_{p,q}))''$, 
due to the KMS condition, and $\mathfrak{A}u^\lambda_{p,q}$ extends to a group of *-automorphisms of $\pi_\omega
(\mathfrak{A}(\mathscr{W}_{p,q}))''$ which coincides with the \textsc{Tomita-Takesaki} modular group $\sigma^{(2
\pi)^{-1}\lambda}_\omega$ (see Appendix \ref{ap2}, in particular Lemma \ref{ap2l2}, page \pageref{ap2l2}, and
Theorem \ref{ap2t13}, page \pageref{ap2t13}). Since we've assume passivity and weak ergodicity in mean 
from an ``elementary'' state $\omega$ over $\mathfrak{A}_{AdS_d}$ with respect to all C*-dynamical systems 
$(\mathfrak{A}(\mathscr{W}_{p,q}),\mathfrak{A}u^\lambda_{p,q})$, $p\ll q\in\mathscr{I}$, it follows that
such an $\omega$ satisfies:

\begin{itemize}
\item The \textsc{Reeh-Schlieder} \emph{property} with respect to the collection of wedges $\mathscr{W}(AdS_d)$, 
i.e., $\omega$ is \emph{cyclic} and \emph{separating} with respect to the \textsc{von Neumann} subalgebras
$\pi_\omega(\mathfrak{A}(\mathscr{W}_{p,q}))''$,\footnote{This property is known by specialists
in Conformal Field Theory as the ``state-operator correspondence''.} and
\item The action of $\mathfrak{A}u^\lambda_{p,q}$ on $\pi_\omega(\mathfrak{A}(\mathscr{W}_{p,q}))''$ is
given precisely by the \textsc{Tomita-Takesaki} modular group $\sigma^{(2\pi)^{-1}\lambda}_{\omega
\restr{\pi_\omega(\mathfrak{A}(\mathscr{W}_{p,q}))''}}$.
\end{itemize}

$\Delta_{p,q}$ is just the modular operator which implements $\mathfrak{A}u^\lambda_{p,q}$, i.e.,
$\pi_\omega(\mathfrak{A}u^{2\pi\lambda}_{p,q}(A))=\Delta_{p,q}^{-i\lambda}\pi_\omega(A)\Delta_{p,q}^{i\lambda}$ 
for all $A\in\mathfrak{A}(\mathscr{W}_{p,q})$. Let us consider now the inclusion of \textsc{von Neumann} 
algebras $\pi_\omega(\mathfrak{A}(\mathscr{W}_{r,q}))''\subset\pi_\omega(\mathfrak{A}(\mathscr{W}_{p,q}))''$,
where $p\ll_{\mathscr{I}}r\ll_{\mathscr{I}}q$ and, hence, $\mathscr{W}_{r,q}\subset\mathscr{W}_{p,q}$ and
$\partial_+\mathscr{W}_{r,q}\subset\partial_+\mathscr{W}_{p,q}$. The inclusion of the respective underlying 
C*-algebras satisfies the following properties, due to the regularity and the local covariance of 
$\mathfrak{A}$:

\begin{enumerate}
\item $\cup_{\lambda\in\mathbb{R}}\mathfrak{A}u^\lambda_{p,q}(\mathfrak{A}(\mathscr{W}_{r,q}))$ is C*-dense
in $\mathfrak{A}(\mathscr{W}_{p,q})$;
\item $\mathfrak{A}u^\lambda_{p,q}(\mathfrak{A}(\mathscr{W}_{r,q}))\subset\mathfrak{A}(\mathscr{W}_{r,q})$
for all $\lambda\geq 0$.
\end{enumerate}

It follows from 1.) and 2.) that the inclusion $\pi_\omega(\mathfrak{A}(\mathscr{W}_{r,q}))''\subset\pi_\omega
(\mathfrak{A}(\mathscr{W}_{p,q}))''$ satisfies the conditions proposed in the work of \textsc{Borchers} and
\textsc{Yngvason} \cite{boryng}. The main results of this work, rephrased within our present context, are:

\begin{theorem}[\cite{boryng}, Theorem 2.1]\label{ch4t1}
Define $\mathfrak{M}\doteq\pi_\omega(\mathfrak{A}(\mathscr{W}_{p,q}))''$, $\mathfrak{N}\doteq\pi_\omega
(\mathfrak{A}(\mathscr{W}_{r,q}))''$, $\Delta^{i\lambda}_{p,q}\doteq T(-2\pi\lambda)$ and $\mathfrak{N}(\lambda)
=\mbox{\upshape Ad}T(\lambda)\mathfrak{N}\doteq T(\lambda)\mathfrak{N}T(-\lambda)$. In this case, 
\begin{enumerate}
\item[(i)] $\mbox{\upshape Ad}\Delta_{r,q}^{iu}\mathfrak{N}(\lambda)=\mathfrak{N}(\varphi(u,\lambda))$, 
where \[\varphi(u,\lambda)=\log(1+e^{-2\pi u}(e^\lambda-1))\] for all $(u,\lambda)\in\mathbb{R}^2$ for which
the right hand side is defined. In particular, $\mbox{\upshape Ad}\Delta_{r,q}^{iu}\mathfrak{M}$\\$\subset
\mathfrak{M}$ for all $u\geq 0$, and $\mathfrak{N}=\cap_{u\geq 0}\mbox{\upshape Ad}\Delta_{r,q}^{iu}\mathfrak{M}$.
\item[(ii)] The operator $G\doteq\log\Delta_{p,q}-\log\Delta_{r,q}$ is nonnegative and essentially 
self-adjoint in a core common to $\log\Delta_{p,q}$ and $\log\Delta_{r,q}$.\footnote{\upshape A common core 
here is a linear subset $D$ dense in $\mathscr{H}_\omega$ such that $D\subset D(\log\Delta_{p,q})\cap{}D(\log
\Delta_{r,q})$ and such that $\overline{\log\Delta_{p,q}\restr{D}}=\log\Delta_{p,q}$ and $\overline{\log
\Delta_{r,q}\restr{D}}=\log\Delta_{r,q}$, where $D(\log\Delta_{p,q}),D(\log\Delta_{r,q})\subset\mathscr{H}_\omega$ 
are respectively the domains of the operators $\log\Delta_{p,q}$ and $\log\Delta_{r,q}$.} The one-parameter
unitary group $\Gamma(\tau)\doteq e^{(2\pi)^{-1}i\tau\bar{G}}$ satisfies $T(\lambda)\Gamma(\tau)T(\lambda)=
\Gamma(e^\lambda\tau)$ and $\mbox{\upshape Ad}\Gamma(\tau)\mathfrak{N}(\lambda)=\mathfrak{N}(\psi(\tau,\lambda))$,
where \[\psi(\tau,\lambda)=\lambda+\log(1+\tau e^{-\lambda})\] for all $(\tau,\lambda)\in\mathbb{R}^2$
for which the right hand side is defined. In particular, $\mbox{\upshape Ad}\Gamma(\tau)\mathfrak{M}$\\$\subset
\mathfrak{M}$ and $\mbox{\upshape Ad}\Gamma(\tau)\mathfrak{N}\subset\mathfrak{N}$ for all $\tau\geq 0$, and 
$\mathfrak{N}=\mbox{\upshape Ad}\Gamma(1)\mathfrak{M}$.~\hfill~$\Box$
\end{enumerate}
\end{theorem}

Summing up, Theorem \ref{ch4t1} not only gives a precise shape to the commutation relations between 
$\Delta^{i\lambda}_{p,q}$ and $\Delta^{iu}_{r,q}$, but also provides a geometrical realization for the action 
of $u\mapsto\mbox{Ad}\Delta^{iu}_{r,q}$ from that of $\lambda\mapsto\mbox{Ad}\Delta^{i\lambda}_{p,q}$ 
without our having to assume the existence of the former. The next two Theorems show that the actions 
of these unitary groups become essentially indistinguishable if we are geometrically sufficiently 
far from $\partial_-\mathscr{W}_{r,q}$:

\begin{theorem}[\cite{boryng}, Theorem 2.2]\label{ch4t2}
Employing the notation of Theorem \ref{ch4t1}, if $A\in\mathfrak{N}(\lambda)$ and $B\in\mathfrak{N}'$, 
we have \[|\langle B\Omega,\Delta^{iu}_{r,q}A\Omega\rangle-\langle B\Omega,T(-2\pi u)A\Omega\rangle|\leq
2\max\{\Vert A\Omega\Vert\Vert B\Omega\Vert,\Vert A^*\Omega\Vert\Vert B^*\Omega\Vert\}.\]\[.\min\left\{
\frac{|2e^{2\pi u}-1|}{e^\lambda-1},1\right\}\] for all $\lambda>0$, $u\in\mathbb{R}$.~\hfill~$\Box$
\end{theorem}

\begin{theorem}[\cite{boryng}, Theorem 2.3]\label{ch4t3}
Employing the notation of Theorem \ref{ch4t1}, we have:
\begin{enumerate}
\item[(i)] \[\lim_{\lambda\rightarrow+\infty}\Vert\Delta^{iu}_{r,q}(\mbox{\upshape Ad}T(\lambda)A)\Psi-T(-2\pi 
u)(\mbox{\upshape Ad}T(\lambda)A)\Psi\Vert=0\] for all $A\in\mathfrak{M}$ and $\Psi\in\mathscr{H}_\omega$, 
with uniform convergence in $u\in(-\infty,u_0]$ for all $u_0<+\infty$;
\item[(ii)] \[\lim_{\lambda\rightarrow+\infty}\Vert\Delta^{iu}_{r,q}\mbox{\upshape Ad}T(\lambda)A-T(-2\pi u)
\mbox{\upshape Ad}T(\lambda)A\Vert=0\] for all $A$ in a strongly dense subalgebra of $\mathfrak{M}$, 
with uniform convergence in $u\in(-\infty,u_0]$ for all $u_0<+\infty$.\hfill~$\Box$
\end{enumerate}
\end{theorem}

Summing up, denoting $\mathscr{D}_{p,q}=I^-(q,\mathscr{M}in(p)\cong\mathbb{R}^{1,d-2})$, it follows that
such groups are essentially indistinguishable from dilations by a factor $e^{-2\pi\lambda}$ around $q$ 
for $\lambda\rightarrow+\infty$. Taking into account the action of the dilation subgroup of 
$\mathscr{R}^{1,d-2}$ in $AdS_d$, we obtain that the modular groups associated to $\omega$ and the
wedges of $AdS_d$ asymptotically implement scaling transformations around points of $\mathscr{I}$, in the
precise sense given by Theorems \ref{ch4t2} and \ref{ch4t3} -- all the above considerations remain valid 
if we invert the time orientation, i.e., exchanging the roles of $p$ and $q$.\\

More in general, inspired in the formalism proposed in \cite{buve1}, we write the following

\begin{definition}\label{ch4d3}
Let $p\ll q\in\mathscr{M}in(r)$ for some $r\in\mathscr{I}$, where it's assumed that $\mathscr{M}in(r)$ 
is endowed with the vector space operations of $\mathbb{R}^{1,d-2}$. The \emph{future scaling algebra} 
$\mathfrak{A}^+(\mathscr{W}_{p,q})$ (resp. \emph{past scaling algebra} $\mathfrak{A}^-(\mathscr{W}_{p,q})$)  
in $\mathscr{W}_{p,q}$ consists in the uniformly norm-limited functions $(0,1]\ni\theta\mapsto
\underline{A}^\pm(\theta)\doteq A^\pm_\theta\in\mathfrak{A}(\mathscr{W}_{p,q})$ satisfying $A^+_1\doteq A^+
\in\mathfrak{A}(\mathscr{W}_{r,s})\Rightarrow A^+_\theta\in\mathfrak{A}(\mathscr{W}_{\theta r+(1-\theta)q,
\theta s+(1-\theta)q})$ (resp. $A^-_1\doteq A^-\in\mathfrak{A}(\mathscr{W}_{r,s})\Rightarrow A^-_\theta\in
\mathfrak{A}(\mathscr{W}_{\theta r+(1-\theta)p,\theta s+(1-\theta)p})$), for all $p\leq_{\mathscr{I}} r
\ll_{\mathscr{I}} s\leq_{\mathscr{I}} q$, $\theta\in(0,1]$. The algebraic operations are defined pointwise, 
and the C*-norm is given by $\Vert\underline{A}^\pm\Vert\doteq\sup_{0<\theta\leq 1}\Vert A^\pm_\theta\Vert$.
\end{definition}

\begin{remark}\label{ch4r2}
We can say that the concept of scaling algebras defines an abstract version of the \emph{renormalization
group} in Local Quantum Physics. A different, cruder definition of scaling algebras associated to wedges 
in $AdS_d$ was proposed by the author in \cite{ribeiro1}.
\end{remark}

Notice that the definition above preserves isotony and local causality of the local subalgebras of 
$\mathfrak{A}(\mathscr{W}_{p,q})$.\\

Given a state $\omega$ in $\mathfrak{A}(\mathscr{W}_{p,q})$, we can perform its \emph{lift} $\omega^\pm_\theta$ 
to $\mathfrak{A}^\pm(\mathscr{W}_{p,q})$ at scale $\theta\in(0,1]$ by defining $\omega^+_\theta(\underline{A}^+)
\doteq\omega(A^+_\theta)$ and $\omega^-_\theta(\underline{A}^-)\doteq\omega(A^-_\theta)$. Obviously, 
$\omega^+_\theta$ defines a state in $\mathfrak{A}^\pm(\mathscr{W}_{p,q})$.\\

Let's consider $s=q$ in Definition \ref{ch4d3}, i.e., we have the inclusion considered in Theorems 
\ref{ch4t1} to \ref{ch4t3}. In this case, we have that $\theta\mapsto A^+_\theta\doteq\mathfrak{A}
u^{f^{-1}_{r,+}(\theta)}_{p,q}(A)$, where $A\in\mathfrak{A}(\mathscr{W}_{r,q})$ and \[f_{r,+}(\lambda)
=\frac{d_{\bar{g}^{(0)}}(u^\lambda_{p,q}(r),q)}{d_{\bar{g}^{(0)}}(r,q)},\,f_{r,-}(\lambda)=\frac{d_{\bar{g}^{(0)}}
(p,u^{-\lambda}_{p,q}(r))}{d_{\bar{g}^{(0)}}(p,r)}\] are bijective, $\mathscr{C}^\infty$ and strictly 
decreasing functions from $[0,+\infty)\ni\lambda$ to $(0,1]\ni\theta$, defines an element of 
$\mathfrak{A}^+(\mathscr{W}_{r,q})$ and, thus, of $\mathfrak{A}^+(\mathscr{W}_{p,q})$. Conversely, we can 
define a family of *-automorphisms $\delta^{\mu,+}_{p,q}$ of $\mathfrak{A}^+(\mathscr{W}_{p,q})$ implementing 
scaling transformations around $q$ by a factor $\mu\in(0,1]$ through formula $(\delta^{\mu,+}_{p,q}
\underline{A}^+)(\theta)\doteq A^+_{\mu\theta}$. Taking into account the commutation relation
\begin{equation}\label{ch4e11}
\theta u^\lambda_{p,q}(r)+(1-\theta)q=u^{\theta\lambda}_{\theta p+(1-\theta)q,q}(\theta r+(1-\theta)q),
\end{equation}
we can induce an action of $\mathfrak{A}u^\lambda_{p,q}$ on $\mathfrak{A}^+(\mathscr{W}_{p,q})$ through 
formula
\begin{equation}\label{ch4e12}
\mathfrak{A}^+u^\lambda_{p,q}(\underline{A}^+)(\theta)\doteq\mathfrak{A}u^{\theta\lambda}_{\theta 
p+(1-\theta)q,q}(A^+_\theta).
\end{equation}

Analogous considerations hold for past scaling algebras by changing notation accordingly, if 
we invert the time orientation, exchange the roles of $r$ and $s$ and substitute $f_{s,-}$ 
for $f_{r,+}$. We assume that the action (\ref{ch4e12}) and its past counterpart are strongly 
continuous respectively in $\mathfrak{A}^+(\mathscr{W}_{p,q})$ and $\mathfrak{A}^-(\mathscr{W}_{p,q})$.\\

The \emph{raison d'\^etre} of the redundant description of the realization of $\mathfrak{A}$ in $AdS_d$ 
provided by the past and future scaling algebras comes from the fact that we can use it to define 
\emph{scaling limits:} let us notice that the net of lifted states $\{\omega^+_\theta:\theta\in(0,1]\}$ 
induced by $\omega$ has always pointwise convergent subnets (i.e., in the *-weak topology of the dual of
$\mathfrak{A}^+(\mathscr{W}_{p,q})$) to states $\omega^+_{0,\iota}$ over $\mathfrak{A}^+(\mathscr{W}_{p,q})$ 
($I\ni\iota$ is some index set), by the \textsc{Banach-Alaoglu} theorem \cite{reedsimon1}. The GNS 
representation $\pi^+_{0,\iota}$ associated to each $\omega^+_{0,\iota}$ of the elements of $\mathfrak{A}^+
(\mathscr{W}_{p,q})$ thus have the interpretation of (weak) scaling limits of the local procedures. 
However, there remains the question of uniqueness of the representations $\pi^+_{0,\iota}$ up to 
\emph{net isomorphisms}, i.e., up to the existence or not of *-isomorphisms $\phi_{\iota,\iota'}$ from 
$\pi^+_{0,\iota}(\mathfrak{A}^+(\mathscr{W}_{p,q}))$ to $\pi^+_{0,\iota'}(\mathfrak{A}^+(\mathscr{W}_{p,q}))$ 
that preserve the local structure -- $\phi_{\iota,\iota'}(\pi^+_{0,\iota}(\mathfrak{A}^+(\mathscr{W}_{r,s})))=
\pi^+_{0,\iota'}(\mathfrak{A}^+(\mathscr{W}_{r,s}))$ for all $p\leq_{\mathscr{I}} r\ll_{\mathscr{I}}s
\leq_{\mathscr{I}}q$, $\iota,\iota'\in I$. Such a question is partially clarified by the following

\begin{proposition}\label{ch4p1}
Suppose that $\omega$ is a state in $\mathfrak{A}_{AdS_d}$, passive and weakly ergodic in mean with respect to
the C*-dynamical systems $(\mathfrak{A}(\mathscr{W}_{p,q}),\mathfrak{A}u^\lambda_{p,q})$, for all pairs 
$(p,q)\in\mathscr{D}(\mathscr{I})$. Then,
\begin{enumerate}
\item[(i)] The limit state $\omega^+_{0,\iota}$ is $\beta$-KMS with $\beta=2\pi$ with respect to the C*-dynamical
system $(\pi^+_{0,\iota}(\mathfrak{A}^+(\mathscr{W}_{p,q})),\mathfrak{A}^+_{0,\iota}u^\lambda_{p,q})$,
where $\mathfrak{A}^+_{0,\iota}u^\lambda_{p,q}(\pi^+_{0,\iota}(A))\doteq\pi^+_{0,\iota}(\mathfrak{A}^+u^\lambda_{p,q}
(A))$. 
\item[(ii)] The scaling limits $\pi^+_{0,\iota}(\mathfrak{A}^+(\mathscr{W}_{r,s}))$ and $\pi^+_{0,\iota}
(\mathfrak{A}^+(u^\lambda_{p,q}(\mathscr{W}_{r,s})))$ are unitarily equivalent, for all $p\leq_{\mathscr{I}} 
r\ll_{\mathscr{I}} s\leq_{\mathscr{I}} q$.
\item[(iii)] If $\omega$ is \emph{primary}, i.e. $\pi_\omega(\mathfrak{A}_{AdS_d})''$ is a factor (see 
Appendix \ref{ap2}, page \pageref{ap2d3}), it then satisfies the following property of \emph{holographic
local definiteness:} given $r\in\mathscr{I}$, we have that \[\bigcap_{{(p,q)\in\mathscr{D}(\mathscr{I}),}
\atop{\mathscr{W}_{p,q}\ni r}}\pi_\omega(\mathfrak{A}(\mathscr{W}_{p,q}))''=\mathbb{C1}.\]
It follows from this property that the sets of scaling limits $\{(\omega')^+_{0,\iota'}:\iota'\in I'\}$ 
of all states $\omega'$ in the local folium of $\omega$ with respect to the collection of wedges (i.e., $\omega'
\restr{\mathfrak{A}(\mathscr{W}_{p,q})}$ defines a normal state over $\pi_\omega(\mathfrak{A}
(\mathscr{W}_{p,q}))''$ for all $(p,q)\in\mathscr{D}(\mathscr{I})$) coincide with $\{\omega^+_{0,\iota}:
\iota\in I\}$. As a consequence of this and of (i), each \textsc{von Neumann} algebra $\pi_\omega(\mathfrak{A}
(\mathscr{W}_{p,q}))''$, if infinite (which is generally the case in Quantum Field Theory), is type III (see 
Appendix \ref{ap2}, Definition \ref{ap2d4}, page \pageref{ap2d4}).
\item[(iv)] Suppose that the linear map from $\pi_\omega(\mathfrak{A}(\mathscr{W}_{p,q}))$ to
$\mathscr{H}_\omega$ given by \[\pi_\omega(A)\mapsto E_{p,q}(K)\pi_\omega(A)\Omega,\] where $E_{p,q}$
is the \textsc{Borel} measure with values in the projections of $\mathscr{H}_\omega$ which defines the 
spectral resolution of the operator $\log\Delta_{p,q}$ and $K$ is a compact set in $\mathbb{R}=\sigma
(\log\Delta_{p,q})$, is a \emph{compact} map, i.e., it takes bounded subsets of the domain to relatively 
compact subsets of the image, for all $(p,q)\in\mathscr{D}(\mathscr{I})$.\footnote{\upshape As compact
linear maps are characterized for being able to be approximated in norm by finite-rank linear maps
with arbitrary precision, such a condition tells us that the spectral density of local excitations
of the reference state doesn't grow too fast with the energy. Such a condition on the phase space 
behaviour of quantum field theories was proposed by \textsc{Haag} and \textsc{Swieca} for the 
(Hamiltonian) generator of time translations in \textsc{Minkowski} spacetime. IT would be interesting 
to know is the compactness condition we've given can be deduced, in the same way that the \textsc{Haag-Swieca}
compactness condition, from other structural conditions which have been explored in the literature of
Algebraic QFT, such as modular nuclearity \cite{haag}.} 
Then the maps $\phi_\iota:\pi^+_{0,\iota}(\underline{A}^+)\mapsto w-\lim_{\kappa}\pi_\omega(\mathfrak{A}
u^{-f^{-1}_{p,+}(\theta_\kappa)}_{p,q}(A^+_{\theta_\kappa}))$, where $\{\omega^+_{\theta_\kappa}\}_\kappa$ is 
a subnet *-weakly converging to $\omega^+_{0,\iota}$, define net isomorphisms to the GNS representation 
of the original net associated to $\omega$. In particular, by virtue of (i), the scaling limits are all
unitarily equivalent, and the modular group of any scaling limit acts asymptotically as the modular 
group associated to $\omega^+_1$
as $\lambda\rightarrow+\infty$.
\end{enumerate}
\begin{quote}{\small\scshape Proof (sketch).\quad}
{\small\upshape 
\begin{enumerate}
\item[(i)] It follows from Theorem 5.3.30 in \cite{bratteli2} that the *-weak limit of a net of $\beta$-KMS 
states is $\beta$-KMS. 
\item[(ii)] It follows immediately from the continuity hypothesis on $\mathfrak{A}^+u^\lambda_{p,q}$ the
existence of a *-isomorphism, which is unitarily implemented due to the existence of cyclic and separating 
vector for both algebras (see Theorem 2.5.32 in \cite{bratteli1} for a proof of the latter fact).
\item[(iii)] The first assertion is obtained by employing \textsc{Rehren} duality: By virtue of (i) and (ii), 
the conformal group of $\mathscr{M}in(r)\cong\mathbb{R}^{1,d-2}$ for all $r\in\mathscr{I}$ is unitarily 
implemented in $\mathscr{H}_\omega$ by the \textsc{Tomita-Takesaki} modular groups associated to $\mathfrak{A}
\circ\rho_{AdS_d}^{-1}(\mathscr{D}_{p,q})=\mathfrak{A}(\mathscr{W}_{p,q})$, and $\omega$ is a conformally invariant
state of the \textsc{Rehren}-dual theory $\mathfrak{A}\circ\rho_{AdS_d}^{-1}$. By virtue of the resulting
commutation relations between the different modular groups, it follows from Proposition 2.3 in \cite{buchsum1} 
that the generators $P^\mu$ of translations in $\mathscr{M}in(r)$ satisfy the \emph{spectral condition}, i.e., 
$\sigma(P)\subset J^+(0,\mathbb{R}^{1,d-2})$ -- summing up, $\omega$ is a \textsc{vacuum}. As $\omega$ is 
primary, it then follows from Theorem 4.6 in \cite{araki2} that $\omega$ is pure and every translationally
invariant element of $\mathscr{H}_\omega$ is proportional to $\Omega$. Let now $Z\in\bigcap_{{(p,q)\in\mathscr{D}
(\mathscr{I}),}\atop{\mathscr{W}_{p,q}\ni r}}\pi_\omega(\mathfrak{A}(\mathscr{W}_{p,q}))''\doteq\mathfrak{N}(r)$. 
In this case, $Z^*\in\mathfrak{N}(r)$ and, if $\mathbb{R}^{1,d-2}\ni x\alpha_x$ is the group of *-automorphisms 
which implements translations in $\mathscr{M}in(s)\ni 0=x(r)$, where $x$ is a Cartesian chart in $\mathscr{M}
in(s)$, then $[Z^*,\alpha_x(Z)]=0$ for all spacelike $x$ and, by weak continuity of $\alpha_x$, also for
$x=0$ and $x$ lightlike. Consider the bounded function $\mathbb{R}\ni t\mapsto\langle\Omega,Z^*\alpha_{te}(Z)
\Omega\rangle$, where $e$ is a lightlike vector in $\mathbb{R}^{1,d-2}$. The \textsc{Fourier} transform of
this function has, due to the spectral condition, support in $\bar{\mathbb{R}}_+$. Due to the lightlike
commutativity established above, such a function coincides with the function $\mathbb{R}\ni t\mapsto\langle
\Omega,\alpha_{te}(Z)Z^*\Omega\rangle$, whose \textsc{Fourier} transform has support in $\bar{\mathbb{R}}_-$. 
It thus follows that the \textsc{Fourier} transform of the former is supported in $\{0\}$ and, since it's 
bounded, it's necessarily constant. Therefore, if $x\mapsto U(x)$ is the unitary implementation of translations
in $\mathscr{H}_\omega$, it follows that $U(te)Z\Omega=Z\Omega$ for all $t\in\mathbb{R}$ and $e$ lightlike, 
hence $U(x)Z\Omega=Z\Omega$ for all $x$. By the uniqueness of $\Omega$, we thus have $Z\Omega=\langle\Omega,
Z\Omega\rangle\Omega$. As $\Omega$ is separating for $\pi_\omega(\mathfrak{A}(\mathscr{W}_{p,q}))''$ for all
$(p,q)\in\mathscr{D}(\mathscr{I})$, we have $Z=\langle\Omega, Z\Omega\rangle\mathbb{1}$, as claimed.
The second assertion follows from the fact that, by virtue of holographic local definiteness, $\lim_{\theta
\searrow 0}\Vert(\omega-\omega')\restr{\mathfrak{A}(\mathscr{W}_{\theta p+(1-\theta)q,q})}\Vert=0$ (this can be
proven in a manner analogous to the second part of the proof of Lemma 4.1 in \cite{buve1}, whose first part is
essentially the proof of the first assertion of (iii) above).
The third assertion follows from \textsc{Driessler}'s criterion (Theorem \ref{ap2t12}, page \pageref{ap2t12}), 
which, in a certain way, was specially devised for our situation.
\item[(iv)] One proves it in a way analogous to Proposition 5.1 in \cite{buve1}. The last assertion then follows
from Theorems \ref{ch4t2} and \ref{ch4t3}.
\end{enumerate}
~\hfill~$\Box$}
\end{quote}
\end{proposition}

\begin{remark}\label{ch4r3}
The analog of Proposition \ref{ch4p1} for past scaling algebras holds, if we invert the time orientation and
substitute $f_{q,-}$ for $f_{p,+}$ in item (iv). Moreover, there is a unitary equivalence between past and future
scaling limits by virtue of item (ii) of this Proposition.
\end{remark}

\begin{remark}\label{ch4r4}
The unitary implementability of conformal transformations in the GNS \textsc{Hilbert} space
of a conformally invariant vacuum by the \textsc{Tomita-Takesaki} modular groups was previously demonstrated 
by \textsc{Brunetti, Guido} and \textsc{Longo} \cite{bglongo1}. Thus, our line of reasoning can be seen as an
``holographic'' version of this result.
Moreover, it follows from Theorem 2.3 (i) in \cite{bglongo1} that the *-representation $\pi_\omega$ satisfies 
\textsc{Haag} \emph{duality} in $\mathscr{I}$ and essential \textsc{Haag} duality in $\mathscr{M}in(r)$ for 
all $r\in\mathscr{I}$ (for the definition and an explanation of the usefulness and meaning of (essential) 
\textsc{Haag} duality within our context, see Subsection \ref{ch4-sector-dig}, page \pageref{ch4-sector-dig}). 
We can make use of the fact that essential \textsc{Haag} duality holds for the scaling limits and refine the 
conclusion of item (iii), proving that $\pi_\omega(\mathfrak{A}(\mathscr{W}_{p,q}))''$ is a type-III${}_1$ 
factor if we employ a different criterion for the structure of this algebra \cite{buve1}. We haven't proven 
this stronger result because we don't need this more refined information in what follows, and 
\textsc{Driessler}'s criterion, besides being simpler, has quite a natural interpretation within our context.
\end{remark}

The equivalence in item (iv) of Proposition \ref{ch4p1} shows that, indeed, (\ref{ch4e8}) and (\ref{ch4e9}) 
act effectively as a condition on scaling limits around points in $\mathscr{I}$, thus revealing the 
relation between this set of hypotheses and the formulation of \textsc{Bertola et al.} \cite{bertobms}, based 
in $k$-point \textsc{Wightman} functions in $AdS_d$. The boundary conditions employed in this work, similar 
to the ones proposed by \textsc{Witten} \cite{witten1} for the AdS/CFT correspondence, are given precisely by
a scaling limit of the $k$-point functions around $\mathscr{I}$. The important item (iii) shows that the
scaling limit is, indeed, a condition on the local folium of $\omega$ and, hence, on the *-ideals $\mbox{Ann}
\omega\restr{\mathfrak{A}(\mathscr{W}_{p,q})}$. Such a conditions is locally covariant by virtue of (ii) and
the First *-Isomorphism Theorem (Theorem \ref{ap2t2}, page \pageref{ap2t2}), thus defining a locally covariant
boundary condition in the sense of Section \ref{ch4-bound}.\\

There is, though, a very important point to be stressed: in the work of \textsc{Bertola et al.},
the scaling limit \emph{defines} the dual quantum theory at the boundary, or, equivalently, the boundary
quantum theory corresponds in this case to the \emph{universality class} of the bulk quantum theory under the
``renormalization group flow'' induced by scaling transformations around points in $\mathscr{I}$), which doesn't 
happen to \textsc{Rehren} duality, which consists in \emph{transplanting} the set that indexes the precosheaf 
which realizes the locally covariant quantum theory $\mathfrak{A}$ in $AdS_d$ by means of the \textsc{Rehren}
bijection. A structural property which illustrates rather well this difference is the \textsc{Borchers}'s 
\emph{timelike tube property:} if $\mathfrak{A}$ is \emph{additive}, and there exists a collection of regions 
$\mathscr{O}_i\subset\mathscr{M}$ covering an open, arbitrarily small neighbourhood of a timelike curve segment
with endpoints $r\ll s\in\mathscr{M}$ such that $I^+(r)\cap{} I^-(s)$ is simply connected, then 
\[\overline{\bigvee_i\mathfrak{A}(\mathscr{O}_i)}=\mathfrak{A}(I^+(r)\cap{} I^-(s)).\] That is, the \emph{germ} 
of local algebras around a timelike curve segment and, in particular, of a timelike submanifold is *-isomorphic 
to the local algebra associated to its causal completion (a proof of this fact for quantum fields in real
analytic spacetimes can be found in \cite{stroh}). Such a result shows that \textsc{Rehren} duality (Definition 
\ref{ch4d2}, page \pageref{ch4d2}) really concerns the germs of algebras around $\mathscr{I}$ (whose definition 
demands, indeed, the imposition of some sort of boundary condition as the one elaborated above 
\footnote{\label{ch4fn1} Actually, we've committed a slight abuse in our definition of \textsc{Borchers}'s
timelike tube property. The correct formulation (which we won't use) involves the local \textsc{von Neumann} 
algebras and weak topology closures, associated to a GNS representation generated by a state $\omega$. Hence, 
the definition also involves properties of $\omega$, or, more precisely, of its local folium.}), thus being
richer than a mere scaling limit. Such a difference also manifests itself in the properties of the local algebras 
of the dual quantum theory at the boundary in both approaches, as discussed by \textsc{Dütsch} and \textsc{Rehren} 
\cite{dure2,rehren5}.\\

Item (iii) of Proposition \ref{ch4p1} and Theorems \ref{ch4t2} and \ref{ch4t3} may, on the other hand, be 
understood as properties of \emph{return to equilibrium}, if we imagine the isometries $u^\lambda_{p,q}$ in 
$AdS_d$ as \textsc{Cauchy} evolutions with boundary conditions at $\mathscr{I}$, in the light of the results 
of Chapter \ref{ch2}. Hence, we have a strict (we could say ``holographic'') relation between \emph{scaling 
limits} around points in $\mathscr{I}$ and \emph{thermalization} of the ``elementary'' states $\omega$ through 
the dynamics generated by $\mathfrak{A}u^\lambda_{p,q}$. In the case of $AdS_d$, this return is trivial, for
we've seen that $\omega$ is a KMS state with respect to this dynamics and, at the same time, $\omega$ is a state over
the \textsc{Rehren}-dual theory which is invariant under scaling transformations. In the next Section, we'll
make use of this more general viewpoint and consider ``elementary'' states which, by virtue of nontrivial 
gravitational effects, return, accordingly, ``nontrivially'' to equilibrium and, equivalently, present only
asymptotic scaling invariance around $\mathscr{I}$.

\section{\label{ch4-rehren}Algebraic holography in AAdS spacetimes}

We've finally arrived at the moment of defining the quantum theories of interest to us in AAdS spacetimes 
$(\mathscr{M},g)$ satisfying the hypotheses of Theorem \ref{ch1t4}, which guarantees that $I^+(p,
\overline{\mathscr{M}})\cap{}I^-(q,\overline{\mathscr{M}})\cap{}\mathscr{I}=I^+(p,\mathscr{I})\cap{} I^-(q,
\mathscr{I})\doteq\mathscr{D}_{p,q}$ for all $(p,q)\in\mathscr{D}(\mathscr{I})$. The most delicate part concerns
the implementation of the asymptotic isometries associated to wedges constructed in Chapter \ref{ch2} (pages 
\pageref{ch2p3}--\pageref{ch2e49}). However, before that, let us write:\\

\begin{definition}\label{ch4d4}
The quantum theory \textsc{Rehren}\emph{-dual} to the realization of $\mathfrak{A}$ in $(\mathscr{M},g)$ 
is the precosheaf $\mathscr{D}_{p,q}\mapsto\mathfrak{A}(\mathscr{D}_{p,q})\doteq\mathfrak{A}(\rho_{(\mathscr{M},
g)}^{-1}(\mathscr{D}_{p,q}))=\mathfrak{A}(\mathscr{W}_{p,q})$ of C*-algebras, indexed by the collection of 
diamonds in $\mathscr{I}$. Here, $\rho_{(\mathscr{M},g)}$ is the \textsc{Rehren} bijection (\ref{ch1e32}), 
introduced in page \pageref{ch1e32}.
\end{definition}

Let us now consider a wedge $\mathscr{W}_{p,q}$ with geodesically convex closure. We'll define, for 
$\lambda\in\mathbb{R}$, $\epsilon>0$ fixed, the regions
\begin{equation}\label{ch4e13}
\mathscr{W}^+_{p,q}(\lambda)\doteq(\lambda^{\bar{g}}_{p,q})^{-1}((\lambda-\epsilon,+\infty))
\cap{}\mathscr{M},\,\mathscr{W}^-_{p,q}(\lambda)\doteq(\lambda^{\bar{g}}_{p,q})^{-1}((-\infty,
\epsilon-\lambda))\cap{}\mathscr{M},
\end{equation}
and, correspondingly in $\mathscr{I}$,
\begin{equation}\label{ch4e14}
\mathscr{D}^+_{p,q}(\lambda)\doteq(\lambda^{\bar{g}}_{p,q})^{-1}((\lambda-\epsilon,+\infty))
\cap{}\mathscr{I},\,\mathscr{D}^-_{p,q}(\lambda)\doteq(\lambda^{\bar{g}}_{p,q})^{-1}((-\infty,
\epsilon-\lambda))\cap{}\mathscr{I}.
\end{equation}

Let $\mathfrak{A}$ be a regular, extended locally covariant quantum theory, satisfying the following 
hypotheses:

\begin{enumerate}
\item[(a)] The realization of $\mathfrak{A}$ in $(\mathscr{M},g)$ is locally causal with respect to the collection 
of wedges $\mathscr{W}(\mathscr{M},g)$;
\item[(b)] $\mathfrak{A}$ is weakly primitively causal with respect to the wedges of $(\mathscr{M},g)$ (thus 
guaranteeing the possibility of imposing a time evolution subject to boundary conditions);
\item[(c)] There exists a state space $\mathfrak{S}$ such that $\mathfrak{S}(\mathscr{M},g)$ possesses
a primary representative $\omega$ satisfying the property of holographic local definiteness (defined in 
item (iii) of Proposition \ref{ch4p1}, page \pageref{ch4p1}) and such that its local folium with respect to the 
causally convex subregions of wedges is contained in $\mathfrak{S}(\mathscr{M},g)$.
\end{enumerate}

In what follows, we've uncluttered the notation by omitting the symbols of restriction of the metric 
$g$ to subregions and those of quotients modulo nontrivial *-ideals which annihilate the propagators 
(\emph{boundary *-ideals}), having always in mind the *-Isomorphism Theorems. It follows from (b) that 
$\mathfrak{A}(\mathscr{W}^\pm_{p,q}(\lambda),g)$ is *-isomorphic to $\mathfrak{A}(\mathscr{W}_{p,q})$ for 
all $\gamma$, $(p,q)\in\mathscr{D}(\mathscr{I})$, modulo boundary *-ideals. Moreover, we can define the 
\emph{germ} $\mathfrak{A}(\partial_\pm\mathscr{W}_{p,q})$ of $\mathfrak{A}(\mathscr{W}_{p,q})$ at the past
and future horizons $\partial_\pm\mathscr{W}_{p,q}$ by the projective limits (modulo boundary *-ideals)
\begin{equation}\label{ch4e15}
\mathfrak{A}(\partial_\pm\mathscr{W}_{p,q})\doteq\ilim{\lambda\nearrow+\infty,\mathfrak{A}}
\mathfrak{A}(\mathscr{W}^\pm_{p,q}(\lambda),g),
\end{equation}
recalling that the equivalence relation which defines the limit is given by the identification
(modulo boundary *-ideals) $A\in\mathfrak{A}(\mathscr{W}^\pm_{p,q}(\lambda),g)\sim(\mathfrak{A}
i_{\mathscr{W}^\pm_{p,q}(\lambda),\mathscr{W}^\pm_{p,q}(\lambda')})^{-1}(A)$, for all $\lambda'\geq\lambda$, 
similarly to the case of germs of local algebras at \textsc{Cauchy} surfaces, defined in Chapter \ref{ch3} 
(see formula (\ref{ch3e8}), page \pageref{ch3e8}).\\

With the above definitions, if $r\in(\lambda^{\bar{g}}_{p,q})^{-1}(\lambda_0)\cap{}\mathscr{I}$
and $r\ll_{\mathscr{I}}s\leq_{\mathscr{I}}q$, obviously $\mathscr{W}_{r,s}\subset\mathscr{W}^+_{p,q}
(\lambda_0)$ and, hence, $\mathscr{W}_{u^\lambda_{p,q}(r),u^\lambda_{p,q}(s)}\subset\mathscr{W}^+_{p,q}
(\lambda_0+\lambda)$ for all $\lambda\geq 0$. 

\begin{definition}\label{ch4d5}
Let $p\ll q\in\mathscr{M}in(r)$ for some $r\in\mathscr{I}$, where it's assumed that $\mathscr{M}in(r)$ 
is is endowed with the vector space operations of $\mathbb{R}^{1,d-2}$. The \emph{future scaling algebra} 
$\mathfrak{A}^+(\mathscr{W}_{p,q})$ (resp. \emph{past scaling algebra} $\mathfrak{A}^-(\mathscr{W}_{p,q})$) 
in $\mathscr{W}_{p,q}$ consists of the uniformly norm-bounded continuous functions $(0,1]\ni\theta\mapsto 
\underline{A}^+(\theta)\doteq A^+_\theta\in\mathfrak{A}(\mathscr{W}^+_{p,q}(f^{-1}_{p,+}(\theta)),g)$ (resp. 
$(0,1]\ni\theta\mapsto\underline{A}^-(\theta)\doteq A^-_\theta\in\mathfrak{A}(\mathscr{W}^-_{p,q}(f^{-1}_{q,-}
(\theta)),g)$) satisfying $A^+_1\doteq A^+\in\mathfrak{A}(\mathscr{W}_{r,s})$, $r\in\mathscr{W}^+_{p,q}
(\lambda_0)$ for some $\lambda_0$ implies $A^+_\theta\in\mathfrak{A}(\mathscr{W}^+_{p,q}(\lambda_0+f_{r,+}^{-1}
(\theta)),g)$ (resp. $A^-_1\doteq A^-\in\mathfrak{A}(\mathscr{W}_{r,s})$, $s\in\mathscr{W}^-_{p,q}(\lambda_0)$ 
for some $\lambda_0$ implies $A^-_\theta\in\mathfrak{A}(\mathscr{W}^-_{p,q}(\lambda_0+f_{s,-}^{-1}(\theta)),
g)$), for all $p\leq_{\mathscr{I}} r\ll_{\mathscr{I}} s\leq_{\mathscr{I}} q$, $\theta\in(0,1]$, where $f_{r,+}$ and 
$f_{s,-}$ are defined in page \pageref{ch4e11}. The algebraic operations are defined pointwise, and the C*-norm 
is given by $\Vert\underline{A}^\pm\Vert\doteq\sup_{0<\theta\leq 1}\Vert A^\pm_\theta\Vert$.
\end{definition}

The more general definition given above copes with the loss of ``spatial'' localization caused by the action
of the \emph{propagators}
\begin{equation}\label{ch4e16}
\alpha^\pm_{\lambda,\lambda'}\doteq(\mathfrak{A}i_{\mathscr{W}^\pm_{p,q}(\lambda'),\mathscr{W}_{p,q}})^{-1}
\circ\mathfrak{A}i_{\mathscr{W}^\pm_{p,q}(\lambda),\mathscr{W}_{p,q}},
\end{equation}
which then can be used to build elements of $\mathfrak{A}^\pm(\mathscr{W}_{p,q})$ (modulo boundary *-ideals) 
in the same way we've employed $\mathfrak{A}u^\lambda_{p,q}$ in the case of $AdS_d$. We suppose, accordingly, 
that the propagators (\ref{ch4e16}) are strongly continuous in the norm of the scaling algebras.\\

We stress now a \emph{very} important point: the images of the elements of the scaling algebras belong to
the \emph{intrinsic} algebras, and not to their images in $\mathfrak{A}(\mathscr{W}_{p,q})$, which identify 
elements which differ by an element of a boundary *-ideal. This eliminates an apparent contradiction 
between the germs of local algebras at the past and future horizons and the principle of 
holographic local definiteness. The maps
\begin{equation}\label{ch4e17}
A^+_\theta\mapsto I^+_{p,q}(\underline{A}^+)(\theta)\doteq\mathfrak{A}i_{\mathscr{W}^+_{p,q}(f^{-1}_{p,+}
(\theta)),\mathscr{W}_{p,q}}(A^+_\theta)
\end{equation}
and
\begin{equation}\label{ch4e18}
A^-_\theta\mapsto I^-_{p,q}(\underline{A}^-)(\theta)\doteq\mathfrak{A}i_{\mathscr{W}^-_{p,q}(f^{-1}_{q,-}
(\theta)),\mathscr{W}_{p,q}}(A^-_\theta)
\end{equation}
define *-morphisms from the scaling algebras to the uniformly bounded continuous functions $(0,1]\ni\theta$
in $\mathfrak{A}(\mathscr{W}_{p,q})$ and preserve the ``time'' localization. Defining the lift 
$\omega^\pm_\theta$ of our reference state $\omega$ to $I^\pm_{p,q}(\mathfrak{A}^\pm(\mathscr{W}_{p,q}))$
at scale $\theta$ in the same manner as in the AdS case, we can finally impose our boundary conditions:

\begin{enumerate}
\item[(d)] \[\mbox{Ann}\mathfrak{S}(\mathscr{W}^\pm_{p,q}(\lambda),g)\doteq\bigcap_{\omega'\in\mathfrak{S}
(\mathscr{W}^\pm_{p,q}(\lambda),g)}\mbox{Ann}\omega'\supset\mbox{Ker}\mathfrak{A}i_{\mathscr{W}^\pm_{p,q}
(\lambda),\mathscr{W}_{p,q}}\] for all $\lambda\in\mathbb{R}$, and the scaling limits $\omega^\pm_{0,\iota}$ 
define GNS representations linked by net *-isomorphisms with respect to the collection of wedges in $(\mathscr{M},g)$, 
and *-isomorphic to the *-representation of $\mathfrak{A}(\partial_\pm\mathscr{W}_{p,q})$ induced by the 
quotient modulo $\mbox{Ann}\mathfrak{S}(\partial_\pm\mathscr{W}^\pm_{p,q})$ (this state space is defined 
by employing the contravariance of $\mathfrak{S}$ and the universality of projective limits). The 
*-automorphic action $(0,1]\ni\mu\mapsto\delta^{\mu,\pm}_{p,q}$ of scaling transformations on $\pi_{0,\iota}
(I^\pm_{p,q}(\mathfrak{A}^\pm(\mathscr{W}_{p,q})))$ is unitarily implemented by the formula
\begin{equation}\label{ch4e19}
\pi_{0,\iota}(\delta^{\exp(\pm\kappa_\pm\lambda),\pm}_{p,q}(I^\pm_{p,q}(\underline{A}^\pm)))=\mbox{Ad}
\Delta_{p,q}^{\frac{i}{2\pi}\lambda}\pi_{0,\iota}(I^\pm_{p,q}(\underline{A}^\pm)),
\end{equation}
where $\Delta_{p,q}$ is the \textsc{Tomita-Takesaki} modular operator associated to $\omega_{0,\iota}$
(which is, hence, essentially unique for all $\iota$), and $\kappa_\pm=\mp 1$ is the asymptotic 
past/future surface gravity of $\mathscr{W}_{p,q}$, obtained in Theorem \ref{ch2t2} (page \pageref{ch2t2}). 
\end{enumerate}

Condition (d), albeit rather complicated, is naturally motivated by Proposition \ref{ch4p1} and
Theorem \ref{ch2t2} in the case of $AdS_d$. Together with (c), it's not only common to all states
in the local folium of $\omega$ but also implies that the \textsc{von Neumann} algebras $\pi_\omega(\mathfrak{A}
(\mathscr{W}_{p,q}))''$ are type III, analogously to item (iii) of Proposition \ref{ch4p1}. The assumed
continuity of the propagators, together with local covariance, guarantees that the scaling limits of
different wedges all live in the same \textsc{Hilbert} space, as in item (ii). The geometrical character 
\emph{at} $\mathscr{I}$ of the action of the modular groups, on its turn, follows from Theorem \ref{ch4t1}. 
By virtue of the asymptotic value we've found for the surface gravities of wedges and the commutation
relations between the different modular groups, resulting from Theorem \ref{ch4t1}, we have that the modular
groups act geometrically exactly as conformal transformations in $\mathscr{I}$, and the scaling limit 
$\omega_{0,\iota}$ defines a conformally invariant vacuum \cite{buchsum1} over the precosheaf $\mathscr{D}_{p,q}
\mapsto\pi_{0,\iota}((I^\pm_{p,q}(\mathfrak{A}^\pm\circ\rho^{-1}_{(\mathscr{M},g)})(\mathscr{D}_{p,q}))''$. 
Summing up, the scaling limit we've defined loses \emph{all geometrical information about} $(\mathscr{M},g)$, 
and could be taken as a quantum theory \textsc{Rehren}-dual to a quantum theory in $AdS_d$. Moreover, the 
implementation of asymptotic isometries only occurs in the scaling limit. We notice here the strong 
kinship of the construction above to the \emph{lightfront holography} proposed by \textsc{Schroer}
\cite{schroer5,schroer6}.\\

We've emphasized a \emph{dynamical} definition as Definition \ref{ch4d5} for the scaling algebras 
in AAdS wedges, instead of the simpler Definition \ref{ch4d3} given in the case of $AdS_d$ (which, in principle,
also applies here) due to several reasons. First, the boundary condition (d) cast after Definition \ref{ch4d5} 
emphasizes the character of \emph{return to equilibrium} dual to the scaling limit, and extends to 
more general contexts (i.e., relatively compact diamonds in causally simple spacetimes), such as the
construction of global time functions in Chapter \ref{ch2}. Second, it opens up the possibility of 
dynamically probing deviations from thermal equilibrium in the same way we probe deviations from
critical (i.e., scale invariant) behaviour by means of, say, operator product expansions \cite{frehertel,
weinberg2,bostel2,buor}.

\section{\label{ch4-sector}Holographic analysis of superselection sectors}

The development which concludes this Thesis consists in showing how the geometrical properties of nontrivial
AAdS spacetimes (i.e., different from $AdS_d$) within the hypotheses employed throughout Chapter \ref{ch1} 
modify in a radical way not only the implementation of \emph{geometrical} symmetries, but also of 
\emph{internal} symmetries, encoded in the structure of superselection sectors. In what follows, we'll make
the following set of hypotheses about the spacetimes and the locally covariant quantum theories of our 
interest:

\begin{enumerate}
\item[(i)] $(\mathscr{M},g)$ is an AAdS spacetime satisfying the hypotheses of Proposition \ref{ch1p1};
\item[(ii)] $\mathfrak{A}$ is an extended locally covariant quantum theory, whose realization
in $(\mathscr{M},g)$ satisfies local causality w.r.. $\mathscr{W}(\mathscr{M})$;
\item[(iii)] We have at our disposal a state space $\mathfrak{S}_0$, called \emph{reference state space}, 
such that any $\omega,\omega'\in\mathfrak{S}_0(\mathscr{W}_{p,q})$ satisfy condition (\ref{ch3e12}) (page 
\pageref{ch3e12}) for all $(p,q)\in\mathscr{D}(\mathscr{I})$. In particular, as we've seen, the \textsc{von 
Neumann} algebras $\pi_\omega(\mathfrak{A}(\mathscr{W}_{p,q}))''$, $\omega\in\mathfrak{S}_0(\mathscr{W}_{p,q})$, 
are all *-isomorphic.
\item[(iv)] $\mathfrak{S}_0(\mathscr{W}_{p,q})$ has a representative $\omega_0$, (also) called \emph{reference
state}, which satisfies:
\begin{itemize}
\item The \textsc{Reeh-Schlieder} property, i.e., the cyclic vector $\Omega_0$ in the GNS \textsc{Hil\-bert} 
space $\mathscr{H}_0\doteq\mathscr{H}_{\omega_0}$ associated to $\omega_0$ is \emph{separating} for $\pi_0
(\mathfrak{A}(\mathscr{W}_{p,q}))''$, where $\pi_0\doteq\pi_{\omega_0}$ is the GNS *-representation associated 
to $\omega_0$;
\item The property of holographic local definiteness;
\item The property (III), i.e., $\pi_0(\mathfrak{A}(\mathscr{W}_{p,q}))''$ is type III for all $(p,q)\in
\mathscr{D}(\mathscr{I})$ (see Appendix \ref{ap2}, Definition \ref{ap2d4}, page \pageref{ap2d4}).
\end{itemize}
\end{enumerate}

Additional conditions will be included from time to time alongside the following developments. Notice that the 
states $\omega_0$ satisfying the boundary conditions (c)--(d) imposed in the previous Section satisfy the
properties listed by hypothesis (iv).

\begin{remark}\label{ch4r5}
Hypotheses (iii) and (iv) on the reference state space $\mathfrak{S}_0$ are similar to the ones adopted by
\textsc{Brunetti} and \textsc{Ruzzi} \cite{bruruzzi} in their locally covariant formulation for the algebraic
theory of superselection sectors. We shan't adopt, though, the full-fledged approach of this work for simplicity, 
since we'll assess only basic elements of this theory in what follows.
\end{remark}

\subsection[A digression]{\label{ch4-sector-dig}A digression. \textsc{Haag} duality and elements of the algebraic theory
of superselection sectors}

The fact that the collection $\mathscr{W}(AdS_d)$ of wedges in $AdS_d$ is closed under the operation 
of taking the causal complement (more precisely, formulae (\ref{ch1e29}) are valid) guarantees that
not only local causality is preserved by the \textsc{Rehren} bijection, but also the possible maximality 
of the local algebras with respect to this property. We'll see in the next Subsection that, due to the geometry 
of the wedges in nontrivial AAdS spacetimes, this maximality acquires contours of a \emph{``no-go''} theorem
about the nontriviality of the local algebras in the bulk of such spacetimes, with implications on the 
structure of superselection sectors. However, before that, we shall formulate in a precise manner this 
maximality hypothesis, and illustrate its role in the determination of this structure.

\begin{definition}\label{ch4d6}
Let $\mathfrak{A}$ be an extended locally covariant quantum theory, $(\mathscr{M},g)\in Obj\mathscr{S}tc_d$, 
$\pi$ a *-representation of $\mathfrak{A}_{\mathscr{M}}$ in the \textsc{Hilbert} space $\mathscr{H}$, and 
$\mathscr{Q}(\mathscr{M},g)$ a collection of causally complete open sets of $(\mathscr{M},g)$. We say that
the realization of $\mathfrak{A}$ in $(\mathscr{M},g)$ is $\pi$-\textsc{Haag}-\emph{dual} with respect to $\mathscr{Q}
(\mathscr{M},g)$ if the \textsc{von Neumann} algebras associated to $\pi$ and localized at elements of 
$\mathscr{Q}(\mathscr{M},g)$ are \emph{maximally} locally causal, i.e., $\pi(\mathfrak{A}(\mathscr{O}))''
=\pi(\mathfrak{A}(\mathscr{O}'))'$, where $\pi(\mathfrak{A}(\mathscr{O}))'$ denotes the commutant
of $\pi(\mathfrak{A}(\mathscr{O}))$ in $\mathscr{H}$ (see in Appendix \ref{ap2} the discussion 
immediately preceding Definition \ref{ap2d3}, page \pageref{ap2d3}) -- recall that local causality of 
$\mathfrak{A}$ only implies $\pi(\mathfrak{A}(\mathscr{O}'))''\subset\pi(\mathfrak{A}(\mathscr{O}))'$. More 
in general, we say that the realization of $\mathfrak{A}$ in $(\mathscr{M},g)$ is \emph{essentially} 
$\pi$-\textsc{Haag}-\emph{dual} with respect to $\mathscr{Q}(\mathscr{M},g)$ if the \textsc{Haag}-\emph{dual} 
extension $\mathscr{Q}(\mathscr{M},g)\ni\mathscr{O}\mapsto\pi(\mathfrak{A}(\mathscr{O}))^d\doteq\pi
(\mathfrak{A}(\mathscr{O}'))'$ of the realization of $\mathfrak{A}$ in $(\mathscr{M},g)$ in the 
representation $\pi$ is \emph{locally causal}, i.e., $\pi(\mathfrak{A}(\mathscr{O}_1))^d\subset(\pi
(\mathfrak{A}(\mathscr{O}_2))^d)'$ for all pairs $\mathscr{O}_1,\mathscr{O}_2\in\mathscr{Q}(\mathscr{M},
g)$ such that $\mathscr{O}_1\perp\mathscr{O}_2$. A *-representation $\pi$ of $\mathfrak{A}_{\mathscr{M}}$ 
is said to be (resp. essentially) \textsc{Haag}-dual if the realization of $\mathfrak{A}$ in $(\mathscr{M},
g)$ is (resp. essentially) $\pi$-\textsc{Haag}-dual.
\end{definition}

\begin{remark}\label{ch4r6}
There is a subtlety in the definition of essential \textsc{Haag} duality: it's immediate to see that,  
if $\mathscr{Q}(\mathscr{M},g)$ is closed under causal complements, i.e., $\mathscr{O}\in\mathscr{Q}
(\mathscr{M},g)\Leftrightarrow\mathscr{O}'\in\mathscr{Q}(\mathscr{M},g)$, then \textsc{Haag} duality
and essential \textsc{Haag} duality with respect to $\mathscr{Q}(\mathscr{M},g)$ coincide. Hence, the interesting 
effects that arise when the realization of $\mathfrak{A}$ in $(\mathscr{M},g)$ is only essentially 
$\pi$-\textsc{Haag}-dual with respect to $\mathscr{Q}(\mathscr{M},g)$ appear only when the latter is \emph{not} 
closed under causal complements. The typical example is $(\mathscr{M},g)=\mathbb{R}^{1,d-1}$ and 
$\mathscr{Q}(\mathbb{R}^{1,d-1})=\{I^+(p)\cap I^+(q):p\ll q\in\mathbb{R}^{1,d-1}\}$.
\end{remark}

The breakdown of \textsc{Haag} duality implies, for instance, the spontaneous breaking of internal 
symmetries in the representation induced by a vacuum in theories of local observables in \textsc{Minkowski} 
spacetime \cite{haag}. The reason, in intuitive terms, is the following: the presence of internal
symmetries, manifest in the structure of the particle multiplets which appear in scattering processes, 
is encoded in the local algebras by means of intertwiners \footnote{Given a group $G$ and two unitary 
representations $G\ni g\mapsto U_1(g),U_2(g)$ of $G$ respectively in \textsc{Hilbert} spaces $\mathscr{H}_1$
and $\mathscr{H}_2$, a \emph{intertwiner} between $U_1$ and $U_2$ is a bounded linear operator 
$T:\mathscr{H}_1\rightarrow\mathscr{H}_2$ such that $TU_1(g)=U_2(g)T$ for all $g\in G$. In particular, such
a definition applies, as it'll be often used, to *-representations of *-algebras.} between the unitary 
representations of the group of internal symmetries. \textsc{Haag} duality guarantees that there are enough 
intertwiners to completely reconstruct the representation theory of this group (= structure of superselection
sectors) and, hence, the group itself -- this result, considered one of the great triumphs of the algebraic 
approach to QFT, was demonstrated by \textsc{Doplicher} and \textsc{Roberts} \cite{araki2,haag}, and 
constitutes an abstract formulation of duality between compact groups and their irreducible representations 
(\textsc{Tannaka-Kre\u{i}n} theorem). \\

If only essential \textsc{Haag} duality holds, it's possible to reconstruct the part of internal symmetries 
which doesn't go under spontaneous breaking by repeating the procedure of \textsc{Doplicher} and 
\textsc{Roberts} for the \textsc{Haag}-dual extension of the local algebras, according to Definition 
\ref{ch4d6}. The procedure used to identify the spontaneously broken part of the internal symmetries is, 
however, a bit more subtle \cite{haag}.\\

Let's now make more precise how the structure of superselection sectors is encoded in the realization
of $\mathfrak{A}$ in $(\mathscr{M},g)$, following the treatment in \cite{araki2} and \cite{guilrv},
to which we refer for details. 
Let $\mathfrak{A}$, $(\mathscr{M},g)$ and $\mathscr{Q}(\mathscr{M},g)$ as in Definition \ref{ch4d6},
and $\pi_0$ a \textsc{Haag}-dual *-representation of $\mathfrak{A}_{\mathscr{M}}$ in the \textsc{Hilbert} 
space $\mathscr{H}_0$, which we'll adopt as reference. We say that a *-representation $\pi$ of 
$\mathfrak{A}_{\mathscr{M}}$ in a \textsc{Hilbert} space $\mathscr{H}$ \emph{satisfies the DHR criterion} 
\footnote{DHR stands for the names of \textsc{Doplicher}, \textsc{Haag} and \textsc{Roberts}.} with respect to
$\mathscr{Q}(\mathscr{M},g)$ if, for all $\mathscr{O}\in\mathscr{Q}(\mathscr{M},g)$, $\pi
\restr{\mathfrak{A}(\mathscr{O}')}$ is unitarily equivalent to $\pi_0\restr{\mathfrak{A}(\mathscr{O}')}$. 
Let, then, $V$ be a unitary operator from $\mathscr{H}$ to $\mathscr{H}_0$ such that $V\pi(A)V^*=\pi_0(A)$ 
for all $A\in\mathfrak{A}(\mathscr{O}')$. Define the following *-morphism from $\pi_0(\mathfrak{A}_{\mathscr{M}})$ 
to $\mathscr{B}(\mathscr{H}_0)$:

\begin{equation}\label{ch4e20}
\rho(\pi_0(A))\doteq V\pi(A)V^*.
\end{equation}

By construction, $\rho\circ\pi_0$ is unitarily equivalent to $\pi$. Assuming, for now, that $\mathscr{Q}
(\mathscr{M},g)$ is \emph{directed} under the partial ordering induced by set inclusions, consider 
$\mathscr{O}_1,\mathscr{O}_2\in\mathscr{Q}(\mathscr{M},g)$ such that $\mathscr{O}_2\supset\mathscr{O},
\mathscr{O}_1$. Then, 
\begin{equation}\label{ch4e21}
[\rho(\pi_0(A)),\rho(\pi_0(B))]=[\rho(\pi_0(A)),\pi_0(B)]=0
\end{equation}
for all $A\in\mathfrak{A}(\mathscr{O}_1)$ and $B\in\mathfrak{A}(\mathscr{O}_2')\subset\mathfrak{A}
(\mathscr{O}')$. Due to \textsc{Haag} duality, we have $\rho(\pi_0(A))\in\pi_0(\mathfrak{A}(\mathscr{O}_2'))'
=\pi_0(\mathfrak{A}(\mathscr{O}_2))''$. It follows from the definition of quasilocal algebra that $\rho\circ
\pi_0(\mathfrak{A}_{\mathscr{M}})\subset\rho(\pi_0(\mathfrak{A}_{\mathscr{M}})'')\subset\pi_0
(\mathfrak{A}_{\mathscr{M}})''$.\\

The argument above is a recurrent line of reasoning in the algebraic theory of superselection sectors. 
With it, we can define the transport of localized *-endomorphisms of $\pi_0(\mathfrak{A}_{\mathscr{M}})''$ 
without the aid of the implementation of a translation group. More precisely, we have the following facts:

\begin{itemize}
\item Given two *-endomorphisms $\rho_1$, $\rho_2$ localized respectively in $\mathscr{O}_1$ and 
$\mathscr{O}_2$ and such that the representations $\rho_1\circ\pi_0$ and $\rho_2\circ\pi_0$ are 
unitarily equivalent, i.e., $\rho_1\circ\pi_0(A)=U\rho_2\circ\pi_0(A)U^*$, then $U\in\pi_0(\mathfrak{A}
(\mathscr{O}_3'))'$, where $\mathscr{O}_3\supset\mathscr{O}_1,\mathscr{O}_2$. That is, *-endomorphisms 
localized in $\mathscr{O}$ can be transported to any other element of $\mathscr{Q}(\mathscr{M},g)$ by means
of \emph{local} intertwiners.
\item The composition $\rho_1\circ\rho_2$ is clearly localized in $\mathscr{O}_3\supset\mathscr{O}_1,
\mathscr{O}_2$, where $\rho_i$ is localized in $\mathscr{O}_i$. Moreover, if $\mathscr{O}_1\perp
\mathscr{O}_2$, then $\rho_1\circ\rho_2=\rho_2\circ\rho_1$.
\end{itemize}

A *-endomorphism $\rho$ is said to be \emph{irreducible} if the *-representation $\rho\circ\pi_0$ is.
In this case, we say that the set of pure states in the folium of $\rho\circ\pi_0$ is the 
\emph{(superselection) sector} associated to $\rho$.\\

Let us denote the \textsc{Banach} space of (local) intertwiners between localized *-en\-do\-mor\-phisms 
$\rho$ and $\sigma$ by $(\rho,\sigma)\doteq\{U:U\rho(A)=\sigma(A)U,\,\forall A\}$. Given $T\in(\rho,
\sigma)$, $T'\in(\rho',\sigma')$ we can define the \emph{(tensor) product} $T\otimes T'\doteq T\rho(T')
\in(\rho\circ\rho',\sigma\circ\sigma')$. We can prove that, if $T$ and $T'$ are causally disjoint, 
then $T\otimes T'=T'\otimes T$. It thus follows that, if $\sigma$ and $\sigma'$ are localized in 
causally disjoint regions, then $(T'\otimes T)^*\circ(T\otimes T')\doteq\epsilon(\rho,\rho')\in(\rho
\circ\rho',\rho'\circ\rho)$ is \emph{independent} of $\sigma,\sigma'$ or $T,T'$. Generalizing to 
$n$ localized *-endomorphisms $\rho_i$, $i=1,\ldots,n$, we obtain intertwiners $\epsilon_p(\rho_1,
\ldots,\rho_n)$ for each permutation $p$ of $\{1,\ldots,n\}$, in such a way that $\epsilon_{p'}(\rho_1,
\ldots,\rho_n)\circ\epsilon_p(\rho_1,\ldots,\rho_n)=\epsilon_{p'\circ p}(\rho_1,\ldots,\rho_n)$. In this
manner, if $\rho_1=\cdots=\rho_n=\rho$ the map $S_n\ni p\mapsto\epsilon_p(\rho,\ldots,\rho)\in\rho^n
(\pi_0(\mathfrak{A_{\mathscr{M}}}))'$ constitutes a unitary representation of the group $S_n$ of 
permutations of $n$ elements in $\rho^n\circ\pi_0$. The operators $\epsilon$ are called \emph{statistics
operators}, for they reflect the spin-statistics relation of the superselection sectors. It's precisely 
these operators, together with additional structures (conjugates, left inverses), who allow one to 
rebuild the group of internal symmetries and the particle multiplets.\\

In the same way that the composition of *-endomorphisms corresponds to the tensor product of 
representations, we can describe subrepresentations and direct sums by \emph{subobjects} and 
\emph{direct sums} of *-endomorphisms: we say that $\sigma$ is a \emph{subobject} of $\rho$ if
there exists an \emph{isometry} $W\in(\sigma,\rho)$ (i.e., $W^*W=\mathbb{1}$ and, hence, $WW^*$ is 
a projection), and $\tau$ is a \emph{direct sum} of $\rho$ and $\sigma$ if there exist isometries 
$V\in(\rho,\tau)$, $W\in(\sigma,\tau)$ such that $VV^*+WW^*=\mathbb{1}$ and, hence, $\tau(.)=V\rho(.)
V^*+W\sigma(.)W^*$. To guarantee that these *-endomorphisms remain localized, we need to invoke the 
so-called \emph{property B of} \textsc{Borchers}, valid, for instance, for type-III factors: given 
$\mathscr{O}_1,\mathscr{O}_2\in\mathscr{Q}(\mathscr{M},g)$ such that $\overline{\mathscr{O}_1}\subset
\mathscr{O}_2$ and a projection $E\in\pi_0(\mathfrak{A}(\mathscr{O}_1))''$, there exists an isometry
$W\in\pi_0(\mathfrak{A}(\mathscr{O}_1))''$ such that $WW^*=E$. In the case of type-III factors, we can 
take $\mathscr{O}_2=\mathscr{O}_1$. \\

When $\mathscr{Q}(\mathscr{M},g)$ is not directed, the *-morphism $\rho$ defined in (\ref{ch4e20})
is defined only in the precosheaf $\mathscr{Q}(\mathscr{M},g)\ni\mathscr{O}\subset\mathscr{O}_1\mapsto
\pi_0(\mathfrak{A}(\mathscr{O}_1))''$, and here satisfies $A\in\mathfrak{A}(\mathscr{O}_1)\Rightarrow
\rho(\pi_0(A))\in\pi_0(\mathfrak{A}(\mathscr{O}_1))''$.

\subsection{\label{ch4-sector-aads}(Essential) \textsc{Haag} duality in (A)AdS spacetimes}

Now, with the adequate language at hand, let us go back to discussing the question raised in the first
paragraph of the previous Subsection. Let us consider and AAdS spacetime $(\mathscr{M},g)$, an extended 
locally covariant quantum theory $\mathfrak{A}$ realized in $(\mathscr{M},g)$ and $\pi_0$ a reference 
*-representation of $\mathfrak{A}_{\mathscr{M}}$, which we assume irreducible for simplicity. Once more, 
$\mathscr{D}(AdS_d)$ as well as $\mathscr{W}(AdS_d)$ are closed under causal complements, due to formulae 
(\ref{ch1e29}) (page \pageref{ch1e29}). Employing the \textsc{Rehren} bijection (\ref{ch1e32}) (page 
\pageref{ch1e32}), it automatically follows that the realization of $\mathfrak{A}$ in $AdS_d$ is 
$\pi_0$-\textsc{Haag}-dual with respect to $\mathscr{W}(AdS_d)$ if and only if the \textsc{Rehren}-dual realization 
$\mathfrak{A}$ in $ESU_{d-1}$ is with respect to $\mathscr{D}(AdS_d)$ -- recall as well that, in both cases, 
\textsc{Haag} duality and essential \textsc{Haag} duality coincide.\\

Let1s see now what happens if $(\mathscr{M},g)$ is a nontrivial AAdS spacetime, satisfying the hypotheses 
of Theorem \ref{ch1t4}, page \pageref{ch1t4}. In this case, $\mathscr{W}(\mathscr{M},g)$ is no longer closed
under causal complements, there thus being a fundamental difference between \textsc{Haag} duality and
essential \textsc{Haag} duality. Nevertheless, still holds the following

\begin{lemma}\label{ch4l1}
If $\pi_0$ is (essentially) \textsc{Haag}-dual with respect to $\mathscr{D}(\mathscr{M},g)$, then  it's 
\textsc{Haag}-dual with respect to $\mathscr{W}(\mathscr{M},g)$ and $\pi_0(\mathfrak{A}(\mathscr{W}_{p,\bar{q}}))''=
\pi_0(\mathfrak{A}(\mathscr{W}_{q,\bar{p}}'))''=\pi_0(\mathfrak{A}(\mathscr{W}_{\bar{p},q})')'$ for all pairs 
$p,q\in\mathscr{I}$ such that $(p,\bar{q})\in\mathscr{D}(\mathscr{I})$, where $\bar{p}$ is the \emph{antipodal} 
of $p$ (see formulae (\ref{ch1e27})--(\ref{ch1e28}) and the discussion preceding them in page \pageref{ch1e27}).
\begin{quote}{\small\scshape Proof.\quad}
{\small\upshape Essential \textsc{Haag} duality with respect to $\mathscr{D}(\mathscr{M},g)$ is not only equivalent to
\textsc{Haag} duality (for the collection of diamonds in $\mathscr{I}$ is closed under causal complements), 
but also implies \textsc{Haag} duality for wedges, for $\pi_0(\mathfrak{A}(\mathscr{D}_{p,\bar{q}}))''=\pi_0
(\mathfrak{A}(\mathscr{W}_{p,\bar{q}}))''\subset\pi_0(\mathfrak{A}(\mathscr{W}_{p,\bar{q}}'))'\subset\pi_0
(\mathfrak{A}(\mathscr{W}_{q,\bar{p}}))'=\pi_0(\mathfrak{A}(\mathscr{D}_{q,\bar{p}}))'=\pi_0(\mathfrak{A}
(\mathscr{D}_{p,\bar{q}}'))'$. The last assertion follows from the identities $\pi_0(\mathfrak{A}
(\mathscr{W}_{p,\bar{q}}))''=\pi_0(\mathfrak{A}(\mathscr{D}_{p,\bar{q}}))''$ $=\pi_0(\mathfrak{A}
(\mathscr{D}_{q,\bar{p}}'))''=\pi_0(\mathfrak{A}(\mathscr{D}_{q,\bar{p}}))'=\pi_0(\mathfrak{A}(\mathscr{W}_{q,
\bar{p}}))'=\pi_0(\mathfrak{A}(\mathscr{W}_{q,\bar{p}}'))''$.~\hfill~$\Box$}
\end{quote}
\end{lemma}

The last expression of Lemma \ref{ch4l1} is trivial for $AdS_d$, but possesses  potentially dramatic 
consequences for nontrivial AAdS spacetimes, for, in this case, $\pi_0(\mathfrak{A}(\mathscr{W}_{p,\bar{q}}))''
\cap\pi_0(\mathfrak{A}(\mathscr{W}_{\bar{p},q}))''\supset\pi_0(\mathfrak{A}(\mathscr{W}_{\bar{p},q}'\cap{}
\mathscr{W}_{p,\bar{q}}'))''$, and the region $\mathscr{W}_{\bar{p},q}'\cap{}\mathscr{W}_{p,\bar{q}}'$ is a nonvoid
open set if $(\mathscr{M},g)$ satisfies the hypotheses of Proposition \ref{ch1p1} (page \pageref{ch1p1}), stronger 
than those of Theorem \ref{ch1t4} (see Figure \ref{ch1f4}, page \pageref{ch1f4}). Supposing that the precosheaf 
of \textsc{von Neumann} algebras $\mathscr{K}(\mathscr{M},g)\ni\mathscr{O}\mapsto\pi_0(\mathfrak{A}(\mathscr{O}))''$ 
satisfies the property of \emph{extended locality}
\begin{equation}\label{ch4e22}
\pi_0(\mathfrak{A}(\mathscr{O}_1))''\cap{}\pi_0(\mathfrak{A}(\mathscr{O}_1))''=\mathbb{C1},\,\forall
\mathscr{O}_1,\mathscr{O}_2\in\mathscr{K}(\mathscr{M},g):\mathscr{O}_1\perp\mathscr{O}_2,
\end{equation}
it then follows that $\pi_0(\mathfrak{A}(\mathscr{W}_{\bar{p},q}'\cap{}\mathscr{W}_{p,\bar{q}}'))''=\mathbb{C1}$
and, hence, $\pi_0(\mathfrak{A}(\mathscr{O}_{r,s}))''=\mathbb{C1}$ for any $r,s\in\mathscr{M}$ such that 
$\mathscr{O}_{p,q}\doteq I^+(r,\mathscr{M})\cap{} I^-(s,\mathscr{M})\subset\mathscr{W}_{\bar{p},q}'
\cap{}\mathscr{W}_{p,\bar{q}}'$. As:

\begin{itemize}
\item Any sufficiently small diamond in $(\mathscr{M},g)$ satisfies the above condition for some pair
$p,q\in\mathscr{I}$ satisfying the conditions of Lemma \ref{ch4l1}, and
\item In \textsc{Minkowski} spacetime $\mathbb{R}^{1,d-1}$, any theory of local observables which is 
additive, translation covariant and locally causal in the GNS representation $\pi_0$ associated to a 
pure vacuum state (i.e., the joint spectrum of the translation generators is contained in $J^+(0,
\mathbb{R}^{1,d-1})$ and the vacuum vector is an eigenvector of the generators with zero eigenvalue) satisfies 
extended locality \cite{landau},
\end{itemize}
we're led to the following ``\emph{`no-go'} theorem'':

\begin{corollary}\label{ch4c1}
Let $(\mathscr{M},g)$ be an AAdS spacetime satisfying the hypotheses of Proposition \ref{ch1p1},
$\mathfrak{A}$ an extended locally covariant quantum theory realized in $(\mathscr{M},g)$ and $\pi_0$ 
an irreducible *-representation of $\mathfrak{A}_{\mathscr{M}}$ satisfying the hypotheses of Lemma \ref{ch4l1}. 
If $\pi_0$ satisfies extended locality (\ref{ch4e22}), then any \textsc{von Neumann} algebra localized in
a sufficiently small open set consists only in multiples of $\mathbb{1}$. If the precosheaf of \textsc{von 
Neumann} algebras $\mathscr{K}(\mathscr{M},g)\ni\mathscr{O}\mapsto\pi_0(\mathfrak{A}(\mathscr{O}))''$ is
moreover additive, then $\pi_0(\mathfrak{A}_{\mathscr{M}})''=\mathbb{C1}$.~\hfill~$\Box$
\end{corollary}

\begin{remark}\label{ch4r7}
There is, to the author's knowledge, no proof of the property of extended locality in curved spacetimes under
hypotheses similar to the ones employed in \textsc{Minkowski} spacetime (i.e., substituting, say, some 
version of the microlocal spectrum condition \cite{brunfek} for the spectral condition of the vacuum). We 
conjecture, though, that such a demonstration is possible for real analytic AAdS spacetimes, along lines
analogous to the proof of the timelike tube property of \textsc{Borchers} (page \pageref{ch4fn1} and
footnote \ref{ch4fn1}).
\end{remark}

We conclude from Corollary \ref{ch4c1} that essential \textsc{Haag} duality with respect to $\mathscr{D}(\mathscr{M},
g)$ implies potentially a trivial quantum theory if combined with bulk additivity, and even a non additive
bulk quantum theory may end up being trivial at sufficiently small scales -- in particular, the procedure of
holographic reconstruction of the realization of $\mathfrak{A}$ in $(\mathscr{M},g)$ from the dual theory 
in the boundary, whose geometrical part was schematized in Subsubsection \ref{ch1-aads-causal-loc} (page 
\pageref{ch1-aads-causal-loc}), is \emph{useless} in both cases!
\footnote{On the other hand, the possibility that $\pi_0$ is \textsc{Haag}-dual and $\pi_0(\mathfrak{A}
(\mathscr{W}_{p,\bar{q}}'\cap{}\mathscr{W}_{q,\bar{p}}'))''=\mathbb{C1}$ is not completely uninteresting
from the physical viewpoint -- as wedges model, in a certain sense, the exterior of black holes, we can 
imagine the region $\mathscr{W}_{p,\bar{q}}'\cap{}\mathscr{W}_{q,\bar{p}}'$ as being ``inside the horizon''. 
From this viewpoint, we see that a quantum theory with the characteristics above is unable to ``see'' physical 
events localized ``inside the event horizon''. As \textsc{Haag} duality with respect to $\mathscr{D}(\mathscr{M},g)$ 
is necessary so that the \textsc{Rehren}-dual quantum theory in the boundary be conformally covariant and
$\pi_0$ be the GNS representation associated to a vacuum state \cite{bglongo1}, we conjecture that such a
(non additive) quantum theory probably doesn't suffer from the \emph{black hole information paradox}, but is
also essentially insensitive to the details of the bulk geometry, in the same way as the scaling limit built 
in the previous Section.} The following hypothesis, though, does lead to nontrivial results:

\begin{enumerate}
\item[(v)] $\pi_0$ is \emph{essentially} \textsc{Haag}-dual with respect to $\mathscr{W}(\mathscr{M},g)$, 
but it's \emph{not} (essentially) \textsc{Haag}-dual with respect to $\mathscr{D}(\mathscr{M},g)$.
\end{enumerate}

The second premise allows us to escape from Corollary \ref{ch4c1}. In particular, in this case the
precosheaf of \textsc{von Neumann} algebras $\mathscr{D}(\mathscr{M},g)\ni\mathscr{D}_{p,\bar{q}}
\mapsto\pi_0(\mathfrak{A}(\mathscr{D}_{q,\bar{p}}))'$ is not locally causal, but the precosheaf
\begin{equation}\label{ch4e23}
\mathscr{W}(\mathscr{M},g)\ni\mathscr{W}_{p,q}\mapsto\mathfrak{B}(\mathscr{W}_{p,q})\doteq\pi_0
(\mathfrak{A}(\mathscr{W}_{p,q}'))'
\end{equation}
is, by hypothesis (v). 

\begin{lemma}\label{ch4l2}
Suppose that $\pi_0$ satisfies hypothesis (v). Then, $\pi_0$ satisfies holographic local definiteness
if and only if the quasilocal algebras $\mathfrak{B}_{\mathscr{P}oi(r)}''$ are \emph{irreducible},
i.e. $\mathfrak{B}_{\mathscr{P}oi(r)}'=\mathbb{C1}$, for all $r\in\mathscr{I}$. In particular,
in this case $\mathfrak{B}_{\mathscr{M}}''$ is irreducible.
\begin{quote}{\small\scshape Proof.\quad}
{\small\upshape It follows immediately from formula \[\bigcap_{{(p,\bar{q})\in\mathscr{D}(\mathscr{I}),}
\atop{\mathscr{W}_{p,\bar{q}}\ni\bar{r}}}\pi_0(\mathfrak{A}(\mathscr{W}_{p,\bar{q}}))''=\bigcap_{{(p,
\bar{q})\in\mathscr{D}(\mathscr{I}),}\atop{\mathscr{W}_{p,\bar{q}}\ni\bar{r}}}\pi_0(\mathfrak{A}
(\mathscr{W}_{q,\bar{p}}'))''=\]\[=\left(\bigcup_{q\ll_{\mathscr{I}}\bar{p}\in\mathscr{M}in(r)}
\mathfrak{B}(\mathscr{W}_{q,\bar{p}})\right)',\] and the fact that \[\bigcap_{{(p,\bar{q})\in
\mathscr{D}(\mathscr{I}),}\atop{\mathscr{W}_{p,\bar{q}}\ni\bar{r}}}\mathscr{W}_{q,\bar{p}}'=
\bigcap_{{(p,\bar{q})\in\mathscr{D}(\mathscr{I}),}\atop{\mathscr{W}_{p,\bar{q}}\ni\bar{r}}}
\mathscr{W}_{p,\bar{q}}=\{\bar{r}\},\] by virtue of Proposition \ref{ch1p1} and Theorem 
\ref{ch1t4}.~\hfill~$\Box$}
\end{quote}
\end{lemma}

In particular, internal symmetries of the precosheaf $\mathscr{W}_{p,q}\mapsto\mathfrak{B}(\mathscr{W}_{p,q})$
cannot be spontaneously broken in $\mathscr{H}_0$ if $\pi_0$ satisfies holographic local definiteness and
hypothesis (v). Notice that this \emph{doesn't} imply that the internal symmetries of the precosheaf 
$\mathscr{W}_{p,q}\mapsto\pi_0(\mathfrak{A}(\mathscr{W}_{p,q}))''$ are not broken.\\

Let us consider, now, a *-representation $\pi$ of $\mathfrak{A}_{\mathscr{M}}$ in the \textsc{Hilbert} space
$\mathscr{H}_\pi$ satisfying 
\begin{equation}\label{ch4e24}
\forall p,q\in\mathscr{D}(\mathscr{I}),\exists V\in\mathscr{B}(\mathscr{H}_0,\mathscr{H}_\pi)
\mbox{ unitary such that }\pi(A)=V\pi_0(A)V^*,\forall A\in\mathfrak{A}(\mathscr{W}_{p,q}').
\end{equation}

In particular, $\pi$ satisfies the DHR criterion with respect to $\mathscr{D}(\mathscr{M},g)$ under the \textsc{Rehren}
bijection. The fact that $\pi_0$ doesn't satisfy essential \textsc{Haag} duality with respect to $\mathscr{D}(\mathscr{M},
g)$ guarantees that the *-representations satisfying (\ref{ch4e24}) constitute a \emph{proper} subclass of
DHR excitations.\\

One should notice that, even from the original viewpoint of the original form of \textsc{Rehren} duality, 
i.e., $(\mathscr{M},g)=AdS_d$, it looks weird at first sight to employ a selection criterion which aims 
at modelling vacuum excitations corresponding to particles, since conformal quantum field theories are not 
elementary particle theories. However, it was demonstrated by \textsc{Buchholz, Mack} and \textsc{Todorov} 
\cite{buchmt} that:

\begin{theorem}[\cite{buchmt}]\label{ch4t4}
If $\pi_0$ is the GNS representation of a conformally invariant vacuum state $\omega_0$ (which is, 
hence, \textsc{Haag}-dual with respect to $\mathscr{D}(AdS_d)$ \cite{bglongo1}), any state $\omega$ with GNS 
representation $\pi_\omega$ with positive energy is a DHR excitation of $\omega_0$, i.e., given any $(p,q)\in
\mathscr{D}(\mathscr{I})$, we have $\pi_\omega\restr{\mathfrak{A}(\mathscr{D}_{p,q}')}$ is unitarily equivalent
to $\pi_0\restr{\mathfrak{A}(\mathscr{D}_{p,q}')}$. 
\begin{quote}{\small\scshape Proof.\quad}
{\small\upshape Consider the following facts: (i) $(p,\bar{q})\in\mathscr{D}(\mathscr{I})$ if and only if 
$(q,\bar{p})\in\mathscr{D}(\mathscr{I})$; (ii) $\omega_0\restr{\pi_0(\mathfrak{A}(\mathscr{D}_{p,q}))''}$ 
is normal and faithful for all $(p,q)\in\mathscr{D}(\mathscr{I})$ (\textsc{Reeh-Schlieder} property).
As the \textsc{Reeh-Schlieder} property also holds if we substitute $\pi$ for $\pi_0$ by virtue of
energy-momentum positivity, it follows that the map $\pi_0(A)\mapsto\pi(A)$ for $A\in\mathfrak{A}
(\mathscr{D}_{p,q})$ extends to a normal *-isomorphism between $\pi_0(\mathfrak{A}(\mathscr{D}_{p,q}))''$ 
and $\pi_\omega(\mathfrak{A}(\mathscr{D}_{p,q}))''$. As the cyclic vectors $\Omega_0$ and $\Omega$ in the
respective GNS \textsc{Hilbert} spaces $\mathscr{H}_0$ and $\mathscr{H}_\omega$ associated to $\omega_0$
and $\omega$ are also separating by the \textsc{Reeh-Schlieder} property, it follows from Theorem 2.5.32
in \cite{bratteli1} (see also Theorem 7.2.9 in \cite{kadring2}) that such a *-isomorphism is implemented
by a unitary operator between the GNS \textsc{Hilbert} spaces of $\omega_0$ and $\omega$. From (i), we finally
see that $\pi$ satisfies the DHR criterion, with $\rho$ localized in the diamond $\mathscr{D}_{p,q}'$.~\hfill~$\Box$}
\end{quote}
\end{theorem}

As the \textsc{Reeh-Schlieder} property is valid for the reference representations $\pi_0$ we've adopted 
for AAdS spacetimes, we see that such a criterion is sufficiently general for our purposes. The endomorphism 
$\rho$ corresponding to $\pi$ and localized, say, in $\mathscr{D}_{p,\bar{q}}$ has the following properties:

\begin{enumerate}
\item By construction, $\rho\circ\pi_0(A)=\pi_0(A)$ for all $A\in\mathfrak{A}(\mathscr{D}_{q,\bar{p}})$;
\item If $(p',q')\in\mathscr{D}(\mathscr{I})$ are such that $p,\bar{q},p',q'\in\mathscr{M}in(r)$ for some 
$r\in\mathscr{I}$ and $B\in\mathfrak{A}(\mathscr{D}_{p',q'})$, then $\rho\circ\pi_0(B)\in\mathfrak{B}
(\mathscr{D}_{p'',q''})$ for $(p'',q'')\in\mathscr{D}(\mathscr{I})$ such that $\mathscr{D}_{p,\bar{q}},
\mathscr{D}_{p',q'}\subset\mathscr{D}_{p'',q''}$ (repeat the argument used in Subsection \ref{ch4-sector-dig}, 
page \pageref{ch4-sector-dig}).
\end{enumerate}

The second property above leads to the following scenario: $\rho\circ\pi_0$ only assumes values in 
$\pi_0(\mathfrak{A}_{\mathscr{M}})$ for $A\in\mathfrak{A}\left(\bigcup_{(p',q')\in\mathscr{D}(\mathscr{I}):
\mathscr{W}_{p',q'}\supset\mathscr{W}_{p,\bar{q}}}\mathscr{W}_{p',q'}\right)$. However, if $\rho'$ is the
endomorphism associated to $\pi$ and localized in $\mathscr{W}_{q,\bar{p}}$, we have that $\rho'$ can
still be obtained from the transport of $\rho$ by means of local intertwiners, for, although there is 
no $(p',q')\in\mathscr{D}(\mathscr{I})$ such that $\mathscr{W}_{p',q'}\supset\mathscr{W}_{p,\bar{q}}\cup
\mathscr{W}_{q,\bar{p}}$, $\rho$ can always be transported to a smaller wedge contained in 
$\mathscr{W}_{q,\bar{p}}$.\\

We can imagine, in view of the possibility of spontaneous breaking of internal symmetries allowed by 
essential \textsc{Haag} duality, that $\pi_0$ is \emph{not} irreducible. More precisely, suppose that the 
(trivially) localized endomorphism $\iota\doteq\mbox{id}_{\pi_0(\mathfrak{A}_{\mathscr{M}})}$ associated to $\pi_0$ 
has nontrivial subobjects, i.e., for each $(p,\bar{q})\in\mathscr{D}(\mathscr{I})$ there is a projection 
$\mathbb{1}\neq E\in\pi_0(\mathfrak{A}(\mathscr{W}_{p,\bar{q}}))'$, a *-endomorphism $\sigma$ of $\pi_0
(\mathfrak{A}_{\mathscr{M}})$ localized in $\mathscr{W}_{p,\bar{q}}$ and an isometry $W\in(\sigma,\iota)$ 
such that $WW^*=E$ (by the property B of \textsc{Borchers}, $W\in\pi_0(\mathfrak{A}(\mathscr{W}_{p,\bar{q}}))'$).
Actually, we only need the existence of $E$ (guaranteed, on its turn, by the non irreducibility of $\pi_0$), 
for the existence of $W$ and, thus, of $\sigma$ ($\sigma(\pi_0(A))\doteq W^*\pi_0(A)W$) follows from the 
property B. Hence, we can also consider the isometry $V$ and the localized endomorphism $\sigma'$ associated 
to the projection $\mathbb{1}-E$. $\sigma\circ\pi_0$ as well as $\sigma'\circ\pi_0$ satisfy essential 
\textsc{Haag} duality with respect to $\mathscr{W}(\mathscr{M},g)$.

\begin{definition}\label{ch4d7}
If $\sigma$ and $\sigma'$ are subobjects of $\iota$ which generate \emph{disjoint} representations, their
unitary equivalence classes are then called \emph{phases} of the reference state $\omega_0$. A phase $\sigma$ 
is said to be \emph{pure} if the corresponding representation is irreducible.
\end{definition}

Let us take the subobjects $\sigma$ and $\sigma'$ of $\iota$ constructed in the previous paragraph,
localized respectively, say, in $\mathscr{D}_{p,\bar{q}}$ and $\mathscr{D}_{q,\bar{p}}$. As in this case 
$\sigma(\pi_0(A))=\sigma'(\pi_0(A))=\pi_0(A)$ if $A\in\mathfrak{A}(\mathscr{W}_{p,\bar{q}}')\cap{}\mathfrak{A}
(\mathscr{W}_{q,\bar{p}}')\supset\mathfrak{A}(\mathscr{W}_{p,\bar{q}}'\cap{}\mathscr{W}_{q,\bar{p}}')$, we can 
consider then the following *-endomorphism $\tau$ of $\pi_0(\mathfrak{A}_{\mathscr{M}})''$, which satisfies 
the following properties:

\begin{enumerate}
\item[(S1)] $\tau\circ\sigma(\pi_0(A))=\sigma(\pi_0(A))$ for all $A\in\mathfrak{A}(\mathscr{W}_{p,\bar{q}})$;
\item[(S2)] $\tau\circ\sigma'(\pi_0(A))=\sigma'(\pi_0(A))$ for all $A\in\mathfrak{A}(\mathscr{W}_{q,\bar{p}})$.
\end{enumerate}

Obviously, we have $\tau\circ\sigma(\pi_0(A))=\tau\circ\sigma'(\pi_0(A))=\tau(\pi_0(A))$ if $A\in\mathfrak{A}
(\mathscr{W}_{p,\bar{q}}'\cap\mathscr{W}_{q,\bar{p}}')$. We can, thus, say that $\tau$ is ``localized'' in 
$\mathscr{W}_{p,\bar{q}}'\cap\mathscr{W}_{q,\bar{p}}'$ and interpolates the phases $\sigma$ and $\sigma'$ 
(recall that, in AAdS spacetimes satisfying, as we've assumed in this Section, the hypotheses of Proposition 
\ref{ch1p1} and, hence, of Theorem \ref{ch1t4}, $\mathscr{W}_{p,\bar{q}}'\cap\mathscr{W}_{q,\bar{p}}'\neq
\varnothing$! See Figure \ref{ch1f4}, page \ref{ch1f4}).

\begin{definition}\label{ch4d8}
An endomorphism $\tau$ which satisfies (S1) and (S2) for some pair of phases $\sigma$, $\sigma'$ of $\omega_0$ 
localized respectively in wedges $\mathscr{W}_{p,\bar{q}}$, $\mathscr{W}_{q,\bar{p}}$ is said to be 
\emph{solitonic}. Then, we say that $\tau$ \emph{separates} or \emph{interpolates} $\sigma$ and $\sigma'$, 
and the unitary equivalence class of $\tau$ is a \emph{soliton}. 
\end{definition}

Notice that an intertwiner which transports $\sigma$ from the wedge $\mathscr{W}_{p,\bar{q}}$ to another,  
sufficiently close wedge $\mathscr{W}_{p',\bar{q'}}$, when applied to $\sigma'$, it transports the latter
from $\mathscr{W}_{q,\bar{p}}$ to $\mathscr{W}_{q',\bar{p'}}$ and, hence, when applied to $\tau$, produces 
a solitonic endomorphism ``localized'' in  $\mathscr{W}_{p',\bar{q'}}'\cap\mathscr{W}_{q',\bar{p'}}'$ in the
sense above. Thus, we're entitled to remove the quotation marks and say that $\tau$ is, indeed,
\emph{localized} in $\mathscr{W}_{p,\bar{q}}'\cap\mathscr{W}_{q,\bar{p}}'$. The composition of solitonic
endomorphisms $\tau,\tau'$ is not always defined, for it demands that the corresponding endomorphisms
representing the interpolated phases $\tau$ and $\tau'$ ``fit together''.

\subsection{\label{ch4-sector-rec}Obstructions to the inversion of \textsc{Rehren} duality}

It remains to discuss the possibility that, starting from a precosheaf $\mathscr{D}_{p,q}\mapsto\mathfrak{A}
(\mathscr{D}_{p,q})$ realizing a locally covariant quantum theory $\mathfrak{A}$ in $(\mathscr{I},\bar{g}^{(0)})$, 
we may construct for each AAdS spacetime $(\mathscr{M},g)$ satisfying the hypotheses of the present Section a
precosheaf $\mathscr{O}\mapsto\mathfrak{A}(\mathscr{O})$ indexed by open, globally hyperbolic and causally 
convex regions $\mathscr{O}\subset\mathscr{M}$, and not only wedges (whose local algebras are obtained by 
defining $\mathfrak{A}(\mathscr{W}_{p,q})\doteq\mathfrak{A}\circ\rho_{(\mathscr{M},g)}(\mathscr{W}_{p,q})$). \\

Let us take a reference state $\omega$ in the realization of $\mathfrak{A}$ in $(\mathscr{M},g)$. If the 
precosheaf of \textsc{von Neumann} algebras $\mathscr{O}\mapsto\pi_\omega(\mathfrak{A}(\mathscr{O}))''$ is
\emph{additive}, we can try to perform this reconstruction by employing the results of Subsubsection 
\ref{ch1-aads-causal-loc} (page \pageref{ch1-aads-causal-loc}) if we define, for a relatively compact and 
geodesically convex $\mathscr{O}_{p,q}\subset\mathscr{M}$ sufficiently small,
\begin{equation}\label{ch4e25}
\mathfrak{B}(\mathscr{O}_{p,q})\doteq\bigcap_{\mathscr{W}_{p,q}\supset\mathscr{O}_{p,q}}
\pi_\omega(\mathfrak{A}(\mathscr{W}_{p,q}))''.
\end{equation}

\begin{proposition}\label{ch4p2}
Suppose that the precosheaf of \textsc{von Neumann} algebras $\mathscr{O}\mapsto\pi_\omega(\mathfrak{A}
(\mathscr{O}))''$ is additive and satisfies \textsc{Haag} duality with respect to wedges. Then $\mathfrak{B}
(\mathscr{O}_{p,q})=\pi_\omega(\mathfrak{A}(\mathscr{O}_{p,q}'))'$.
\begin{quote}{\small\scshape Proof.\quad}
{\small\upshape Notice that the operation which takes a \textsc{von Neumann} algebra to its commutant
satisfies the same properties that of the operation of causal complement in causally complete regions
(see Appendix \ref{ap1}, end of Subsection \ref{ap1-glob-causal}, page \pageref{ap1-glob-infty}), if we 
substitute the algebra generated by the union for the union of algebras. Then it follows from (\ref{ch4e25}) 
that $\mathfrak{B}(\mathscr{O}_{p,q})\doteq\bigcap_{\mathscr{W}_{p,q}\supset\mathscr{O}_{p,q}}\pi_\omega
(\mathfrak{A}(\mathscr{W}_{p,q}'))'=\left(\bigvee_{\mathscr{W}_{p,q}\supset\mathscr{O}_{p,q}}\pi_\omega
(\mathfrak{A}(\mathscr{W}_{p,q})')''\right)'=\pi_\omega\left(\mathfrak{A}\left(\bigcup_{\mathscr{W}_{p,q}
\supset\mathscr{O}_{p,q}}\mathscr{W}_{p,q}'\right)\right)'=\pi_\omega(\mathfrak{A}(\mathscr{O}_{p,q}'))'$.
~\hfill~$\Box$}
\end{quote}
\end{proposition}

If the precosheaf $\mathscr{O}\mapsto\pi_\omega(\mathfrak{A}(\mathscr{O}))''$ is \textsc{Haag}-dual, 
then it's also additive and, thus, the reconstruction procedure based upon (\ref{ch4e25}) can be taken
to completion in a successful way. If we assume, on the other hand, only \emph{essential} \textsc{Haag}
duality, the \textsc{Haag}-dual precosheaf $\mathscr{O}\mapsto\pi_\omega(\mathfrak{A}(\mathscr{O}'))'$ 
\emph{doesn't need} to be additive. Having in sight the scenario of spontaneous breaking of internal 
symmetries, we can give a natural physical interpretation for this phenomenon: the \textsc{Haag}-dual 
net circumvents the spontaneous breaking of internal symmetries by including precisely those observables 
corresponding to extended physical procedures, localized, say, around domain walls, just like the soliton 
sectors built in the previous Subsection. Such a situation is analogous to ordered phases in quantum spin 
models in Statistical Mechanics, where \textsc{Kramers-Wannier} duality produces a gas of ``contours around
pure-phase domains'', which, on its turn, finds itself in the symmetric (disordered) phase \cite{kadanoff2,
minlos}. If we imagine the region $\mathscr{W}_{p,\bar{q}}'\cap\mathscr{W}_{q,\bar{p}}'$ ($\neq\varnothing$!), 
sketched in Figure \ref{ch1f4} (page \ref{ch1f4}) as the rough localization of a \emph{domain wall}, 
i.e., a maximal (that is, which isn't properly included into any other submanifold of the same codimension), 
acausal codimension-two submanifolds of $\mathscr{M}$, it's quite tempting to interpret the solitons in 
Definition \ref{ch4d8} in the light of the AdS/CFT correspondence as \emph{D-branes}, for, as we've seen, 
the former share essentially the same properties of the objects of the same name in string theory 
\cite{zwiebach}. \\

We moreover notice that gravitationally induced superselection sectors have already been previously 
considered in the literature, in particular related to the superselection rule for electric charge
\cite{ashsen,sorkin}. Such sectors enjoy a rather large amount of universality, and are induced even
in free quantum field theories.\\

We emphasize that the soliton sectors we've built \emph{don't exist} in $AdS_d$, for in this case \textsc{Haag} 
duality and essential \textsc{Haag} duality with respect to $\mathscr{W}((\mathscr{M},g)=AdS_d)$, as we've seen, coincide
(this, of course, is directly related to the fact that $\mathscr{W}_{p,\bar{q}}'\cap{}\mathscr{W}_{q,\bar{p}}'=
\varnothing$ in $AdS_d$), and the direct sum of reference *-representations satisfying \textsc{Haag} duality
doesn't necessarily, but the direct sum of\emph{essentially} \textsc{Haag}-dual reference *-representations is 
essentially \textsc{Haag}-dual \cite{rob2}. More in general, \textsc{Haag} duality with respect to wedges leads, by 
Lemma \ref{ch4l2}, to the irreducibility of $\pi_0$, making impossible the existence of soliton sectors even if
we manage to escape Corollary \ref{ch4c1} and have nontrivial observables in $\mathscr{O}\subset\mathscr{M}$ 
arbitrarily small for $(\mathscr{M},g)\neq AdS_d$ (in $AdS_d$, there is no problem in assuming \textsc{Haag}
duality, and in this case the original \textsc{Rehren} duality allows a one-to-one correspondence). Hence, the 
presence of soliton sectors seems to essentially render useless or impossible (modulo the validity of the 
property of extended locality) the unequivocal reconstruction of the bulk quantum theory by means of the 
prescription (\ref{ch4e25}) in nontrivial AAdS geometries. We speculate that an alternative, more indirect 
possibility, for at least a partial reconstruction, albeit beyond the mere specification of the localization 
of observables in wedges by the \textsc{Rehren} bijection, would be to elaborate an operator product expansion
around the scaling limit, as suggested at the end of Section \ref{ch4-rehren}.

\cleardoublepage

\thisfancyput(0in,-5.4in){\includegraphics{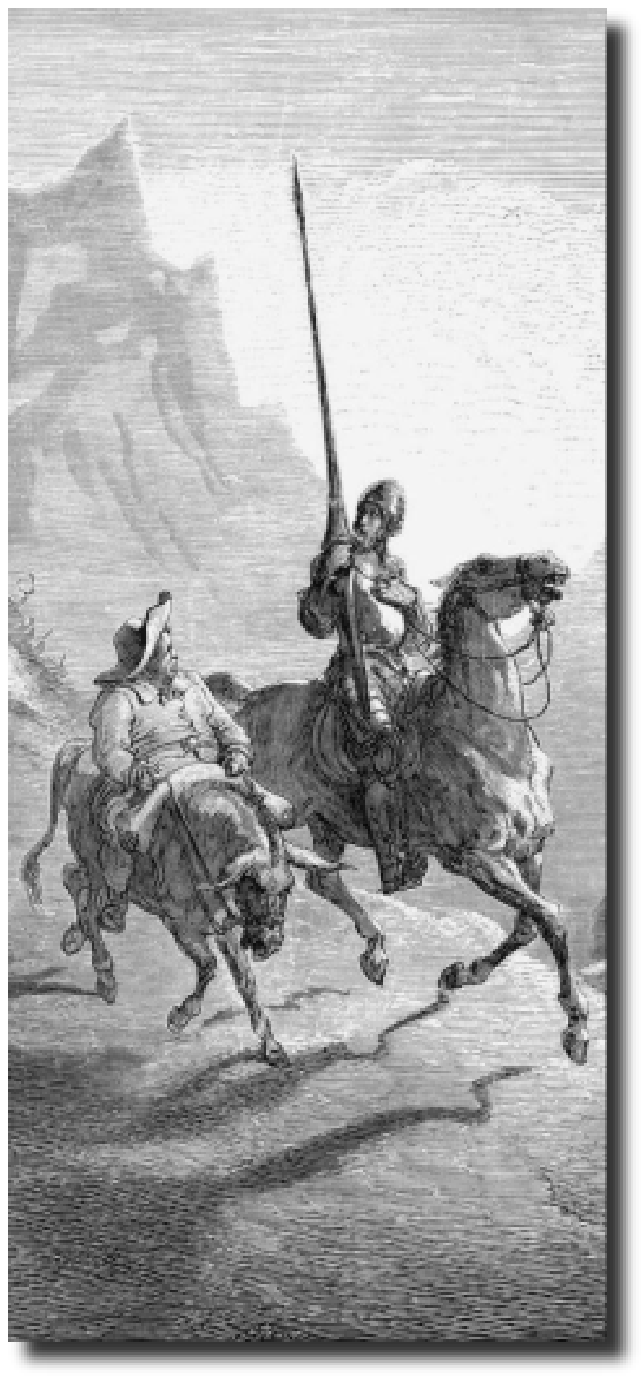}}

\epigraphhead[170]{\epigraph{\qquad --- ?`C\'omo puede ser eso? ---
respondi\'o don Quijote ---. ?`Tan de esencia de la historia es 
saber las cabras que han pasado por estenso, que si se yerra una 
del n\'umero no puedes seguir adelante con la historia?

\qquad --- No, se\~nor, en ninguna manera --- respondi\'o Sancho ---;
porque as\'{\i} como yo pregunt\'e a vuestra merced que me dijeste cu\'antas
cabras hab\'{\i}an pasado, y me respondi\'o que no sab\'{\i}a, en aquel mesmo 
instante se me fue a m\'{\i} de la memoria cuanto que quedaba por decir,
y a fe que era de mucha virtud y contiento.

\qquad --- ?`De modo --- dijo don Quijote --- que ya la historia
es acabada?

\qquad --- Tan acabada es como mi madre --- dijo Sancho.

\qquad --- D\'{\i}gote de verdad --- respondi\'o don Quijote --- que t\'u has
contado unas de las m\'as nuevas consejas, cuento o historia que nadie
pudo pensar en el mundo, y que tal modo de contarla ni dejarla jam\'as
se podr\'a ver ni habr\'a visto en toda la vida, aunque no esperaba yo 
otra cosa de tu buen discurso; mas no me maravillo, ques quiz\'a estos
golpes que no cesan te deben de tener turbado el entendimiento.\endnotemark[12]}
{\textsc{Miguel de Cervantes Saavedra}\\ \emph{El ingenioso hidalgo 
don Quijote de la Mancha}, Primero Libro, Cap. XX}}
\part{A synthesis}

\chapter*{\label{ch5}Coda. Concluding remarks}
\addcontentsline{toc}{chapter}{Coda}

\epigraph{\begin{verse}[7cm] \emph{Go, go, go, said the bird: human kind \\
Cannot bear very much reality. \\ Time past and time future \\
What might have been and what has been \\ Point to one end, which is
always present.}\end{verse}~\vspace*{-13pt}}{\textsc{T. 
S. Eliot} \\ ``Burnt Norton'' (\emph{Four Quartets}, 1943), I}

The present work was able to reveal quite a rich structure underlying the AdS/CFT correspondence, 
of which we've been able only to scratch the surface. \\

The main contributions of the present work, as a whole, can be listed as follows:

\begin{itemize}
\item One has specified the essential aspects underlying \textsc{Rehren} duality. 
In particular, the geometrical and causal character of the change of localization proposed by 
\textsc{Rehren} was outlined from the global viewpoint (Chapter \ref{ch1}) as well as the local 
one (i.e., in a neighbourhood of $\mathscr{I}$ -- Chapter \ref{ch2}). 
\item The elaborate machinery of global Lorentzian geometry developed from two different
viewpoints -- kinematic (causal structure, conformal infinity) and dynamical (time evolution, 
return to equilibrium) -- allowed one to determine sufficiently robust conditions on the geometry 
of AAdS spacetimes so as certain basic properties of \textsc{Rehren} duality are preserved and 
others are modified in a relatively clear way, as seen in Theorem \ref{ch1t4} (page \pageref{ch1t4}), 
in Proposition \ref{ch1p1} (page \pageref{ch1p1}) and in Theorem \ref{ch1t9} (page \pageref{ch1t9}).
\item We proposed a geometrically intrinsic manner of constructing global time functions
associated to relatively compact diamonds in causally simple spacetimes and AAdS wedges, which 
admits a formulation of the (asymptotic) zeroth and second laws of black hole dynamics (Section 
\ref{ch2-thermo}, page \pageref{ch2-thermo}ff.).
\item At the quantum level, we implemented an instance of the principle of local covariance, 
proposed by \textsc{Brunetti, Fredenhagen} and \textsc{Verch} (Chapter \ref{ch3}), in a situation 
which demands the imposition of boundary conditions (see Definition \ref{ch4d1}, page \pageref{ch4d1},
and the discussion that follows it) adapted to the geometry of (A)AdS spacetimes, with the help 
of the algebraic formalism developed by \textsc{Sommer} and the prototype case of $AdS_d$
(Proposition \ref{ch4p1}, page \pageref{ch4p1}).
\item We proposed a formalism of ``scaling algebras'' around points at the conformal infinity
of AAdS spacetimes which, aided by the powerful structural results of \textsc{Borchers} 
and \textsc{Yngvason} (Theorems (\ref{ch4t1}) to (\ref{ch4t3}), page \pageref{ch4t1}ff.)
and by the geometrical viewpoint of Section \ref{ch2-thermo}, not only permitted a precise 
determination of boundary conditions for physically relevant states, but also established
a close relation between return to thermal equilibrium by the action of  asymptotic isometries
and scaling limits (condition (d), page \pageref{ch4e19} and the discussion that follows).
\item We pointed out how soliton excitations similar to D-branes (Definition \ref{ch4d8}, page
\pageref{ch4d8}) may arise in the case of nontrivial AAdS geometries, relating them to a 
nontrivial ``vacuum'' structure (Definition \ref{ch4d7}, page \pageref{ch4d7}) and to a potential
impossibility of reconstructing the bulk quantum theory from its \textsc{Rehren} dual
by means of the prescription (\ref{ch4e25}) (page \pageref{ch4e25}).
\end{itemize}

A problem of great interest in the literature on Lorentzian geometry concerns the stability
and rigidity of AAdS spacetimes given conditions similar to the ones assumed in Theorem \ref{ch1t4} 
and in Proposition \ref{ch1p1}. An important partial result was recently obtained by \textsc{Anderson} 
\cite{anderson5}, which tells us that, given (i) a global time function $t$ in an asymptotically 
simple, geodesically complete AAdS spacetime $(\mathscr{M},g)$ satisfying 
a technical condition of unique continuation of solutions of the linearized \textsc{Einstein} 
equations across conformal infinity, (ii) a sequence of \textsc{Cauchy} data defined at distinct 
levels $t^{-1}(t_i)$ of $t$, such that $t_i\stackrel{i\rightarrow\infty}{\longrightarrow}
+\infty$, and tending to \emph{stationary} \textsc{Cauchy} data, then $(\mathscr{M},g)$ is
\emph{globally} stationary. Although this is not a strict rigidity result, it illustrates
how the boundary conditions at conformal infinity act so as to prevent the dispersion of perturbations
of \textsc{Cauchy} data. We are, however, far from a result about the global nonlinear stability
of (A)AdS spacetimes with the same precision as the one due to \textsc{Christodoulou} and 
\textsc{Klainerman} \cite{christoklai} for \textsc{Minkowski} spacetime (see also the Remark \ref{ch2d1}, 
page \pageref{ch2d1}, for more considerations about this issue). Another open question 
is the feasibility of demonstrating the rigidity of AdS spacetimes under (absence of) gravitational 
time delay of null geodesics \cite{pagesw} (see Figure \ref{ch1f3}, page \pageref{ch1f3}). 
Such questions interest us to the extent that answering them would tell us how natural are our 
geometrical assumptions from the gravitational viewpoint.\\

The construction of global time functions for relatively compact diamonds we've obtained
can, we believe, be further deepened, in the direction of asymptotic analogs of the other
laws of black hole dynamics, thus giving us details on the ``fluctuations'' of
geometrical quantities around the scaling limit at the tips of the diamonds, hence seeking
an analog (of a generalization) of the first law of black hole dynamics for diamonds. From
the viewpoint of AAdS wedges, it interests us to relate some quantity of this kind with the
rescaled electric part $E_{ab}$ of the \textsc{Weyl} tensor, which enters as part of the 
``initial data'' for the \textsc{Fefferman-Graham} expansion of the metric around infinity.\\

The most glaring limitation of the present work is the total absence of nontrivial examples, 
concerning geometry as well as (even free) QFT models. The geometrical issue, as seen above, 
involves finding nontrivial examples of nonsingular AAdS spacetimes, which is very difficult 
without a general result as in the case of zero cosmological constant. Regarding the quantum
part, a future, quite compelling direction of investigation would be to check the boundary 
condition (\ref{ch4e19}), at least for free fields, in terms of more familiar premises. Doing
this for general spacetimes involves plunging into the arsenal of microlocal analysis \cite{hormander1,
duhorm,duistermaat,hormander3} in order to build in a precise way the \textsc{Green} functions. \\

The construction of an \emph{operator product expansion} within our extension of \textsc{Rehren} 
duality represents a challenge of its own, albeit of the highest interest, for it might 
potentially reveal relations between deviations from thermal equilibrium in AAdS wedges and 
geometrical quantities such as $E_{ab}$. An example which motivates such a conjecture is the 
relation between the \emph{poles} of retarded two-point functions of the dual quantum theory at 
the boundary, which determine the properties of return to equilibrium, and \emph{quasinormal modes}
of AAdS black holes \cite{horhub,sonstar}. Such a relation should have an analog for wedges and
free fields in AAdS.\\

Last but not least, we obviously presented in Subsection \ref{ch4-sector-aads} (page 
\pageref{ch4-sector-aads}ff.) only the bare bones of the properties of soliton sectors of the
bulk quantum theory associated to pairs of phases (satisfying the DHR criterion) of its \textsc{Rehren}
dual at the boundary. The results presented above certainly deserve a deeper scrutiny, along the 
lines of the works of \textsc{Müger} \cite{muger1,muger2} and \textsc{Schlingemann} \cite{schlinge}
in the general case, and along the lines of the works of \textsc{Ashtekar} and \textsc{Sen} \cite{ashsen}  
and \textsc{Sorkin} \cite{sorkin} for the construction of examples, combining the techniques 
of these papers with the aforementioned tools of microlocal analysis.

\epigraphhead{}

\addcontentsline{toc}{part}{Appendices}
\part*{Appendices}

\begin{appendix}

\chapter{\label{ap1}Elements of Lorentzian geometry}
\chaptermark{Lorentzian geometry}

We shall present here the necessary definitions and the statement of some useful results in Lorentzian
geometry. Our references are \cite{beemee,hawkellis,oneill,wald2}.

\section{\label{ap1-loc}Local theory}

Let $\mathscr{M}$ be a $d$-dimensional manifold, $d\geq 2$. A \emph{semi-Riemannian} metric of 
index $p$ is a section $g$ of $T^*\!\mathscr{M}\otimes T^*\!\mathscr{M}$ such that $g(r)$ is a 
nondegenerate symmetric bilinear form of index $p$, $\forall r\in\mathscr{M}$. The pair $(\mathscr{M},g)$ 
is said to be a \emph{semi-Riemannian manifold} -- for $p=0$, $(\mathscr{M},g)$ is simply a Riemannian
manifold. If $p=1$, we say that $(\mathscr{M},g)$ is a \emph{Lorentzian manifold}. \\

For all cases, we'll denote by $\nabla_a$ the \textsc{Levi-Civita} connection associated to $g$ -- 
recall that a (\textsc{Koszul}) \emph{connection} in a vector bundle $\mathscr{E}\stackrel{p}{
\longrightarrow}\mathscr{M}$ is a map $\nabla:\Gamma^\infty(\mathscr{M},T\!\mathscr{M})\times
\Gamma^\infty(\mathscr{M},\mathscr{E})\rightarrow\Gamma^\infty(\mathscr{M},\mathscr{E})$, 
$\mathscr{C}^\infty(\mathscr{M})$-linear in the first variable and, for all $X\in\Gamma^\infty
(\mathscr{M},T\!\mathscr{M})$, $\nabla_X\cdot$ is a \emph{derivation} on the $\mathscr{C}^\infty
(\mathscr{M})$-module $\Gamma^\infty(\mathscr{M},\mathscr{E})$, i.e., 
\begin{equation}\label{ap1e0}
\nabla_X(fS)=\langle df,X\rangle S+f\nabla_XS,\,\forall S.
\end{equation} 

Due to (\ref{ap1e0}), the difference between two connections $\nabla$ and $\tilde{\nabla}$ is a  
$\mathscr{C}^\infty(\mathscr{M})$-bilinear map with values in $\Gamma^\infty(\mathscr{M},\mathscr{E})$, i.e., 
is a section of $T\!\mathscr{M}\otimes\mathscr{E}^*\otimes\mathscr{E}$. A connection $\nabla$  
functorially induces a unique connection in all tensor bundles $\mathscr{E}^{(r,s)}=(\otimes^r
\mathscr{E})\otimes(\otimes^s\mathscr{E}^*)$ ($\mathscr{E}*$ is the \emph{dual} bundle to $\mathscr{E}$) 
by means of the \textsc{Leibniz} rule: \[\nabla_X(S_1\otimes S_2)=(\nabla_XS_1)\otimes S_2+S_1\otimes
(\nabla_X S_2),\,(\nabla_XS')(S)+S'(\nabla_XS)=\langle dS'(S),X\rangle.\] In the case $\mathscr{E}=T\!
\mathscr{M}$, we say that $\nabla$ is \emph{symmetric} is the \emph{torsion} tensor $T(X,Y)\doteq
\nabla_XY-\nabla_YX-[X,Y]$ vanishes for all $X,Y$ ($[X,Y]=\lie{X}Y$ denotes the \textsc{Lie} bracket of $X$ 
with $Y$). In a semi-Riemannian manifold $(\mathscr{M},g)$, a \textsc{Levi-Civita} connection is the 
(unique!) symmetrical connection $\nabla$ in $T\!\mathscr{M}$ which satisfies $\nabla_X g=0$ for all $X$. 
We use, for the latter, the notation $\nabla_a$ to denote its tensor character.\\

The \textsc{Riemann} curvature tensor $\mbox{Riem}(g)$ associated to $g$ (more precisely, associated to 
$\nabla_a$) is given by $2\nabla_{[a}\nabla_{b]}X_c=\mbox{Riem}(g)^d_{abc}X_d$, for all $X_a$. We denote 
yet the \textsc{Ricci} tensor and the scalar curvature associated to $g$ respectively by $\mbox{Ric}(g)_{ab}
\doteq\mbox{Riem}(g)^{c}_{acb}$ and $R(g)=g^{ab}\mbox{Ric}(g)_{ab}$, where $g^{ab}\doteq(g^{-1})^{ab}$, $g^{ab}
g_{bc}=\delta^a_c$. The \textsc{Weyl} tensor $C(g)$ associated to $g$ is given, for $d\geq 3$, by
\begin{equation}\label{ap1e1}
C_{abcd}=\mbox{Riem}_{abcd}-\frac{2}{d-2}(g_{a[c}\mbox{Ric}_{d]b}-g_{b[c}\mbox{Ric}_{a]d})+\frac{2}{(d-1)
(d-2)}R_{a[c}g_{d]b}.
\end{equation}

This expression algebraically vanishes for $d\leq 3$, and, in this case, $\mbox{Riem}(g)$ is completely
determined by $\mbox{Ric}(g)$, through formula (\ref{ap1e1}). We adopt in this formula, as well as in 
several places throughout the present work, a slight abuse of notation by omitting the specification of $g$ for
the \textsc{Weyl} tensor. We'll do it whenever such a practice doesn't cause confusion, to alleviate the
notation. \\

Given a semi-Riemannian manifold $(\mathscr{M},g)$, $p\in\mathscr{M}$ we say that $X(p)\in T_p
\mathscr{M}$ is \emph{timelike}, \emph{spacelike}, \emph{causal} or \emph{lightlike} (\emph{null}) if, 
respectively, $g(X,X)(p)<0$, $>0$, $\leq 0$ or $=0$. A vector field $X$ is timelike, spacelike, 
causal or lightlike / null if $X(p)$ is for all $p\in\mathscr{M}$. We define analogously timelike, 
spacelike, causal and lightlike / null covectors and covector fields by exchanging $g$ with $g^{-1}$.  
Such a characteristic of a (co)vector field is said to be its \emph{causal character}. The raising and
lowering of indices by $g,g^{-1}$ doesn't modify the causal character.\\

A piecewise $\mathscr{C}^\infty$ curve $\gamma$ is said to be \emph{timelike, spacelike, causal} or 
\emph{null} if, in each $\mathscr{C}^\infty$ component $\gamma_\alpha$, $\alpha=1,\ldots,k$, the tangent 
vector $\dot{\gamma}^a$ satisfies respectively $\dot{\gamma}_\alpha^a\dot{\gamma}_{\alpha a}<0$, 
$\dot{\gamma}_\alpha^a\dot{\gamma}_{\alpha a}>0$, $\dot{\gamma}_\alpha^a\dot{\gamma}_{\alpha a}\geq 0$ or 
$\dot{\gamma}_\alpha^a\dot{\gamma}_{\alpha a}=0$. A causal $\gamma$ is \emph{past} or \emph{future} directed 
if, for $t^a$ denoting a timelike vector field which determines the time orientation of $(\mathscr{M},g_{ab})$, 
we have respectively $t^a\dot{\gamma}_{\alpha a}>0$ or $t^a\dot{\gamma}_{\alpha a}<0$ along each component 
$\gamma_\alpha$. \\

A $\mathscr{C}^\infty$ curve $\gamma$ is a \emph{geodesic} if its tangent vector is covariantly 
constant: $\dot{\gamma}^a\nabla_a\dot{\gamma}_b=0$. Thus, a geodesic segment is uniquely determined  
by the choice of a point $p$ of $\mathscr{M}$ and a tangent vector at $p$, by the theorem of existence  
and uniqueness of solutions of ordinary differential equations on manifolds. Given $\mathscr{O}_1\subset
\mathscr{O}\subset\mathscr{M}$, we say that $\mathscr{O}_1$ is \emph{geodesically convex} with respect to 
$\mathscr{O}$ if, given any $p,q\in\mathscr{O}_1$, there exists a unique geodesic segment contained in 
$\mathscr{O}$ linking $p$ to $q$. Any $p\in\mathscr{M}$ has a geodesically convex neighbourhood.
More in general, a neighbourhood $\mathscr{U}$ of $p\in\mathscr{M}$ is said to be \emph{geodesically normal} 
if for all $q\in\mathscr{U}$ there exists a unique geodesic linking $p$ to $q$ (i.e., we don't assume that
this is true for $q,q'\neq p$). For such neighbourhoods, we can write the following system of coordinates: 
the \emph{exponential map}

\begin{equation}\label{ap1e2}
\exp_p:{\mathscr{V}\subset T_p\mathscr{M}\rightarrow\mathscr{U}}\atop{X\mapsto\exp_p(X)=\gamma_X(1)},
\end{equation}
where $\gamma_X:[0,1]\rightarrow\mathscr{M}$ is the (unique) geodesic segment such that $\gamma(0)=p$ and
$\dot{\gamma}(0)=X$, is a diffeomorphism for $\mathscr{U}$ sufficiently small. In this case, $\exp^{-1}_p
(\mathscr{U})=\mathscr{V}$ is a \emph{star-shaped} neighbourhood of the origin in $T_p\mathscr{M}\cong
\mathbb{R}^d$ (i.e., given $q\in\mathscr{V}$, we have $tq\in\mathscr{V}$ for all $t\in[0,1]$), whose Cartesian
coordinates $(\exp^{-1}_p)^\mu\doteq x^\mu:\mathscr{U}\rightarrow\mathscr{V}$ are denominated \emph{(geodesic) 
normal} coordinates. The metric in $\mathscr{U}$, expressed by means of the latter, becomes \cite{bergm,spivak2}
\begin{eqnarray}\label{ap1e3}
g_{\mu\nu}(p) & = & (g\circ x^{-1})_{\mu\nu}(0)=\eta_{\mu\nu},\,\partial_\mu(g\circ x^{-1})(0)=0;\\
g_{\mu\nu}(\exp_p(x)) & = & \eta_{\mu\nu}-\frac{1}{3}\mbox{Riem}(g\circ x^{-1})(0)_{\mu\rho\nu\sigma}x^\rho 
x^\sigma+o(x^2);\label{ap1e4}\\ & \Downarrow & \nonumber\\
\sqrt{|\det(g\circ x^{-1})|}(\exp_p(x)) & = & 1-\frac{1}{6}\mbox{Ric}(g\circ x^{-1})(0)_{\rho
\sigma}x^\rho x^\sigma+o(x^2).\label{ap1e5}
\end{eqnarray}

The formulae (\ref{ap1e3})--(\ref{ap1e4}), already present in embryonic form in the celebrated 
\emph{Ha\-bi\-li\-ta\-tions\-schrift} of \textsc{B. Riemann}, were proven in complete generality by 
\textsc{\'E. Cartan}.\\

In the case of Lorentzian manifolds, we say that a hypersurface (i.e., a codimension-one submanifold) is 
\emph{timelike}, \emph{spacelike} or \emph{lightlike} / \emph{null} if a normal vector $n^a$ at each point 
is respectively spacelike, timelike or null. Timelike and spacelike hypersurfaces, when endowed with the
pullback $g^{(0)}$ of $g$ under the inclusion map (sometimes called \emph{first fundamental form} -- see the 
paragraph after the next), are respectively Riemannian and Lorentzian manifolds by themselves. The null case 
is exceptional, for then a normal is also tangent to the hypersurface, and the pullback of $g$ results in a 
\emph{degenerate} symmetric bilinear form, for the one-dimensional subspace of null tangent vectors has 
eigenvalue zero -- the remaining directions are necessarily spacelike, for there are no two linearly 
independent vectors which are mutually orthogonal with respect to $g$ and such that one is lightlike and the other 
is causal.\\

Let $\Sigma$ be a \emph{nondegenerate} hypersurface in a Lorentzian manifold $(\mathscr{M},g)$ 
(i.e., $\Sigma$ is either timelike or spacelike), with unit normal $n^a$ and induced metric $h_{ab}=g_{ab}-
g(n,n)n_an_b$ (such that $h(X,X)=g(X,X)$ if $X\in T\Sigma$, and $h(n,.)=0$). We can define, for some 
$\epsilon>0$, geodesic normal coordinates $(p,\epsilon')$ in a tubular neighbourhood $\mathscr{U}$ 
of $\Sigma$ in $\mathscr{M}$ by writing $\Sigma\times\{0\}\doteq\Sigma$ and, for $p\in\Sigma$ and $\epsilon'$ 
satisfying $|\epsilon'|<\epsilon$ (beware that $\epsilon$ may depend on $p$, unless $\Sigma$ is compact), 
writing $(p,\epsilon')=\gamma(\epsilon')$, where $\gamma$ is the unique geodesic with $\gamma(0)=p$ and 
$\dot{\gamma}(0)^a=n^a(p)$ -- with this, we can extend $n^a$ from $\Sigma$ to $\mathscr{U}$, where it still
holds that $g(n,n)=\pm 1$ (we use the same notation for the extension of $n$, since there is no confusion). 
As, in this case, we see that $g(n,.)=d\phi$, where $\phi:\mathscr{U}\rightarrow\mathbb{R}$ is given by 
$\phi((p,\epsilon')=\epsilon'$, it follows that $\nabla_a n_b=\nabla_b n_a$ and $n^a\nabla_a n_b=n^a\nabla_b 
n_a=0$. \\

We'll make use of the following routine to relate the intrinsic (given by $h$) and extrinsic (given by $g$) 
geometries of $\Sigma$: let $\nabla^{(0)}$ be the \textsc{Levi-Civita} connection associated to the restriction
$g^{(0)}$ of $g$ to $\Sigma$ (= pullback of $g$ under the natural inclusion map / embedding $i^{\Sigma,\mathscr{M}}$ 
of $\Sigma$ into $\mathscr{M}$), let's identify vector fields $X,Y$ in $\Sigma$ with the vector fields in
$\mathscr{U}$ obtained by means of parallel transport of $X$ and $Y$ along the geodesics normal to $\Sigma$. 
Hence, we can identify the connection $\nabla^{(0)}$ with the component of $\nabla_XY$ \emph{tangential} to 
$\Sigma$, i.e., \[\nabla^{(0)}_XY=(g^{-1}h)\nabla_XY,\] where $(g^{-1}h)^a_b=g^{ac}h_{bc}=h^a_b$ is the projector 
onto the subspace tangent to the foliation. The component of $\nabla_XY$ \emph{normal} to $\Sigma$, given by 
$g(\nabla_XY,n)=-X^aY^b\nabla_an_b$, defines the \emph{extrinsic curvature} (or \emph{second fundamental form}) 
$K_{ab}\doteq-\frac{1}{2}(\lie{n} h)_{ab}=-\nabla_an_b$. Notice that, due to the above considerations, $K$ is 
symmetric and, indeed, a tensor \emph{in} $\Sigma$. The desired relation is, thus, given by the \textsc{Gauss} 
(\ref{ap1e6}) and \textsc{Codazzi-Mainardi} \emph{equations} (\ref{ap1e7})
\begin{eqnarray}
\label{ap1e6} h^q_ah^r_bh^s_ch^t_d\mbox{Riem}(g)_{qrst} & = & \mbox{Riem}
(h)_{abcd}-2g(n,n)K_{a[c}K_{d]b},\\
\label{ap1e7} h^c_bn^d\mbox{Ric}(g)_{cd} & = & 2h^c_{[a}D_{|c|}K^a_{b]}.
\end{eqnarray}

The lightlike case can be treated in the following manner: taking a lightlike hypersurface $\mathscr{S}$, 
we see that the null geodesics constructed as in the nondegenerate case above are, at the same time, 
\emph{normal} and \emph{tangent} to $\mathscr{S}$, and generate the latter, in the sense that it's possible 
to choose an affine parametrization $\lambda$ common to all these geodesics (using, for instance, an 
auxiliary Riemannian metric in $\mathscr{M}$ to normalize the lightlike directions), in such a way that 
constant-$\lambda$ submanifolds are spacelike, codimension-two submanifolds of $\mathscr{M}$, with null 
normal and tangent to $\mathscr{S}$ denoted by $k^a=(\frac{d}{d\lambda})^a$. It's possible, in this situation, 
to impose the following Riemannian structure in $\mathscr{S}$: at each $p\in\mathscr{S}$, whose affine 
parameter assumes the value $\lambda_p$, we say that $X(p),Y(p)\in T_p\mathscr{S}$ are equivalent if $X(p)-
Y(p)$ is lightlike, denoting such an equivalence class by $\hat{X}(p)(=\hat{Y}(p))$. That is, we take at each 
$p$ the quotient $\widehat{T_p\mathscr{S}}\ni\hat{X}(p)$ of $T_p\mathscr{S}$ modulo its degenerate 
one-dimensional subspace. The metric $\hat{h}$ induced in $T\mathscr{S}$ by the quotient is a (nondegenerate) 
Riemannian metric with one dimension less, which is identified at each constant-$\lambda$ submanifold with the
metric induced directly by $g$. Thus, we can define the \emph{null extrinsic curvature} $\hat{K}=-\frac{1}{2}
\frac{d}{d\lambda}\hat{h}$.\\ 

We can visualize a nondegenerate $\Sigma$ hypersurface locally as a parametrization of the family of geodesics 
normal to $\Sigma$. We can, more generally, consider a family (or \emph{congruence}) of geodesics with the
same causal character, going through an open set $\mathscr{O}\subset\mathscr{M}$, such that each $p\in\mathscr{O}$
is crossed by a single geodesic of the family (locally, any vector field $X^a$ generate a family of curves with 
this property, which are geodesics if and only if $X^a\nabla_a X^b=0$). The difference with respect to the case of 
hypersurfaces is that the distribution of subspaces of $T_p\mathscr{O}$ normal to the congruence at each $p$ 
need not be integrable, that is, it needs not to generate a family of hypersurfaces normal to the congruence. 
We can, nevertheless, yet in this case define the \emph{first fundamental form} of the congruence in the 
nondegenerate case $h_{ab}\doteq g_{ab}-g(X,X)X_a X_b$ and in the lightlike case $\hat{h}_{ab}\doteq(g_{ab}-X_a 
X_b)/^\wedge$, and the \emph{(null) second fundamental form} $\accentset{(\wedge)}{K}_{ab}=\nabla_b 
X_a(/^\wedge)$,\footnote{We use, for congruences, a sign convention \emph{opposite} to the one employed for
hypersurfaces when we define the second fundamental form.} where $X$ is the (co)vector field tangent to the 
congruence, which is parametrized in such a manner that $X$ is normalized at $\pm 1$ resp. in the spacelike /
timelike case (resp. normalized at $1$ with respect to an auxiliary Riemannian metric in the lightlike case). \\

It's immediate to see that, in the nondegenerate case, $K$ acts on the subspaces normal to the congruence, 
for $X^aK_{ab}=\frac{1}{2}\nabla_b{X^a X_a}=0$ and $X^b K_{ab}=X^b\nabla_b X_a=0$. In the null case, $K$ 
(without hats) acts nontrivially in the spacelike normal subspace \emph{and} in the other null direction 
normal to this subspace, whose tangent vector field of the same time orientation that of $X^a$ we denote by
$\bar{X}^a$. This vector field locally generates another congruence of null geodesics, naturally associated 
to the first -- namely, it's the \emph{only} other congruence which shares the same spacelike normal subspaces. 
the normalization adopted for $\bar{X}^a$ is given by $g(X,\bar{X})=-2$.\\

Unlike as in the case of hypersurfaces, however, $K_{ab}$ is not necessarily a \emph{symmetric} tensor: 
$\accentset{(\wedge)}{\omega}_{ab}\doteq\accentset{(\wedge)}{K}_{[ab]}$ (called \emph{(null) twist} of the
congruence) expresses precisely the failure of the distribution of subspaces normal to the congruence to be 
integrable. By the \textsc{Frobenius} theorem \cite{wald2}, the congruence is normal to a family of 
hypersurfaces if and only if $X_{[a}K_{bc]}\equiv 0$. The symmetric part can be decomposed in terms of the 
trace with respect to $\accentset{(\wedge)}{h}$, called \emph{(null) expansion} (or \emph{(null) mean curvature}) 
$\theta\doteq\accentset{(\wedge)}{K}_{ab}\accentset{(\wedge)}{h}^{ab}$ (the hat in the null case is 
unnecessary for the first member for it's a scalar), and the trace-free part, denominated \emph{(null) shear}  
$\accentset{(\wedge)}{\sigma}_{ab}\doteq\accentset{(\wedge)}{K}_{(ab)}-\frac{1}{d-1}\accentset{(\wedge)}{\theta}\,
.\,\accentset{(\wedge)}{h}_{ab}$ (a factor $(d-2)$ must be substituted for the factor $(d-1)$ in the null case, 
because of the dimensional ``reduction'' due to the quotient modulo the equivalence relation associated to 
${}^\wedge$). Summing up, 
\begin{equation}\label{ap1e8}
\accentset{(\wedge)}{K}_{ab}=\accentset{(\wedge)}{\omega}_{ab}+\accentset{(\wedge)}{\sigma}_{ab}
+\frac{1}{d-1(2)}\theta\accentset{(\wedge)}{h}_{ab}.
\end{equation}

The manner the geometry normal to the congruence changes along the parallel transport is given by the matrix 
\textsc{Riccati} equation
\begin{equation}\label{ap1e9}
X^c\nabla_c K_{ab}=-K^c_b K_{ac}+\mbox{Riem}(g)_{cbad}X^c X^d.
\end{equation}

Let us now take the quotient of (\ref{ap1e9}) in the null case, which is the case of interest in the present 
work, and decompose the result in terms of the expansion, shear and twist in the left hand side. Denoting by
$\lambda$ the common affine parametrization of the congruence, we have:

\begin{eqnarray}
\label{ap1e10} \frac{d\theta}{d\lambda} & = & -\frac{1}{d-2}\theta^2-\hat{\sigma}^{ab}\hat{
\sigma}_{ab}+\hat{\omega}^{ab}\hat{\omega}_{ab}-R_{cd}X^c X^d;\\
\label{ap1e11} X^c\nabla_c\hat{\sigma}_{ab} & = & -\theta\hat{\sigma}_{ab}+C(g)_{cbad}X^c 
X^d;\footnotemark\\
\label{ap1e12} X^c\nabla_c\hat{\omega}_{ab} & = & -\theta\hat{\omega}_{ab}.
\end{eqnarray}
\footnotetext{$C(g)_{cbad}X^c X^d$ is a symmetric tensor which assumes a unique value at each equivalence
class associated to ${}^\wedge$.}

Equation (\ref{ap1e10}) is the celebrated \textsc{Raychaudhuri} \emph{equation}.\\

Any one-parameter subfamily of geodesics $\gamma_\lambda$, $-\epsilon<\lambda<\epsilon$, $\epsilon>0$ of a 
congruence produces a vector field $Y\doteq\frac{d}{d\lambda}\restr{\lambda=0}$ in $\gamma_0$, which denotes 
the relative displacement a geodesic ``infinitesimally close to $\gamma_0$''. The relative acceleration of
two such geodesics is given by the \emph{equation of geodesic deviation} (also called \textsc{Jacobi}
\emph{equation})
\begin{equation}\label{ap1e13}
X^c\nabla_c(X^b\nabla_b Y^a)=-R^a_{cbd}X^c X^d Y^b.
\end{equation}

More in general, we can define a \textsc{Jacobi} \emph{field} over a geodesic $\gamma$ as a solution $Y$ of 
(\ref{ap1e13}) with $X=\dot{\gamma}$. We say that a causal geodesic $\gamma$ \emph{has a pair of conjugate
points} $p=\gamma(\lambda_1),q=\gamma(\lambda_2)$, $\lambda_1<\lambda_2$, if there exists a \textsc{Jacobi} 
field $Y$ which vanishes nowhere in $\gamma((\lambda_1,\lambda_2))$ but does so at $p$ and $q$. Geometrically, 
this means that the geodesics emanating from $p$ and ``infinitesimally close'' to $\gamma$ tend to focus at $q$.\\

To visualize this phenomenon in the nondegenerate case, let us take an orthonormal frame $e^a_i(\lambda)$, 
$i=0,\ldots,d-1$ parallelly propagated along $\gamma$, such that $e^a_0(\lambda)=X^a(\lambda)$. Taking 
$\lambda_1=0$ for simplicity, we have $Y^a(0)=0$ and, hence, \[Y^i(\lambda)=\sum^{d-1}_{j=1}A^i_j(\lambda)
\frac{dY^j}{d\lambda}(0),\] where the matrix $A^i_j$ expressed in terms of the chosen frame satisfies 
\[\frac{d^2 A^i_j}{d\lambda^2}=-R^i_{klm}X^k X^m A^l_j.\] As \[\frac{dY^i}{d\lambda}=e^i_bX^a\nabla_a Y^b=e^i_b
Y^a\nabla_a X^b=\sum^{d-1}_{j=1}K^i_jY^j,\] we have \[K^i_j=\left[\frac{dA}{d\lambda}\right]^i_k[A^{-1}]^k_j,\] 
and thus $\theta=\mbox{tr}[K]=\frac{1}{\det A}\frac{d}{d\lambda}(\det A)$. Hence, $\det A\rightarrow 0$, which 
implies that $A$ possesses an eigenspace normal to $X^a$ with eigenvalue zero and, thus, there exists a  
\textsc{Jacobi} field which vanishes at $p$ and $q$, if and only if $\theta\rightarrow-\infty$ at $q$. The 
lightlike case is treated by adopting a frame parallelly propagated along $\gamma$ such that $e^a_0=X^a$, $e^a_1
=\bar{X}^a$, and $e^a_2,\ldots,e^a_{d-1}$ an orthonormal set such that $g(e_0,e_i)=g(e_1,e_i)=0$, $i=2,\ldots,d-1$. 
The formulae above don't involve $e^a_1$ in the latter case.

\section{\label{ap1-glob}Global theory}

By a \emph{spacetime} one understands a \emph{time orientable} Lorentzian manifold $(\mathscr{M},g)$ , i.e., 
there exists a timelike vector field $\mathscr{C}^\infty$ $T^a$ in $\mathscr{M}$ which vanishes nowhere (in 
particular, we can choose $g(T,T)=-1$). \\

\subsection{\label{ap1-glob-causal}Causal structure, geodesic completeness}

Let $(\mathscr{M},g_{ab})$ be a spacetime, $\mathscr{O}\subset\mathscr{M}$, $p\in\mathscr{O}$. 
The \emph{chronological} (resp. \emph{causal}) \emph{future} of $p$ with respect to $\mathscr{O}$, denoted by 
$I^+(p,\mathscr{O})$ (resp. $I^+(p,\mathscr{O})$) is given by
\begin{eqnarray}
I^+(p,\mathscr{O}) & \dot{=} & \{x\in\mathscr{O}:\exists\gamma:[0,a]\stackrel{\mathscr{C}^\infty}{
\longrightarrow}\mathscr{O}\;\mbox{future directed,}\nonumber\\ & & \mbox{timelike such that}\;\gamma(0)
=p,\,\gamma(a)=x\};\label{ap1e14}\\
J^+(p,\mathscr{O}) & \dot{=} & \{x\in\mathscr{O}:x=p\;\mbox{or}\;\exists\gamma:[0,a]\stackrel{
\mathscr{C}^\infty}{\longrightarrow}\mathscr{U}\;\mbox{future directed,}\nonumber\\ & & \mbox{causal such
that}\;\gamma(0)=p,\,\gamma(a)=x\}\label{ap1e15}.
\end{eqnarray}

Exchanging future with past, one defines in a dual way the \emph{chronological} (resp. 
\emph{causal}) \emph{past} $I^-(p,\mathscr{O})$ (resp. $J^-(p,\mathscr{O})$) of $p$ with respect to $\mathscr{O}$. 
It follows from these definitions that $I^\pm(p,\mathscr{O})$ is open and $int(J^\pm(p,\mathscr{O}))=I^\pm
(p,\mathscr{O})$. A set $\mathscr{O}_1\subset\mathscr{O}$ is said to be \emph{causally convex} with respect to
$\mathscr{O}$ if, given any $p\leq_\mathscr{O}q\in\mathscr{O}_1$, we have $J^+(p,\mathscr{O})\cap J^-(q,
\mathscr{O})\subset\mathscr{O}_1$.\\

Using the sets above, we can define chronology and causality relations between two points. Let $p,q\in
\mathscr{O}\subset\mathscr{M}$. We say that $p$ \emph{chronologically} (resp. \emph{causally}) 
\emph{precedes} $q$ with respect to $\mathscr{O}$ if $p\in I^-(q,\mathscr{O})$ (resp. $p\in J^-(q,\mathscr{O})$). 
We denote this relation by $p\ll_\mathscr{O}q$ (resp. $p\leq_\mathscr{O}q$). Equivalently, we say in this
case that $q$ \emph{chronologically} (resp. \emph{causally}) \emph{succeeds} $p$ with respect to $\mathscr{O}$, 
with the notation $q\gg_\mathscr{O}p$ (resp. $q\geq_\mathscr{O}p$). If $p\leq_\mathscr{O}q$ and $p\neq q$, 
we write $p<_\mathscr{O}q$, or, dually, $q>_\mathscr{O}p$. If $p\neq q$, $p\not\ll_\mathscr{O}q$ and $p\not
\gg_\mathscr{O}q$ (resp. $p\nless_\mathscr{O}q$ and $p\ngtr_\mathscr{O}q$), we say that $p$ and $q$ are 
\emph{chronologically} (resp. \emph{causally}) \emph{disjoint} -- in this case, we write $p\curlywedge_\mathscr{O} 
q$ (resp. $p\perp_\mathscr{O}q$). All chronology and causality relations defined above, as well as the
chronological or causal past and future, are defined for nonunit and nonvoid sets in an obvious manner. If 
$p\neq q\in\mathscr{O}\subset\mathscr{M}$ implies $p\curlywedge_\mathscr{O} q$ (resp. $p\perp_\mathscr{O}q$), 
we say that $\mathscr{O}$ is \emph{achronal} (resp. \emph{acausal}) with respect to $\mathscr{O}$.\\

Some consequences of the definitions of the previous paragraph are:

\begin{itemize}
\item $\ll_\mathscr{O}$ is an open relation, that is, 
\begin{eqnarray}
p\ll_\mathscr{O}q\,\Rightarrow\,\exists\mathscr{O}_1,\mathscr{O}_2\subset\mathscr{M}\mbox{ open 
such that}\label{ap1e16}\\p\in\mathscr{O}_1,q\in\mathscr{O}_2\,\mbox{and}\,\mathscr{O}_1
\cap{}\mathscr{O}\ll_\mathscr{O}\mathscr{O}_2\cap{}\mathscr{O}.\nonumber
\end{eqnarray}
\item\begin{equation}\label{ap1e17}
J^+(J^+(p,\mathscr{O}),\mathscr{O})=J^+(p,\mathscr{O}),\mbox{ i.e., }p\leq_\mathscr{O}q\mbox{ and }
q\leq_\mathscr{O}r\,\Rightarrow\,p\leq_\mathscr{O}r.
\end{equation}
\item\begin{eqnarray} I^+(I^+(p,\mathscr{O}),\mathscr{O})=I^+(J^+(p,\mathscr{O}),\mathscr{O})=
J^+(I^+(p,\mathscr{O}),\mathscr{O})=I^+(p,\mathscr{O}),\nonumber\\
\mbox{ i.e., }(p\ll_\mathscr{O}q\mbox{ and }q\ll_\mathscr{O}r)\mbox{ or }(p\ll_\mathscr{O}q\mbox{ and }
q\leq_\mathscr{O}r)\mbox{ or }(p\leq_\mathscr{O}q\mbox{ and}\label{ap1e18}\\q\ll_\mathscr{O}r)\,
\Rightarrow\,p\ll_\mathscr{O}r.\nonumber
\end{eqnarray}
\item\begin{equation}\label{ap1e19}
\overline{I^+(\mathscr{O}_1,\mathscr{O}_2)}=\overline{J^+(\mathscr{O}_1,\mathscr{O}_2)},\,\forall
\mathscr{O}_1\subset\mathscr{O}_2,
\end{equation} where the closure is taken in the relative topology of $\mathscr{O}_2$.
\item\begin{equation}\label{ap1e20}
\partial I^+(\mathscr{O}_1,\mathscr{O}_2)=\partial J^+(\mathscr{O}_1,\mathscr{O}_2),\,\forall
\mathscr{O}_1\subset\mathscr{O}_2,
\end{equation} where the boundary is taken in the relative topology of $\mathscr{O}_2$.
\end{itemize}

The set defined in (\ref{ap1e20}) is called \emph{future achronal boundary} of $\mathscr{O}_1$ with respect to
$\mathscr{O}_2$, and constitutes an achronal, topological submanifold of $\mathscr{M}$ such that any 
$p\in\partial I^+(\mathscr{O}_1,\mathscr{O}_2)$ belongs to a (necessarily unique) geodesic segment, which
is achronal with respect to $\mathscr{O}_2$ and contained in $\partial I^+(\mathscr{O}_1,\mathscr{O}_2)$, which is 
either past inextendible or has a past endpoint in $\mathscr{O}_1$. Such geodesics are denominated 
\emph{generators} of $\partial I^+(\mathscr{O}_1,\mathscr{O}_2)$. Notice that (\ref{ap1e16})--(\ref{ap1e20}) 
still hold if we exchange future with past.\\

One of the most powerful tools in the analysis of the global geometry of Lorentzian manifolds is the 
\emph{Lorentzian distance}. Let $p\leq_\mathscr{M}q\in\mathscr{M}$, and denote by $\Omega_{p,q}$ the space 
of future directed, causal piecewise $\mathscr{C}^\infty$ curves linking $p$ to $q$ (this space is empty if 
$p\nleq_\mathscr{M}q$). $\Omega_{p,q}$ inherits the following topology from the space $\mbox{Path}(\mathscr{M})$ 
of $\mathscr{C}^0$ curves in $\mathscr{M}$: we say that a sequence of piecewise $\mathscr{C}^\infty$ curves 
$\{\gamma_n:[0,\lambda]\rightarrow\mathscr{M}\}$ converges to $\gamma:[0,\lambda]\rightarrow\mathscr{M}$ if 
$\gamma_n(0)\rightarrow\gamma(0)$ and, given any open set $\mathscr{O}\subset\mathscr{M}$ containing $\gamma
([0,\lambda])$, there exists $N\in\mathbb{Z}_+$ such that $\gamma_n([0,\lambda])\subset\mathscr{O}$ for all 
$n\geq N$. This topology is called $\mathscr{C}^0$ \emph{topology}. Consider $\lambda\in\Omega_{p,q}$ 
parametrized by $\lambda\in[\lambda_0=0,\lambda_k]$, with $\mathscr{C}^\infty$ components $\gamma_i$, $i=1,
\ldots,k$, defined respectively in the intervals $[\lambda_{i-1},\lambda_i]$, $\lambda_0<\lambda_1<\cdots<
\lambda_k$. The Lorentzian arc length of $\gamma$ is given by 
\begin{equation}\label{ap1e21}
L_g(\gamma)\doteq\sum^k_{i=1}\int^{\lambda_i}_{\lambda_{i-1}}\sqrt{-g(\dot{\gamma}_i
(\lambda),\dot{\gamma}_i(\lambda))}d\lambda.
\end{equation}

For timelike curves, which correspond to trajectories of observers, (\ref{ap1e21}) is also said to be 
the \emph{proper time} of the trajectory.

\begin{definition}\label{ap1d1}
The \emph{Lorentzian distance} in $(\mathscr{M},g)$ is the function $d_g:\mathscr{M}\times\mathscr{M}
\rightarrow\bar{\mathbb{R}}_+\cup{}\{+\infty\}$ given by 
\begin{equation}\label{ap1e22}
(p,q)\mapsto d_g(p,q)=\left\{\begin{array}{rl}
\sup_{\gamma\in\Omega{p,q}}L_g(\gamma) & \mbox{ if }p\leq_\mathscr{M} q;\\ 0 & \mbox{ otherwise.}
\end{array}\right.
\end{equation}
\end{definition}

It follows immediately from the definition that $d_g$ satisfies a \emph{reverse} triangular inequality: 
if $p\leq_\mathscr{M}r\leq_\mathscr{M}q$, then \[d_g(p,q)\geq d_g(p,r)+d_g(r,q).\] where the value $d_g(p,q)
=+\infty$ may be obtained is in the extremal \textsc{Reissner-Nord\-str\"om} spacetime (see \cite{beemee}). In
the case $p=q$, we have $d_g(p,p)=0$ or $d_g(p,p)=+\infty$. $d_g$ is lower semicontinuous for all $(p,q)$ 
where $d_g(p,q)<+\infty$. We say that $\gamma\in\Omega_{p,q}$ is \emph{maximal} if $L(\gamma)=d_g(p,q)$. 
One can prove \cite{beemee} that, if $\gamma$ is maximal, then it's a ($\mathscr{C}^\infty$) causal geodesic,
up to reparametrization. In particular, a maximal null geodesic is \emph{achronal}. On the other hand, if
a, say, future causal geodesic $\gamma$ possesses a pair of conjugate points $p\leq q$, then any points 
$p',q'$ in $\gamma$ with $p'$ in the past of $p$ and $q'$ in the future of $q$, we have that the segments 
of $\gamma$ which link $p'$ to $q$ and $p$ to $q'$ no longer maximize the Lorentzian arc length between
these pairs of points -- in particular, if $\gamma$ is lightlike, it ceases to be achronal if sufficiently 
extended. The same occurs if a causal geodesic, \emph{distinct from} $\gamma$, links $p$ to $q$.\\

Now we recapitulate some causality conditions one can impose on $(\mathscr{M},g)$. We say that $(\mathscr{M},
g)$ is \emph{chronological} (resp. \emph{causal}) if there is no $p\in\mathscr{M}$ such that $p\ll_\mathscr{M}
p$ (resp. $p<_\mathscr{M}p$) -- equivalently, $(\mathscr{M},g)$ is chronological if and only if $d_g(p,p)=0$, 
and causal if and only if $\Omega_{p,p}=\varnothing$, for all $p\in\mathscr{M}$. We say that $(\mathscr{M},g)$ 
is \emph{strongly causal} if, for all $p\in\mathscr{M}$, there exist arbitrarily small open neighbourhoods 
$\mathscr{O}$ of $p$ such that no causal curve has a disconnected intersection with $\mathscr{O}$. This implies 
that no compact set completely contains a past or future intextendible causal curve (i.e., such that it's not 
properly contained in any other curve), and, in particular, that $(\mathscr{M},g)$ is causal. \\

A particularly interesting class of spacetimes is the one which admits a foliation by acausal hypersurfaces, 
for these admit a global notion time evolution (possibly supplemented by boundary conditions). Namely, we
say that $t\in\mathscr{C}^0(\mathscr{M})$ is a \emph{global time function} if $t\circ\gamma$ is strictly 
increasing for any future directed causal curve $\gamma$. If $t$ is $\mathscr{C}^1$, we equivalently say
that $t$ is a global time function if $dt$ is a future directed timelike covector and $t(\mathscr{M})=
\mathbb{R}$. In this case, $t^{-1}(\tau)$ is a spacelike hypersurface for all $\tau\in\mathbb{R}$, and 
any $p\in\mathscr{M}$ belongs to $t^{-1}(\tau)$ for precisely one value of $\tau$. As $dt$ induces the 
same time orientation as adopted for $(\mathscr{M},g)$, it follows that $t^{-1}(\tau)$ is acausal for all
$\tau\in\mathbb{R}$. A spacetime which admits a (continuous) global time function is said to be \emph{stably 
causal}. The name comes from the fact that such spacetimes can be equivalently defined by the following 
property: there exists a timelike covector field $T_a$ such that the Lorentzian metric $g_{ab}-\lambda T_a T_b$ 
is chronological for all $\lambda\in[0,1)$. In our case, $T=dt$; conversely, given $T_a$ as above, it's 
possible to build a (continuous) global time function \cite{geroch1,wald2}. The question of how to construct 
global time $\mathscr{C}^\infty$ functions in this context was partially answered by \textsc{Seifert} 
\cite{seifert} and \textsc{Dieckmann} \cite{dieckmann}, and completely recently by \textsc{Bernal} and 
\textsc{Sánchez} \cite{bernsan1,bernsan2,bernsan3}, thus eliminating the necessity of assuming 
differentiability of global time functions in the definition of stable causality. Hence, as a differentiable
$t$ has no critical points, all constant-$t$ hypersurfaces are diffeomorphic among each other.\\

Finally, we say that a strongly causal spacetime $(\mathscr{M},g)$ is \emph{causally simple} if, for all
$p\in\mathscr{M}$ one has that $J^\pm(p,\mathscr{M})$ is closed, or, equivalently, $J^\pm(p,\mathscr{M})
\smallsetminus I^\pm(p,\mathscr{M})=\partial I^\pm(p,\mathscr{M})$, and \emph{globally hyperbolic} if, for
all $p,q\in\mathscr{V}$, the set $J^-(p,\mathscr{M})\cap J^+(q,\mathscr{M})$ is compact if nonvoid.\footnote{Very 
recently, \textsc{Bernal} and \textsc{Sánchez} \cite{bernsan4} showed that we can substitute ``causal'' for 
``strongly causal'' in the definition of global hyperbolicity given here.} Equivalently, $(\mathscr{M},g)$ is
globally hyperbolic if and only if $\Omega_{p,q}$ is \emph{compact} in the $\mathscr{C}^0$ topology if nonvoid. 
Thus, it follows that, in a globally hyperbolic spacetime, $d_g$ is continuous and $d_g(p,q)<+\infty,\,
\forall p,q\in\mathscr{M}$. Global hyperbolicity clearly implies simple causality, which, on its turn, implies 
stable causality (for the latter assertion, see \cite{beemee}). All above conditions can be defined for subsets 
of $\mathscr{M}$.\\

Let now $\mathscr{S}\subset\mathscr{M}$ be closed and achronal. The \emph{future} (resp. \emph{past}) 
\emph{domain of dependence} of $\mathscr{S}$, denoted by $D^+(\mathscr{S})$ (resp. $D^-(\mathscr{S})$) 
consists of the set
\begin{eqnarray}
D^{+/-}(\mathscr{S})\dot{=}\{p\in\mathscr{M}:\forall\mbox{ past/future inextendible, causal}\gamma:[0,a)
\longrightarrow\mathscr{M}\nonumber\\\mbox{such that }\gamma(0)=p,\,\exists 
b<a\mbox{ such that }\gamma(b)\in\mathscr{I}\}.\label{ap1e23}
\end{eqnarray}

$D(\mathscr{S})\doteq D^+(\mathscr{S})\cup D^-(\mathscr{S})$ is denoted \emph{domain of dependence} or 
\textsc{Cauchy} \emph{development} of $\mathscr{S}$. The \emph{edge} of $\mathscr{S}$ (notation: 
$\dot{\mathscr{S}}$) consists of the points $p\in\mathscr{S}$ such that \emph{any} open neighbourhood of $p$ 
has points $q\in I^-(p),\,r\in I^+(p)$ and a timelike curve $\gamma$ linking $q$ to $r$ with empty intersection 
with $\mathscr{S}$. If $\dot{\mathscr{S}}=\varnothing$ then $\mathscr{S}$ is a codimension-one embedded 
topological submanifold of $\mathscr{M}$. A domain of dependence possesses the following properties:

\begin{itemize}
\item The set $int(D(\mathscr{S}))$ is globally hyperbolic (in this case, we say that $S$ is a \textsc{Cauchy} 
\emph{surface} for $int(D(\mathscr{S}))$), as is $D(\mathscr{S})$ if $\dot{\mathscr{S}}=\varnothing$. 
Conversely, one can prove that any globally hyperbolic region has a \textsc{Cauchy} surface -- moreover, one 
can build a global time function $t$ such that $t^{-1}(\tau)$ is a \textsc{Cauchy} surface for all $\tau\in
\mathbb{R}$.
\item The achronal, closed set $H^+(\mathscr{S})\doteq \overline{D^+(\mathscr{S})}\smallsetminus I^-(D^+
(\mathscr{S}))$, called \emph{future} \textsc{Cauchy} \emph{horizon} of $\mathscr{S}$ has the following 
property: any $p\in H^+(\mathscr{S})$ is contained in a (necessarily unique) achronal, null geodesic segment
contained in $H^+(\mathscr{S})$ which is either past inextendible or possesses a past endpoint in $\dot{
\mathscr{S}}$. An analogous definition exists for $H^-(\mathscr{S})$, the \emph{past} \textsc{Cauchy} 
\emph{horizon} of $\mathscr{S}$; 
\item The \textsc{Cauchy} \emph{horizon} $H(\mathscr{S})\doteq H^+(\mathscr{S})\cup H^-(\mathscr{S})$ equals 
$\partial D(\mathscr{S})$.
\end{itemize}

The main characteristic of a globally hyperbolic spacetime is that it's ``dynamically closed'', i.e., the 
\textsc{Cauchy} problem for any equation of motion which propagates initial data in a locally causal manner
(i.e., the support of the solution at each instant grows in time with speed less than that of the light), as, 
for instance, the wave equation, is \emph{well posed}, i.e., it possesses a unique solution, which depends 
continuously on the initial data, for the solution associated to initial data with compact support in the
time-zero hypersurface has spatially compact support at every other instant. This eliminates the need for 
boundary conditions, which doesn't necessarily happen in the more general case of stably causal or causally 
simple spacetimes \cite{wald1}.\\

On geodesic completeness: we say that a geodesic $\gamma$ is \emph{past, future complete} or simply 
\emph{complete} if, respectively, its affine parameter extends to $\mathbb{R}_+$, $\mathbb{R}_-$ or 
$\mathbb{R}$.\footnote{Here the words ``past'' and ``future'', unlike elsewhere, refer only to the affine 
parameter, rather than to any time orientation whatsoever, for they also apply, as shown afterwards, to 
spacelike geodesics.} $(\mathscr{M},g)$ is said to be \emph{spacelike, timelike} or \emph{null geodesically 
complete} if, respectively, any spacelike, timelike or null geodesic is complete. The counterexamples collected 
and constructed by \textsc{Geroch} in \cite{geroch0} show that the three definitions are logically independent, 
that is, none of these implies the other. Timelike and/or null geodesic incompleteness are often used criteria 
to determine the presence of singularities, for they cannot be circumvented by some sort of \textsc{Cauchy}
completion with respect to, say, some distance function which generates the manifold topology by means of
paracompactness (which implies metrizability \cite{dugundji}).\\

To conclude, we must consider two important notions for theories of local observables: the \emph{causal 
complement} of an open set $\mathscr{O}$ with respect to $\mathscr{O}\supset\mathscr{O}_1$ is the set $(\mathscr{O}_1)
'_\mathscr{O}\doteq\mbox{int}\{p\perp_\mathscr{O}\mathscr{O}_1\}$, and the \emph{causal completion} of 
$\mathscr{O}_1$ with respect to $\mathscr{O}$ is given by $(\mathscr{O}_1)''_\mathscr{O}\equiv(\mathscr{O}_1)'_\mathscr{O})
'_\mathscr{O}$.\footnote{Such definitions differ from the ones adopted in \cite{ribeiro2}, where we don't assume 
$\mathscr{O}$ open. The differences involved, though, don't imply any modification in the proofs of the results 
of \cite{ribeiro2}.} If $(\mathscr{O}_1)''_\mathscr{O}=\mathscr{O}_1$, we say that $\mathscr{O}_1$ is \emph{causally 
complete} with respect to $\mathscr{O}$. It follows from these definitions that:

\begin{itemize}
\item $(\mathscr{O}_1)''_\mathscr{O}$ is the smallest causally complete open set with respect to $\mathscr{O}$ which 
contains $\mathscr{O}_1$. Notice here the importance of our taking $\mathscr{O}_1$ \emph{open} in our definition 
of causal complement -- if, for instance,  $\mathscr{O}_1=\{p\}$ or, more in general, a discrete achronal set, 
we have $(\mathscr{O}_1)''_{\mathscr{O}}=\varnothing$. See later on how we'll do to define the operations of 
causal complement for sets which are not necessarily open, for which the property above \emph{doesn't} 
obligatorily hold.
\item $(\mathscr{O}_1)'''_\mathscr{O}=(\mathscr{O}_1)'_\mathscr{O}$;
\item $\mathscr{O}_1\subset\mathscr{O}_2\Rightarrow(\mathscr{O}_1)'_\mathscr{O}\supset
(\mathscr{O}_2)'_\mathscr{O}$; 
\item $(\cup{}_\alpha\mathscr{O}_\alpha)'_\mathscr{O}=\cap{}_\alpha(\mathscr{O}_\alpha)'_\mathscr{O}\,
\Rightarrow\,(\cup{}_\alpha(\mathscr{O}_\alpha)'_\mathscr{O})'_\mathscr{O}=\cap{}_\alpha
(\mathscr{O}_\alpha)''_\mathscr{O}$ and, hence, any intersection of causally complete sets is causally complete 
-- we shall use these formulae to extend the definition of causal complement to arbitrary sets.
\item Any \emph{diamond} $\mathscr{O}_{p,q}\doteq I^-(p,\mathscr{O})\cap I^+(q,\mathscr{O})=(\{p,
q\})''_{\mathscr{O}}$, $p\ll_\mathscr{O}q$, is causally complete.
\end{itemize}

Throughout the present work, we assume that all our spacetimes are \emph{strongly causal}, unless otherwise stated.\\

\begin{remark}\label{ap1r1}
If $\mathscr{O}=\mathscr{M}$, the part of the notations of this Subsection indicating $\mathscr{O}$ can be
omitted.
\end{remark}

\subsection{\label{ap1-glob-infty}Conformal infinity}

\begin{definition}\label{ap1d2} 
The \emph{conformal infinity} or \emph{conformal boundary} of a $d$-dimensional spacetime $(\mathscr{M},g)$ is
a $d-1$-dimensional spacetime $(\mathscr{I},\bar{g}^{(0)})$ such that there exists a $d$-dimensional Lorentzian 
manifold $(\overline{\mathscr{M}},\bar{g})$ with boundary (the \emph{conformal closure} or \emph{conformal
completion} of $(\mathscr{M},g)$) satisfying:
\begin{itemize}
\item $\mathscr{I}\equiv\partial\overline{\mathscr{M}}$; there exists a diffeomorphism $\Phi$ of $\mathscr{M}$ 
onto $\Phi(\mathscr{M})\doteq\overline{\mathscr{M}}\smallsetminus\partial\overline{\mathscr{M}}$;
\item $\bar{g}^{(0)}$ is the semi-Riemannian metric (possibly degenerate) induced by $\bar{g}$ in $\mathscr{I}$;
\item There is a \emph{conformal} or \textsc{Weyl} \emph{factor}, there is, a positive $\mathscr{C}^\infty$ 
function $z$ in $\mathscr{M}$, which admits a $\mathscr{C}^\infty$ extension to $\overline{\mathscr{M}}$ such that 
$z\restr{\mathscr{I}}\equiv 0$ and $dz\restr{\mathscr{I}}\neq 0$ in $\mathscr{I}$ (the \textsc{Weyl} factor and 
its extension to the conformal closure are always denoted by the same symbol, since there is no confusion here) 
satisfying $\bar{g}=z^2 g$. We denote respectively by $\bar{\nabla}_a$ and ${}^{(0)}\bar{\nabla}_a$ the 
\textsc{Levi-Civita} connections associated to $\bar{g}$ and $\bar{g}^{(0)}$.
\end{itemize}
\end{definition}

We'll now show what becomes of the formulae for $\bar{\nabla}_a$, $\mbox{Riem}(\bar{g})$ and $\mbox{Ric}(\bar{g})$ 
in $\mathscr{M}$. First, recall that any two linear connections $\nabla_a$ and $\bar{\nabla}_a$ differ in its
action on the $\mathscr{C}^\infty$-module $\Gamma(\mathscr{M},T^*\!\mathscr{M})$ by a tensor $C^c_{ab}$ of rank 
$(1,2)$, i.e., 
\begin{equation}\label{ap1e24}
\bar{\nabla}_aX_b=\nabla_aX_b-C^c_{ab}X_c.
\end{equation}

Hence, starting from the \textsc{Levi-Civita} conditions \[\nabla_ag_{bc}=0,\,\nabla_a\nabla_bf=\nabla_b
\nabla_af,\,\forall f\in\mathscr{C}^\infty(\mathscr{M}),\] we obtain \[C^c_{ab}=\frac{1}{2}\bar{g}^{cd}
(\nabla_a\bar{g}_{bd}+\nabla_b\bar{g}_{ad}-\nabla_d\bar{g}_{ab}).\footnotemark\] 
\footnotetext{Specializing this formula to a local chart with domain endowed with the \textsc{Minkowski}
metric, we thus obtain the usual expression usual in local coordinates for the \textsc{Christoffel} symbols.} 
In particular, taking $g$ and $\bar{g}$ as in Definition \ref{ap1d2}, it follows $\nabla_a\bar{g}_{bc}=2z
g_{bc}\nabla_az$ and, hence, 
\begin{equation}\label{ap1e25}
C^c_{ab}=\frac{1}{z}g^{cd}(g_{bd}\nabla_az+g_{ad}\nabla_bz-g_{ab}\nabla_dz).
\end{equation}

Geodesics with respect to $\nabla_a$ in general are no longer so with respect to $\bar{\nabla}_a$, for \[X^a
\nabla_a X^b=0\;\Rightarrow\;X^a\bar{\nabla}_aX^b=X^aC^b_{ac}X^c=\frac{2}{z}X^bX^c\nabla_cz-\frac{g(X,X)}{z}g^{bd}
\nabla_dz.\] However, if $X^a=\dot{\gamma}^a(\lambda)$ is null with respect to $g$ (and, thus, $\bar{g}$), where $\lambda$ 
is the original affine parametrization of $\gamma$, choosing $\bar{\lambda}=e^{2\log z(\gamma(\lambda))}\lambda=z^2
\lambda$, it follows that \[\frac{d^2}{d\bar{\lambda}^2}\gamma(\bar{\lambda})=0,\] i.e., $\gamma$ is a null
geodesic with respect to $\bar{g}$ \emph{up to reparametrization}. The relations involving the \textsc{Riemann} and 
\textsc{Ricci} tensors become 
(taking $z^{-1}\nabla_a z=\nabla_a\log z$)
\begin{eqnarray}\label{ap1e26}
\mbox{Riem}(\bar{g})^d_{abc} & = & \mbox{Riem}(g)^d_{abc}-2\nabla_{[a}C^d_{b]c}+2C^e_{c[a}C^d_{b]e} = \\ 
& = & \mbox{Riem}(g)^d_{abc}+2\delta^d_{[a}\nabla_{b]}\nabla_c\log z-2g^{de}g_{c[a}\nabla_{b]}\nabla_e
\log z+\nonumber\\ & & +2(\nabla_{[a}\log z)\delta^d_{b]}\nabla_c\log z-2(\nabla_{[a}\log z)g_{b]c}
g^{df}\nabla_f\log z+\nonumber\\ & & -2g_{c[a}\delta^d_{b]}g^{ef}(\nabla_e\log z)\nabla_f\log z\nonumber
\end{eqnarray}
and (contracting the indices $d$ and $b$ in (\ref{ap1e22}))
\begin{eqnarray}\label{ap1e27}
\mbox{Ric}(\bar{g})_{ac} & = & \mbox{Ric}(g)_{ac}-\frac{d-2}{z}\nabla_a\nabla_cz-\frac{1}{z}g_{ac}
g^{de}\nabla_d\nabla_ez+\\ & & +2\frac{d-2}{z^2}(\nabla_az)\nabla_cz-\frac{d-3}{z^2}g_{ac}g^{de}
(\nabla_dz)\nabla_ez.\nonumber
\end{eqnarray}

It's convenient to have at our disposal the inverse form of (\ref{ap1e27}), employed in Chapter 
\ref{ch2}. Exchanging the roles of $g$ and $\bar{g}$ and taking $z\mapsto z^{-1}$, we arrive at
\begin{equation}\label{ap1e28}
\mbox{Ric}(g)_{ac}=\mbox{Ric}(\bar{g})_{ac}+\frac{d-2}{z}\bar{\nabla}_a\bar{\nabla}_cz+\bar{g}_{ac}
\bar{g}^{de}\left(\frac{1}{z}\bar{\nabla}_d\bar{\nabla}_ez-\frac{d-1}{z^2}(\bar{\nabla}_dz)
\bar{\nabla}_ez\right).
\end{equation}

It follows from formulae (\ref{ap1e1}), (\ref{ap1e26}) and (\ref{ap1e27}) that the \textsc{Weyl} tensor is 
\emph{invariant} under conformal changes of the metric, i.e., \[C(\bar{g})^d_{abc}=C(g)^d_{abc},\,\mbox{or }
C(\bar{g})_{abcd}=z^2C(g)_{abcd},\] where it's understood that the raising and lowering of indices of tensors 
built uniquely from a metric $g$ must be made with $g$ (other cases are treated individually according to the
context).\\

Returning to $\mathscr{I}$, we point out that it's always possible to choose $z$ in such a way that 
$\bar{\nabla}_a\bar{\nabla}_bz\restr{\mathscr{I}}=0$, with no restrictions on the chosen representative of
the conformal class of $\bar{g}^{(0)}$ (a proof of this fact in the case that $g$ satisfies the \textsc{Einstein}
equations without matter is presented in Section \ref{ch2-fefgra}). It's customary to call the spacetime 
$(\mathscr{M},g)$ simply the \emph{bulk}, and the corresponding conformal infinity $(\mathscr{I},\bar{g}^{(0)})$,
the \emph{boundary}. \\

An important characteristic of conformal changes of a Lorentzian metric is that the causal structure is an 
invariant of the conformal structure associated to this metric. In particular, the conformal infinity is also
called \emph{null infinity}, due to the invariance of the concept of null geodesics under conformal changes 
of the metric, as seen above. This is based upon the idea that complete null geodesics may have, from the
viewpoint of $(\overline{\mathscr{M}},\bar{g})$, past and future endpoints (in the sense of ``ideal points'' 
of \textsc{Geroch, Kronheimer} and \textsc{Penrose} \cite{gerokp}) in $\mathscr{I}$. However, one must notice 
that $\mathscr{I}$ need \emph{not} be the ``infinity'' for \emph{any} complete null geodesic. A case in which 
this certainly occurs is when $(\mathscr{M},g)$ is strongly causal and $(\overline{\mathscr{M}},\bar{g})$ is 
compact (example: \textsc{Minkowski} spacetime); however, if $(\overline{\mathscr{M}},\bar{g})$ is not 
compact, it may happen that some null geodesics are unable to reach $\mathscr{I}$. This motivates the following

\begin{definition}\label{ap1d3} 
Let $(\mathscr{M},g)$ be a $d$-dimensional spacetime with conformal infinity $(\mathscr{I},$\\$\bar{g}^{(0)})$. 
We say that $(\mathscr{M},g)$ is \emph{asymptotically simple} if any null geodesic in $(\mathscr{M},g)$ has a 
unique extension to $(\overline{\mathscr{M}},\bar{g})$ such that $\mathscr{I}$ contains precisely both endpoints
of the latter.
\end{definition}

Obviously, this is only possible if $(\mathscr{M},g)$ is null geodesically complete. Actually, in the case 
that $(\mathscr{I},\bar{g}^{(0)})$ is timelike (and, hence, a spacetime in its own right), one can say more, 
justifying the name ``asymptotically simple'':

\begin{theorem}
\label{ap1t1} {\upshape\quad If $(\mathscr{M},g)$ is asymptotically simple with timelike conformal infinity, 
then it's causally simple.}
\begin{quote}{\small\scshape Proof.\quad}
{\small\upshape First, notice that if $p,q\in\mathscr{M}$ is such that $p\not\ll_{\overline{
\mathscr{M}}}q$, then $p\not\ll_{\mathscr{M}}q$ (a past counterpart holds by an analogous argument), 
for a timelike curve in $\overline{\mathscr{M}}$ linking $p$ to $q$ can always be slightly deformed 
so as to result in a timelike curve \emph{contained in $\mathscr{M}$} and linking $p$ to $q$. Now, 
suppose that $p\in\partial I^-(q,\mathscr{M})$ and $p\notin J^-(q,\mathscr{M})$. By the argument above, 
we have $p\in\partial I^-(q,\overline{\mathscr{M}})$. Moreover, by hypothesis, a null generator $\gamma$ 
of $\partial I^-(q,\mathscr{M})$ needs to attain its future endpoint at infinity without crossing $q$ 
before that. Let $r$ be such an endpoint. Then, $r\in\partial I^-(q,\overline{\mathscr{M}})$ as the 
latter set is closed. Since the infinity is totally geodesic \cite{ribeiro2}, $\gamma$ must hit it
transversally and, thus, any causal extension of $\gamma$ must be broken \footnote{Recall that a
continuous curve segment is said to be \emph{broken} if it's piecewise $\mathscr{C}^\infty$ but not 
$\mathscr{C}^1$ in a finite set.}. Hence, if we extend $\gamma$ slightly towards the future by a 
null generator segment $\gamma'$ in $\partial I^-(q,\overline{\mathscr{M}})$ crossing $r$ (by,
say, adjusting the affine parameter $t$ of $\gamma'$ equal to zero at $r$ and extending it up to 
$t=\epsilon>0$), then there exists a timelike curve in $\overline{\mathscr{M}}$ linking $p$ to 
$\gamma'(\epsilon)$ \cite{hawkellis}, which violates the achronality of $\partial I^-(q,\overline{\mathscr{M}})$. 
Repeat the argument exchanging future with past.\hfill$\Box$}\end{quote}
\end{theorem}

An asymptotically simple spacetime need not, though, be globally hyperbolic -- a prime example is 
the AdS spacetime.

\chapter{\label{ap2}Elements of operator algebras}
\chaptermark{Operator algebras}

The formalism of operator algebras is a rather powerful language for the discussion of 
structural aspects of Quantum Physics which are independent of the representation of operators in 
some fixed \textsc{Hilbert} space, which is absolutely fundamental in the relativistic case (i.e., 
Quantum Field Theory) and, more in general, in the study of systems with an \emph{infinite}
number of degrees of freedom (thermodynamic limits in Quantum Statistical Mechanics, etc.).\\

C*-algebras and \textsc{von Neumann} algebras (to be seen in Section \ref{ap2-cwalg}) are 
adequate to the discussion of \emph{bounded} operators. We can, when considering unbounded 
operators, make use of continuous functional calculus (spectral theorem) when the latter applies 
and treat only bounded functions of these operators, hence avoiding technical problems relative
to the determination of domains. When we discuss field operators, as well as their distributional
character, it's however convenient to have at our disposal an \emph{algebraic} framework for 
dealing with such unbounded operators \emph{directly}, albeit in a manner independent of 
representations. For such, we present a reasonably detailed study of *-algebras in Section 
\ref{ap2-staralg}, before we introduce C*-algebras or \textsc{von Neumann} algebras, to show 
that several known results in the context of C*-algebras remain valid or suffer only minor 
technical modifications in the context of *-algebras. This allows us to treat the cases of 
algebras of observables and of fields more or less in parallel. For the latter, we'll present the 
formalism proposed by \textsc{Wightman}, \textsc{Borchers} and \textsc{Uhlmann} in Section 
\ref{ap2-borchers}.\\

We limit ourselves to demonstrate results whose proof is simple from the analytical viewpoint and 
at the same time instructive, and those which are required in the present work in ``nonstandard'' form. 
Our main references for this Appendix are \cite{bratteli1,bratteli2}, and we employ as 
auxiliary references \cite{murphy,pct,tak1,tak2}. For functional analytic aspects, we adopt 
\cite{reedsimon1,rudin}.

\section{\label{ap2-staralg}*-Algebras}

This Section is dedicated to purely algebraic aspects. We shall begin by defining the notion of *-algebra 
and some of its properties.

\begin{definition}\label{ap2d1}
An (associative) \emph{*-algebra} is a vector space $\mathfrak{F}$ over $\mathbb{C}$, endowed 
with an associative, distributive \emph{product} $x,y\mapsto xy$ which commute with scalar 
multiplication, and with an \emph{involution} $x\mapsto x^*$ which satisfies $x^{**}=x$, $(xy)^*=y^*
x^*$ and $(\alpha x+\beta y)^*=\bar{\alpha}x^*+\bar{\beta}y^*$, for all $x,y\in\mathfrak{F}$, 
$\alpha,\beta\in\mathbb{C}$. We say that $\mathfrak{F}$ is \emph{unital} if there exists an  
\emph{identity} element $\mathbb{1}\in\mathfrak{F}$ such that $\mathbb{1}x=x\mathbb{1}=x$, for
all $x\in\mathfrak{F}$ (in this case, $\mathbb{1}$ is unique).
\end{definition}

The \emph{centre} of $\mathfrak{F}$ is given by $\mathfrak{Z}(\mathfrak{F})\doteq\{x\in\mathfrak{F}:
[x,y]=0,\,\forall y\in\mathfrak{F}\}$, where $[x,y]\doteq xy-yx$ denotes the \emph{commutator} of $x$ 
and $y$. A *-algebra $\mathfrak{F}$ is \emph{Abelian} if $\mathfrak{Z}(\mathfrak{F})=\mathfrak{F}$; 
at the other extreme, we say that $\mathfrak{F}$ is \emph{factorial} or \emph{primary} if $\mathfrak{Z}
(\mathfrak{F})=\{0\}$. 
A \emph{*-subalgebra} $\mathfrak{G}\subset\mathfrak{F}$ is a vector subspace closed under product 
and involution. Examples of *-subalgebras of $\mathfrak{F}$ are $\mathfrak{Z}(\mathfrak{F})$ and, more in 
general, the \emph{(relative) commutant} or \emph{normalizer} of a *-subalgebra $\mathfrak{G}\subset
\mathfrak{F}$, given by $\mathfrak{G}'=\{x\in\mathfrak{F}:[x,y]=0,\forall y\in\mathfrak{G}\}$.\\

We can adjoin an identity $\mathbb{1}$ to a nonunital *-algebra $\mathfrak{F}$ in the following way: 
let $\widetilde{\mathfrak{F}}=\mathfrak{F}\oplus\mathbb{C1}\ni(x,\lambda)$ with product $(x,
\lambda)(y,\mu)\doteq(xy+\mu x+\lambda y,\lambda\mu)$ and involution $(x,\lambda)^*\doteq(x^*,
\bar{\lambda})$. $\mathfrak{F}$ is naturally identified with the *-subalgebra $(\mathfrak{F},0)$, and 
we have $\mathbb{1}=(0,1)$.\\

Let $\mathfrak{F}$, $\mathfrak{G}$ be *-algebras. A \emph{*-morphism} is a linear map $\pi:
\mathfrak{F}\rightarrow\mathfrak{G}$ which preserves the product and the involution. If $\pi$ is 
respectively one-to-one, onto, bijective, we say that $\pi$ is a \emph{*-monomorphism, *-epimorphism, 
*-isomorphism} -- we write $\mathfrak{F}\cong\mathfrak{G}$ if there exists a *-isomorphism from 
$\mathfrak{F}$ to $\mathfrak{G}$. If $\mathfrak{F}=\mathfrak{G}$, we say that $\pi$ is a 
\emph{*-endomorphism}, and a \emph{*-automorphism} if $\pi$, in this case, is bijective. A 
\emph{*-representation} of $\mathfrak{F}$ in a \textsc{Hilbert} space $\mathscr{H}$ \footnote{\label{fninv}
We demand a \textsc{Hilbert} space structure to avoid worrying about the definition of involution in 
$\mbox{End}_{\mathbb{C}}\mathscr{H}$. A definition of *-representation by the latter's elements 
(i.e., without reference to the inner product) would involve having at our disposal a 
definition of involution, at least in the image of the *-algebra in $\mbox{End}_{\mathbb{C}}
\mathscr{H}$.} is a *-morphism 
$\pi:\mathfrak{F}\rightarrow\mathscr{B}(\mathscr{H})$. We say, then that $\pi$ is \emph{faithful} if 
$\pi$ is a *-monomorphism, and \emph{irreducible} (abbreviated \emph{irrep}) if the only subspaces 
of $X$ invariant under the action of $\pi(\mathfrak{F})$ are $\{0\}$ and $X$ -- by the \textsc{Schur} lemma, 
$\pi$ is irreducible if and only if $\{a\in\mathscr{B}(\mathscr{X}):[a,\pi(x)]=0,\,\forall x\in\mathfrak{F}\}
=\{0\}$.\\

A *-ideal of $\mathfrak{F}$ is a subspace $\mathfrak{I}\subset\mathfrak{F}$ \emph{invariant under} 
${}^*$, such that $x\in\mathfrak{F},y\in\mathfrak{I}$ implies $xy\in\mathfrak{I}$. By *-invariance, 
it follows automatically that $(x^* y^*)^*=yx\in\mathfrak{I}$. Notice that, on the other hand, left
\emph{ideals} (not necessarily *-invariant) of $\mathfrak{F}$ are not necessarily right ideals, and 
vice-versa. $\mathfrak{F}$ is \emph{simple} if it doesn't possess nontrivial *-ideals (i.e., 
different from $\{0\}$ and $\mathfrak{F}$) -- any nontrivial simple representation $\pi$ of 
$\mathfrak{F}$ simples is faithful, for $\mbox{ker}\pi$ is a *-ideal of $\mathfrak{F}$.\\

Given a *-ideal $\mathfrak{I}\subset\mathfrak{F}$, the \emph{quotient *-algebra} $\mathfrak{F}/\mathfrak{I}$ 
of $\mathfrak{F}$ modulo $\mathfrak{I}$ has as elements the cosets $[x]=x+\mathfrak{I}\subset\mathfrak{F}$, 
where $[x]=[y]$ if $x-y\in\mathfrak{I}$. For instance, if $\pi:\mathfrak{F}\rightarrow\mathfrak{G}$ is a 
*-morphism, it follows that $\mathfrak{F}/\mbox{ker}\pi\cong\pi(\mathfrak{F})\subset\mathfrak{G}$. The 
quotient possesses the following universal property: given a *-algebra $\mathfrak{G}$ and a *-morphism $\rho:
\mathfrak{F}\rightarrow\mathfrak{G}$ such that $\mbox{ker}\rho\subset\mathscr{I}$, there exists a unique 
*-morphism $[\rho]:\mathfrak{F}/\mathfrak{I}\rightarrow\mathfrak{G}$ such that $\rho(x)=[\rho]([x])$ for all 
$x\in\mathfrak{F}$. A *-endomorphism $\pi$ of $\mathfrak{F}$ induces a \emph{*-endomorphism} $[\pi]:
\mathfrak{F}/\mathfrak{I}\ni[x]\mapsto\pi([x])=\pi(x)+\pi(\mathscr{I})$ if and only if $\mathfrak{I}$ is 
\emph{invariant} under $\pi$, i.e., $\pi(\mathfrak{I})\subset\mathfrak{I}$. In particular, $[\pi]$ is an 
injective *-endomorphism of $\mathfrak{F}/\mathfrak{I}$ if and only if, in addition, $\mbox{ker}\pi\subset
\mathfrak{I}$ and $\pi(\mathfrak{I})supseteqq\mathfrak{I}$.\\

We state below four fundamental theorems on *-algebras --  the latter three are the so-called
\emph{*-Isomorphism Theorems}, analogous to the Isomorphism Theorems for rings and modules (the 
incorporation of the operation of involution ${}^*$ is immediate and can be checked directly).

\begin{theorem}[Fundamental Theorem of *-Morphisms]\label{ap2t1}
Let $\mathfrak{F}$ and $\mathfrak{G}$ be two *-algebras, and $\rho:\mathfrak{F}\rightarrow\mathfrak{G}$ a 
*-morphism. Then the induced *-morphism $\tilde{\rho}:\mathfrak{F}/\mbox{Ker}\rho\rightarrow\rho(\mathfrak{F})
\subset\mathfrak{G}$ is a *-isomorphism.
\begin{quote}{\small\scshape Proof.\quad}
{\small\upshape Immediate.~\hfill~$\Box$}
\end{quote}
\end{theorem}

Theorem \ref{ap2t1} can be generalized, resulting in the

\begin{theorem}[First *-Isomorphism Theorem \footnotemark]\label{ap2t2}
Let $\mathfrak{F}$ and $\mathfrak{G}$ two *-algebras, \\and $\rho:\mathfrak{F}\rightarrow\mathfrak{G}$
a *-epimorphism. Then, $\mathfrak{H}\mapsto\rho(\mathfrak{H})$ establishes a one-to-one correspondence 
between the additive subgroups $\mathfrak{H}\supset\mbox{Ker}\rho$ of $\mathfrak{F}$ and the additive 
subgroups of $\mathfrak{G}$. Under this  correspondence, $\mathfrak{H}$ is a *-subalgebra (resp. *-ideal) 
of $\mathfrak{F}$ if and only if $\rho(\mathfrak{H})$ is a *-subalgebra (resp.*-ideal) of $\mathfrak{G}$. 
Moreover, if $\mathfrak{I}\supset\mbox{Ker}\rho$ is a *-ideal, then the map \[x+\mathfrak{I}\mapsto\rho(x)+
\rho(\mathfrak{I})\] is a *-isomorphism of $\mathfrak{F}/\mathfrak{I}$ onto $\mathfrak{G}/\rho(\mathfrak{I})$.
\begin{quote}{\small\scshape Proof.\quad}
{\small\upshape The first assertion follows immediately from Theorem \ref{ap2t2}. The remaining one is 
demonstrated in a manner analogous to the case of rings -- see the proof of Theorem 2.6 of \cite{jacobson},
pages 107--108.~\hfill~$\Box$}
\end{quote}
\end{theorem}
\footnotetext{There seems to be no agreement in the literature on the coining ``First Isomorphism Theorem'', 
neither for rings and modules, nor for groups. Some authors attribute this name to the analog of Theorem 
\ref{ap2t1}, and others, to the analog of Theorem \ref{ap2t2}, as we do here, for Theorem \ref{ap2t2} can
be understood, as said above, as a generalization of Theorem \ref{ap2t1}, and more accordingly with the 
Second and Third (*-)Isomorphism Theorems.}

\begin{theorem}[Second *-Isomorphism Theorem]\label{ap2t3}
Let $\mathfrak{F}$ be a *-algebra, $\mathfrak{G}$ a *-subalgebra of $\mathfrak{F}$, and $\mathfrak{I}$
a *-ideal of $\mathfrak{F}$. Then $\mathfrak{G}+\mathfrak{I}$ is a *-subalgebra of $\mathfrak{F}$, 
$\mathfrak{G}\cap{}\mathfrak{I}$ is a *-ideal of $\mathfrak{G}$, $\mathfrak{I}$ is a *-ideal of
$\mathfrak{G}+\mathfrak{I}$, and $(\mathfrak{G}+\mathfrak{I})/\mathfrak{I}\cong\mathfrak{G}/(\mathfrak{G}
\cap{}\mathfrak{I})$.
\begin{quote}{\small\scshape Proof.\quad}
{\small\upshape Analogous to the case of rings -- see \cite{jacobson}, page 108.~\hfill~$\Box$}
\end{quote}
\end{theorem}

\begin{theorem}[Third *-Isomorphism Theorem]\label{ap2t4}
Let $\mathfrak{F}$ be a *-algebra, and $\mathfrak{I}\subset\mathfrak{J}$ two *-ideals of $\mathfrak{F}$. 
Then $\mathfrak{I}$ is a *-ideal of $\mathfrak{J}$, and $\mathfrak{F}/\mathfrak{J}\cong(\mathfrak{F}/
\mathfrak{I})/(\mathfrak{J}/\mathfrak{I})$.
\begin{quote}{\small\scshape Proof.\quad}
{\small\upshape Follows immediately from Theorem \ref{ap2t2}.~\hfill~$\Box$}
\end{quote}
\end{theorem}

The demonstration of the analogs of Theorems \ref{ap2t2} and \ref{ap2t3} for rings consist only in the 
construction of the isomorphisms and explicit checking of the relevant properties. 

\begin{remark}\label{ap2r1}
In the case of topological *-algebras, in particular the examples to be discussed throughout this 
Appendix, Theorems \ref{ap2t1} to \ref{ap2t4} not only at the algebraic level but also at the topological 
level, as far as one adds the adjective \emph{``closed''} (in the respective topologies) to the terms 
``subgroup'', ``*-subalgebra'' and ``*-ideal'' -- the continuity of the *-morphisms is automatic in all 
cases, by virtue of the C* property of the (semi)norms (see Definitions \ref{ap2d2} and \ref{ap2d7}). In
this case, given a (semi)norm $\Vert.\Vert$ in a *-algebra $\mathfrak{F}$, the corresponding (semi)norm 
in the quotient *-algebra $\mathfrak{F}/\mathfrak{I}$, where $\mathfrak{I}\subset\mathfrak{F}$ is a *-ideal, 
is given by \[\mathfrak{F}/\mathfrak{I}\ni[x]\mapsto\Vert[x]\Vert\doteq\inf_{y\in\mathfrak{I}}\Vert x+y\Vert.\] 
Completeness of this norm (or separating family of seminorms) can be checked directly; for the verification 
of the C* property in the case of C*-algebras, see Proposition 2.2.19 of \cite{bratteli1}.
\end{remark}

A \emph{state} $\omega$ over a unital *-algebra $\mathfrak{F}$ is a \emph{positive} linear functional on 
$\mathfrak{F}$, i.e., $\omega(x^*x)\geq 0$ for all $x\in\mathfrak{F}$, which satisfies $\omega(\mathbb{1})=1$.
\footnote{If there is a topology in $\mathfrak{F}$ at our disposal, we can define the normalization condition 
of a state even in the absence of a unit.} We say that $\omega$ is \emph{faithful} if $\omega(x^*x)>0$ for all 
$x\neq 0$ (the reason behind the name will emerge soon). The set of states over $\mathfrak{F}$ is \emph{convex}: 
if $\omega_1$ and $\omega_2$ are states, then $\lambda\omega_1+(1-\lambda)\omega_2$ is as well, for all $\lambda
\in(0,1)$. If $\omega$ doesn't admit a decomposition $\omega=\lambda\omega_1+(1-\lambda)\omega_2$ for $\omega_1,
\omega_2\neq\omega$, $\lambda\in(0,1)$, we say that $\omega$ is \emph{pure} (equivalently, $\omega$ is pure if
it cannot be written as the sum of two positive linear functionals $\phi_1,\phi_2\notin\mathbb{R}_+\omega$). 
Otherwise, we say that $\omega$ is \emph{mixed}. We denote the set of all states over $\mathfrak{F}$ by 
$\mathscr{S}_{\mathfrak{F}}$.\\

The positivity of $\omega$ leads to the following, crucial property:

\begin{lemma}[\textsc{Cauchy-Schwarz} inequality]\label{ap2l1}
Let $\phi$ be a positive linear functional over a (not necessarily unital) *-algebra $\mathfrak{F}$. Then, (i) 
$\phi(x^*y)=\overline{\phi(y^*x)}$ (in particular, if $\mathfrak{F}$ is unital, $\phi(x^*)=\overline{\phi(x)}$) 
and (ii) $|\phi(x^*y)|^2\leq\phi(x^*x)\phi(y^*y)$, for all $x,y\in\mathfrak{F}$ (in particular, if $\mathfrak{F}$ 
is unital, $|\phi(x)|^2\leq\phi(\mathbb{1})\phi(x^*x)$).
\begin{quote}{\small\scshape Proof (sketch).\quad}
{\small\upshape For all $\lambda\in\mathbb{C}$, we have $\phi((\lambda x+y)^*(\lambda x+y))\geq 0$. By linearity, 
\[|\lambda|^2\phi(x^*x)+\bar{\lambda}\phi(x^*y)+\lambda\phi(y^*x)+\phi(y^*y)\geq 0.\] The reality of the first 
member of the inequality above implies (i), whence it follows we can now take $\lambda\in\mathbb{R}$.
Choosing an appropriate interval for $\lambda$, one proves (ii).~\hfill~$\Box$}
\end{quote}
\end{lemma}

With Lemma \ref{ap2l1} in our possession, we can easily build a representation of $\mathfrak{F}$ by densely 
defined linear operators in a \textsc{Hilbert} space $\mathscr{H}_\phi$ naturally associated to a positive 
linear functional $\phi$. First, notice that the \emph{annihilator} $\mbox{Ann}\phi$ of $\phi$, given by 
$\mbox{Ann}\phi=\{x\in\mathfrak{F}:\phi(x^*x)=0\}$, is a left ideal of $\mathfrak{F}$. Indeed, $\mbox{Ann}\phi$ 
is a vector subspace of $\mathfrak{F}$, due to the linearity of $\phi$ and Lemma \ref{ap2l1}. Moreover, given 
$x\in\mbox{Ann}\phi,y\in\mathfrak{F}$, we have $\phi((yx)^*yx)^2\leq\phi(x^*x)\phi((x^*y^*y)(y^*yx))=0$. Consider 
now the quotient space $\mathfrak{F}/\mbox{Ann}\phi\ni[x],[y]$, which admits the scalar product given by 
$\langle[x],[y]\rangle\doteq\phi(x^*y)$. Completing $\mathfrak{F}/\mbox{Ann}\phi\ni[x],[y]$ with respect to the norm 
$\Vert.\Vert$ associated to $\langle.,.\rangle$, we obtain the \textsc{Hilbert} space $\mathscr{H}_\phi\doteq
\overline{(\mathfrak{F}/\mbox{Ann}\phi)^{\Vert.\Vert}}$. \\

Let us consider the operator $\tilde{\pi}_\phi(x)$ associated to $x$, defined in the dense linear subspace 
$\mathfrak{F}/\mbox{Ann}\phi$ by $\tilde{\pi}_\phi(x)[y]\doteq[xy]$. As the adjoint of $\tilde{\pi}_\phi(x)$ 
is given by $\tilde{\pi}_\phi(x)^*\restr{\mathfrak{F}/\mbox{\scriptsize Ann}\phi}=\tilde{\pi}_\phi(x^*)$ and 
both are densely defined, it follows that $\tilde{\pi}_\phi(x)$ is closable \cite{reedsimon1}, with closure 
$\overline{\pi_\phi(x)}=\pi_\phi(x)^{**}\subset\pi_\phi(x^*)^*$ (recall that, given densely defined linear
operators $A,B$ with respective domains $D(A),D(B)$, $A\subset B$ denotes that $B$ is an \emph{extension} of 
$A$, i.e., $D(B)\supset D(A)$ and $B\restr{D(A)}=A$) Thus, $\pi_\phi(x)=\overline{\pi_\phi(x)}=\pi_\phi(x)^{**}$ 
defines a representation of $\mathfrak{F}$ by densely defined, closed linear operators $\mathscr{H}_\phi$. The 
densely defined antilinear operator given by $S_0:[x]\mapsto[x*]$ will be crucial in the development of the
\textsc{Tomita-Takesaki} modular theory later on.\\

Notice that the representation $\pi_\phi$ is faithful if and only if $\phi(x^*x)>0$ for all $x\neq 0$. If
$\mathfrak{F}$ is \emph{unital} and $\omega$ is a \emph{state}, then $\phi_\omega$ is faithful if and only if 
$\omega$ is faithful, justifying the name we've given. For $\mathfrak{F}$ unital, there is yet a special
element of $\mathscr{H}_\omega$, given by $\Omega=[\mathbb{1}]$. Obviously, $\Vert\Omega\Vert=1$ if $\omega$ 
is a state; more importantly, it follows that the set $\pi_\omega(\mathfrak{F})\Omega=\mathfrak{F}/\mbox{Ann}
\phi$ is dense in $\mathscr{H}_\omega$, i.e., $\Omega$ is a \emph{cyclic} vector for $\pi_\omega(\mathfrak{F})$.
\footnote{In the case of C*-algebras, it's possible to construct a cyclic vector even in the absence of a
unit \cite{murphy}.} In this case, the representation built above, denoted by the triple $(\mathscr{H}_\omega,
\pi_\omega,\Omega)$, is called \emph{cyclic} or \textsc{Wightman}\emph{-GNS representation} of $\mathfrak{F}$ 
associated to the positive linear functional $\omega$.\footnote{GNS calls for the names of \textsc{Gel'fand} 
and \textsc{Naimark} \cite{gelfand} and \textsc{Segal} \cite{segal}, which proposed the construction above in 
the case that $\mathfrak{F}$ is a C*-algebra -- see Section \ref{ap2-cwalg}. \textsc{Wightman}'s name comes 
along as well for the construction above is the core of its Theorem of reconstruction of quantum fields from 
a hierarchy of $k$-point functions \cite{pct}. We shall have more to say about all this in Section 
\ref{ap2-borchers}.} $\pi_\omega$ is faithful if and only if $\Omega$ is not only cyclic but also 
\emph{separating} for $\mathfrak{F}$, i.e., $\pi_\omega(x)\Omega=0\,\Rightarrow\,x=0.$\\

The \textsc{Wightman}-GNS representation of a \emph{unital} *-algebra $\mathfrak{F}$ associated to a positive 
linear functional $\phi$ is unique in the following sense: let $\pi$ be a representation of $\mathfrak{F}$ by
densely defined, closed linear operators in a \textsc{Hilbert} space $\mathscr{H}$, such that:

\begin{itemize}
\item There exists a domain $\mathscr{D}$, common to all $\pi(x)$'s, dense in $\mathscr{H}$ and such that 
$\pi\mathfrak{F}\mathscr{D}\subset\mathscr{D}$;
\item Thee exists a vector $\Phi'\in\mathscr{D}$ cyclic for $\pi(\mathfrak{F})$ such that $\pi(\mathfrak{F})
\Phi'=\mathscr{D}$ and $\langle\Phi',\pi(x)\Phi'\rangle$\\$=\phi(x)$ for all $x\in\mathfrak{F}$.
\end{itemize}

We can then define the linear operator $U:\mathfrak{F}/\mbox{Ann}\phi\rightarrow\mathscr{H}$ by $U[x]\doteq
\pi(x)\Phi'$. $U$ is densely defined and satisfies $\langle U[x],U[y]\rangle=\phi(x^*y)=\langle\pi(x)\Phi',
\pi(y)\Phi'\rangle$ for all $x,y\in\mathfrak{F}$, that is, $U$ is an isometry. By the BLT Theorem 
\cite{reedsimon1}, $U$ extends uniquely to an isometry $U$ of $\mathscr{H}_\phi$ into $\mathscr{H}$, 
injective and with dense range. Invoking once more the BLT Theorem, we have that $U$ is a bijection, and
its inverse $U^-1=U^*$ is also an isometry. That is, the representations $\pi_\phi$ and $\pi$ are unitarily 
equivalent.\footnote{In the case that $\mathfrak{F}$ is a C*-algebra, the uniqueness of the GNS representation, 
given by $U$, still holds even if $\mathfrak{F}$ doesn't have a unit, but in this case the cyclic vector 
$\pi_\phi$ doesn't necessarily belongs to $\mathfrak{F}/\mbox{Ann}\phi$, and we must analogously impose only 
that $\pi(\mathfrak{F})\Phi'\subset\mathscr{D}$ is dense in $\mathscr{H}$.} Another important consequence 
is that, if $\Phi=[\mathbb{1}]$ is the cyclic vector of $\pi_\phi$, then $U\Phi=\Phi'$. An immediate consequence
of this fact, which we can state as a Corollary, is:

\begin{corollary}\label{ap2c1}
If $\alpha$ is a *-automorphism of a unital *-algebra $\mathfrak{F}$ and $\phi$, a  positive linear functional 
such that $\phi\circ\alpha=\phi$, then there exists a unique unitary operator $U_\alpha$ in $\mathscr{H}_\phi$ 
such that $\pi_\phi(\alpha(x))=U\pi_\phi(x)U^{-1}$.~\hfill~$\Box$
\end{corollary}

That is, a sufficient condition for a *-automorphism $\alpha$ to be unitarily implementable in $\mathscr{H}_\phi$ 
is that $\phi$ is invariant under the action of $\alpha$. This is the main mechanism of implementation of 
symmetries in quantum theory. Finally, there is a close relation between the purity of $\omega$ and a ``weak'' 
counterpart of the notion of irreducibility for $\pi_\omega$:

\begin{theorem}\label{ap2t5}
Let $\omega$ be a state over a unital *-algebra $\mathfrak{F}$. Define the \emph{weak commutant} \[\pi_\omega
(\mathfrak{F})_w'\doteq\{T\in\mathscr{B}(\mathscr{H}_\omega):\langle T^*[x],\pi_\omega(y)[z]\rangle=\langle
\pi_\omega(y^*)[x],T[z]\rangle,\,\forall x,y,z\in\mathfrak{F}\}\] of $\pi_\omega(\mathfrak{F})$; $\pi_\omega
(\mathfrak{F})_w'$ is a *-invariant linear subspace of $\mathscr{B}(\mathscr{H}_\omega)$. Then $\pi_\omega
(\mathfrak{F})_w'=\mathbb{C1}$ if and only if $\omega$ is pure. Moreover, there exists a one-to-one 
correspondence $\omega_T\leftrightarrow T$ between positive linear functionals $\omega_T$ satisfying $\omega_T
\leq\omega$ and positive elements $T$ of $pi_\omega(\mathfrak{F})_w'$ satisfying $\Vert T\Vert\leq 1$.
\begin{quote}{\small\scshape Proof.\quad}
{\small\upshape $(\Rightarrow)$ Suppose that $\omega$ is mixed. Then there exists a positive linear functional 
$\omega'$ such that $\omega'(x^*x)\leq\omega(x^*x)$, for all $x$, and such that $\omega'$ is not a multiple 
of $\omega$. Applying the \textsc{Cauchy-Schwarz} inequality, we have $|\omega'(x^*y)|^2\leq\Vert\pi_\omega(x)
\Omega\Vert^2\Vert\pi_\omega(y)\Vert^2$. Hence, $\pi_\omega(x)\Omega\times\pi_\omega(y)\Omega\mapsto\omega'(x^*y)$ 
defines a densely defined, bounded sesquilinear form on $\mathscr{H}_\omega$. Thus, there exists a unique bounded
operator $T$ in $\mathscr{H}_\omega$ such that $\langle\pi_\omega(x)\Omega,T\pi_\omega(y)\Omega\rangle=\omega'(x^*y)$, 
and $T$ is not a multiple of $\mathbb{1}$. Moreover, $0\leq\langle\pi_\omega(x)\Omega,T\pi_\omega(x)\Omega\rangle
\leq\Vert\pi_\omega(x)\Vert^2$ and thus $0\leq T\leq\mathbb{1}$. However, in this case, \[\langle\pi_\omega(x)
\Omega,T\pi_\omega(y)\pi_\omega(x)\Omega\rangle=\omega'(x^*yz)=\omega'((y^*x)^*z)=\langle\pi_\omega(y^*)\pi_\omega(x)
\Omega,T\pi_\omega(z)\Omega\rangle\] for all $x,y,z\in\mathfrak{F}$, proving that $T$ is a nontrivial element of 
$\pi_\omega(\mathfrak{F})_w'$.

$(\Leftarrow)$ Suppose that $T\in\pi_\omega(\mathfrak{F})_w',\,T\notin\mathbb{C1}$. Then, $T^*$ and, thus, 
$T+T^*$ belong to $\pi_\omega(\mathfrak{F})_w'$, being the latter self-adjoint and nontrivial. Define $S\doteq
\lambda(2\Vert T+T^*\Vert\mathbb{1}+(T+T^*))\in\pi_\omega(\mathfrak{F})_w'$, where $\lambda>0$ is chosen such 
that $\Vert S\Vert<1$. Then, $0<S<\mathbb{1}$, and hence there exists a spectral projector $P$ of $S$ such 
that $0<P<\mathbb{1}$ and $P\in\pi_\omega(\mathfrak{F})_w'$. Consider the linear functional $\omega'(x)=\langle 
P\Omega,\pi_\omega(x)\Omega\rangle$. $\omega'$ is positive, for \[\omega'(x^*x)=\langle P\Omega,\pi_\omega(x^*)
\pi_\omega(x)\Omega\rangle=\langle\pi_\omega(x)\Omega,P\pi_\omega(x)\Omega\rangle=\langle P\pi_\omega(x)\Omega,
\pi_\omega(x)\Omega\rangle\geq 0.\] Finally, \[\omega(x^*x)-\omega'(x^*x)=\langle\pi_\omega(x)\Omega,P\pi_\omega(x)
\Omega\rangle\geq 0,\] showing that $\omega$ is mixed.

The final assertion final follows automatically from the arguments above.~\hfill~$\Box$}
\end{quote}
\end{theorem}

\section{\label{ap2-cwalg}C*-algebras and \textsc{von Neumann} algebras}

\subsection{\label{ap2-cwalg-c}C*-algebras}

Let's now give a sufficient condition so as to make the closed linear operators $\pi_\phi(x)$ built by means of
the \textsc{Wightman}-GNS representation in Section \ref{ap2-staralg} \emph{bounded}. If we can find for all $x$ 
a $C(x)>0$ such that $\phi((xa)^*xa)\leq C(x)^2\phi(a^*a)$ for all $a\in\mathfrak{F}$ \emph{and any positive
  linear functional} $\phi$ (i.e., $C$ doesn't depend on $\phi$), it then follows that $\Vert\pi_\phi(x)[a]\Vert
\leq C(x)\Vert[a]\Vert$. Hence, by the BLT Theorem, $\pi_\phi(x)$ admits a unique extension to $\mathscr{H}_\phi$, 
and $\Vert\pi_\phi(x)\Psi\Vert\leq C(x)\Vert\Psi\Vert$ for all $\Psi\in\mathscr{H}_\phi$. In this case, $\pi_\phi
(x^*)=\pi_\phi(x)^*$ and, thus, $\pi_\phi$ defines a \emph{*-representation} of $\mathfrak{F}$ by bounded operators
in $\mathscr{H}_\phi$.\\

The most obvious manner of obtaining $C$ as in the previous paragraph is to impose a \emph{norm} $\Vert.\Vert$ 
in $\mathfrak{F}$ such that the product and the involution of $\mathfrak{F}$ are continuous with respect to this norm, 
defining a \emph{normed *-algebra}. In particular, we can choose a norm equivalent to the original such that 
$\Vert xy\Vert\leq\Vert x\Vert\Vert y\Vert$ and, hence, $\Vert\mathbb{1}\Vert=1$ if $\mathfrak{F}$ is unital 
(such a property is assumed to hold throughout the whole of the present work). We'll see that $C(x)\doteq\Vert 
x\Vert$ satisfies the condition in the previous paragraph if:

\begin{enumerate}
\item $\mathfrak{F}$ is \emph{complete} with respect to $\Vert.\Vert$ (then we say that $\mathfrak{F}$ is a 
\textsc{Banach} \emph{*-algebra}), and 
\item $\Vert x^*x\Vert=\Vert x\Vert^2$. In this case, it follows that $\Vert x\Vert^2,\Vert x^*\Vert^2\leq
\Vert x\Vert.\Vert x^*\Vert$ and, thus, $\Vert x\Vert=\Vert x^*\Vert$. 
\end{enumerate}

\begin{definition}\label{ap2d2}
A \emph{C*-algebra} $\mathfrak{A}$ is a \textsc{Banach} *-algebra whose norm satisfies the \emph{C* condition} 
$\Vert x^*x\Vert=\Vert x\Vert^2$.\footnotemark
\end{definition}
\footnotetext{\label{cstarnorm} Actually, it's not necessary to impose the condition $\Vert xy\Vert\leq
\Vert x\Vert\Vert y\Vert$ or even continuity of the product and of the involution as axioms defining 
a C*-algebra. \textsc{Araki} and \textsc{Elliott} \cite{arelliott} showed that the first condition, and, hence, 
the condition $\Vert x\Vert=\Vert x^*\Vert$ follow from the remaining axioms (completeness of the norm, *-algebra 
structure and the C* condition).}

A typical example of C*-algebra is the unital *-algebra $\mathscr{B}(\mathscr{H})$ of bounded operators in a
complex \textsc{Hilbert} space $\mathscr{H}$. Let's now cite two frequently used results, which moreover show
the strength of Definition \ref{ap2d2}.

\begin{proposition}\label{ap2p1}
Any *-morphism $\rho$ from a unital \textsc{Banach} *-algebra $\mathfrak{B}$ to a C*-algebra $\mathfrak{C}$ 
satisfies $\Vert\rho(x)\Vert\leq\Vert x\Vert$ for all $x\in\mathfrak{B}$. If $\mathfrak{B}$ is a C*-algebra, 
then $\rho(\mathfrak{B})$ is a C*-subalgebra of $\mathfrak{C}$, and $\Vert\rho(x)\Vert=\Vert x\Vert$ for all 
$x$ if and only if $\rho$ is a *-monomorphism. 
\begin{quote}{\small\scshape Proof.\quad}
{\small\upshape See Propositions 2.3.1 and 2.3.3 of \cite{bratteli1}.~\hfill~$\Box$}
\end{quote}
\end{proposition}

\begin{proposition}\label{ap2p2}
Any positive linear functional $\phi$ over a C*-algebra $\mathfrak{A}$ (not necessarily unital) satisfies the 
following properties:
\begin{itemize}
\item[(i)] $\phi(x^*)=\overline{\phi(x)}$;
\item[(ii)] $|\phi(x)|^2\leq\Vert\phi\Vert\phi(x^*x)$ (in particular, $\Vert\phi\Vert=\phi
(\mathbb{1})$ if $\mathfrak{A}$ is unital);
\item[(iii)] $|\phi(x^*yx)|\leq\phi(x^*x)\Vert y\Vert$;
\item[(iv)] $\Vert\phi\Vert=\sup\{\phi(x^*x):\Vert x\Vert=1\}$.
\end{itemize}
\begin{quote}{\small\scshape Proof.\quad}
{\small\upshape See Proposition 2.3.11 of \cite{bratteli1}.~\hfill~$\Box$}
\end{quote}
\end{proposition}

It follows from Proposition \ref{ap2p2}(iii) that $\phi((xy)^*xy)\leq\Vert x\Vert^2\phi(y^*y)$. Hence, $C(x)
=\Vert x\Vert$ satisfies the estimate in the first paragraph of this Section, and the GNS representation
$\pi_\phi$ is, thus, a *-representation of $\mathfrak{A}$ by elements of the C*-algebra $\mathscr{B}
(\mathscr{H}_\phi)$. The apparently particular example of C*-algebra given by $\pi_\phi(\mathfrak{A})$, is
actually the most general possible, for \textsc{Gel'fand} and \textsc{Naimark} proved \cite{gelfand} that
any C*-algebra is *-isomorphic to a C*-subalgebra of $\mathscr{B}(\mathscr{H})$ for some \textsc{Hilbert} 
space $\mathscr{H}$. More precisely, the \emph{universal representation} \[\mathscr{H}_u=\bigoplus_{\omega
\in\mathscr{S}_{\mathfrak{A}}}\mathscr{H}_\omega,\,\pi_u=\bigoplus_{\omega\in\mathscr{S}_{\mathfrak{A}}}\pi_\omega\] 
is faithful, for one can prove \cite{bratteli1} that, given $x\in\mathscr{A}$, there exists a pure state 
$\omega_x$ such that $\omega_x(x^*x)=\Vert x\Vert^2$. As $\pi_{\omega_x}$ in this case is irreducible (see next 
paragraph), it follows that $\pi_u$ has, among its irreducible subrepresentations, $\pi_{\omega_x}$ for all 
$x$. Hence, $\pi_u(x)\neq 0$ for all $x\neq 0$.\\

Finally, let us consider the following class of elements of a C*-algebra $\mathfrak{A}$. We say that $x\in
\mathfrak{A}$ is \emph{positive} if it's self-adjoint and its spectrum $\sigma(x)$ satisfies $\sigma(x)\subset
\bar{\mathbb{R}}_+$, or, equivalently, can be written as $x=y^*y,\,y\in\mathfrak{A}$.\footnote{At first sight, 
we could define positivity directly in a *-algebra in Section \ref{ap2-staralg} by employing this condition. 
We avoid this explicit terminology in this more general case (except, of course, in the definition of 
positive linear functionals and states over a *-algebra), for the \textsc{Wightman}-GNS representation of 
of a ``positive'' element in this sense defines a symmetric and closed operator, but not necessarily a 
self-adjoint one. Thus, the intuitive notion of positive operator as referring to the non negativity of the 
spectrum is lost for *-algebras, which could lead to dangerous misunderstandings.} The positive elements of 
$\mathfrak{A}$ form a convex cone $\mathfrak{A}_+$ which satisfies $\mathfrak{A}_+\cap(-\mathfrak{A}_+)=\{0\}$, 
and any self-adjoint element $x\in\mathfrak{A}$ can be written as the orthogonal decomposition $x=x_+-x_-$, 
where $x_\pm=\frac{1}{2}(x\pm|x|)\in\mathfrak{A}_+$, $|x|=\sqrt{x^*x}$ and $x_+ x_-=0$. We can, thus, impose 
the following ordering $\leq$ to the self-adjoint elements of $\mathfrak{A}$, given by $x\leq y\Leftrightarrow 
y-x\in\mathfrak{A}_+$. We see immediately that $A\leq B$ and $B\leq A$ imply $A=B$, and, by the spectral radius
formula \cite{reedsimon1,rudin}, it follows that $\mathfrak{A}_+\ni x\leq\Vert x\Vert\mathbb{1}$. Hence, when
we refer ourselves to increasing or decreasing subsets of $\mathfrak{A}_+$ and to upper limits, suprema, lower
limits or infima of subsets of $\mathfrak{A}_+$, it'll be always with respect to the ordering $\leq$ defined above.

\subsection{\label{ap2-cwalg-w}\textsc{von Neumann} algebras}

Let us now consider a state $\omega$ over $\mathfrak{A}$. In this case, the weak commutant $\pi_\omega
(\mathfrak{A})_w'$ is a C*-algebra, and coincides with the \emph{commutant} $\pi_\omega(\mathfrak{A})'\doteq
\{T\in\mathscr{B}(\mathscr{H}_\omega):[T,\pi_\omega(A)]\doteq T\pi_\omega(A)-\pi_\omega(A)T=0,\,\forall A\in
\mathfrak{A}\}$ of $\pi_\omega(\mathfrak{A})$ in $\mathscr{B}(\mathscr{H}_\omega)$. We can then strengthen the 
conclusion of Theorem \ref{ap2t5}: $\omega$ is pure if and only if $\pi_\omega$ is irreducible. Moreover, 
$\pi_\omega(\mathfrak{A})'$ is a C*-algebra of quite a special kind, for $\pi_\omega(\mathfrak{A})'=(\pi_\omega
(\mathfrak{A})')''$.

\begin{definition}\label{ap2d3}
Let $\mathfrak{R}$ be a unital C*-algebra, concretely realized as a C*-subalgebra of $\mathscr{B}(\mathscr{H})$ 
for some \textsc{Hilbert} space $\mathscr{H}$. We say that $\mathfrak{A}$ is a \textsc{von Neumann} \emph{algebra} 
if $\mathfrak{R}=\mathfrak{R}''$.
\end{definition}

We say that a primary \textsc{von Neumann} algebra $\mathfrak{R}$, i.e., $\mathfrak{Z}(\mathfrak{F})\doteq
\mathfrak{F}\cap{}\mathfrak{F}'=\mathbb{C1}$, is simply a \emph{factor}. The \emph{Double Commutant Theorem} 
of \textsc{von Neumann} tells us that a nontrivial *-subalgebra $\mathfrak{R}$ of $\mathscr{B}(\mathscr{H})$ 
is \textsc{von Neumann} algebra if and only if $\mathfrak{R}$ is closed in the strong, weak and 
$\sigma$\emph{-weak} (also called \emph{ultraweak}) topologies of $\mathscr{B}(\mathscr{H})$. The latter 
is given by the seminorms $\sum^\infty_{i=1}|\langle\Psi_i,.\Phi_i\rangle|$, where $\{\Psi_i\}_i,\{\Phi_i\}_i
\subset\mathscr{H}$ satisfy $\sum^\infty_{i=1}\Vert\Psi_i\Vert^2,\sum^\infty_{i=1}\Vert\Phi_i\Vert^2<\infty$. 
Moreover, given a C*-subalgebra $\mathfrak{A}\in\mathscr{B}(\mathscr{H})$, we have that $\mathfrak{A}''$ is
the closure of $\mathfrak{A}$ in the strong, weak and $\sigma$-weak topologies. We cite tho results about the
latter topology:

\begin{proposition}\label{ap2p3}
Let $\mathfrak{R}$ be a \textsc{von Neumann} algebra, and $\mathfrak{I}$ a $\sigma$-weakly closed two-sided
ideal in $\mathfrak{R}$. Then $\mathfrak{I}$ is a *-ideal, and there exists a projection $E\in\mathfrak{Z}
(\mathfrak{R})$ such that $\mathfrak{I}=E\mathfrak{R}$.
\begin{quote}{\small\scshape Proof.\quad}
{\small\upshape See Proposition 2.4.22 of \cite{bratteli1}.~\hfill~$\Box$}
\end{quote}
\end{proposition}

\begin{theorem}\label{ap2t6}
Given two \textsc{von Neumann} algebras $\mathfrak{R}$ and $\mathfrak{S}$, a *-morphism from $\mathfrak{R}$ 
to $\mathfrak{S}$ is necessarily $\sigma$-weakly continuous.
\begin{quote}{\small\scshape Proof.\quad}
{\small\upshape See Theorem 2.4.23 of \cite{bratteli1}.~\hfill~$\Box$}
\end{quote}
\end{theorem}

If $\mathfrak{A}$ is a C*-subalgebra of $\mathscr{B}(\mathscr{H})$, one sees that $\mathfrak{A}$ is 
dense in $\mathfrak{A}''$ in the strong, weak and $\sigma$-weak topologies of $\mathscr{B}(\mathscr{H})$.\\

Notice that, if an element $T\in\mathscr{B}(\mathscr{H})$ commutes with $x=x^*\in\mathfrak{R}$, it 
commutes with all spectral projections of $x$. Hence, this projections belong to $\mathfrak{R}$. 
Employing the decomposition of an element of $\mathscr{B}(\mathscr{H})$ in a linear combination of 
self-adjoint elements, it follows that the linear hull of the projections of $\mathfrak{R}$ is strongly 
dense in $\mathfrak{R}$; making use of the decomposition in a linear combination of unitary
elements \cite{reedsimon1}, one can show that $x\in\mathfrak{R}$ if and only if the partial isometry 
$u$ and the positive element $|x|$ occurring in the polar decomposition $x=u|x|$ of $x$, as well as all 
spectral projectors of $|x|$, belong to $\mathfrak{R}$ \cite{bratteli1}. More in general, given a densely 
defined, closed linear operator $T$ in $\mathscr{H}$, we say that $T$ is \emph{affiliated} to $\mathfrak{R}$ 
if $\mathfrak{R}'D(T)\subseteq D(T)$ and $Tx\supseteq xT$ for all $x\in\mathfrak{R}'$. Equivalently, $T$ is
affiliated to $\mathfrak{R}$ if and only if the partial isometry $U$ and the spectral projections of the
positive operator $|T|$ occurring in the polar decomposition $T=U|T|$ of $T$ belong to $\mathfrak{R}$.\\

Given a \textsc{von Neumann} algebra $\mathfrak{R}$, its \emph{predual} $\mathfrak{R}_*$ consists of the 
$\sigma$-weakly continuous linear functionals on $\mathfrak{R}$. This is a \textsc{Banach} subspace of the 
dual $\mathfrak{R}^*$ of the \textsc{Banach} space $\mathfrak{R}$, and the latter coincides precisely with 
the dual $(\mathfrak{R}_*)^*$ of $\mathfrak{R}_*$ \cite{bratteli1}. The typical example if $\mathscr{B}
(\mathscr{H})$ ($\mathscr{H}$ separable), whose predual is given by the ideal of trace class operators 
$\mathscr{L}_1(\mathscr{H})$, and the dual pairing is given by the trace in $\mathscr{H}$. We have, in 
general, the following structural result, due to \textsc{Sakai}:

\begin{theorem}[Sakai \upshape\cite{tak1}]\label{ap2t7}
A C*-algebra $\mathfrak{A}$ s *-isomorphic to a \textsc{von Neumann} algebra if and only if it's the dual 
of some \textsc{Banach} space.~\hfill~$\Box$
\end{theorem}

Due to Theorem \ref{ap2t7}, it's customary to call such ``abstract'' \textsc{von Neumann} algebras 
\emph{W*-algebras} ($W$ corresponds to \emph{``weak''}, denoting closure in the weak topology), but here 
we'll use both terms alternately when dealing with such C*-algebras. We see also that the topology most 
naturally associated to a \textsc{von Neumann} algebra is the \emph{$\sigma$-weak} topology, and it's 
this one we refer to when, for instance, we employ the ``topologized'' version of Theorems \ref{ap2t1} to 
\ref{ap2t4} in the context of W*-algebras, according to Remark \ref{ap2r1}. However, as one can see from 
Proposition 2.4.2 of \cite{bratteli1}, the product operation is separately $\sigma$-weakly continuous in
both variables, but it's \emph{not jointly} $\sigma$-weakly continuous is the W*-algebra in question is
faithfully realized in an \emph{infinite}-dimensional \textsc{Hilbert} space.\\

Given a C*-algebra $\mathscr{A}$ and a state $\omega$, we have a \textsc{von Neumann} algebra in 
$\mathscr{H}_\omega$ naturally associated to this pair, given by $\pi_\omega(\mathfrak{A})''$. More in 
general, it's possible to show that the double dual $\mathfrak{A}^{**}$ of $\mathfrak{A}$ is a C*-algebra 
*-isomorphic to $\pi_u(\mathfrak{A})''$, called \emph{(universal) enveloping W*-algebra} of $\mathfrak{A}$. 
This is the smallest W*-algebra containing $\mathfrak{A}$, and is *-isomorphic to $\pi_\omega(\mathfrak{A})''$ 
if $\omega$ is faithful. A state $\omega$ over a \textsc{von Neumann} algebra $\mathfrak{R}$ in $\mathscr{H}$ 
is called \emph{normal} if there exists a trace class element $\rho\in\mathscr{B}(\mathscr{H})$ such that 
$\mbox{Tr}(\rho)=1$ and $\omega(x)=\mbox{Tr}(\rho x)$. Any state $\omega$ over a C*-algebra $\mathfrak{A}$ 
admits a unique normal extension to $\mathfrak{A}^{**}$, which is faithful if and only if $\omega$ is 
\cite{tak1}.\\

Normal states possess the following characterization:

\begin{theorem}\label{ap2t8}
Let $\omega$ be a state over a \textsc{von Neumann} algebra $\mathfrak{R}$ in the \textsc{Hilbert} space
$\mathfrak{H}$. Then the following are equivalent:
\begin{itemize}
\item[(i)\quad] $\omega$ is normal;
\item[(ii)\quad] $\omega$ is $\sigma$-weakly continuous;
\item[(iii)\quad] Given any increasing net $\{A_\alpha\}\subset\mathfrak{R}_+$ bounded from above, one has 
$\omega(\sup_\alpha A_\alpha)=\sup_\alpha\omega(A_\alpha)$.
\end{itemize}
\begin{quote}{\small\scshape Proof.\quad}
{\small\upshape See Theorem 2.4.21 of \cite{bratteli1}.~\hfill~$\Box$}
\end{quote}
\end{theorem}

A comment about the condition (iii) of Theorem \ref{ap2t8} is in order. It's not obvious that, given an
increasing net $\{A_\alpha\}\subset\mathfrak{R}_+$ bounded from above, there exists $\sup_\alpha A_\alpha\in
\mathfrak{R}$. That this is nevertheless true, it can be seen as follows: let $\mathfrak{R}_\alpha$ be the 
closure of $\{A_\beta:\beta>\alpha\}$ in the weak topology of $\mathscr{B}(\mathscr{H})$. As the closed unit
ball $\mathscr{B}(\mathscr{H})_1$ of the latter is weakly compact,\footnote{Notice that any $\mathscr{B}
(\mathscr{H})$ is a \textsc{von Neumann} algebra and, as such, it's the dual of a \textsc{Banach} space -- 
the latter, in this case, is formed by its trace class elements \cite{bratteli1}. The assertion thus
follows from the \textsc{Banach-Alaoglu} theorem \cite{reedsimon1}.} it follows from the cone structure 
of $\mathfrak{R}_+$ that there exists $A\in\cap_\alpha\mathfrak{R}_\alpha$. For all $A_\alpha$, $\{B\in
\mathscr{B}(\mathscr{H})_+:B\geq A_\alpha\}$ is $\sigma$-weakly closed and contains $\mathfrak{R}_\alpha$, 
hence $A\geq A_\alpha$ for all $\alpha$. Any element $B$ majorizing $\{A_\alpha\}$ majorizes its weak closure 
and, thus, $B\geq A$. Hence, $A=\sup_\alpha A_\alpha$. Moreover, $A_\alpha\rightarrow A$ in the $\sigma$-weak
topology, for the latter coincides with the weak topology in $\mathscr{B}(\mathscr{H})_1$ \cite{bratteli1}, 
thus establishing (ii)$\Rightarrow$(iii) in Theorem \ref{ap2t8}. We say in general that a positive linear 
functional $\phi$ is a \textsc{von Neumann} algebra $\mathfrak{R}$ is \emph{normal} if it satisfies condition 
(iii) in Theorem \ref{ap2t8}.\footnote{It was proved by \textsc{Kadison} \cite{tak1} that a C*-algebra 
$\mathfrak{A}$ is a W*-algebra if and only if any increasing net $\{A_\alpha\}\subset\mathfrak{A}_+$ 
bounded from above has a supremum $\sup_\alpha A_\alpha\in\mathfrak{A}$ and the stated that satisfy condition 
(iii) in Theorem \ref{ap2t8} separate elements in $\mathfrak{A}_+$, i.e., given $0\neq x\in\mathfrak{A}_+$, 
there exists a normal state $\omega$ such that $\omega(A)\neq 0$.}\\

We conclude this Subsection with the classification of \textsc{von Neumann} algebras introduced by 
\textsc{Murray} and \textsc{von Neumann}: let \[\mathscr{P}(\mathfrak{R})\doteq\{E\in\mathfrak{R}:
E=E^*=E^2\}\] the set of \emph{projections} of the \textsc{von Neumann} algebra $\mathfrak{R}$.
$\mathscr{P}(\mathfrak{R})$ has a partial ordering $\leq$ inherited from $\mathfrak{R}_+$, and given any 
$E,F\in\mathscr{P}(\mathfrak{R})$, there exists the \emph{complement} $E^\perp\doteq\mathbb{1}-E$ of $E$, 
the \emph{infimum} $E\wedge F\doteq s-\lim_{n\rightarrow\infty}(EF)^n$ (by $s-\lim$ it's understood the limit 
in the strong topology of $\mathfrak{R}$) and the \emph{supremum} $E\vee F\doteq(E^\perp\wedge F^\perp)^\perp$ 
of $E$ and $F$ with respect to the ordering $\leq$ \footnote{Given a partially ordered set $(S,\leq)$, the 
\emph{infimum} of $x,y\in S$ is an element $x\wedge y\in S$ such that $x\wedge y\leq x,y$ and $z\leq x,y
\Rightarrow z\leq x\wedge y$, the \emph{supremum} of $x,y\in S$ is an element $x\vee y\in S$ such that $x,y
\leq x\vee y$ and $x,y\leq w\Rightarrow x\vee y\leq w$, and a \emph{complement} operation is a bijection 
${}^\perp:S\rightarrow S$ which satisfies $(x^\perp)^\perp=x$, $(x\wedge y)^\perp=x^\perp\vee y^\perp$ and 
$(x\vee y)^\perp=x^\perp\wedge y^\perp$. If any pair $x,y\in S$ possesses a supremum and an infimum, we
say that $(S,\leq,\vee,\wedge)$ is a \emph{lattice} -- in particular, any lattice is \emph{directed}, i.e., 
given $x,y\in S$, there exists $z\geq x,y$. If, moreover, $S$ has a complement operation ${}^\perp$, then
we say that $(S,\leq,\vee,\wedge,{}^\perp)$ is an \emph{orthocomplemented lattice}.} -- that is, 
$(\mathscr{P}(\mathfrak{R}),\leq,\vee,\wedge,{}^\perp)$ is an orthocomplemented lattice. \\

We introduce the following equivalence relation in $\mathscr{P}(\mathfrak{R})$: given $E,F\in\mathscr{P}
(\mathfrak{R})$, we say that $E\sim F$ if there exists $W\in\mathfrak{R}$ such that $W^*W=E$ and $WW^*=F$ 
(i.e., $W$ is a \emph{partial isometry} with source $E$ and target $F$). We say yet that $E$ is a 
\emph{subprojection} of $F$ if $E\leq F$ ($E$ is \emph{proper} if moreover $E\neq F$), and define the
following order relation in $\mathscr{P}(\mathfrak{R})/\sim\ni[E],[F]$: $[E]\neq[F]$ if $E$ is equivalent 
to a subprojection of $F$ (the definition clearly doesn't depend on the choice of representative). 
Employing a faithful *-representation of $\mathfrak{R}$ in a \textsc{Hilbert} space $\mathscr{H}$, it's
easy to prove that $E\leq F$ if and only if $EF=E$, justifying our terminology. Moreover, using the
one-to-one correspondence $E\mapsto\mbox{Ran}E$ which exists between elements of $\mathscr{P}(\mathscr{B}
(\mathscr{H}))$ and closed linear subspaces of $\mathscr{H}$, we see that \[\mbox{dim Ran}E\leq+\infty
\mbox{ if and only if there is no }0\neq F\lneq E\mbox{ such that }F\sim E.\] We can then \emph{define} 
abstractly that $E\in\mathscr{P}(\mathfrak{R})$ is \emph{finite} if it satisfies the condition above, which 
obviously doesn't depend on the realization of $\mathfrak{R}$ in $\mathscr{H}$. We say otherwise that $E$ 
is \emph{infinite}.\\

\begin{definition}[Murray-von Neumann]\label{ap2d4}
Let $\mathfrak{R}$ be a \textsc{von Neumann} algebra. A function $d:\mathscr{P}(\mathfrak{R})\rightarrow
\bar{\mathbb{R}}_+\cup\{+\infty\}$ is said to be a \emph{dimensional functions} if $d(E)=0\Rightarrow E=0$, 
$E\sim F\Rightarrow d(E)=d(F)$ and $E\leq F^\perp\Rightarrow d(E\vee F)=d(E+F)=d(E)+d(F)$ -- such properties 
determine $d$ up to a multiple $\lambda\in\mathbb{R}\smallsetminus\{0\}$, and $d(E)<+\infty$ if and only if 
$E$ is finite. We say that $\mathfrak{R}$ is \emph{finite} if $d(\mathbb{1})<+\infty$, and \emph{infinite} 
otherwise. $\mathfrak{R}$ is \emph{semifinite} if any $E\in\mathscr{P}(\mathfrak{R})$ possesses a finite 
subprojection $F\neq 0$, and \emph{purely infinite} if any $E\in\mathscr{P}(\mathfrak{R})$ is infinite.\\

If $\mathfrak{R}$ is a factor, we say that $\mathfrak{R}$ is:
\begin{itemize}
\item\emph{Type I${}_n$} if $\mbox{\upshape Ran}d=\lambda\{0,1,\ldots,n\}$, $n=0,\ldots,+\infty$ 
(equivalently, $\mathfrak{R}$ is type I if it has a minimal projection $E\neq 0$, i.e., $F\neq 
E\Rightarrow F=E$). $\mathfrak{R}$ is finite if and only if $n<+\infty$;
\item\emph{Type II} if it's semifinite, but not type I. $\mathfrak{R}$ is, then, \emph{type II${}_1$} 
if $\mbox{\upshape Ran}d=\lambda[0,1]$ (equivalently, if $\mathfrak{R}$ is finite) and \emph{type 
II${}_\infty$} if $\mbox{\upshape Ran}d=\bar{\mathbb{R}}_+\cup\{+\infty\}$;
\item\emph{Type III} if it's purely infinite, i.e., $\mbox{\upshape Ran}d=\{0,+\infty\}$.
\end{itemize}
\end{definition}

Several of the more advanced results about \textsc{von Neumann} algebras make use of the concept of 
\emph{weight}, which generalizes the concept of positive linear functional. Given a W*-algebra $\mathfrak{R}$, 
a \emph{weight} in $\mathfrak{R}$ is a function $\omega:\mathfrak{R}_+\rightarrow\bar{\mathbb{R}}_+\cup\{+
\infty\}$ (where we set $0.+\infty=0$) which satisfies (i) $\omega(A)+\omega(B)=\omega(A+B)$ and (ii) $\omega
(\alpha A)=\alpha\omega(A)$, for all $A,B\in\mathfrak{R}_+$, $\alpha\in\bar{\mathbb{R}}_+$. A weight $\tau$ is
said to be \emph{tracial} (or simply a \emph{trace}) if $\omega(A^*A)=\omega(AA^*)$, for all $A$. Writing 
$\mathfrak{R}_{\omega+}\doteq\{A\in\mathfrak{R}_+:\omega(A)<+\infty\}$ and $\mathfrak{L}_\omega\doteq\{A\in
\mathfrak{R}:\omega(A^*A)<+\infty\}$, we have immediately that $\mathfrak{R}_{\omega+}$ is an \emph{hereditary} 
cone in $\mathfrak{R}_+$, i.e., $0\leq A\leq B\in\mathfrak{R}_{\omega+}\Rightarrow A\in\mathfrak{R}_{\omega+}$, 
and $\mathfrak{R}_\omega\doteq\mbox{span}_{\mathbb{C}}\mathfrak{R}_{\omega+}=\mathfrak{L}_\omega^*\mathfrak{L}_\omega$. 
Hence, the weight $\omega$ extends by linearity to a positive linear functional in the *-algebra 
$\mathfrak{R}_\omega$ -- in particular, if $\omega$ is a trace, then $\omega(AB)=\omega(BA)$ for all $A,B\in
\mathfrak{R}_\omega$. We say that a weight $\omega$ is \emph{normal} if it satisfies condition (iii) of 
Theorem \ref{ap2t8}, \emph{faithful} if $\omega\restr{\mathfrak{R}_\omega}$ is faithful, and \emph{semifinite} 
if $\mathfrak{R}_\omega$ is $\sigma$-weakly dense in $\mathfrak{R}$. It's possible to prove that \emph{any} 
\textsc{von Neumann} algebra $\mathfrak{R}$ admits a normal, faithful semifinite weight, and that $\mathfrak{R}$ 
is semifinite if and only if it admits a normal, faithful semifinite \emph{trace} $\tau$ \cite{tak1,tak2} -- 
the reciprocal of the last assertion is immediate, for if $E$ is a projection in $\mathfrak{R}_\tau$, then $E$ 
is necessarily finite. \\

$\sigma$\emph{-finite} \textsc{von Neumann} algebras $\mathfrak{R}$, i.e., such that any mutually orthogonal
subset $S\in\mathscr{P}(\mathscr{R})$ (i.e., which satisfies $E\leq F^{\perp}$ for all $E,F\in S$) is 
\emph{countable}, can be decomposed in a ``direct integral'' of factors, in terms of the set of central
projections $\mathfrak{Z}(\mathfrak{R})\cap{}\mathscr{P}(\mathscr{R})$. $\mathfrak{R}$ is $\sigma$-finite
if and only if it admits a normal, faithful state $\omega$ \cite{bratteli1}. Thus, one can say that, for such
algebras, the classification task is reduced to the classification of factors. A refinement of the classification 
given by Definition \ref{ap2d4}, for type-III factors, will be presented succinctly at the end of Section 
\ref{ap2-cwalg-totak}.

\subsection{\label{ap2-cwalg-totak}\textsc{Tomita-Takesaki} modular theory}

We're now in a position to discuss a structure typical of \textsc{von Neumann} algebras, extremely useful 
and powerful. Consider a C*-algebra $\mathfrak{A}$ and a \emph{faithful} state $\omega$. We then identify 
the enveloping W*-algebra $\mathfrak{A}^{**}=\pi_\omega(\mathfrak{A})''\doteq\mathfrak{R}$ in the 
\textsc{Hilbert} space $\mathscr{H}_\omega\doteq\mathscr{H}$, which possesses, on its turn, a cyclic and 
separating vector $\Omega$. Let's now define the densely defined antilinear operators $S_0$ (already 
mentioned \emph{en passant} in Section \ref{ap2-staralg}) and $F_0$ as \[D(S_0)=\mathfrak{R}\Omega,\,S_0 
A\Omega\doteq A^*\Omega;\;D(F_0)=\mathfrak{R}'\Omega,\,S_0 A'\Omega\doteq A'^*\Omega\] ($F_0$ is densely 
defined, for a vector $\Phi$ is cyclic for a \textsc{von Neumann} algebra $\mathfrak{S}$ if and only if 
it's separating for $\mathfrak{S}'$). It follows that $S_0\subseteq F^*_0$ and $F_0\subseteq S^*_0$, and, 
thus, $S_0$ as well as $F_0$ are closable. One can yet prove that \cite{bratteli1}
\begin{equation}\label{ap2e1}
F^*_0=\bar{S}_0\doteq S\mbox{ and }S^*_0=\bar{F}_0\doteq F.
\end{equation}

\begin{definition}\label{ap2d5}
Let $J$ be the \emph{anti-unitary} operator (i.e., $J$ is antilinear, invertible and such that $\langle 
J\Phi,J\Psi\rangle=\overline{\langle\Phi,\Psi\rangle}=\langle\Psi,\Phi\rangle$ for all $\Phi,\Psi
\in\mathscr{H}$) and $\Delta$ the positive operator of the polar decomposition $S=J\Delta^{\frac{1}{2}}$ of 
$S$. We say that $\Delta$ is the \emph{modular operator}, and $J$, the \emph{modular conjugation}, associated
to $(\mathfrak{R},\Omega)$.
\end{definition}

The name has its origin in harmonic analysis: when we define the involution of the convolution *-algebra 
of compactly supported continuous functions on a locally compact topological group, we need to introduce 
the \textsc{Radon-Nik\'odym} derivative of the right-invariant \textsc{Haar} measure with respect to the left-invariant
\textsc{Haar} measure, called \emph{modular function} (this function is equal to 1 for Abelian or compact
groups). It follows from (\ref{ap2e1}) that
\begin{equation}\label{ap2e2}
\Delta=FS,\,\Delta^{-1}=SF,\,F=J\Delta^{-\frac{1}{2}},\,J=J^*\,(\mbox{and, hence, }J^2=\mathbb{1})
\mbox{ and }\Delta^{-\frac{1}{2}}=J\Delta^{\frac{1}{2}}J.
\end{equation}

The most important result involving $J$ and $\Delta$ is the

\begin{theorem}[Tomita-Takesaki \upshape\cite{bratteli1,tak2}]\label{ap2t9}
Let $\mathfrak{R}$ be a \textsc{von Neumann} algebra in $\mathscr{H}$ with cyclic and separating vector 
$\Omega$, and $\Delta$ and $J$ respectively the modular operator and the modular conjugation associated to 
$(\mathfrak{R},\Omega)$. Then, \[J\mathfrak{R}J=\mathfrak{R}'\mbox{ and }\Delta^{it}\mathfrak{R}
\Delta^{-it}=\mathfrak{R},\,\forall t\in\mathbb{R}.\]~\hfill~$\Box$
\end{theorem}

In particular, $t\mapsto\Delta^{it}$ defines a strongly continuous, one-parameter groups of unitary operators 
in $\mathscr{H}$, with generator $\log\Delta$, defined by the spectral theorem, which, by Theorem \ref{ap2t9}, 
implements a $\sigma$-weakly continuous group of *-automorphisms of $\mathfrak{R}$ and of $\mathfrak{R}'$ 
(in the latter, the sign of $t$ is changed), called \emph{modular group}. For $\mathfrak{R}=\pi_\omega
(\mathfrak{A})''$, where $\mathfrak{A}$ is a C*-algebra, and $\Omega$ the cyclic and separating GNS vector 
associated to a state $\omega$, we write, then, for each $x\in\mathfrak{A}$:

\begin{eqnarray}
\label{ap2e3} & j(x)\doteq\pi^{-1}_\omega(J\pi_\omega(x)J),\mbox{ and} & \\
\label{ap2e4} & \sigma^\omega_t(x)\doteq\pi^{-1}_\omega(\Delta^{it}\pi_\omega(x)\Delta^{-it}). & 
\end{eqnarray}

Notice that the formulae (\ref{ap2e3}) and (\ref{ap2e4}) assume values in $\mathfrak{A}^{**}$, 
i.e., $\pi^{-1}_\omega$ denotes, in a slight but innocuous abuse of notation, its unique $\sigma$-weakly 
continuous extension to $\pi_\omega(\mathfrak{A})''$. A formula of the highest importance which follows from
the definition of the modular objects is:

\begin{eqnarray}
 & \langle\Delta^{\frac{1}{2}}\pi_\omega(x)\Omega,\Delta^{\frac{1}{2}}\pi_\omega(y)\Omega\rangle=
\langle J\pi_\omega(x^*)\Omega,J\pi_\omega(y^*)\Omega\rangle=\langle \pi_\omega(y^*)\Omega,
\pi_\omega(x^*)\Omega\rangle & \nonumber\\ & \Downarrow & \nonumber\\\label{ap2e5} & \omega
(\sigma^\omega_{i/2}(x)\sigma^\omega_{-i/2}(y))=\omega(yx). &
\end{eqnarray}

The \textsc{Tomita-Takesaki} modular theory can be extended to arbitrary \textsc{von Neumann} algebras 
if we employ a normal, faithful semifinite weight instead of a normal, faithful state. The proofs, however, 
are considerably more complicated.\\

Finally, there is a rigid relation between the modular groups associated to two normal, faithful states (or, 
more in general, semifinite weights) $\phi,\omega$ over a \textsc{von Neumann} algebra $\mathfrak{R}$, 
given by the

\begin{theorem}[Connes]\label{ap2t10}
There exists a one-parameter family $t\mapsto\Gamma_t\doteq(D\phi:D\omega)_t$ of unitary elements of 
$\mathfrak{R}$ which satisfies $\sigma^\phi_t(A)=\Gamma_t\sigma^\omega_t(A)\Gamma^*_t$ and the cocycle 
relation $\Gamma_{t+s}=\Gamma_t\sigma^\omega_t(\Gamma_s)$. In particular, if $\phi=\omega(U.U^*)$, where 
$U\in\mathfrak{R}$ is a unitary element, then necessarily $\Gamma_t=U^*\sigma^\omega_t(U)$.
\begin{quote}{\small\scshape Proof.\quad}
{\small\upshape See Theorem 2.7.16 of \cite{bratteli1} and Theorem 5.3.34 of \cite{bratteli2}.
~\hfill~$\Box$}
\end{quote}
\end{theorem}

The expression $S(\phi|\omega)\doteq\frac{d}{dt}\omega((D\phi:D\omega)_t)\restr{t=+0}$ also defines the 
\emph{relative entropy} of $\phi$ with respect to $\omega$ \cite{araki3,bratteli2,ohyapetz}. The cocycles defined in 
Theorem \ref{ap2t10} are of fundamental importance in the classification of type-III factors performed by 
\textsc{Connes} \cite{bratteli1,tak2}. Let $\mathfrak{R}$ be a factor ($\sigma$-finite, for simplicity), and 
$\mathbb{R}\ni u\mapsto\alpha_g$ a one-parameter, $\sigma$-weakly continuous group of *-automorphisms of 
$\mathfrak{R}$, so that it's meaningful to talk about the $\sigma$\emph{-weak integral} \[L^1(\mathbb{R})\ni 
f\mapsto\alpha_f(A)\doteq\int f(t)\alpha_t(A)dt,\,A\in\mathfrak{R}.\] That is, the functional $\mathfrak{R}_*
\ni\phi\mapsto\phi(\alpha_f(A))\doteq\int f(t)\phi(\alpha_t(A))dt$ defines an element of $(\mathfrak{R}_*)^*
=\mathfrak{R}$ for each $A\in\mathfrak{R}$, $f\in L^1(\mathbb{R})$. Define the \emph{fixed-point subalgebra} 
$\mathfrak{R}^\alpha\doteq\{A\in\mathfrak{R}:\alpha_t(A)=A,\,\forall t\in\mathbb{R}\}$. The \textsc{Arveson} 
\emph{spectrum} of $\alpha_.$ is given by the intersection of the supports of the $\mathfrak{R}$-valued
tempered distributions $f\mapsto\alpha_f(A)$ through $A\in\mathfrak{R}$, or, equivalently, \[\sigma(\alpha)
\doteq\mathbb{R}\smallsetminus\bigcup\{U\subset\mathbb{R}\mbox{ open: }\alpha_f(A)=0,\,\forall A\in
\mathfrak{R},f\in L^1(\mathbb{R}),\,\mbox{supp}f\subset U\},\] and the \textsc{Connes} \emph{spectrum} is
given by \[\Gamma(\alpha)\doteq\bigcap_{E\in\mathscr{P}(\mathfrak{R})\cap{}\mathfrak{R}^\alpha}\sigma(\alpha
\restr{\mathfrak{R}E}).\] Let's now consider the modular group $\alpha_.=\sigma^\omega_.$ associated to the
normal, faithful state $\omega$. As the spectrum of a self-adjoint operator is invariant under the adjoint
action of a unitary operator \cite{reedsimon1}, it follows from Theorem \ref{ap2t10} that $\Gamma(\mathfrak{R})
\doteq\Gamma(\sigma^\omega)$ doesn't depend on $\omega$. One can yet prove \cite{bratteli1,tak2} that 
$\mbox{exp}\Gamma(\mathfrak{R})=S(\mathfrak{R})\smallsetminus\{0\}$ is a multiplicative subgroup of 
$\mathbb{R}_+$, where $S(\mathfrak{R})=\bigcap_{\omega\in\mathfrak{R}_*,\,\mbox{\scriptsize Ann}\omega=\{0\}}
\sigma(\Delta_\omega)$ and $\Delta_\omega$ is the \textsc{Tomita-Takesaki} modular operator associated to 
$\omega$, and that $\Gamma(\mathfrak{R})=\{0\}$ if and only if $\mathfrak{R}$ is \emph{semifinite}. Hence 
it follows that the remaining possibilities for $S(\mathfrak{R})$ by virtue of its multiplicative structure 
\begin{itemize}
\item $S(\mathfrak{R})=\{0,1\}$;
\item $S(\mathfrak{R})=\{0\}\cup\{\lambda^n:n\in\mathbb{Z}\},\,\lambda\in(0,1)$;
\item $S(\mathfrak{R})=\bar{\mathbb{R}}_+$,
\end{itemize}
can only be realized for $\mathfrak{R}$ type III.

\begin{definition}[Connes]\label{ap2d6}
We say that $\mathfrak{R}$ is \emph{type III${}_0$} if $S(\mathfrak{R})=\{0,1\}$, \emph{type III${}_\lambda$}, 
$\lambda\in(0,1)$, if $S(\mathfrak{R})=\{0\}\cup\{\lambda^n:n\in\mathbb{Z}\},\,\lambda\in(0,1)$, and \emph{type 
III${}_1$} if $S(\mathfrak{R})=\bar{\mathbb{R}}_+$.
\end{definition}

Of particular interest in Quantum Field Theory are the hyperfinite factors (a \textsc{von Neumann} algebra 
$\mathfrak{R}$ is \emph{hyperfinite} or \emph{injective} if it's generated by the union of a sequence of matrix 
algebras of increasing dimension) of type $III_1$. These typically appear as thermodynamic limits of local
algebras associated to a net of local observables, such as the \textsc{Araki-Woods} factor, given by the
thermodynamic limit of algebras of observables of spin systems at finite temperature \cite{bratteli2}. It was 
proven by \textsc{Haagerup}, completing former partial results of \textsc{Connes}, that the \textsc{Araki-Woods} 
factor is the \emph{unique} hyperfinite, type-III${}_1$ factor up to *-isomorphism \cite{tak3}. We finalize this 
Section with two useful criteria to determine if $\mathfrak{R}$ is type III.

\begin{theorem}\label{ap2t11}
Given a normal, faithful state $\omega$, $\mathfrak{R}$ is type III if and only if $\sigma^\omega_.
\restr{\mathfrak{R}E}$ is \emph{not} a group of \emph{internal} automorphisms (i.e., there is no $\sigma$-weakly 
continuous group $\mathbb{R}\ni t\mapsto U_t$ of unitaries of $\mathfrak{R}$ such that $\sigma^\omega_t(.)=U^*_t.
U_t$, $\forall t\in\mathbb{R}$) for \emph{any} $E\in\mathscr{P}(\mathfrak{R})\cap{}\mathfrak{Z}(\mathfrak{R})$.
\begin{quote}{\small\scshape Proof.\quad}
{\small\upshape See Theorem 2.7.17 of \cite{bratteli1} and the discussion preceding Theorem 5.3.35 of 
\cite{bratteli2}.
~\hfill~$\Box$}
\end{quote}
\end{theorem}

The second criterion is due to \textsc{Driessler} \cite{driessler2}:

\begin{theorem}[\upshape\cite{driessler2}]\label{ap2t12}
Let $\mathscr{H}$ be a separable \textsc{Hilbert} space, $\mathfrak{R}$ a \textsc{von Neumann} subalgebra of 
$\mathscr{B}(\mathscr{H})$ and $\Omega$ a cyclic and separating for $\mathfrak{R}$. Suppose there exists an
infinite \textsc{von Neumann} subalgebra $\mathscr{R}_1\subset\mathfrak{R}$ and $\{\alpha_n\}_{n\in\mathbb{Z}_+}$ 
a sequence of *-automorphisms of $\mathscr{B}(\mathscr{H})$ satisfying:
\begin{enumerate}
\item $\alpha_n(\mathfrak{R}_1)\subset\mathfrak{R}_1$, for all $n$;
\item $w-\lim_{n\rightarrow\infty}\alpha_n(A_1)=\omega(A_1)\mathbb{1}$, for all $A_1\in
\mathfrak{R}_1$, $0\not\equiv\omega\in\mathfrak{R}_{1*}$;
\item $s-\lim_{n\rightarrow\infty}[A,\alpha_n(A_1)]=\{0\}$, for all $A\in\mathfrak{R}$,
$A_1\in\mathfrak{R}_1$.
\end{enumerate}
Then $\mathfrak{R}$ is type III.
\begin{quote}{\small\scshape Proof.\quad}
{\small\upshape Suppose that $P\neq 0$ is a finite projection. In this case, $P\mathfrak{R}P$ is finite and, as
we've seen at the end of Subsection \ref{ap2-cwalg-w}, possesses a normal, faithful semifinite weight $\tau$. As 
$P\Omega$ is cyclic and separating for $P\mathfrak{R}P$ in $P\mathscr{H}$, it follows from the analog of Theorem 
2.5.31 in \cite{bratteli1} for weights \cite{tak2} that $\tau$ is implemented by a single tracial vector $\xi\in 
P\mathscr{H}$. We obtain, then, that \[\lim_n\langle\xi,\alpha_n(A)P\alpha_n(B)\xi\rangle\stackrel{3.)}{=}\lim_n
\langle\xi,\alpha_n(AB)\xi\rangle\stackrel{2.)}{=}\langle\xi,AB\xi\rangle=\langle\xi,BA\xi\rangle.\] The limit 
above thus defines a normal, faithful and \emph{finite} trace $\tau_1$ in $\mathfrak{M}_1$, i.e., $\tau_1(A)<+
\infty$ for all $A\in\mathfrak{R}_1$, implying that $\mathfrak{R}_1$ is finite, which is absurd.~\hfill~$\Box$}
\end{quote}
\end{theorem}

\subsection{\label{ap2-cwalg-kms}C*- e W*-dynamical systems. KMS condition}

We shall now arrive at a physical interpretation for the formula (\ref{ap2e5}). First, notice that,
given $A\in\mathfrak{R}$, the $\sigma$-weak integrals
\begin{equation}\label{ap2e6}
A_n=\sqrt{\frac{n}{\pi}}\int\sigma^\omega_t(A)e^{-nt^2}dt,\,n\in\mathbb{Z}_+
\end{equation}
\[\left(\mbox{i.e., }\phi(A_n)=\sqrt{\frac{n}{\pi}}\int\phi(\sigma^\omega_t(A))e^{-nt^2}dt\mbox{ for all }\phi\in
\mathfrak{F}_*\right)\] define uniquely a sequence of elements $(A_n)$ which $\sigma$-weakly converge to $A$ 
(see Proposition 2.5.18 of \cite{bratteli1}), and such that each $A_n$ is an \emph{analytic element} for 
$\sigma^\omega_t$, i.e., there exists $\lambda>0$ and a function $f_{A_n}:\mathbb{R}+i(-\lambda,\lambda)
\rightarrow\mathfrak{R}$ such that $f(t)=\sigma^\omega_t(A)$ for all $t\in\mathbb{R}$ and $z\mapsto\phi(f(z))$ 
is analytic in $\mathbb{R}+i(-\lambda,\lambda)$ for all $\phi\in\mathfrak{R}_*$. Hence, $\sigma^\omega_t$ has a
$\sigma$-weakly dense set of analytic elements $B$ in $\mathfrak{R}$. For such, we write $f_B(z)=\sigma^\omega_z(B)$.\\

Let's return to (\ref{ap2e5}). Choosing $A=\mathbb{1}$ and $B$ an analytic element of $\mathfrak{R}$ in the
strip $\mathbb{R}+i(-1/2,1/2)$ (this can always be obtained by an appropriate rescaling of $f_B$), we have that
the function $F(z)=\omega(\sigma^\omega_{z+i/2}(B))$ is continuous and bounded in $\mathbb{R}+i[-1,0]$, analytic 
in $\mathbb{R}+i(-1,0)$, and \emph{periodic} with period 1 in $\Im z$, due to (\ref{ap2e5}). Thus, $F$ admits
a \emph{bounded} analytic extension to all $\mathbb{C}$ (inclusive in $\mathbb{R}+i\mathbb{Z}$, by 
\textsc{Schwarz}'s reflection principle). Hence, by the \textsc{Liouville} theorem, $F$ must be constant, i.e., $F(z)
=\omega(B)$ for all $z\in\mathbb{C}$, in particular for $z=t\in\mathbb{R}$. Since the analytic elements are dense 
in $\mathfrak{R}$, it follows that $\omega$ is \emph{invariant} under $\sigma^\omega_t$.\\

More in general, given $A,B\in\mathfrak{R}$, we can choose sequences $(A_n),(B_n)$ such that $\Vert A_n\Vert\leq
\Vert A\Vert$, $\Vert B_n\Vert\leq\Vert B\Vert$ and $A^{(*)}_n\Omega\rightarrow A^{(*)}\Omega$, $B^{(*)}_n\Omega
\rightarrow B^{(*)}\Omega$ (this is always possible by \textsc{Kaplansky}'s density theorem -- see 
\cite{bratteli1}, Theorem 2.4.16). In this case, it follows from (\ref{ap2e5}) and \textsc{Phragm\'en-Lindelöf}'s 
Three Line Theorem (see Theorem 6.4 in \cite{lang1}) that $\omega(A_n f_{B_n}(z))$ converges uniformly in 
$\mathbb{R}+i[-1/2,1/2]$ to a function $F'_{A,B}$, continuous and bounded in $\mathbb{R}+i[-1/2,1/2]$ and analytic 
in $\mathbb{R}+i(-1/2,1/2)$ (see Proposition 5.3.7 in \cite{bratteli2} for a more detailed proof), such that, for
all $t\in\mathbb{R}$, 
\begin{equation}\label{ap2e7}
F_{A,B}'(t+i/2)=\omega(A\sigma^\omega_t(B))\mbox{ and }F_{A,B}'(t-i/2)=\omega(\sigma^\omega_t(B)A).
\end{equation}

We can recast the assertion above in in the following form: let $f\in\mathscr{S}(\mathbb{R})$ such that $\hat{f}
\in\mathscr{C}^\infty_c(\mathbb{R})$. In this case, there exists $R>0$ such that $\mbox{supp}\hat{f}\subset[-R,R]$ 
and, hence, by the \textsc{Paley-Wiener} theorem \cite{hormander3}, $f$ is an entire function which satisfies 
$|f(x+iy)|=O(x^{-\infty}e^{R|y|})$ for all $x,y\in\mathbb{R}$ (indeed, this Theorem also tells us that this estimate 
completely characterizes the $f$'s which satisfy the hypothesis above). We can then rewrite (\ref{ap2e7}) as
\begin{equation}\label{ap2e8}
\int f(t+i/2)\omega(A\sigma^\omega_t(B))dt=\int f(t-i/2)\omega(\sigma^\omega_t(B)A)dt,\,\forall 
f:\hat{f}\in\mathscr{C}^\infty_c(\mathbb{R}).
\end{equation}

\begin{definition}\label{ap2d7}
Let $\mathfrak{A}$ be a C*-algebra and $\mathbb{R}\ni t\mapsto\alpha_t$ a one-parameter group of
*-automorphisms of $\mathfrak{A}$, which we assume strongly continuous for simplicity, i.e., $\Vert\alpha_t(x)-
x\Vert\stackrel{t\rightarrow 0}{\longrightarrow}0$ for all $x\in\mathfrak{A}$. The pair $(\mathfrak{A},\alpha_t)$ 
is said to be a \emph{C*-dynamical system}. If $\mathfrak{A}$ is a \textsc{von Neumann} algebra and $\alpha$ is, 
instead, $\sigma$-weakly continuous, we say that the pair $(\mathfrak{A},\alpha_t)$ is a \emph{W*-dynamical system}. 
\end{definition}

Notice that, if $\mathfrak{A}$ is a C*-subalgebra of $\mathscr{B}(\mathscr{H})$ for some $\mathscr{H}$, the 
C*-dynamical system $(\mathfrak{A},\alpha_t)$ extends uniquely to the W*-dynamical system $(\mathfrak{A}'',
\alpha_t)$, due to Theorem \ref{ap2t6}.

\begin{definition}\label{ap2d8}
Let $(\mathfrak{A},\alpha_t)$ be a C*- (resp. W*-)dynamical system, $\beta\in\mathbb{R}\cup\{\pm\infty\}$. 
A (resp. normal) state $\omega$ over $\mathfrak{A}$ is said to be a \emph{$(\alpha,\beta)$-KMS state} if
it satisfies the \emph{condition of} \textsc{Kubo, Martin} \emph{and} \textsc{Schwinger}: given $x,y\in
\mathfrak{A}$, there exists a function $F_{x,y}$, continuous and bounded in$\mathbb{R}+(\mbox{sgn}\beta)i[0,
|\beta|]$ and analytic in $\mathbb{R}+(\mbox{sgn}\beta)i(0,|\beta|)$, such that \[F_{x,y}(t)=\omega(x\alpha_t
(y))\mbox{ and }F_{x,y}(t+i\beta)=\omega(\alpha_t(y)x).\] In the case $\beta=0$, $\omega$ satisfies $\omega(xy)=
\omega(yx)$ for all $x,y$; then we say that $\omega$ is a \emph{tracial} state. In the case $\beta=+\infty$, 
we say that $\omega$ is a \emph{ground} state.
\end{definition}

It follows from considerations similar to the ones made for the modular group that a $(\alpha,\beta)$-KMS state
$\omega$ is invariant under $\alpha_t$, and, by the uniqueness of the GNS representation, there exists a unique
implementation of $\alpha_t$ by unitary operators $U_\alpha(t)$ in $\mathscr{H}_\omega$. A consequence of this
is that the (unique) normal extension of a $(\alpha,\beta)$-KMS state over $\mathfrak{A}$ to $\mathfrak{A}^{**}$ 
is also a $(\alpha,\beta)$-KMS state for the W*-dynamical system $(\mathfrak{A}^{**},\alpha_t)$, where $\alpha_t$ 
denotes its $\sigma$-weakly continuous extension. \\

The importance of this concept owes to the fact that the KMS condition characterizes \emph{thermodynamic 
equilibrium} of the state $\omega$ at temperature $T=(k_B\beta)^{-1}$, where $k_B$ is the \textsc{Boltzmann}
constant. Indeed, in quantum systems with a finite number of degrees of freedom, the KMS condition characterizes 
the \textsc{Gibbs} states $\omega_\beta=\frac{\mbox{Tr}(e^{-\beta H}.)}{\mbox{Tr}(e^{-\beta H})}$, where $H$ is the
Hamiltonian operator which generates $\alpha_t$. The KMS condition, however, survives the thermodynamic limit, 
which doesn't happen to the characterization of equilibrium based upon \textsc{Gibbs} states. Its original
formulation, given by \textsc{Kubo, Martin} and \textsc{Schwinger}, was based upon \textsc{Green} functions 
associated to creation and annihilation operators, being later rephrased for C*-dynamical systems in the version 
given by Definition \ref{ap2d8} by \textsc{Haag}, \textsc{Hugenholtz} and \textsc{Winnink} \cite{hhw}.\\

Equivalently, by the considerations above, $\omega$ is a $(\alpha,\beta)$-KMS state if and only if, for all $f$
such that $\hat{f}\in\mathscr{C}^\infty_c(\mathbb{R})$,
\begin{equation}\label{ap2e9}
\int f(t)\omega(x\alpha_t(y))dt=\int f(t+i\beta)\omega(\alpha_t(y)x)dt,\,\forall x,y\in\mathfrak{A}.
\end{equation}

The case $\beta=+\infty$ is illuminated by the following consideration: the left hand side of (\ref{ap2e9}) is
nothing more than the distributional \textsc{Fourier} transform of $\omega(x\alpha_t(y))$. If $\mbox{supp}\hat{f}
\subset\mathbb{R}_-$, it follows that $f(t+i\beta)$ decays exponentially as $\beta\rightarrow+\infty$, and, thus, 
the right hand side of (\ref{ap2e9}) tends to zero in this limit. Hence, we conclude that the support of the 
distributional \textsc{Fourier} transform of $\pi_\omega(\alpha_t(x))\Omega$ is contained in $\bar{\mathbb{R}}_+$, 
justifying the name \emph{``ground state''} for $\omega$.\\

For $0<\beta<+\infty$, a $(\alpha,\beta)$-KMS state $\omega$ satisfies the following ``fluc\-tuation\--dis\-sipation
relation'': $\omega(x^*\alpha_t(x))$ and $\omega(\alpha_t(x)x^*)$ are positive distributions; by the
\textsc{Boch\-ner-Schwartz} theorem \cite{reedsimon2}, the distributional \textsc{Fourier} transforms \[\hat{\mu}_x
(\hat{f})\doteq\int f(t)\omega(x^*\alpha_t(x))dt,\,\hat{\nu}_x(\hat{f})\doteq\int f(t)\omega(\alpha_t(x)x^*)dt\] 
define \textsc{Radon} measures over $\mathbb{R}$. A manner of seeing this is to use the spectral decomposition
of the unitary operators $U_\alpha(t)$ which implement $\alpha_t$. We can, then, see that $\hat{\mu}_x$ and 
$\hat{\nu}_x$ are absolutely continuous one with respect to the other, and the \textsc{Radon-Nik\'odym} derivative of 
$\hat{\mu}_x$ with respect to $\hat{\nu}_x$ is given by 
\begin{equation}\label{ap2e10}
\frac{d\hat{\mu}_x}{d\hat{\nu}_x}(p)=e^{-\beta p},\mbox{ or, formally, }d\hat{\mu}_x(p)=e^{-\beta 
p}d\hat{\nu}_x(p).
\end{equation}

If we consider the signed measures $\hat{\delta}_{x,y}$ and $\hat{\phi}_{x,y}$ given respectively by the 
\textsc{Fourier} transforms of $\omega([x,\alpha_t(y)])$ and $\omega(\{x,\alpha_t(y)\})$ ($\{a,b\}\doteq ab+ba$ 
denotes the \emph{anticommutator} of $a,b\in\mathfrak{A}$), we have 
\begin{equation}\label{ap2e11}
d\hat{\delta}_{x,y}(p)=\tanh\left(\frac{\beta p}{2}\right)d\hat{\phi}_{x,y}(p),
\end{equation}
which is precisely the fluctuation-dissipation relation which characterizes the KMS condition, relating 
bounded perturbations of the dynamics to fluctuations of measurements of observables along time. Other relations,
equivalent to (\ref{ap2e10}) and (\ref{ap2e11}) can be analogously obtained among other \textsc{Green} functions
associates to $\omega(x^*\alpha_t(y))$.\\

Finally, we see that any normal, faithful state $\omega$ over a \textsc{von Neumann} algebra $\mathfrak{R}$ is
a $(\sigma^\omega_{-\beta t},\beta)$-KMS state.\footnote{Amusingly, the work of \textsc{Haag, Hugenholtz} and 
\textsc{Winnink} \cite{hhw} preceded by little the original, unpublished work of \textsc{Tomita} \cite{tomita}, 
and, in a way, motivated the inception of modular theory, as it can be seen by the pioneering exposition of 
\textsc{Takesaki} \cite{tak0}.} This example, in a certain sense, is the most general case possible: conversely, 
given a $(\alpha,\beta)$-KMS state $\omega$ over a C*-algebra $\mathfrak{A}$, it follows that 

\begin{lemma}\label{ap2l2}
$\Omega$ is \emph{separating} for $\pi_\omega(\mathfrak{A})''$.
\begin{quote}{\small\scshape Proof.\quad}
{\small\upshape First, notice that, if $\mathfrak{A}$ is a C*-subalgebra of $\mathscr{B}(\mathscr{H})$ for some
\textsc{Hilbert} space $\mathscr{H}$ and $\omega=\langle\Omega,.\Omega\rangle$ is a state over $\mathfrak{A}$ 
such that $\Omega$ is cyclic for $\mathfrak{A}$ and $\mbox{Ann}\omega$ is *-invariant, then $\Omega$ is 
separating for $\mathfrak{A}$: given $x$ such that $\omega(x^*x)=0$, we have $yx\Omega=0$ for all $y\in\mathfrak{A}$. 
However, in this case, $(yx)^*\Omega=x^*y^*\Omega=0$. By the cyclicity of $\Omega$, it follows that $x=0$. 
Returning to our hypotheses, consider now the C*-subalgebra $\mathfrak{R}=\pi_\omega(\mathfrak{A})''$ of 
$\mathscr{B}(\mathscr{H}_\omega)$. Obviously, $\Omega$ is cyclic for $\mathfrak{R}$. Taking $A\in\mathfrak{R}$ 
such that $\omega(A^*A)=0$, consider the function $F_{A^*,A}(z)$ as in Definition \ref{ap2d8}. Then, $F_{A^*,A}(t)
=\omega(A^*\alpha_t(A))=\langle A\Omega,U_\alpha(t)A\Omega\rangle=0$. By the Edge-of-the-Wedge Theorem \cite{pct}, 
it follows that $F_{A^*,A}(z)=0$ for $z\in\mathbb{R}+i[0,\beta]$. But, in this case, $F_{A^*,A}(i\beta)=\omega(AA^*)
=0$. Hence, by the argument above, $A=0$.~\hfill~$\Box$}
\end{quote}
\end{lemma}
 
Consider now the following situation: given a W*-dynamical system $(\mathfrak{R},\alpha_t)$ and a normal, $(\alpha,
\beta)$-KMS state, it follows from Lemma \ref{ap2l2} that $\omega$ is faithful in $\pi_\omega(\mathfrak{R})$. 
However, $\mbox{Ann}\omega$ is then a $\sigma$-weakly closed two-sided ideal. Invoking Proposition \ref{ap2p3}, 
there exists a projection $E\in\mathfrak{R}\cap\mathfrak{R}'$ such that $\mbox{Ann}\omega=(\mathbb{1}-E)\mathfrak{R}$. 
Hence, $\omega(\mathbb{1}-E)=\{0\}$ and $\omega$ is faithful in $\mathfrak{R}E$. Moreover, \[\omega(AE)=\omega(A)=
\langle\Omega,\pi_\omega(A)\Omega\rangle.\] We conclude this Subsection with the following, surprising result of 
\textsc{Takesaki}, which follows form the modular theory presented in Subsection \ref{ap2-cwalg-totak}:

\begin{theorem}[Takesaki \upshape\cite{bratteli2,tak2}]\label{ap2t13}
Let $\mathfrak{R}$ be a \textsc{von Neumann} algebra and $\omega$ a normal state over $\mathfrak{R}$. Then the
following are equivalent:
\begin{enumerate}
\item $\omega$ is faithful in $\pi_\omega(\mathfrak{R})$, i.e., there exists a projection $E\in\mathfrak{R}
\cap\mathfrak{R}'$ such that $\omega(\mathbb{1}-E)=0$ and $\omega\restr{\mathfrak{R}E}$ is faithful.
\item There exists a $\sigma$-weakly continuous group $\alpha_t$ of *-automorphisms of $\mathfrak{R}$ such that 
$\omega$ is a KMS state for the W*-dynamical system $(\mathfrak{R},\alpha_t)$.
\end{enumerate}
Moreover, if the conditions above are satisfied, it follows that $\alpha_t(E)=E$, and $\alpha_t\restr{\mathfrak{R}
E}$ coincides with the modular group of $\mathfrak{R}E$ associated to $\omega$. In particular, $\alpha_t
\restr{\mathfrak{R}E}$ is uniquely determined by $\omega$.
\begin{quote}{\small\scshape Proof.\quad}
{\small\upshape See Theorem 5.3.10 in \cite{bratteli2}.~\hfill~$\Box$}
\end{quote}
\end{theorem}

\section{\label{ap2-borchers}Locally convex *-algebras and \textsc{Borchers-Uhlmann}
algebras}

We recall, following \cite{rudin} and \cite{reedsimon1}, that a topological vector space (always here 
assumed \textsc{Hausdorff}) $X$ (always over $\mathbb{C}$, unless otherwise stated) is \emph{locally convex} if
it admits a fundamental system of open neighbourhoods $\mathscr{V}=\{V_\alpha\ni 0\}_{\alpha\in I}$ such that
each of these neighbourhoods is convex, i.e., $tx+(1-t)y\in V_\alpha$ for all $x,y\in V_\alpha$, $t\in (0,1)$. 
Equivalently, a topological vector space is locally convex if it admits a family $\{\Vert.\Vert_\beta\}_{\beta
\in J}$ of \emph{seminorms} (i.e., maps $\Vert.\Vert_\alpha:X\rightarrow\bar{R}_+$ which satisfy the same 
properties of a norm, except possibly for $\Vert x\Vert=0\Rightarrow x=0$) which \emph{separates points} in $X$, 
i.e., given $x\neq y\in X$, there exists $\beta\in J$ such that $\Vert x\Vert_\beta\neq\Vert y\Vert_\beta$. 
A system of convex open neighbourhoods of the origin corresponding to this family of seminorms is given, for
instance, by $V_{\beta_1,\ldots,\beta_k,n}=\{x\in X:\Vert x\Vert_{\beta_i}<n^{-1}_i,\,\forall i=1,\ldots,k,\,n_i\in
\mathbb{Z}_+\}$.  Conversely, notice that any locally convex vector space admits a fundamental system of open, 
convex and \emph{balanced} neighbourhoods (i.e., $e^{it}V_\alpha=V_\alpha$, for all $t\in[0,2\pi)$). Defining 
$\Vert x\Vert_\beta\doteq\inf\{t>0:t^{-1}x\in V_\beta\}$, it follows from the continuity of scalar multiplication 
and the fact that $V_\beta$ is balanced that $\Vert.\Vert_\beta$ defines a continuous seminorm -- notice that 
$\Vert V_\beta\Vert_\beta\subset[0,1)$, for $V_\beta$ is open. Moreover, the family $\{\Vert.\Vert_\beta\}_{\beta
\in J}$ of seminorms separates points in $X$, since for each $x\neq 0$ there exists $\beta$ such that $x\notin 
V_\beta$ and, in particular, $\Vert x\Vert_\beta\geq 1$.\\

Our representative example of a locally convex topological vector space will be $\mathscr{D}(\mathscr{O})=
\mathscr{C}^\infty_c(\mathscr{O})$, $\mathscr{O}\subset R^d$ open, whose locally convex topology is defined in 
the following way: this space is given by the union of subspaces \[\bigcup^\infty_{n=1}\mathscr{D}_{K_n}
(\mathscr{O}),\] where $\mathscr{D}_{K_n}(\mathscr{O})\doteq\{f\in\mathscr{D}(\mathscr{O}):\mbox{supp}f\subset 
K_n\}$ has a locally convex topology given by the seminorms \[\Vert f\Vert_{k,K_n}=\sum_{|\alpha|\leq k}\sup_{K_n}|
\partial^{(\alpha)}f|,\] and $K_n=\bar{K}_n\Subset\mathscr{O}$ satisfies $K_n\Subset K_{n+1}$ and$\cup_n K_n=
\mathscr{O}$ (i.e., the collection $\{K_n\}_{n\in\mathbb{Z}_+}$ is an \emph{exhaustion} of $\mathscr{O}$), 
implying $\mathscr{D}_{K_n}(\mathscr{O})\subseteq\mathscr{D}_{K_{n+1}}(\mathscr{O})$ in a way that the topology
of $\mathscr{D}_{K_n}(\mathscr{O})$ is clearly induced by the topology of $\mathscr{D}_{K_{n+1}}(\mathscr{O})$. 
The topology of $\mathscr{D}(\mathscr{O})$ is the finest topology such that the inclusions $\mathscr{D}_{K_n}
(\mathscr{O})\hookrightarrow\mathscr{D}(\mathscr{O})$ are continuous for all $n$, called \emph{inductive
limit} of (the topologies of) the collection $(\mathscr{D}_{K_n}(\mathscr{O}))_{n\in\mathbb{Z}_+}$.\\

A separating collection of seminorms which defines the topology of an inductive limit is not (notice!) 
necessarily countable, though the collection of seminorms given above is (i.e., the latter is not always 
sufficient to generate the topology of an inductive limit). If, however, $\mathscr{O}$ is replaced by $K$ 
\emph{compact} and/or the support requirements are abandoned, we enumerate the family of seminorms given 
above as $\Vert.\Vert_k$, $k=1,2,...$. Defining \[d(x,y)\doteq\sum^\infty_{k=1}\frac{\Vert x-y\Vert_k}{2^k
(1+\Vert x-y\Vert_k)},\] we have that $d(.,.)$ defines a translation-invariant metric which generates the 
topology of the space. We have, nevertheless, that $\mathscr{C}^\infty_{(c)}(\mathscr{O})$ and 
$\mathscr{C}^\infty_{(c)}(K)$ are \emph{sequentially complete}, i.e., any \textsc{Cauchy} sequence (for general
topological vector spaces, a sequence $(x_n)$ is said to be \textsc{Cauchy} if, for any neighbourhood $V$ 
of the origin, there exists $N\in\mathbb{Z}_+$ such that $x_n-x_m\in V$ for all $n,m>N$) converges. In 
particular, $\mathscr{C}^\infty(\mathscr{O})$ and $\mathscr{C}^\infty_{(c)}(K)$ are complete metric spaces 
(\textsc{Fr\'echet} \emph{spaces}). The topology of the respective topological distribution duals is equivalent 
to the topology of pointwise convergence of sequences, being that this topology is also sequentially complete.\\

We can now give the

\begin{definition}\label{ap2d9}
A \emph{locally convex *-algebra} is a *-algebra $\mathfrak{F}$ which, as a vector space, is locally convex, 
and such that the operations of product and involution are continuous in this topology. We demand that the 
seminorms $\Vert.\Vert_\alpha$ of $\mathfrak{F}$ satisfy the C* property, i.e., $\Vert x^*x\Vert_\alpha=\Vert 
x\Vert^2_\alpha$, for all $\alpha$.
\end{definition}

Let's now build the example which interests us the most, the \textsc{Borchers-Uhlmann} algebra \cite{borchers1,
uhlmann,borchers2}. The space $\mathscr{D}(\mathscr{O})$ is \emph{nuclear}: there exists a unique locally convex
topology in the algebraic tensor product $\mathscr{D}(\mathscr{O})\otimes\mathscr{D}(\mathscr{O})\subset
\mathscr{D}(\mathscr{O}\times\mathscr{O})$ which extends the topology of $\mathscr{D}(\mathscr{O})$. The 
\textsc{Cauchy} completion of $\mathscr{D}(\mathscr{O})\otimes\mathscr{D}(\mathscr{O})$ in this topology is 
precisely $\mathscr{D}(\mathscr{O}\times\mathscr{O})$. Dually, the \emph{Kernel Theorem} of \textsc{L. Schwartz} 
\cite{hormander3} asserts that to any continuous linear map $K:\mathscr{D}(\mathscr{O})\rightarrow\mathscr{D}'
(\mathscr{O})$ corresponds a unique $u\in\mathscr{D}'(\mathscr{O}\times\mathscr{O})$ such that $(K\phi)(\psi)=
u(\phi\otimes\psi)$.\\

\begin{definition}\label{ap2d10}
The \textsc{Borchers-Uhlmann} algebra associated to $\mathscr{O}$ is the unital, locally convex *-algebra 
$\mathfrak{F}(\mathscr{O})$ given by \[\mathfrak{F}(\mathscr{O})\doteq\sum^\infty_{k=0}\otimes^k
\mathscr{C}^\infty_c(\mathscr{O}),\] where $\otimes^0\mathscr{C}^\infty_c(\mathscr{O})=\mathbb{C1}$ denotes the 
adjunction of the unit $\mathbb{1}$. More precisely, it's the $\bar{\mathbb{Z}}_+$-graded algebra composed of 
elements $f=(f_{(0)},f_{(1)},\ldots)$, where $f_{(k)}\in\otimes^k\mathscr{C}^\infty_c(\mathscr{O})$ and $f_{(k)}=0$ 
except for a finite set of values of $k$. The algebraic operations are given by:
\begin{itemize}
\item $\alpha f+\beta g=(\alpha f_{(0)}+\beta g_{(0)},\alpha f_{(1)}+\beta g_{(1)},\ldots,)$;
\item $fg=f\otimes g=(f_{(0)}g_{(0)},f_{(1)}g_{(0)}+f_{(0)}g_{(1)},\ldots,\sum_{i+j=k}f_{(i)}
\otimes g_{(j)},\ldots)$;
\item $f^*=(f^*_{(0)},f^*_{(1)},\ldots)$, where $f^*_{(0)}=\overline{f_{(0)}}$ and, for $f_{(k)}
=f^1_{(k)}(x_1)\cdots f^k_{(k)}(x_k)$, $x_i\in\mathscr{O}$, $i=1,\ldots,k$, we define $f^*_{(k)}
=\overline{f^k_{(k)}(x_1)}\cdots\overline{f^1_{(k)}(x_k)}$, extending the operation to $\otimes^k
\mathscr{C}^\infty_c(\mathscr{O})$ by linearity.
\end{itemize}
The seminorms of $\mathfrak{F}(\mathscr{O})$ are the ones inherited from $\mathscr{C}^\infty_c
(\mathscr{O})$. It follows immediately from the \textsc{Leibniz} rule that these seminorms satisfy the C* property.
\end{definition}

A state $\omega$ over $\mathfrak{F}(\mathscr{O})$ is given by a hierarchy of distributions $\omega_n\in
\mathscr{D}'(\mathscr{O}^n)$, $\omega_0(f_{(0)}\mathbb{1})=f_{(0)}$, subject to the requirement of 
\emph{positivity}: $\sum^k_{n,m=0}\omega_{n+m}(f^*_{(n)}\otimes f_{(m)})\geq 0$, $\forall f_{(n)}\in\otimes^n
\mathscr{C}^\infty_c(\mathscr{O})$, $n=0,\ldots,k$, $k\in\bar{\mathbb{Z}}_+$. Notice that positivity is enough
to guarantee the weak continuity of $\omega_n$, and, indeed, a good deal of regularity is guaranteed \emph{a priori} 
for $\omega_n$. One demands directly that the $\omega_n$'s define distributions to include the possibility
of $\omega$ not being a positive functional (a situation in which this occurs is the case of gauge fields).\\

The \textsc{Wightman}-GNS representation of $\mathfrak{F}(\mathscr{O})$ associated to $\omega$ produces 
a distribution $\phi$ in $\mathscr{C}^\infty_c(\mathscr{O})$ with values in the set of closed linear operators
(densely) defined in $\mathfrak{F}(\mathscr{O})/\mbox{Ann}\omega\subset\mathscr{H}_\omega$. $\phi$ satisfies 
the hermiticity condition $\phi(\bar{f})^*=\phi(f)$.\\

Everything we've done above goes without essential changes from $\mathbb{R}^d$ to any $d$-dimensional manifold
$\mathscr{M}$. Moreover, this example is the base for us to construct the \textsc{Borchers-Uhlmann} algebra 
associated to fields with arbitrary spin, which will be sketchly done as follows. Consider a vector bundle 
$\mathscr{E}\stackrel{p}{\longrightarrow}\mathscr{M}$ with fiber $E\cong\mathbb{R}^D$. Given $\mathscr{O}
\subset\mathscr{M}$ open, we take \[\mathfrak{F}(\mathscr{O})\doteq \sum^\infty_{k=0}\otimes^k\Gamma^\infty_c
(\mathscr{O},\mathscr{E}).\] In this case, the $k$-point function $\omega_k$ associated to a state $\omega$ 
assumes values in $\otimes^k E$, and the field obtained is a $D$-tuple of operator-valued distributions.\\

We call here attention to the extremely important fact that the \textsc{Borchers-Uhl\-mann} algebra contains only 
\emph{kinematic} information about a theory, determining the support, the tensor character and the global 
internal symmetries of the field (multiplet). \emph{All} the \emph{dynamical} information of the theory 
(equations of motion, commutation relations, spectrum, etc.) is contained in $\omega$. We can alternatively 
define $\phi$ by taking the quotient of $\mathfrak{F}(\mathscr{O})$ modulo some *-ideal $\mathfrak{I}$ (we 
treat several instances of this scenario in Chapters \ref{ch3} and \ref{ch4}). In this case, a \emph{bona 
fide} state $\omega$ over the field $\phi$ must satisfy $\mbox{Ann}\omega\supset\mathfrak{I}$, for it
follows from the Third *-Isomorphism Theorem (Theorem \ref{ap2t4}), supplemented by Remark \ref{ap2r1}, that 
$(\mathfrak{F}(\mathscr{O})/\mathfrak{I})/(\mbox{Ann}\omega/\mathfrak{I})\cong\mathfrak{F}(\mathscr{O})/\mbox{Ann}
\omega$.\\

One can construct, under special circumstances (as, for instance, in the case of a quantum field theory 
satisfying the \textsc{Wightman} axioms), a version of the \textsc{Tomita-Takesaki} modular theory for 
*-algebras such as the \textsc{Borchers-Uhlmann} algebra. This problem, addressed for the first time by 
\textsc{Bisognano} and \textsc{Wichmann} \cite{bw1,bw2}, is attacked in a systematic fashion in \cite{inoue}.

\chapter{\label{ap3}Categories and functors}

Here, we shall introduce in a minimalistic way the needed concepts from the
theory of categories and functors, a language devised with the purpose of organizing
the structures that appear in the different areas of mathematics. Our reference
for this Appendix is the classic text \cite{maclane} of \textsc{Mac Lane}, one of the
fathers of the concept.\\

A brief comment about the use of set theory made when one speaks about categories
is in order here. Strictly speaking, the concept of category \footnote{More
precisely, of \emph{metacategory} -- see Definition \ref{ap3d1}.} (Definitions 
\ref{ap3d1} and \ref{ap3d2}) makes no mention whatsoever to any of the axioms of set 
theory. Indeed, it's common (and quite useful) to consider ``large'' categories, i.e., 
such that the objects and arrows (morphisms) constitute aggregates more general than sets, 
such as \emph{classes} \footnote{The concept of class makes itself necessary every time 
one tries to build ``sets'' with cardinality bigger than the one of the universe of the
set theory adopted (naïve or following the axioms of \textsc{Zermelo-Fraenkel}). The
paradigmatic example, pointed by the first time by \textsc{B. Russell}, is the class 
of subsets of the universe.}. According to the system of set-theoretical axioms of 
\textsc{G\"odel} and \textsc{Bernays}, the notion of class is realized by all positive
instances of a certain propositional function that defines the properties of the objects 
of the class. For our purposes, it'll suffice to say that a class is a \emph{set} if each
one of the positive instances which forms the former reduces itself to a \emph{single} 
element of the universe. Summing up, sets are ``small'' aggregates (i.e., elements of 
the universe) and classes, ``large'' aggregates.\footnote{\label{largecat} 
There are certain categories whose objects may not constitute a class, as in the case 
of categories of functors (see Section \ref{ap3-functor}).} We won't need more general
set-theoretical concepts.

\section{\label{ap3-rud}Rudiments}

\begin{definition}\label{ap3d1}
A \emph{metacategory} consists of:

\begin{enumerate}
\item \emph{objects} $a,b,c,\ldots$;
\item \emph{arrows} (or \emph{morphisms}) $f,g,h,\ldots$; 
\item \emph{domain} (or \emph{source}) $D(.)$ 
and \emph{counterdomain} (or \emph{target}) $CD(.)$ operations on arrows, 
resulting in the objects $D(f),CD(f),D(g),CD(g),\ldots$ (it's customary
to employ the notation \[f:a\longrightarrow b\mbox{\quad or \quad}
a\stackrel{f}{\longrightarrow}b,\mbox{\quad for \quad} a=D(f),b=CD(f));\]
\item an \emph{identity} $id_.\equiv\mathbb{1}_.$ operation on objects, 
resulting in the arrows $\mathbb{1}_a,\mathbb{1}_b,\ldots$ with 
$D(\mathbb{1}_a)=CD(\mathbb{1}_a)$;
\item and a \emph{composition law} $.\circ.$ on
ordered pairs of arrows $(f,g)$ with $D(f)=CD(g)$, resulting in the arrows
$f\circ g$ with $D(f\circ g)=D(g)$ and $CD(f\circ g)=CD(f)$,
\end{enumerate}

with the operations above subject to the following axioms:

\begin{description}
\item[\mdseries\scshape Cat1 (associativity):\quad] For any objects and
arrows arranged in the form \[a\stackrel{f}{\longrightarrow}b\stackrel{g}
{\longrightarrow}c\stackrel{h}{\longrightarrow}d,\] we have $h\circ
(g\circ f)=(h\circ g)\circ f$;
\item[\mdseries\scshape Cat2 (unitary law):\quad] For any arrow $f$,
we have $\mathbb{1}_{CD(f)}\circ f=f\circ\mathbb{1}_{D(f)}=f$.\footnotemark
\end{description}
\end{definition}
\footnotetext{\label{arrowcat} Since this axiom determines $\mathbb{1}_a$ 
completely for any object $a$, it's sometimes convenient to identify $a$ 
with the arrow $\mathbb{1}_a$. In particular, this makes possible to define a
metacategory just by means of arrows (see \cite{maclane} for a detailed
discussion).}

For our purposes, it'll be enough to consider a category as the following 
realization of a metacategory within set theory (see footnote \ref{largecat}):

\begin{definition}\label{ap3d2}
A \emph{category} $\mathscr{C}$ consists of a \emph{class} of objects $Obj\mathscr{C}$
(sometimes identified, in abuse of notation, with $\mathscr{C}$), and a 
\emph{class} of arrows $Arr\mathscr{C}$ \footnotemark, endowed with operations satisfying
the axioms of a metacategory, \emph{such that the subclasses $Hom_{\mathscr{C}}(a,b)
\doteq\{f:D(f)=a,CD(f)=b\}$ are sets, for any pair $a,b\in Obj\mathscr{C}$.}
\end{definition}
\footnotetext{This, by virtue of the discussion in footnote \ref{arrowcat},
can be considered as a superclass of $Obj\mathscr{C}$. Under this point of view,
it would be more adequate to identify $\mathscr{C}$ with $Arr\mathscr{C}$, but here
we surrender to current practice.}

A \emph{subcategory} $\mathscr{D}$ of $\mathscr{C}$ consists of a category
whose objects form a subclass $Obj\mathscr{D}$ of $Obj\mathscr{C}$ and the arrows
form a subclass $Arr\mathscr{D}$ of $Arr\mathscr{C}$, such that the operations of 
$\mathscr{D}$ are inherited from $\mathscr{C}$. $\mathscr{D}$ is a \emph{full} 
subcategory if $Hom_{\mathscr{D}}(a,b)=Hom_{\mathscr{C}}(a,b),\,\forall a,b\in
\mathscr{D}$. We also say that a category $\mathscr{C}$ is \emph{small} 
if $Obj\mathscr{C}$ (and, thus, $Arr\mathscr{C}$) is a set. We can associate 
$\mathscr{C}$ to the \emph{opposite category} $\mathscr{C}^{op}$ such that
 $Obj\mathscr{C}^{op}=Obj\mathscr{C}$ and $Hom_{\mathscr{C}^{op}}(a,b)=
Hom_\mathscr{C}(b,a)$, with the \emph{inverse} composition law $\circ$ with respect to 
$\mathscr{C}$.\\

Since the arrows abstract from the category $\mathscr{S}et$ of sets the 
notion of functions between sets, it's natural to seek the categorical abstraction 
injective, surjective and bijective functions. We say that an arrow $f$ is 
\emph{surjective}, \emph{epic} or simply an \emph{epi} (resp. \emph{injective}, 
\emph{monic} or a \emph{mono}) if, given two arrows $g_1,g_2\in Hom_{\mathscr{C}}
(CD(f),b)$ (resp. $h_1,h_2\in Hom_{\mathscr{C}}(a,D(f))$) such that $g_1\circ f=
g_2\circ f$ (resp. $f\circ h_1=f\circ h_2$), we have $g_1=g_2$ (resp. $h_1=h_2$) 
\footnote{\label{epimono} One haven't pursued a definition of epic (resp. monic) 
arrows by means of right (resp. left) inverses, for, although such definition
applies to $\mathscr{S}et$, the same fails to happen, for instance, to the category 
$\mathscr{G}rp$ of groups -- epimorphisms possesses a \emph{set-theoretical} right
inverse, for it may not be possible to choose an \emph{homomorphism} with that
property. The definition we gave, on the other hand, embraces all cases of
interest.}. An arrow $e$ is \emph{right} (resp. \emph{left}) \emph{invertible}
if there exists an arrow $e'$ such that $D(e')=CD(e)$, $CD(e')=D(e)$ 
and satisfying $e\circ e'=\mathbb{1}_{CD(e)}$ (resp. $e'\circ e=\mathbb{1}_{D(e)}$), 
and simply \emph{invertible} (or an \emph{isomorphism}) if both 
properties are satisfied \emph{and any right inverse is a left one and vice-versa}. 
In this case, the inverse is unique, which does not happen to the previous cases. 
We also say in this case that $D(e)$ and $CD(e)$ are \emph{isomorphic}. A 
\emph{groupoid} is a small category such that any arrow is an isomorphism.\\

Let $f$ be an epi and $g$ a mono. If $f\circ g=\mathbb{1}_{D(g)}$, it's customary to say
that $g$ is a \emph{section} of $f$ (in analogy with sections of fiber bundles), and $f$ a 
\emph{retraction} of $g$. We also here say that $f$ is a \emph{split epi} and $g$, a
\emph{split mono}. In this case, $h=g\circ f:D(f)\longrightarrow D(f)$ is an 
\emph{idempotent} arrow (or a \emph{projection}), i.e., $h^2\doteq h\circ h=h$. Conversely, 
an idempotent arrow $h\in Hom_{\mathscr{C}}(a,a)$ is said to be \emph{split} by an object $b$ 
if there exist an epi $f:a\longrightarrow b$ and a mono $g:b\longrightarrow a$ such that 
$f\circ g =\mathbb{1}_b$ and $g\circ f=h$.\\

Finally, let us move towards some examples, which will also be used to fix the
notation for some standard categories which shall be frequently used:

\begin{description}
\item[\textrm{$\mathbf{1}$:}\quad] Category endowed with a single object $a$
and a single arrow $\mathbb{1}_a$;
\item[\textrm{$\mathbf{2}$:}\quad] Category endowed with only two objects $a,b$ and 
three arrows $\mathbb{1}_a$, $\mathbb{1}_b$, $a\stackrel{f}{\longrightarrow}b$;
\item[\textrm{$U$:}\quad] (Small) \emph{discrete} category, where $Obj U=U$ is a
set, $\mbox{Hom}_U(a,b)=\varnothing$ if $a\neq b$, and $\mbox{Hom}_U(a,a)\{\mathbb{1}_a\}$
(summing up, we can also write $U=ObjU=ArrU$); 
\item[\textrm{$\mathscr{C}\times\mathscr{D}$:}\quad] \emph{Product} of the categories
$\mathscr{C}$ and $\mathscr{D}$ -- $Obj\mathscr{C}\times\mathscr{D}\ni(a,b)$, where
$a\in Obj\mathscr{C}$ and $b\in Obj\mathscr{D}$, $\mbox{Hom}_{\mathscr{C}\times\mathscr{D}}
((a,b),(c,d))\ni (f:a\rightarrow c,g:b\rightarrow d)$, $\mathbb{1}_{(a,b)}=(\mathbb{1}_a,
\mathbb{1}_b)$ and $(f,g)\circ(f',g')=(f\circ f',g\circ g')$;
\item[\textrm{$\mathscr{C}^I$:}\quad] \emph{Cartesian power} of a category
$\mathscr{C}$ to the set $I$ -- formally, $Obj\mathscr{C}\ni a:I\rightarrow Obj
\mathscr{C}$ ($a$ is then said to be the \emph{product} of the $a(\alpha)$'s), 
$\mbox{Hom}_{\mathscr{C}^I}(a,b)=\{f:I\ni\alpha\mapsto f(\alpha)\in\mbox{Hom}_{\mathscr{C}}
(a(\alpha),b(\alpha))\}$, $\mathbb{1}_a:I\ni\alpha\mapsto\mathbb{1}_{a(\alpha)}$ and $(f\circ 
g):I\ni\alpha\mapsto f(\alpha)\circ g(\alpha)$;
\item[\textrm{$(U,\leq)$:}\quad] Partially ordered set -- $Obj(U,\leq)=U$ is the underlying
set, and $\mathscr{Hom}_{(U,\leq)}(a,b)$, if nonvoid, possesses a single arrow 
$\leq$, also denoted by $a\leq b$, satisfying the usual properties of an order 
relation, here suitably rephrased in categorical language: $\mathbb{1}_a=a\leq a$ and 
$(b\leq c)\circ(a\leq b)=a\leq c$. In this case, $Arr(U,\leq)\doteq\leq$ is said to be a 
\emph{partial order} in $U$. Moreover, we say that $(U,\leq)$ is: \emph{directed} if, 
given $a,b\in U$, there exists $c\in\ U$ such that $a\leq c$ and $b\leq c$; and \emph{totally 
ordered} if for any $a,b\in U$ necessarily $a\leq b$ or $b\leq a$. The \textbf{Zorn} lemma 
here is stated in the following way: any full, totally ordered subcategory $\mathscr{C}$ of
$(U,\leq)$ possesses a \emph{terminal object} $u$, i.e., for any $a\in\mathscr{C}$ there 
exists an \emph{unique} arrow $a\leq u$ (dually, $a\in Obj\mathscr{C}$ is an \emph{initial} 
object if for any $b\in Obj\mathscr{C}$ there exists an \emph{unique} arrow $a\rightarrow b$). 
An example of a directed set is the power set, ordered by inclusions $(U,\leq)=(P(X),\subseteq)$, 
of the set $X$ -- in particular, this directed set possesses a terminal object $X$ and initial 
objects $\{x\}$, $x\in X$;
\item[\textrm{$\mathscr{S}et$:}\quad] Sets, whose arrows are functions;
\item[\textrm{$\mathscr{G}rp$:}\quad] Groups, whose arrows are homomorphisms;
\item[\textrm{$\mathscr{T}op$:}\quad] Topological spaces, whose arrows are
continuous functions;
\item[\textrm{$\mathscr{T}vs$:}\quad] Topological vector spaces (assumed in the text Hausdorff 
and locally convex -- see Section \ref{ap2-borchers}), whose arrows are linear continuous maps.
\end{description}

Other categories shall be introduced in the text as needed.

\section{\label{ap3-functor}Functors and natural transformations}

\begin{definition}\label{ap3d3}
Let $\mathscr{A}$ and $\mathscr{B}$ be categories. A \emph{covariant} (resp. 
\emph{contravariant}) \emph{functor} from $\mathscr{A}$ to $\mathscr{B}$ is a rule 
$\mathfrak{F}$ which associates to each $a\in Obj\mathscr{A}$ an object $\mathfrak{F}a
\in Obj\mathscr{B}$ and to each $f\in Arr\mathscr{A}$ an arrow $\mathfrak{F}f\in 
Arr\mathscr{B}$ satisfying $D(\mathfrak{F}f)=\mathfrak{F}D(f)$ (resp. 
$D(\mathfrak{F}f)=\mathfrak{F}CD(f)$), $CD(\mathfrak{F}f)=\mathfrak{F}CD(f)$ 
(resp. $CD(\mathfrak{F}f)=\mathfrak{F}D(f)$) and such that, for $a,b,c\in Obj
\mathscr{A}$ and $f,g\in Arr\mathscr{A}$ satisfying $a\stackrel{f}{\longrightarrow}
b\stackrel{g}{\longrightarrow}c$, we have \[\mathfrak{F}(g\circ f)=\mathfrak{F}g
\circ\mathfrak{F}f\,(\mbox{resp. }\mathfrak{F}(g\circ f)=\mathfrak{F}f\circ
\mathfrak{F}g).\] It's customary to write a functor as an ``arrow between categories'', 
i.e., $\mathfrak{F}:\mathscr{A}\rightarrow\mathscr{B}$ or even $\mathscr{A}
\stackrel{\mathfrak{F}}{\longrightarrow}\mathscr{B}$.
\end{definition}

We can say, equivalently, that a contravariant functor $\mathfrak{F}:\mathscr{A}
\rightarrow\mathscr{B}$ is a covariant functor $\mathfrak{F}:\mathscr{A}
\rightarrow\mathscr{B}^{op}$. Hence, when we say functor, we just mean ``covariant 
functor''.\\

$\mathfrak{F}$ is said to be an \emph{equivalence of categories} if it's invertible, 
i.e., there exists a functor $\mathfrak{F}^{-1}$ such that $\mathfrak{F}^{-1}\circ
\mathfrak{F}=\mathfrak{id}_\mathscr{A}$ and $\mathfrak{F}\circ\mathfrak{F}^{-1}=
\mathfrak{id}_\mathscr{B}$, where $\mathscr{C}\stackrel{\mathfrak{id}_\mathscr{C}}{
\longrightarrow}\mathscr{C}$ is the \emph{identity} functor associated to the category
$\mathscr{C}$ (i.e., $\mathfrak{id}_\mathscr{C}a=a$, $\mathfrak{id}_\mathscr{C}f=f$, 
$\forall a\in Obj\mathscr{C}$, $f\in Arr\mathscr{C}$). Another example is the functor 
$\mathfrak{i}_{\mathscr{C},\mathscr{D}}$ given by the inclusion of a category $\mathscr{C}$ 
into a supercategory $\mathscr{D}$ of $\mathscr{C}$.\\

We shall give now some examples of functors which will be of great use to us:

\begin{itemize}
\item Let $\mathscr{C}$ be a category and $(U,\leq)$ a partially ordered set.
A functor $\mathfrak{F}:(U,\leq)\rightarrow\mathscr{C}^{op}$ (resp. $\mathfrak{G}:(U,\leq)
\rightarrow\mathscr{C}$) is said to be a \emph{pre(co)sheaf} (of objects of $\mathscr{C}$) in 
$U$.\footnote{Actually, the usual definition of pre(co)sheaf in $U$ consists of a functor 
$\mathscr{F}:(P(U),\subseteq)\rightarrow\mathscr{C}$ or, more generally, $\mathfrak{F}:(\tau(U),
\subseteq)\rightarrow\mathscr{C}$, where $\tau$ is a topology (possibly discrete, as in the previous 
case) in $U$. It's not difficult to see that such a definition is superseded by the one we've given 
above.} If $(U,\leq)$ is directed, we say that $\mathfrak{F}$ is an \emph{inverse system} in 
$\mathscr{C}$ indexed by $U$, and $\mathfrak{G}$, a \emph{net} (of objects of $\mathscr{C}$) 
in $U$, or a \emph{direct system} in $\mathscr{C}$ indexed by $U$;
\item The Cartesian power of $\mathscr{C}$ to $I$ can be understood as a functor
$\mathfrak{F}:I\rightarrow\mathscr{C}$, where $I$ is here understood as a (discrete) category;
\item Let $I$ be a set and $\mathscr{C}$ a category. The \emph{diagonal functor} $\Delta:
\mathscr{C}\rightarrow\mathscr{C}^I$ is given by $Obj\mathscr{C}\ni a\mapsto(\Delta a:I\ni\alpha
\mapsto a)$ and $Arr\mathscr{C}\ni f\mapsto(\Delta f:I\ni\alpha\mapsto f)$;
\item Given a functor $\mathfrak{F}:\mathscr{C}\rightarrow\mathscr{D}$ and $d\in Obj\mathscr{D}$, 
an \emph{universal arrow} from $d$ to $\mathfrak{F}$ (resp. from $\mathfrak{F}$ to $d$) consists 
in a pair $(u,f)$, where $u\in Obj\mathscr{C}$ and $f:d\rightarrow\mathfrak{F}u$ (resp. $f:\mathfrak{F}
u\rightarrow d$) which satisfy the following \emph{universal property:} given any $c\in 
Obj\mathscr{C}$, $g:d\rightarrow\mathfrak{F}c$ (resp. $g:\mathfrak{F}c\rightarrow d$), there exists 
\emph{a single arrow} $g':u\rightarrow c$ (resp. $g':c\rightarrow u$) such that $\mathfrak{F}g'
\circ f=g$ (resp. $f\circ\mathfrak{F}g'=g$). In this case, we say that $(u,f)$ is a universal element
of the functor $\mbox{Hom}_{\mathscr{D}}(d,\mathfrak{F}.):\mathscr{C}\rightarrow\mathscr{S}et$ 
(the action of this functor on arrows is given by $\mbox{Hom}_{\mathscr{D}}(d,\mathfrak{F}.):(g:a\rightarrow 
b)\mapsto(\mbox{Hom}_{\mathscr{D}}(d,\mathfrak{F}a)\ni h\mapsto\mathfrak{F}g\circ h\in\mbox{Hom}
(d,\mathfrak{F}b))$) -- more generally, if $\mathfrak{H}:\mathscr{C}\rightarrow\mathscr{S}et$, an
\emph{universal object} of $\mathfrak{H}$ consists of the pair $(u,e)$, where $u\in Obj\mathscr{C}$ and 
$e\in\mathfrak{H}u$ are such that for each $a\in Obj\mathscr{C}$, $x\in\mathfrak{H}a$ there exists 
\emph{a single arrow} $f:u\rightarrow a$ such that $(\mathfrak{H}f)(e)=x$. Conversely, interpreting $e$
as $e:\{*\}\ni *\mapsto(\mathfrak{H}u)(*)\doteq e$, we see that $(u,e)$ is a universal arrow of $*$
in $\mathfrak{H}$. The dual case cited parallelly above consists only in exchanging $\mathscr{D}$ with 
$\mathscr{D}^{op}$. Examples of universal arrows comprise:
\begin{itemize}
\item \emph{Generators} of free groups ($\mathscr{C}=\mathscr{G}rp$, $\mathscr{D}=\mathscr{S}et$ e 
$\mathfrak{F}$ identifies a group with the underlying set): given $X\in Obj\mathscr{S}et$,
let $u$ be the free group generated by $X$ and $f$ identifying $x\in X$ with the corresponding element
of $u$;
\item \emph{Coproducts} ($\mathscr{D}=\mathscr{C}^I$, where $I$ is a set, and $\mathfrak{F}=\Delta$): 
given $a:I\ni \alpha\mapsto a(\alpha)$ and $(u,f)$ and universal arrow from $a$ to $\Delta$, we then say 
that $u$ is the \emph{coproduct} or \emph{direct sum} of the $a(\alpha)$'s, sometimes denoted $u\doteq
\coprod_{\alpha\in I}a(\alpha)$ or $u\doteq\bigoplus_{\alpha\in I}a(\alpha)$, and $f(\alpha)$, the 
\emph{injection} of $a(\alpha)$ in $u$;
\item \emph{Products} ($\mathscr{D}=\mathscr{C}^I$, where $I$ is a set, and $\mathfrak{F}=\Delta$): 
given $a:I\ni \alpha\mapsto a(\alpha)$ and $(u,f)$ a universal arrow from $\Delta$ to $a$, we then say 
that $u$ is the \emph{product} of the $a(\alpha)$'s, sometimes denoted $u\doteq\prod_{\alpha\in I}a(\alpha)$, 
and $f(\alpha)$, the \emph{projection} of $u$ in $a(\alpha)$;
\item \emph{Completion} of topological spaces, free tensor algebras, etc.,
\end{itemize}
and of universal objects:
\begin{itemize}
\item \emph{Quotient} $S/\sim$ of a set $S$ modulo an equivalence relation $\sim$ ($\mathscr{C}=
\mathscr{S}et$, $\mathfrak{H}:X\mapsto\mathfrak{H}X=\{f:S\rightarrow X:s,s'\in S,\,s\sim s\Rightarrow
f(s)=f(s')\}$);
\item \emph{Tensor product} of two vector spaces (particular case of the previous example), etc..
\end{itemize}
\item Given a category $\mathscr{C}$, a directed set $(U,\leq)$ and a direct (resp. inverse) system 
$\mathfrak{F}$ in $\mathscr{C}$ indexed by $U$, the \emph{inductive} (resp. \emph{projective}) limit, 
or simply \emph{colimit} (resp. \emph{limit}) of $a\in Obj\mathscr{C}^U$ is simply a universal arrow
$(u,f)$ from $a$ to $\Delta$ (resp. from $\Delta$ to $a$), such that $f(\alpha')=f(\alpha)\circ
\mathfrak{F}(\alpha'\leq\alpha)$ (resp. $f(\alpha')=\mathfrak{F}(\alpha'
\leq\alpha)\circ f(\alpha)$) -- we denote 
\begin{equation}\label{ap3e1}
u\doteq\dlim{\alpha,\mathfrak{F}}a(\alpha)\mbox{ (resp. }u\doteq\ilim{\alpha,\mathfrak{F}}a(\alpha)). 
\end{equation}
\end{itemize}

\begin{definition}\label{ap3d4}
Let $\mathscr{A}$ and $\mathscr{B}$ be categories, and the functors \footnotemark 
$\:\mathfrak{F},\mathfrak{G}:\mathscr{A}\rightarrow\mathscr{B}$. A
\emph{natural transformation} $\Phi$ from $\mathfrak{F}$ to $\mathfrak{G}$
is a rule that associates to each $a\in Obj\mathscr{A}$ an arrow $\Phi_a
\in Arr\mathscr{B}$ satisfying $D(\Phi_a)=\mathfrak{F}a$, $CD(\Phi_a)=
\mathfrak{G}a$ and such that, for each $f:a\rightarrow b$, $a,b\in Obj\mathscr{A}$,
we have \[\Phi_a\circ\mathfrak{F}f=\mathfrak{G}f\circ\Phi_b.\]
\end{definition}
\footnotetext{Using the notion of opposite category, we can extend the definition 
to contravariant functors.}

A natural transformation $\phi$ between functors $\mathfrak{F}$, $\mathfrak{G}$
is said to be a \emph{natural equivalence} between $\mathfrak{F}$ and $\mathfrak{G}$ 
if there exists a natural transformation $\phi^{-1}$ from $\mathfrak{G}$ to $\mathfrak{F}$ 
such that $\phi^{-1}_a\circ\phi_a=\phi_a\circ\phi^{-1}_a=\mathbb{1}_a$. For instance, a 
functor $\mathfrak{F}$ from a category $\mathscr{C}$ to a supercategory $\mathscr{D}$ of 
$\mathscr{C}$ is said to be \emph{forgetful} id it's naturally equivalent to the inclusion 
functor $\mathscr{C}$ into $\mathscr{D}$, i.e., $\mathfrak{F}$ ``forgets'' part of the 
structure of $\mathscr{C}$. \\

Since the composition of natural transformations is clearly associative, and there always
exists, for each functor $\mathfrak{F}$, an identity $\mathbb{1}_\mathfrak{F}$
given by $(\mathbb{1}_\mathfrak{F})_a\doteq\mathbb{1}_{\mathfrak{F}a}$,
we can formally consider the \emph{functor category} $\mathscr{C}^\mathscr{D}$,
with $Obj\mathscr{C}^\mathscr{D}=\{\mathfrak{F}:\mathscr{D}\rightarrow\mathscr{C}\}$
and $Arr\mathscr{C}^\mathscr{D}$ consisting of all natural transformations between functors 
from $\mathscr{D}$ to $\mathscr{C}$, \emph{modulo} the set-theoretical caveats discussed
at the beginning of this Appendix, in particular footnote \ref{largecat}. An example of such
a category is, given a category $\mathscr{A}$, the \emph{category of arrows} $\mathscr{A}^\mathbf{2}$.

\chapter{\label{ap4}Rudiments of homotopy}

We introduce here some basic definitions and results which are employed at several points
of Chapter \ref{ch1}. Our reference for the matter is Part I (Sections 1--7) of \cite{greenhar}. 
For applications related to differentiable manifolds (isotopies, etc.), we follow \cite{hirsch}.

\section{\label{ap4-basics}Homotopic maps}

\begin{definition}\label{ap4d1}
Given two topological spaces $X,Y$ and two continuous maps $f,g:X\rightarrow Y$, we say that 
$f$ is \emph{homotopic} to $g$ if there exists a (jointly) continuous map $F:X\times[0,1]
\rightarrow Y$ such that $F(x,0)=f(x)$ and $F(x,1)=g(x)$, for all $x\in X$ (such a $F$ is said to 
be a \emph{homotopy} between $f$ and $g$). In particular, if $A\subset X$ and $f\restr{A}=g\restr{A}$, 
we say that $f$ is \emph{homotopic to} $g$ \emph{relative to} $A$ if we can choose the homotopy
$F$ in a way that $F(x,t)=f(x)=g(x)$ for all $t\in[0,1]$ -- in this case, we say that $F$
is a \emph{homotopy between} $f$ \emph{and} $g$ \emph{relative to} $A$.
\end{definition}

We say that $X$ is \emph{contractible} if the identity map $\mbox{id}_X:X\ni x\mapsto x$ is 
homotopic to the constant map $f_p(x):X\ni x\mapsto p$ for some $p\in X$. $X$ is contractible if 
and only if, given any topological space $Y$, then any two continuous maps $f,g:Y\rightarrow 
X$ are homotopic -- the direct implication is immediate, and the inverse implication is obtained through 
the homotopy $G:Y\times [0,1]\rightarrow X$ given by $G(y,t)=F(f(y),2t)$ ($t\in[0,\frac{1}{2}]$) and 
$G(y,t)=F(g(y),1-2t)$ ($t\in[\frac{1}{2},1]$), where $F$ is a homotopy between $\mbox{id}_X$ and $f_p$. 
As a consequence of this, a contractible topological space $X$ is always \emph{pathwise connected}, i.e., 
given any two points $x,y\in X$, there exists a continuous curve linking $x$ to $y$ -- namely, 
$\gamma:[0,1]\rightarrow X$ given by $\gamma(t)=F(x,2t)$ ($t\in[0,\frac{1}{2}]$) and $\gamma(t)=F(y,1-2t)$ 
($t\in[\frac{1}{2},1]$). \\

A contractible topological space $X$ is also \emph{simply connected}, i.e., any two continuous
curves $\gamma,\gamma':[0,1]\ni s\rightarrow X$ such that $\gamma(0)=\gamma'(0)$ and $\gamma(1)=\gamma'(1)$
are \emph{fixed-endpoint homotopic}, i.e., homotopic relative to $\{0,1\}$.\\

Let us now consider two differentiable manifolds $\mathscr{M},\mathscr{N}$. Given two maps
$\mathscr{C}^\infty$ $f,g:\mathscr{M}\rightarrow\mathscr{N}$ and a homotopy $F$ between $f$ and $g$, 
it's always possible to regularize $F$ in such a way that $F$ is a $\mathscr{C}^\infty$ map from
$\mathscr{M}\times [0,1]$ to $\mathscr{N}$. The same holds if $f=g$ in a subset $U\subset\mathscr{M}$
and $F$ is a homotopy between $f$ and $g$ relative to $U$. Hence, once the concept of homotopy is
transported from the $\mathscr{C}^0$ category to the $\mathscr{C}^\infty$ category, we can refine it
in the following way:

\begin{itemize}
\item If $F(.,t)$ is an embedding of $\mathscr{M}$ into $\mathscr{N}$ for all $t\in[0,1]$ (and, in 
particular, $f$ and $g$ are as well), we say that $F$ is an \emph{isotopy} between $f$ and $g$. 
A typical example of isotopy is given by a foliation of $K\subset\mathscr{N}$ by diffeomorphic copies
of $\mathscr{M}$ -- in this case, $K=F(\mathscr{M},[0,1])$ and $F(\mathscr{M},t_1)
\cap{}F(\mathscr{M},t_2)=\varnothing$ if $t_1\neq t_2$.
\item If $\mathscr{M}=\mathscr{N}$, $F(.,t)$ is a diffeomorphism for all $t\in[0,1]$ and $f=
\mbox{id}_{\mathscr{M}}$, we say that $F$ is a \emph{diffeotopy} or \emph{ambient isotopy}.
An example of diffeotopy is the restriction of an one-parameter group of diffeomorphisms $\mathscr{M}
\times\mathbb{R}\ni(x,t)\mapsto\phi_t(x)$ (i.e., $\phi_{t_1+t_2}=\phi_{t_1}\circ\phi_{t_2}$ and $\phi_0
=\mbox{id}_{\mathscr{M}}$) to $\mathscr{M}\times[0,1]$ -- in this case, $g=\phi_1$.
\end{itemize}

\section{\label{ap4-cov}The fundamental group and covering spaces}

\begin{definition}\label{ap4d2}
Let $X$ be a topological space and $x_0\in X$, called \emph{base}. The \emph{loop space} 
$\Omega(X,x_0)$ of $X$ based on $x_0$ consists of the continuous curves
$\gamma:[0,1]\rightarrow X$ such that $\gamma(0)=\gamma(1)=x_0$. Consider: the 
equivalence relation $\sim$ in $\Omega(X,x_0)$ \[\gamma_1\sim\gamma_2\Leftrightarrow
\gamma_1\mbox{ is fixed-endpoint homotopic to }\gamma_2,\] the operations of
\begin{itemize}
\item\emph{Product:} $\gamma_1,\gamma_2\in\Omega(X,x_0)\rightarrow \gamma_1\gamma_2$
given by
\begin{equation}\label{ap4e1}
\gamma_1\gamma_2(t)=\left\{\begin{array}{lr} \gamma_1(2t) & (t\in[0,\frac{1}{2}])\\
\gamma_2(2t-1) & (t\in[\frac{1}{2},1])\end{array}\right., e
\end{equation}
\item\emph{Inversion:} $\gamma^{-1}(t)=\gamma(1-t)$
\end{itemize}
and the \emph{identity loop} $\gamma_0(t)\equiv x_0$ in $\Omega(X,x_0)$. The \emph{fundamental
group} (or \emph{first homotopy group}) of $X$ based on $x_0$ is set-theoretically given by 
$\pi_1(X,x_0)=\Omega(X,x_0)/\sim\ni[\gamma]$, with group operations $[\gamma_1][\gamma_2]
\doteq[\gamma_1\gamma_2]$, $[\gamma]^{-1}\doteq[\gamma^{-1}]$ and identity $\mathbb{1}\doteq[\gamma_0]$.
\end{definition}

Clearly, the pathwise connected component $X$ to which $x_0$ belongs is simply 
connected if and only if $\pi_1(X,x_0)=\{\mathbb{1}\}$. If $X$ is pathwise connected, then 
$\pi_1(X,x_1)$ is isomorphic to $\pi_1(X,x_2)$ for all $x_1,x_2\in X$, and then we can define 
the \emph{fundamental group} $\pi_1(X)\doteq\pi_1(X,x_0)$ of $X$.

\begin{definition}\label{ap4d3}
Given two topological spaces $X,Y$, a map $\phi:Y\rightarrow X$ is said to be a \emph{covering 
map} if each $x\in X$ possesses an open neighbourhood $U\ni x$ such that
$\phi^{-1}(U)=\dot{\cup_\alpha} V_\alpha$, where $\{V_\alpha\}$ is a collection of \emph{disjoint} 
open subsets ($\dot{\cup}$ denotes disjoint union) of $Y$ such that $\phi\restr{V_\alpha}$ 
is a homeomorphism such that $F(V_\alpha)=U$, for all $\alpha$. In this case,
we say that $Y$ is a \emph{covering (space)} of $X$ -- we also employ the condensed notation
$Y\stackrel{\phi}{\longrightarrow}X$. $Y$ is said to be \emph{universal} if it's pathwise and
simply connected.
\end{definition}

If $Y\stackrel{\phi}{\longrightarrow}X$ is a covering of the topological space $X$, it follows
immediately from Definition \ref{ap4d3} that:

\begin{enumerate}
\item[(i)] $\phi^{-1}(x)$ is discrete for all $x\in X$,
\item[(ii)] $\phi$ is a local homeomorphism, and
\item[(iii)] $\phi$ is surjective, and the topology of $X$ is precisely the quotient topology 
of $E$ modulo the equivalence relation $y_1\sim y_2\Leftrightarrow \phi(y_1)=\phi(y_2)$.
\end{enumerate}

The importance of the universal covering follows from the fact that, despite having quite a
simple global structure, it encodes in a rather convenient way the fundamental group of the
topological space it covers. Defining the group of \emph{covering} (or \emph{deck}) \emph{transformations} 
$G(Y,X)$ of an universal covering $Y$ of $X$ as the group of homeomorphisms $f$ from $Y$ to itself
such that $\phi\circ f=\phi$, one can prove \cite{greenhar} that $G(Y,X)\cong\pi_1(X)$. \\

Moreover, the universal covering $Y$ of $X$ is essentially \emph{unique}: given any
other universal covering $Y'\stackrel{\phi'}{\longrightarrow}X$, there exists a homeomorphism 
$f:Y'\rightarrow Y$ such that $\phi\circ f=\phi'$. We can then speak about \emph{the} universal 
covering of $X$. Conversely, given a group $G$ of homeomorphisms of a pathwise and simply
connected topological space $Y$ which acts \emph{properly discontinuously}, i.e., 
for all $y\in Y$ there exists an open neighbourhood $V\ni y$ such that $f(V)\cap V=\varnothing$ 
for all $f\in G$, it follows that $\pi_1(Y/G)\cong G$ -- notice that, above, $G(Y,X)$ acts properly 
discontinuously in $Y$. We say then that $X=Y/G$ is the \emph{fundamental domain} of $Y$ with
respect to $G$.

\end{appendix}

\bibliography{tese}
\bibliographystyle{plr}

\cleardoublepage

\pagestyle{plain}

\addcontentsline{toc}{chapter}{Notes and translations of (some) epigraphs}
\chaptermark{Notes and translations of (some) epigraphs}

\endnotetext[1]{``See what things are in themselves, distinguishing matter, cause, end.''}
\endnotetext[2]{The literal translation is \emph{``To myself''}. However, the ``official'' Latin
translation for the title ended up being the name by which these personal writings, 
originally in Greek, of the Roman emperor have become more well known -- \emph{``Meditations''}.}
\endnotetext[3]{
``We only listen to the questions for which we are able to find an answer.''}
\endnotetext[4]{\emph{``The Gay Science''}.}
\endnotetext[5]{Editora Melhoramentos, 1986.}
\endnotetext[6]{\begin{quotation}\qquad ``The truth is that when his mind was completely gone, he
had the strangest thought any lunatic in the world ever had, which was that it seemed reasonable and
necessary to him, both for the sake of his honor and as a service to the nation, to become a knight
errant and travel the world with his armor and his horse to seek adventures and engage in everything
he had read that knights errant engaged in, righting all manner of wrongs and, by seizing the 
opportunity and placing himself in danger and ending those wrongs, winning eternal renown and
everlasting fame. The poor man imagined himself already wearing the crown, won by the valor of his
arm, of the empire of Trebizond at the very least; and so it was that with these exceedingly agreeable
thoughts, and carried away by the extraordinary pleasure he took in them, he hastened to put into
effect what he so fervently desired. And the first thing he did was to attempt to clean some armor
that had belonged to his great-grandfathers and, stained with rust and covered with mildew, had 
spent many long years stored and forgotten in a corner.''
\end{quotation} (Tr. Edith Grossman)}
\endnotetext[7]{\emph{``Don Quixote''}. Ecco/Harper Collins, 2003.}
\endnotetext[8]{Editora Nova Fronteira, 1988.}
\endnotetext[9]{\begin{quotation} \qquad `` `It seems to me,' said Don Quixote, `there is no human
history in the world that does not have its ups and downs, especially those that deal with chivalry;
they cannot be filled with nothing but successful exploits.' '' 
\end{quotation} (Tr. Edith Grossman)}
\endnotetext[10]{``There are two ways to reach Despina: by ship or by camel. The
city shows itself in a different way to who arrives by land or by sea. (...)
Each city receives the form of the desert which it opposes; Thus do the cameleer
and the sailor see Despina, city of confine between two deserts.'' (Tr. Pedro
Lauridsen Ribeiro)}
\endnotetext[11]{\emph{```The cities and the desire 3' (The invisible cities)''}.}
\endnotetext[12]{\begin{quotation} \qquad `` `How can that be?' responded Don
Quixote. `Is it so essential to the story to know the exact number of goats that
have crossed that a mistake in the count means you cannot continue the tale?'

\qquad `No, Se\~nor, I can't.' responded Sancho, `because as soon as I asked 
your grace to tell me how many goats had crossed, and you said you didn't know, at
that very moment I forgot everything I had left to say, and, by my faith, it was
very interesting and pleasing.'

\qquad `Do you mean to say that the story is finished?' said Don Quixote.

\qquad `As finished as my mother.' said Sancho.

\qquad `I tell you truthfully,' responded Don Quixote, `that you have told one
of the strangest tales, stories or histories that anyone in the world ever
thought of, and this manner of telling it and then stopping it is something I
shall never see, and have never seen, in my life, although I expected nothing else
from your intellect; but I am not surprised, for perhaps the sound of the pounding,
which has not ceased, has clouded your understanding.' ''
\end{quotation} (Tr. Edith Grossman)}

\theendnotes

\vspace*{\fill}

\begin{footnotesize}
The engravings in the frontispieces of each Part are clippings, done by the Author,
from illustrations of \textsc{Gustave Dor\'e} for \emph{``El ingenioso hidalgo 
don Quijote de la Mancha''} (Primero Libro), of \textsc{Miguel de Cervantes 
Saavedra}. The illustrations, originally drawn with feather tip, were 
wood-engraved by \textsc{H\'eliodore Pisan}, and then employed in the
edition annotated and translated to French by \textsc{Louis Viardot}, 
published by Hachette in 1863 (republished in 1978). Since they have been published 
for the first time more than 100 years ago, these illustrations are in public domain, 
and can be found at Wikimedia Commons (\verb|http://commons.wikimedia.org/wiki/Don_Quixote|)
in electronic format.
\end{footnotesize}

\cleardoublepage

\addcontentsline{toc}{chapter}{Erratum}
\chaptermark{Erratum}
\chapter*{Erratum for the previous versions}

In the original, Portuguese version and in the previous version of the English translation
of this thesis, there was a completely incorrect construction of a measure from the Lorentzian
distance by means of \textsc{Carathéodory}'s procedure in Subsection \ref{ch2-thermo-diam}, 
which (thankfully!) showed up to be irrelevant for our purposes. Thus, this part was removed
and replaced by appropriate remarks and a correct alternative argument which relies only on
Proposition \ref{ch2p2}. We starkly warn the reader and heartily apologise for such errors 
when reading either the original Portuguese version of the previous English version in the 
arXiv.
\vspace*{0.5cm}
\begin{flushright}
The Author\\
Toru\'{n}, June 2008
\end{flushright}

\end{document}